\documentclass[11pt]{article}

\usepackage{amsmath}
\usepackage{amsfonts}
\usepackage{amssymb}
\usepackage{dsfont}
\usepackage{hhline}
\usepackage{wrapfig}
\usepackage{graphicx}
\def\be{\begin{eqnarray}}
\def\ee{\end{eqnarray}}
\def\nn{\nonumber}
\def\brl{\left|}
\def\brr{\right>}
\def\ckl{\left<}

\def\p{\partial}
\def\dim{\mathrm{dim\,}}
\def\Tr{\mathrm{Tr\,}}
\def\Pexp{\mathrm{Pexp\,}}

\newtheorem{step}{Step}[section]

\numberwithin{equation}{section}
\title{On R-matrix approaches to knot invariants}
\author{\textbf{A.Anokhina}\footnote{ {\small \textit{ITEP, Moscow, Russia}};
anokhina@itep.ru}\date{ }}

\begin{document}

\maketitle

\vspace{-4cm}

\begin{center}
\hfill ITEP/TH-49/14\\
\end{center}

\vspace{2cm}

\begin{abstract}
We present an elementary introduction to one of the most important today knot theory
approaches \cite{ReshTur}, which gives rise to a representation for a class of knot polynomials in terms of quantum groups. Historically, the approach was at the same time developed from the state model approach \cite{Kauff} and from the braid group approach \cite{JonesBr}, and we consider the both approaches and there relation to each other and to the $\mathcal{R}$-matrix approach in details with help of the simple explicit examples. We also discuss various kind of motivation for referring to the above approaches as to the \emph{physical} approaches in knot theory \cite{KauffTB}. For instance, we concern a highly inspiring QFT interpretation for a knot polynomial \cite{Witt}.
\end{abstract}

\tableofcontents
\section{Introduction\label{sec:intr}}
The text is a \textbf{review} of various question concerning the $\mathcal{R}$-matrix approaches to the knot invariants. We hope that the text may be useful both for the first acquaintance with the notions and methods we discuss and for clarifying certain subtle points, the simplest examples we study through the text providing good illustrations to the ones. The text is also addressed to anyone who wanders what do the knots to do with the physics. This question definitely desires asking since, as a matter of the fact, up to day knot theory is considered as a chapter of physics at the same extend as a chapter of mathematics. Referring to \cite{KauffTB} for a much broader presentation of the subject, we discuss here some points, which are especially close to the main questions we consider.

Our main task is to provide a pedagogical introduction to the $\mathcal{R}$-matrix representation for the HOMFLY polynomials \cite{ReshTur}, in the version developed and used in \cite{MorSm,MMM1,MMM2,IMMM1,IMMM2,MM1,AMMM1,AMMM2,AMMM3,IMMM3,AM1}. Trying to keep a rather elementary level of presentation and mostly restricting ourselves by the simplest examples, we go in each presented examples into various details and subtleties, which are usually omit in literature. We intentionally do not start from presenting a completed construction, doing what a naive person would do and indicating one pitfall after another instead. As a preface, we discuss in details the representation of the (uncolored) HOMFLY polynomial in term of the matrix representation of the braid group, or, more precisely, of the Hecke algebra \cite{KauffTB}. Although literally replicating the $\mathcal{R}$-matrix representation for the same polynomials, this Hecke algebra representation was known even before \cite{JonesBr}, and we find it instructive to derive this representation in particular cases from the first principles. Apart from that, the braid group approach to the knot polynomials is the most simple and common way of relating them to observables in various physical models, since the braid group is an extension of the permutation group \cite{Pras,KauffTB}. This is the content of sec.\ref{sec:braids}, and we pass to the very $\mathcal{R}$-matrix approach in following sec.\ref{sec:Rmat}.

As a related subject, we would like to discuss a highly intriguing
problem of correspondence between the HOMFLY polynomial and
Chern-Simons Wilson average \cite{Witt}, which naturally arises in
the context of the $\mathcal{R}$-matrix representation for the
former ones \cite{MorRos,MorSm}. Although the very correspondence is
considered as a matter of fact, many problems concerning it remain
unresolved \cite{KauffTB,LabNew,MorRos,MorSm,AlMMM,MM2}. For
instance, a problem of deriving the $\mathcal{R}$-matrix formalism
from the perturbative Chern-Simons theory in the temporial gauge
\cite{MorRos, MorSm} is especially interesting in the context of our
discussion. Although the very problem is left beyond the scope of
our present text, we do what seems to us the only way of approaching
to this problem, as well as to numerous other problems of the same
kind. Namely, we take the standpoint of skeptics and try to answer
the question why the above correspondence is believed to be true.
The discussion on this subject is presented in sec.\ref{sec:QFT}. We
also try to give some sort of physical intuition about the simplest
properties of the very Chern-Simons theory in the same section.

The construction, which serves a kind of bridge between the $\mathcal{R}$-matrix representation for knot invariants and quantum field theory, is referred to a \emph{state model} representation of the knot polynomials, first mentioned under this name by L.Kauffman \cite{Kauff}. A knot polynomial is in fact presented with help of this construction as an observable in a \emph{topological quantum field theory} \cite{Atiyah}, although the theory is defined in rather abstract terms. We outline this construction in sec.\ref{sec:stmod}, illustrating how several different versions of the construction enable calculating a knot polynomial in a simple particular case. Our aim here is two-fold. First, we wish to illuminate the formalism, which might be treated as a rigorous definition of a topological quantum field theory, a knot polynomial as an observable. Second, we explain here the first part of the $\mathcal{R}$-approach, which is in fact a particular case of the state model approach.

As an introduction, we say a few general words on the knot invariants of our interest in sec.\ref{sec:knmath}. Then, before addressing to the main presentation, we discuss (in sec.\ref{sec:topphys}) some simple examples the pure topological problems arising in various fields of physics, the ``non realistic'' topological theories might turning out to be a useful tool for studying the ones.

\section{Knots and knot polynomials\label{sec:knmath}}

This section is a brief review of the knot theory quantities the following presentation refers to. We also try to formulate what kind of questions is asked about these quantities.

A knot theory is one of the most ancient fields of mathematics.
Naively speaking, this science tries to answer the question how to
describe all embeddings of a circle in the three-dimensional space
that can not be continuously transformed into each other. In a wide
sense, this is still an open problem, although huge knot tables with
various built in computer interfaces
\cite{katlas,indknot,indlink,knsc} are available now. In particular,
\cite{katlas}, which provides the detailed description of about 800
in a sense the simplest knots, is often referred to as \emph{a knot
Zoo}.

First of all, to determine the table item relevant to a given curve
in the three-dimension space is a separate, generally involved
problem. But it may be even more important to set and to study
various questions going beyond the plane enumeration, like how
complicated is a given knot, or how alike or different are two
distinct knots, or how to describe \emph{discontinuous}
transformations acting of various knots, or whether one can split
the set of all knots into subsets of in some sense similar knots, or
whether is there a natural way to select an infinite series of knots
with the members being enumerated by a single increasing parameter,
etc.

A powerful tool in studying this kind of questions is a notion of a
\emph{knot invariant}. By definition, it is a quantity which
coincides for any pair of the closed three-dimensional curves that
are continuously transformed into each other. Note that the inverse
is generally wrong. Moreover, a single knot invariant which enables
one to distinguish all invariants is unknown yet. Values of many
different invariants calculated for a vast amount of knots can be
found in \cite{katlas,indknot,indlink,knsc}.

Hence, each knot is associated with a number of quantities, which
are various knot invariants, which may be though of as coordinates
on the set of knots and used to compare different knots in one or
another sense. On the other hand, most of the known invariants can
be explicitly evaluated for an arbitrary three-dimensional curve,
and one can at least conclude that two curves are \emph{not}
continuously transformed one into the other if they have a different
value of the same invariant.

\be
\begin{array}{|c|c|c|c|c|c|c|}
\hline
\mbox{Name of the}&\mbox{Group}&\mbox{Represen-}&\mbox{Formal}&\multicolumn{2}{|c|}{\mbox{Knot}}&\mbox{Polynomial}\\
\cline{5-6}
\mbox{polynomial}&&\mbox{tation}&\mbox{variables}&\mbox{Name}&\begin{array}{c}\mbox{Rolfsen}\\ \mbox{notation}\end{array}&\\
\hline
\mbox{Jones}&SU(2)&\square&q&\begin{array}{c}\mbox{Trefoil,}\\\mbox{fig.}\ref{fig:tref}\end{array}&3_1&-q^{-8}+q^{-6}+q^{-2}\\
\cline{5-7}
&&&&\begin{array}{c}\mbox{Figure-}\\\mbox{eight}\end{array}&4_1&q^4-q^2+1-q^{-2}+q^{-4}\\
\hline
\mbox{HOMFLY}&SU(N)&\square&A,q&&3_1&-A^{-4}+A^{-2}\left(q^2+q^{-2}\right)\\
\cline{5-7}
&&&&&4_1&1+q^2\left(A+A^{-1}\right)-\left(q^2+q^{-2}\right)\\
\hline \mbox{Alexander}
&SU(0)&\square&q&&3_1&q^2-1+q^{-2}\\
\cline{5-7}
&&&&&4_1&q^2+1-q^{-2}\\
\hline
\mbox{Kauffman}&SO(N)&\square&a,q&&3_1&a^2\left(q^2+q^{-2}\right)-a^4\left(q^2-1+q^{-2}\right)\\
&&&&&&+\left(-a^3+a^5\right)\left(q-q^{-1}\right)\\
\hline
\mbox{Colored}&SU(2)&\square\!\square&q&&3_1&q^{-4}+q^{-10}-q^{-14}+q^{-16}-\\
\mbox{Jones}&&&&&&-q^{-18}-q^{-20}+q^{-22}\\
\hline
\end{array}
\label{tab:knpols} \ee

Especially interesting examples of knot invariants present
themselves \emph{knot polynomials}, which are not just numbers but
(Laurent) polynomials in some formal variable(s). The first
discovered knot polynomial is \emph{Alexander polynomial}, which was
already known in 1928 \cite{AlPol}. Today this invariant is thought
of as a very rude one. Nevertheless, Alexander polynomials of all
the prime knots (i.e., the knots not reduced to ``simpler'' knots,
see \cite{Pras} for the exact definition) with the crossing number
(the minimum number of self-crossings in a knot planar projection)
no more than 8, 36 knots altogether including the unknot, are all
different \cite{ChumDuzMost}, and these knots hence can be
enumerated by their Alexander polynomials.

The next polynomial invariant, the \emph{Jones polynomial}, was
introduced only in 1984  \cite{JPol}. Jones polynomials enables one
to distinguish already all prime knots with no more than 9 crossings
\cite{Adams}. E.g., the Jones polynomial for the prime knot  $6_1$
in the Rolfsen table (with $6$ crossings) differs from the Jones
polynomial of the knot $9_{146}$ (with $9$ crossings), which has the
same Alexander polynomial. Unlike that, the knot $10_{132}$ with
$10$ crossings has the same Jones polynomial as the knot $5_1$ with
$5$ crossings.

A short time after the Jones polynomial was introduced, in 1985,
several groups of researches, namely, P.Freyd and D.Yetter, J.
Hoste, W. B. R. Lickorish and K. Millet, A. Ocneany \cite{HOMFLY},
J. H. Przytycki and P. Traczyk \cite{PT}, independently discovered a
\emph{HOMFLY polynomial}, which is a generalization both of the
Jones and Alexander polynomials (the full name of the invariant,
\emph{HOMFLY-PT}, is an abbreviation of the eight enumerated names).
Although the HOMFLY polynomial sometimes enables to distinguish the
knots that are knot distinguished by the Jones polynomial, e.g., the
prime knot $8_9$ and the composite knot $4_1\sharp 4_1$ (see
\cite{Pras} for the definition) \cite{BoHyun}, the two polynomials
have alike ``distinguishable'' powers; in particular, the first pair
of knots with the same Jones polynomial, $5_1$ and $10_{132}$, has
the same HOMFLY polynomial as well. The real interest to the HOMFLY
polynomial was caused by completely different reasons, the present
text may be considered as a review of such ones.

Finally, the invariant which was afterwards called a \emph{Kauffman
polynomial} was first mentioned in 1987 \cite{Kauff}, and was
noticed in the same paper that all the above enumerated knot
polynomials can be defined in a similar way, which we discuss in
details in sec.\ref{sec:stmod}.

The next development of the subject consisted in introducing of the
\emph{colored} first Jones and then HOMFLY and Kauffman polynomials.
In the first papers (for instance, in \cite{Morton}), the colored
polynomials are introduced with help of the \emph{satellite} knots,
which are obtained from the original knot by substituting the
original curve with a braid. For braids with a given number of
strands, such braids form a finite-dimensional linear space, the
dimension being small for a small number of strands. E.g., there are
just 2 linearly independent HOMFLY polynomials among those for the
various satellites with 2-strands braids (if the linear combinations
with coefficients depending on the variable $q$ but not on the
variable $A$ are considered), 3 ones for 3-strand braids, 5 ones for
4-strand braids, and 7 ones for 5-strand braids (however, the
dimensions grow then faster and faster). The colored polynomials can
be introduced as a certain distinguished basis in the space of the
satellite knot polynomials, but they were the spaces themselves,
which we treated in the early papers. Probably the most remarkable
result in studying the colored polynomials from this standpoint is
due to Morton, who demonstrated \cite{Morton} that the simplest pair
of the \emph{Mutant knots}, the so called \emph{Kinoshita-Terasaka}
knot (11n42 in the \emph{Hoste-Thistlethwaite} table) and the Conway
knot (11n34) \cite{katlas}, are \emph{not} distinguished by all
colored Jones polynomials, as well as by the colored HOMFLY
polynomials for the two-strand satellites, but \emph{are}
distinguished by the colored HOMFLY polynomials for the three-strand
satellites. The result of Morton was recently confirmed by an
explicit calculation of the colored HOMFLY polynomials
\cite{IndsMut}, with help of on the $\emph{R}$-matrix approach
\cite{ReshTur}, which is just the main subject of the present text.

However, the sense and properties of the colored polynomials become
much more transparent from the standpoint of their other definition.
The one relies on the construction of our main interest. Although we
discuss this construction in details for plain (uncolored) HOMFLY
polynomials only, this is sufficient to present the very idea, the
color entering just as one of the parameters in the construction.

Examples of the various knot polynomials mentioned above are
presented in table\ref{tab:knpols}. The second and third columns
announce the sense of these polynomials from the standpoint of the
construction we discuss starting form sec.\ref{sec:stmod}.

However, a description of knots with help of their invariants does
not go beyond a formal enumeration, unless some underlying
structures in knot invariants are studied. Although very different
such structures were fruitfully researched for a long time
\cite{Pras}, a nice description of the ``space of all knots'', which
the scientists dream about since ancient times, still lacks for.

A fresh wind came in theory of knot invariants in late 80-s, when it
was noticed that some knot invariants can be introduced in terms and
explicitly calculated with tools, which was originally developed in
quantum physics. This observation inspired some new approaches to
knot invariants, which are now referred to as ``physical''
approaches.

These approaches turned out to be extremely fruitful themselves, but
even more inspiring is the idea of relating a knot to a state of a
physical system, the set of various knots could being described that
as the Hilbert space the system \cite{Adams}. The approaches to knot
invariants discussed in the present text are just of this kind.


\bigskip

\section{Examples of physical problems reduced to topological problems\label{sec:topphys}}
The knot theory approach, which we are mostly interested in and discuss in details in sec.\ref{sec:braids} and\ref{sec:Rmat}, refers to a class of the so called \emph{physical} approaches in knot theory. We briefly review some ideas underlining this name in sec.\ref{sec:QFT} and \ref{sec:stmod}. However, the very idea of studying the topological questions, i.e., of identifying any
two objects related by a continuous transformation at the first
glance seems rather exotic from the physical standpoint. By this reason, we start
from recalling several natural ways of pure topological questions
arising in various physical problems.

One may enumerate several bright examples of physical phenomena known as topological effects. The list starts from the \emph{Aharonov-Bohm effect}
\cite{LL3}, proceeds with the \emph{monopole} solution in the
Weinberg-Salam model \cite{PesSch}, \emph{instantonic solutions} of
the Yang-Mills equations \cite{BPST} and the Polyakov conjecture on
these solutions being responsible for the \emph{confinement}
\cite{Pol}, and includes various experimentally observed
topological quasiparticles in a solid matter, among them
\emph{Abrikosov vortices}, which are responsible for the high
temperature superconductivity and \emph{anyons} in graphene
\cite{anyons}, which probably explain the fractional quantum Hall
effect \cite{qHall} and might enable us with a quantum computer
\cite{Kitaev}. Discovering of these phenomena gave rise to a
separate subject that studies topological effects in gauge theories
\cite{Rub}. Denied that, we intentionally concentrate on another
kind of examples, in which topological problems arise in much more
regular ways.

\subsection{Adiabatic transformations theory\label{sec:adiab}}
\subsubsection{Adiabatic transformations of a physical pendulum}
\begin{wrapfigure}{r}{180pt}
\begin{picture}(180,150)(-90,-75)
\put(0,-75){\vector(0,1){150}}\put(-90,0){\vector(1,0){180}}
\put(-60,-60){\line(0,1){120}}
\put(60,-60){\line(0,1){120}}
\qbezier(0,50)(30,50)(35,40)\qbezier(35,40)(40,30)(60,30)
\qbezier(0,50)(-30,50)(-35,40)\qbezier(-35,40)(-40,30)(-60,30)
\qbezier(0,-50)(30,-50)(35,-40)\qbezier(35,-40)(40,-30)(60,-30)
\qbezier(0,-50)(-30,-50)(-35,-40)\qbezier(-35,-40)(-40,-30)(-60,-30)
\qbezier(0,30)(60,30)(60,0) \qbezier(0,30)(-60,30)(-60,0)
\qbezier(0,-30)(60,-30)(60,0) \qbezier(0,-30)(-60,-30)(-60,0)
\qbezier(0,20)(40,20)(40,0) \qbezier(0,20)(-40,20)(-40,0)
\qbezier(0,-20)(40,-20)(40,0) \qbezier(0,-20)(-40,-20)(-40,0)
\put(0,20){\vector(1,0){1}}\put(0,30){\vector(1,0){1}}\put(0,50){\vector(1,0){1}}
\put(0,-20){\vector(-1,0){1}}\put(0,-30){\vector(-1,0){1}}\put(0,-50){\vector(-1,0){1}}
\put(10,10){\footnotesize{I}}\put(10,35){\footnotesize{IV}}\put(10,55){\footnotesize{II}}
\put(10,-40){\footnotesize{V}}\put(10,-60){\footnotesize{III}}
\put(62,3){$\pi$}\put(-78,2){$-\pi$}
\put(85,5){$\phi$}\put(3,72){$\dot{\phi}$}
\end{picture}
\caption{Various kinds of phase trajectories for a physical
pendulum. \label{fig:pend}}
\end{wrapfigure}
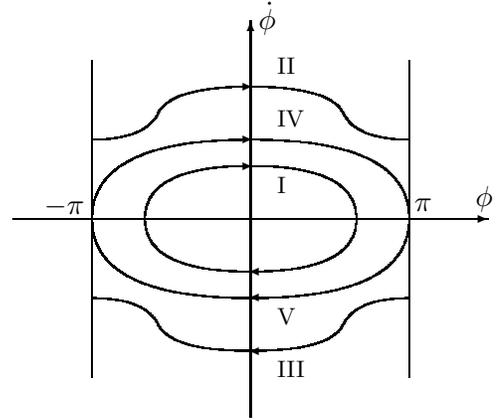

Topology studies a question whether two objects (curves, surfaces,
etc.) can be transformed one into the other by a continuous
transformation. A similar kind of questions arises in physics,
namely in theory of adiabatic transformations \cite{Arn, LL3}, where
one studies a question whether two states of a physical system with
variable parameters can be turned into each other by an
\emph{adiabatic}, i.e., in a sense infinitely slow (see further),
variation of the system parameters. A closely related question is
whether one of the states can be considered as a perturbation over
the other one.

\subsubsection{Idea of an adiabatic transformation}
Consider a physical system with variable parameters.
The equations of motion of the system depend then these
parameters. A solution of these equations depends then on the same
parameters, as well as their solutions. If one interests just in motion of the system for given
values of the parameters, one should just substitute these values in
the solution. However, if one wanders for a motion of the system as
the parameters \emph{vary} in time, one must solve different
\emph{differential} equation, with the coefficients that were
constant now vary, and generally obtain the completely different
function as a solution. Yet, if parameters vary ``enough slow'' (see
the explicit example below), the solution is obtained from a similar
one for unvarying system by plain substitution of the constant
parameters with the corresponding functions. Such ``enough slow''
transformations of a physical system are referred to as
\emph{adiabatic} transformations.

\paragraph{Example}
E.g., the equation of motion of a frequency alternating oscillator
\be \ddot x+\omega(t)x^2=0,\ \
\omega(t)=\omega_0\big(1+\lambda\cos(\Omega t)\big), \ee has a
frequency modulated oscillation \be
x(t)=A\cos\big(\omega(t)t+\phi\big) \ee as a general solution if the
modulation frequency is negligibly small compared to the natural
frequency of the system, \be \ddot
x(t)=\frac{d}{dt}\Big\{-\big(\omega(t)+\dot\omega(t)t\big)A\sin\big(\omega(t)t+\phi\big)\big\}
\stackrel{\Omega/\omega\rightarrow0}{\longrightarrow}
-\omega^2(t)x(t), \ee even if the modulation depth $\lambda$ is not
small.

\bigskip

The basic ideas of the adiabatic transformation theory, as well as their relation to the topology questions is well illustrated by the following simple example.

A motion of a physical pendulum is described by the second order
equation \be \ddot\phi+\omega^2\sin\phi=0,\label{pend2} \ee which,
one time integrated, gives the energy conserving law \be
\dot\phi^2-2\omega^2\cos\phi=\frac{2E}{ml^2}=const.\label{pend1} \ee
Constraint (\ref{pend1}), in turn, can be considered as equation of
the pendulum phase trajectory in the phase plane $(\phi, p\equiv
m\dot\phi)$ (fig.\ref{fig:pend}). There are then five different
kinds of phase trajectories, corresponding to the five kinds of the
pendulum motion. Namely, phase trajectory $I$ corresponds to the
pendulum swinging, never performing a complete turnover, while phase
trajectory $II$ corresponds to the pendulum turning over in the
counter-clock-wise direction, its mirror image w.r.t. the $\phi$
axis $III$ corresponds to the turning over in the clock-wise
direction. There are infinitely many phase trajectories of all the
three kinds, and the two phase trajectories corresponding to the
limiting cases, $IV$ and its mirror image $V$, represent,
respectively, a counter-clock and a clock-wise turn-overs, performed
by the pendulum for infinite time.

A pendulum motion of one kind can not be turned in a \emph{finite time} into a motion of
another kind by an \emph{adiabatic} transformation, i.e., by an infinitely
slow variation of the pendulum energy or period. E.g., if a pendulum performs small oscillations and one pules the wire in, slowly enough for the pendulum motion being approximated at each time point by oscillation of the pendulum with a constant frequency, the pendulum energy increases (as may be shown \cite{Arn}), and naively may be turned arbitrarily large. However, the period decreases at the same time, running at the infinity as the pendulum approaches to the critical trajectory with $E=2mgl$, as one can see from the explicit formula
\be
T=2\tau\left|\int_{-\phi_{max}}^{\phi_{max}}\frac{d\phi}{1+\lambda\cos\phi}\right|,\ \ \cos\phi_{max}=-\lambda^{-1}\le\cos\phi\le 1,\\
|\lambda|>1 \Leftrightarrow E<mgl,\ \lambda=\frac{mgl}{2E},\ \tau=\sqrt{\frac{ml^2}{2E}},\nn
\ee
which follows from (\ref{pend1}) straightforwardly. The energy being enough for a turn-over, the dependance of the period on the energy takes different form,
\be
T=\tau\int_0^{2\pi}\frac{d\phi}{1+\lambda\cos\phi},\ \ \
|\lambda|>1 \Leftrightarrow E>mgl.
\ee
This property of the two described pendulum motions matches the corresponding phase trajectories (fig.\ref{fig:pend}-I and II, respectively)
can not being turned one into the other by a continuous
transformation. Consequently, a turn-over of a pendulum can
\emph{not} be obtained as a perturbation over harmonic oscillation,
because as long as a polynomial in a perturbation parameter
reasonably approximates form of the trajectory, the trajectory
continuously depends on the parameter.

\subsubsection{Adiabatic transformations and discrete degrees of freedom}
In other words, one can one associate each motion of the pendulum
(except for the two limiting cases) with a number $+1$, $0$, or
$-1$, which equals the divided by $2\pi$ phase increment for the
period. This number can be considered as a \emph{discrete degree of
freedom}, which can not be changed by continuous transformations of
the system parameters. Note that this approach is valid only in the
adiabatic limit, where the phase trajectories at each time point are
the same to those of the system with parameters kept constant.

Note that coinciding of the discrete parameter value is not
sufficient for two motions of the system being related by an
adiabatic transformation. For instance, two for two pendulum motions
in the considered example being related in such a way, the quantity
that equals the energy times the period must be the same for these
motions \cite{Arn}. However, coincidence of all discrete degrees of
freedom is necessary for two motions being related by an adiabatic
transformation.

\subsubsection{Adiabatic transformations of quantum systems}
This above approach is widely applied in wave and quantum mechanics,
when one often deals with the discrete specters. For instance, one
can \emph{not} obtain a wave function of a particle in the delta
potential $V(x)=\frac{\hbar^2}{2mb}\delta(x)$ from such one in the
double-delta potential
$V(x)=\frac{\hbar^2}{2mb}\Big(\delta(x)+\delta(x-a)\Big)$ by an
adiabatic transformation that sends the parameter $a$ to infinity,
since there is just one discrete energy level
$E_0=-\frac{\hbar^2}{2mb^2}$ in the former case, while there are the
two ones, $E_{s,a}=E_0\mp\frac{\Delta}{2}$ with
$\Delta=\frac{e^{-\frac{a}{b}}}{mb^2}$ (for $\frac{a}{b}\gg 1$), in
the latter case. The wave functions of the corresponding confined
states are given by the symmetric and antisymmetric functions,
respectively, \be
\psi_{s,a}(x)=const\cdot\left\{\begin{array}{c}e^{-\kappa x},\
x<-a,\\e^{\kappa(x-2a)}+\pm e^{\kappa(-x-2a)},\ -a<x<a\\ \pm
e^{\kappa x},\ x>a \end{array}\right.,\ \kappa=\sqrt{-2mE}. \ee The
linear combination of this functions \be \psi(t)=const\cdot
e^{-i\frac{E_0}{\hbar}t}\left(e^{-i\frac{\Delta}{\hbar}t}\psi_s+e^{i\frac{\Delta}{\hbar}t\psi_a}\right)
\ee describes the particle passing from the one delta-hole to the
other, since \be \psi(0)=const\cdot\left\{\begin{array}{c}e^{-\kappa
x},\ x<-a,\\e^{\kappa(x-2a)},\ -a<x<a\\  0,\ x>a
\end{array}\right.\approx const\cdot e^{-\kappa|x+a|},\ \
\psi(\frac{\pi}{2\Delta})\approx const\cdot e^{-\kappa|x-a|}. \ee
This example also illustrates the close relation between obtaining
an effect as a result of an adiabatic transformation and studying
the effect in perturbation theory. In this particular case, neither
the separation of the energy levels, nor the tunneling solution can
be obtained in perturbation theory, since the distance between the
levels $2\Delta$ is proportional to $e^{-\kappa a}$ so that its
perturbation series in $\frac{1}{a}$ identically vanishes. By this
reason a particle tunneling from one hole to the other is often
referred to as a \emph{non-perturbative} effect.

\begin{wrapfigure}{r}{120pt}
\begin{picture}(120,75)(-60,-40)
\unitlength=0.5mm
\put(0,-30){\vector(0,1){75}}\put(-45,0){\vector(1,0){90}}
{\linethickness{0.6mm}
\put(-20,0){\line(1,0){40}}\put(0,15){\line(0,1){27}}\put(0,-15){\line(0,-1){15}}}
\qbezier(-27,0)(-27,3)(0,3)\qbezier(0,3)(27,3)(27,0)
\qbezier(-27,0)(-27,-3)(0,-3)\qbezier(0,-3)(27,-3)(27,0)
\qbezier(0,10)(-3,10)(-3,45)\qbezier(0,10)(3,10)(3,45)
\put(-20,0){\circle*{2}}\put(20,0){\circle*{2}}\put(0,15){\circle*{2}}\put(0,-15){\circle*{2}}
\put(-8,15){$\small{i}$}\put(-12,-15){$\small{-i}$}
\put(-28,-12){$\small{-a}$}\put(20,-12){$\small{a}$}
\put(5,15){$\small{B}$}\put(15,5){$\small{A}$}
\put(5,42){$\small{\Im z}$}\put(40,2){$\small{\Re z}$}
\end{picture}
\caption{Integration contours in (\ref{Rscph}) winding over $A$ and
$B$ cycles on the torus ($a=\sqrt{\frac{1-\lambda}{1+\lambda}}$).
\label{fig:ABcont}}
\end{wrapfigure}
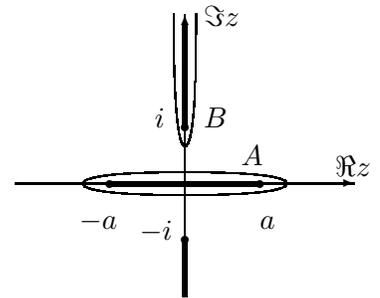

\subsubsection{Adiabatic transformations in statistical physics}
The notion of an adiabatic transformation is of grate invariance in
statistical physics \cite{LL5}, where energies of discrete levels
but not the filling numbers are changed in adiabatic processes.
Adiabatic condition consists then in the characteristic time of the
process being much bigger than the inverse smallest distance between
the energy levels.

\subsection{Analytic continuation approach\label{sec:ancon}}
Another source of the topological questions in physics is a common
method, applied in many chapters of physics and mathematics
\cite{LL3,LL5, ElKr, BogSh, VolTer, PesSch, Shab,Klem}.

\subsubsection{Geometric sense of the analytic continuation}~The technique we mean consists in the \emph{analytic continuation}
of a function $f(x)$ from real values of the argument $x$ to the
complex plane $z$ \cite{Shab}. The resulting function $f(z)$ by
definition satisfies the Cauchy-Riemann condition
$\p_{\bar{z}}f(z)=0$ everywhere but the special points. If the
function $f(x)$ is a rational function of $x$, then the function
$f(z)$ is defined on the \emph{Riemann sphere} (a complex plain
coupled to the point $z=\infty$) without the points where the
function $f(z)$ has the poles. A basic fact of the complex analysis
is that an integral $\int dz f(z)$ depends in this case not on the
particular form of the contour but on the winding numbers of the
contour on the poles. A case of an algebraic function $f(x)$ is more
complicated. The function $f(z)$ is determined then on the several
sheets of the complex plain with cuts, or, equivalently, on a
non-trivial \emph{Riemann surface}. The value of the same integral
this time depends also on whether the contour encircles a cut, or,
equivalently, on whether the contour contains a non-contractible
cycle on a Riemann surface (see the explicit example below).  Other
words, the value of the integral is in the both cases conserved as
the contour is transformed continuously never passing through the
special points. Hence, a physical quantity, which was originally
given by the integral $\int dx f(x)$ is now presented, in the above
sense, as a topological invariant. Evaluation of these kind of
integrals and studying of their properties is often performed with
help of topology methods \cite{Klem}.

\subsubsection{Semiclassical wave function as integral of a meromorphic form over a Riemann surface and geometric sense of the Bohr-Sommerfeld rule.}
A good and rather simple illustration of the very approaches and of the topological terms it refers  to is provided by the semiclassical approximation in quantum mechanics.

A semiclassical wave function of an energy $E$ stationary state of a
quantum particle in the potential $V(x)$ is expressed (in the lowest
approximation) as \be \psi_{sc}(a)=\psi(0)\exp\left(\int^a_0
p(x)dx\right),\ \ p=\sqrt{2m\big(E-V(x)\big)}. \ee The obtained
integral can be treated as an integral of the complex function
$p(z)$ analytically continued to the corresponding Riemann surface,
the integration contour being selected properly.

In particular, if the potential has the form
$V(x)=\frac{\hbar^2}{2mb^2}\cos^2(\frac{x}{b})$, phase of the wave
function can be expressed as
\be-i\log\phi_{cs}=\frac{b\sqrt{2mE}}{\hbar}\left(1+\lambda\frac{\p}{\p\lambda}\right)J(\lambda),\
\ \lambda=\frac{\hbar^2}{2mEb^2},\ee with the integral \be
J(\lambda)\equiv\int\frac{d\xi}{\sqrt{1-\lambda\cos^2\xi}},\label{Rscph}
\ee  the integral being brought to an integral of the algebraic
function \be
J(\lambda)=\int\frac{dz}{\sqrt{(1+z^2)\left((1+\lambda)z^2-(1-\lambda)\right)}}\label{Cscph}
\ee by introducing the variable $z=\tan\frac{\xi}{2}$. The integrand
can be analytically continued to the two sheets of the complex plane
glued along the two cuts, the segment of the real axis
$\left(-\sqrt{\frac{1-\lambda}{1+\lambda}},\sqrt{\frac{1-\lambda}{1+\lambda}}\right)$
and the part of the imaginary axis composed of the segments
$(-i\infty,-i)$ and $(i,i\infty)$ can be chosen as the ones (we
suppose that $\lambda>1$), see fig.\ref{fig:ABcont}. Integral (\ref{Rscph}) taken over the
classically forbidden area, which coincides with one of the selected
cuts selects, equals then the half of integral (\ref{Cscph}) over
closed contour $A$ in fig\ref{fig:ABcont}, which encircles the first cut. This integral gives the phase
increment of the quantum particle as it passes along a closed
contour, so that the semiclassical wave function is defined
unambiguously provided the value of the integral being an integer
multiple of $2\pi i$. This requirement constraints the energy of a
stationary state and is known as the Bohr-Sommerfeld quantization
rule \footnote{However, the semiclassical approximation to the wave
function of a \emph{non-stationary} state is not generally a
single-valued complex function; the phenomenon is known as
\emph{Stocks phenomenon} \cite{ElKr}.}.

On the other hand, integral (\ref{Cscph}) can be equivalently
considered as an integral over the torus (which is the Riemann
surface for the analytically continued integrand) of a meromorphic
one-form. The corresponding contour integral is non-vanishing if the
integration contour winds over a torus cycle, yielding then the
corresponding period of the torus. Such presentation is useful for
deriving the differential equations (Ward identities) for this
integral, since there is just a three-dimensional linear space of
meromorphic one-forms on a torus. In addition, since the considered
contour integral enters the Bohr-Sommerfeld quantization rule, the
energy of the stationary state can be expressed this way via certain
geometric quantities, such as \emph{theta functions} \cite{Klem}.
Being rather a textbook subject \cite{LL3, BogSh, VolTer}, this
approach is still rather popular in QCD phenomenology \cite{UnCut}.

\paragraph{Dispersion relations method.}
The above outlined analytic continuation approach is a basis for the widely used method of
\emph{dispersion relations} (see, e.g., \cite{LL3},\cite{LL5} and \cite{BogSh}
for quantum mechanics, statistical physics and QFT examples, respectively). Roughly
speaking, the method consists in relating an average of some
function to a sum of the function values in certain points, putting
the integral of the function over the real axis is equal to a proper
sum over residues of the analytically continued function. In other
version of the approach, a value of a function in a certain point is
treated as the corresponding residue, $f(w)=\frac{1}{2\pi i}\oint
\frac{f(z)dz}{z-w}$, and then the integration contour is pulled to
the area of the $w$ plain where the function $f(w)$ admits a
perturbative expansion in $w$. The value of a function $f(w)$ for a
$w$ having a physical sense is then approximated by a perturbation
series in a non-physical area, provided that one adds to the series
the certain terms accounting the contour passing through the special
points, when being deformed.

\subsection{Ward identities giving rise to a topological quantum field theory\label{sec:topWard}}
One more way of associating physical quantities with topological
invariants come from the following widely used approach
\cite{Pras}. First, a smooth object (a curve, a surface, etc.) is
associated with a certain graph. By construction, the graph remains
one and the same as the object is subjected to transformations that
form a large subclass of arbitrary continuous transformations. The
remaining transformation correspond to certain operations with the
graph, each operation being presented as a composition of several
elementary ones.

In particular,  various topological invariants are in the most cases
calculated not with help of the smooth object itself, but with help
of the associated graph. It is then necessary and sufficient to care
that two graphs related by each of elementary operations give one
and the same value of the invariant.

One the other  hand, various correlating functions in a quantum
field theory are often associated with the Feynman diagrams.
Constraints on these functions, which are referred to as \emph{Ward
identities}, can be formulated then as equalities of correlators for
the Feynman diagrams related by a certain operation, or, more
generally, as conditions of vanishing of proper linear combinations
of such diagrams.

Coinciding of values of a topological invariant calculated with help
of two graphs related by an elementary operation (and thus
corresponding to the objects that can be continuously deformed one
onto the other) may be then looked at as a Ward identity in a
quantum field theory. A topological invariant may be, in turn,
considered as value of the correlating function in this model.
Hence, any construction that enables to calculate values of some
topological invariant for all objects of a given kind, i.e., for all
knots, may be considered as a definition of a quantum field theory,
with topological invariance constraints as Ward identities. Constructions of this kind are  referred to as \emph{topological quantum field theories} (TQFT) \cite{Atiyah}.

Surprisingly or not, Ward identities of this kind are not exotic but
may arise rather naturally. In particular, one of elementary
equivalence relations of the knot diagrams, the most important one
in the approach we study, gives rise to the constraint known as the
\emph{Yang-Baxter equation}, which coincides with one of the
constraints on the permutation group generators \cite{Pras} (see
sec.\ref{sec:braids} as well). The operators satisfying Yang-Baxter
equation were first applied to the inverse scattering problem in
quantum mechanics \cite{InvScatt}, and they arise in a large class
of physical models, including spin chains and ice-type models in
statistical physics \cite{Baxt} and two-dimensional conformal field
theory \cite{DiFr}.

\section{Towards QFT interpretation of knot invariants\label{sec:QFT}}
In the above section, we outlined how a topological invariant may be
considered as a quantity related to a Feynman diagram of some
abstract quantum field theory. For instance, a class of knot
invariants possess various representations of this kind (see
sec.\ref{sec:stmod}), and the main content of the present text is
devoted to one of them. However, connection between knot invariants
and quantum observables is not exhausted by this point. In this section, we discuss a highly inspiring interpretation, or rather two different interpretations \emph{a la} \cite{Witt,Kaul,Inds1,Inds2,Inds3,KaulLeq,GMMM} and \cite{GuadMarMin1,GuadMarMin2,AlLab,FrohKing,LabPer,LabNew}, respectively, of knot invariants as exactly computable observables in a quantum field theory.

\paragraph{Structure of the section.}
We start from formulating the precise statement about the relation
of knot polynomials and observables in a topological quantum field
theory in sec.\ref{sec:WZW}. We outline then the idea of the
reasoning, which was presented in \cite{Witt} as a derivation of
this statement and developed in the subsequent works \cite{Kaul,Inds1,Inds2,Inds3,Inds6,Inds10,KaulLeq,GuJock,MMM4,MMM5,GMMM}
into a new representation for a class of knot polynomials. We
complete the section with formulating an essential question, which
arise as the result. Sec.\ref{sec:CSWav}-\ref{sec:Konts} are devoted
to the discussion of this question.

In sec.\ref{sec:CSWav}, we write out the basic notions of the
particular TQFT the statement under discussion refers to. We also
attempt to give some geometrical and physical intuition about this
theory properties. Sec.\ref{sec:lnum} contains the detailed
presentation of the simplest explicit example of the discussed
correspondence between the knot (more precisely, link) invariant and
the TQFT observable. In fact, this example was well known  (although
rare mentioned in the published papers; one may cite \cite{Schwarz}
as an exception) even before \cite{Witt} and essentially motivated
the presented there conjecture. In the following sections we address
to a more general case of the knot polynomial --- TQFT observable
correspondence. A significant argument in the sake of the
correspondence is that properties of the certain knot invariants
match the properties of the two important in QFT quantities. The
first them is \emph{ordered exponential}, with help of which an
\emph{evolution operator} in QFT is introduced  \cite{PesSch}. The
second quantity is the \emph{Gaussian average} of the operator
product, on which, for instance, the perturbative definition of the
path integral relies on \cite{PesSch}. By this reason, we start the
discussion from sec.\ref{sec:WAgen}, where we recall the needed
definitions and formulate the properties of the ordered exponential
and gaussian average referring to the question under discussion.
Next comes sec.\ref{sec:CSfram}, in which we discuss the
regularization problem for the particular TQFT quantity
corresponding to the knot polynomial, which can be cured by the
\emph{framing procedure} \cite{GuadMarMin1}. We start from providing
the simplest illustration to the framing procedure, briefly
discussing some subtleties concerning a general case afterwards. We
also comment (in sec.\ref{sec:CSfram}) a part of the framing
procedure as of the argument in the \emph{sake} of the knot
invariant-TQFT observable correspondence under discussion. Finally,
sec.\ref{sec:Konts} contains the main part of the discussion. In the
section, we discuss the particular properties of the certain knot
invariants, which motivate identifying them with the perturbative
contributions to a Gaussian average of an ordered exponential, thus
interpreting them as TQFT observables. One of the subtleties arising
here is that the original observation referred to the knot
invariants are \emph{other} knot invariants than \cite{Witt} refers
to. The relation of them two is also known \cite{ChumDuzMost} and
discussed in the same section.

\subsection{Knot invariants as observables in Wess-Zumino-Witten theory and as axiomatically defined exact Wilson averages\label{sec:WZW}}
A modern sight on the QFT sense of knot invariants is concentrated
in the statement that
\begin{itemize}
\item{A HOMFLY polynomial is a Wilson average in the Chern-Simons theory.}
\end{itemize}
A HOMFLY polynomial \cite{Pras} is the particular case of knot polynomial
mentioned in sec.\ref{sec:knmath} and sec.\ref{sec:stskein}. This is the knot polynomial of our main interest. The
above statement was literally presented and considered in details in
famous paper by Witten \cite{Witt}. However, various considerations
underlying this statement were already widely discussed by that time
(unfortunately, these discussions mostly remained unpublished; see,
e.g., \cite{Schwarz} and \cite{Atiyah}), and the paper itself did
not put  the end in the subject but rather gave rise to numerous
studies of the question \cite{Kaul,Inds1,Inds2,Inds3,KaulLeq,GuadMarMin1,GuadMarMin2,
FrohKing,AlLab,LabPer,LabNew,MorRos,MorSm,MM2} which continue up
to day (one more highly intriguing story is presented in
\cite{MatModLoc}). The discussed correspondence of the knot
polynomials to the QFT observables thus do not reduce to a singe
once proved theorem, rather being an entire subject including many
different and highly intertwined ideas.

\subsubsection{Visual presentation of the construction}
In the present section, we briefly discuss a QFT interpretation of
the knot invariants, as it is presented in \cite{Witt}. The there established correspondence of the knot polynomials relies on the statement
\begin{itemize}
\item{Skein relations for HOMFLY polynomials coincide with the known relations between Wess-Zumino-Witten conformal blocks.}
\end{itemize}
However, the presented in the paper reasoning can be presented rather visually, what may be instructive both for extending the established correspondence to the case of the colored HOMFLY polynomials and for developing the presented construction explicit computational procedure. We  sketch this visual presentation below.

\paragraph{Encircled crossing on a knot diagram as a projection of the four-punctured sphere.} The idea of the construction is presented in
\cite{Witt} rather visually. Namely, an encircled crossing on a knot
planar projection may be looked at as projection of the sphere, two
arcs of the original curve inside. These arcs can be continuously
transformed, newer intersecting, into two lines on the sphere,
pairwise connecting the four intersection points of the sphere with
the curve. One may treat the obtained configuration as a Riemann
sphere with the four selected points and two cuts. Such sphere is a
domain of analytic in certain regions, but generally multi-valued
complex functions, like $\log\frac{(z-a)(z-b)}{(z-c)(z-d)}$, the
cuts connecting the point $a$ with the points $c$, and the point $b$
with the $d$ \cite{Shab}. A certain matrix-valued generalization
of such functions is know as \emph{Wess-Zumino-Witten conformal
block} \cite{KZ:WZW}, and this is the quantity, which is associated
with described sphere in the construction under discussion.

\paragraph{Inverting a crossing as continuous move of the punctures}
Relating the WZW conformal blocks to the knot invariants (namely, to
the HOMFLY polynomials) relies then on the both quantities
satisfying the same system of the defining equations. More
precisely, the WZW blocks on the sphere with the given four selected
points form a three-dimensional linear space, in analogy with the
meromorphic functions with the poles in the given points. In
particular, three WZW conformal blocks related to variously made
cuts satisfy the linear relation,
which, when projected on a plain, reproduces the defining constraint
on the Jones polynomials (see sec.\ref{sec:stmod} for details)
\be
\begin{array}{ccccccccc}
&\begin{picture}(40,60)(0,-28)\put(0,0){$\left(q-q^{-1}\right)\times$}\end{picture}&
\begin{picture}(60,80)(-30,-30)
\put(0,0){\circle{38}}
\put(14,14){\circle*{2}}\put(-14,-14){\circle*{2}}
\put(14,-14){\circle*{2}}\put(-14,14){\circle*{2}}
\qbezier(-14,-14)(0,0)(-14,14)\qbezier(14,-14)(0,0)(14,14)
\put(-14,14){\vector(-1,1){1}}\put(14,14){\vector(1,1){1}}
\put(-22,-21){\footnotesize{D}}\put(-23,14){\footnotesize{C}}
\put(16,-21){\footnotesize{B}}\put(15,14){\footnotesize{A}}
\end{picture}
&\begin{picture}(10,60)(0,-28)\put(0,0){$=$}\end{picture}
&\begin{picture}(10,60)(0,-28)\put(0,0){$q^{-2}\,\times$}\end{picture}&
\begin{picture}(60,80)(-30,-30)
\put(0,0){\circle{38}}
\qbezier(-14,14)(7,7)(14,-14)
\put(-14,14){\circle*{2}}\put(14,-14){\circle*{2}}
\put(-14,-14){\circle*{2}}\put(14,14){\circle*{2}}
\put(-14,14){\vector(-3,1){1}}\put(14,14){\vector(3,1){1}}
\put(-13,-9){\circle*{1}}
\put(-12,-4){\circle*{1}}
\put(-10,0){\circle*{1}}
\put(-8,4){\circle*{1}}
\put(-5,7){\circle*{1}}
\put(0,10){\circle*{1}}
\put(4,12){\circle*{1}}
\put(9,13){\circle*{1}}
\put(-22,-21){\footnotesize{D}}\put(-23,14){\footnotesize{A}}
\put(16,-21){\footnotesize{B}}\put(15,14){\footnotesize{C}}
\end{picture}
&\begin{picture}(10,60)(0,-28)\put(0,0){$-$}\end{picture}
&\begin{picture}(5,60)(0,-28)\put(0,0){$q^2\,\times$}\end{picture}&
\begin{picture}(60,80)(-30,-30)
\put(0,0){\circle{38}}
\qbezier(14,14)(-7,7)(-14,-14)
\put(14,14){\circle*{2}}\put(-14,-14){\circle*{2}}
\put(14,-14){\circle*{2}}\put(-14,14){\circle*{2}}
\put(-14,14){\vector(-3,1){1}}\put(14,14){\vector(3,1){1}}
\put(13,-9){\circle*{1}}
\put(12,-4){\circle*{1}}
\put(10,0){\circle*{1}}
\put(8,4){\circle*{1}}
\put(5,7){\circle*{1}}
\put(0,10){\circle*{1}}
\put(-4,12){\circle*{1}}
\put(-9,13){\circle*{1}}
\put(-22,-21){\footnotesize{D}}\put(-23,14){\footnotesize{A}}
\put(16,-21){\footnotesize{B}}\put(15,14){\footnotesize{C}}
\end{picture}\\[2mm]
&&\mathds{1}&&&\hat\Omega_{AB}&&&\hat\Omega^{-1}_{AB}\\
\multicolumn{9}{c}{\mbox{Skein relations for the Jones polynomial as the projection}}\\
\multicolumn{9}{c}{\mbox{of a three-dimensional figure.}}\\[2mm]
\multicolumn{9}{c}{\mbox{The operator }\Omega_{AB}\mbox{ moves the points }A\mbox{ and }B}\\
\multicolumn{9}{c}{\mbox{The operator }\mbox{ continuously on the each other positions}}\\
\multicolumn{9}{c}{\mbox{in the selected direction, the cuts attached.}}
\end{array}
\ee
The Jones polynomial is argued then to be a ``contraction'' of the WZW
conformal blocks. The precise definition of this contraction, as
well as the explicit formulas for the conformal blocks are not
involved in the approach discussed.

\subsubsection{Knot polynomials as invariants of the conformal blocks}
Presented in \cite{Witt} construction, although being implicit
itself, gave rise to a new approach to calculating knot polynomials
\cite{Kaul,Inds1,Inds2,Inds3,KaulLeq,GMMM}. We briefly outline this approach below.

\paragraph{Cutting of the knot with punctured spheres.}
A knot is composed of ``elementary
pieces'', each piece being constraint by a topological sphere with
$2k$ punctures pairwise connected by $k$ segments of the
original curve. The punctures of the spheres are matched by the
parallel lines in the external space area. It is know then that a
knot can be obtained from the unknot with help of certain
``elementary transformations'' of the ``elementary pieces'', which
consist in intertwining of the curve segments inside the spheres
(see the above papers and references therein for the corresponding
theorems).

\paragraph{Relating a punctured sphere to a linear operator.}
The $k$ line segments composing each ``elementary
piece'' can be continuously transformed into $k$ cuts, pairwise
connecting the $2k$ punctures on the constraining this piece
sphere. Intertwining the segments corresponds then to moving the
punctures continuously on the positions of each other, together
with the attached cuts. A sphere with punctures and cuts
corresponding to a WZW conformal block, such interchangings of the
punctures positions correspond to certain transformations of the WZW
conformal blocks. Unlike the conformal blocks themselves,
operators of these transformations can be calculated explicitly and
rather effectively, the corresponding technology being developed in
\cite{Kaul,Inds1,Inds2,Inds3,Inds6,Inds10,KaulLeq,GuJock,MMM4,MMM5,GMMM}. If the unknot corresponds to the unity
operator, then a knot corresponds to an operator product, and a knot
polynomial is obtained in this approach as a matrix element of the
operator product.

\paragraph{Case of the colored HOMFLY polynomials.}
A more general WZW conformal block is related to a punctured topological
sphere (or to a higher genus curve), a Lie group
representation being associated with each puncture. In particular, the conformal blocks
related to the HOMFLY polynomials are in the fundamental
representation of the $SU(N)$ group; in particular, for $N=2$ the Jones polynomial is obtained. The next conjecture
was that the colored HOMFLY polynomials (see
sec.\ref{sec:knmath}) are related to the WZW conformal blocks for higher
representations of the same group in a similar way (although
the reasoning of \cite{Witt} can not by straightforwardly extended to the case
since the colored polynomials do not possess an implicit definition
generalizing the skein relations definition of the plain
polynomials).

\paragraph{The conformal blocks approach as the state model approach.}
The approach outlined above provides a representation of a knot
invariant of the same kind as the representation we consider (see
sec.\ref{sec:stmod}), and as the representation that arises in the
Kontsevich integral method (see sec.\ref{sec:Konts}), which is
closely related to interpretation of knot invariants from the
standpoint of the perturbative Chern-Simons theory. Moreover, all
three approaches use the so called $R$-matrix (see sec.\ref{sec:Rmat}) as
one of the elementary operators. For instance, $R$-matrix in the WZW
approach arises as, roughly speaking, a square root of the conformal
block monodromy operator.

\subsubsection{Wess-Zumino-Witten conformal blocks and classical Chern-Simons fields}
Finally, a knot polynomial represented as sketched above, is
related in \cite{Witt} to an observable (precisely, to Wilson
averages, see sec.\ref{sec:CSperp}) in a three-dimensional gauge
theory by means of one more implicit step. Namely,
\begin{itemize}
\item{Wess-Zumino-Witten \emph{conformal blocks} are in a
one-to-one correspondence with the curvature free three dimensional
fields inside the sphere, the fields being singular along the curve
arcs.}
\end{itemize}
Such fields may be considered as solutions of the classical
equations of motions (with the field sources along the arcs) for a
certain three-dimensional action, which is called \emph{Chern-Simons
action} \cite{CS,CSphys} (see also sec.\ref{sec:CSperp}). This
correspondence is one of inspiration sources for associating knot
invariants with observables in a topological quantum field theory.
We briefly discuss some points of the subject in the next section.

\subsubsection{Towards relating the knot invariants to the perturbatively computed Wilson averages in the Lagrangian Chern-Simons theory\label{sec:CSperp}}
As we briefly discussed in the last section,  it is the
\emph{implicit} correspondence between the HOMFLY polynomials and
Wilson averages, which is suggested in \cite{Witt}. A problem of
obtaining the knot polynomial as an observable in the Lagrangian
Chern-Simons theory remains then open, being intensively studied by
other researches soon afterwards \cite{AlLab, GuadMarMin1,
GuadMarMin2, FrohKing, LabPer, LabNew}. These studies came to the
same relation between the knot polynomials and the Chern-Simons
Wilson averages as the one stated in \cite{Witt}, approaching it
from a completely different direction. The correspondence of the two
quantities is now established \emph{explicitly}, but only
\emph{perturbatively}. Namely, a perturbation series for the
Chern-Simons Wilson average is compared term-wise with the expansion
for the HOMFLY polynomial in the logarithm of a formal variable.
Apart from that, a variant of this approach also gives rise to an
operator-contraction presentation for the knot invariants, similarly
to the WZW approach and to the approach we follow.

\subsection{Wilson averages in the Chern-Simons theory\label{sec:CSWav}}
In this section, we recall the form of the Chern-Simons action and
the definition of the Wilson average, giving some comments about a
physical sense of these quantities as well.

\subsubsection{Basic definitions}
\paragraph{Chern-Simons action}
In the Lagrangian approach, the three-dimensional Chern-Simons
theory is defined by the cubic action \be \int
d^3x\,\mathrm{Tr}_{\mathrm{adj}}\,\left\{\epsilon^{ijk}\left(A_i\p_j
A_k+\textstyle{\frac{2gi}{3}}A_iA_jA_k\right)\right\} \equiv S_{CS},
\label{CSnab}\ \ee where $A_{\mu}(x)$ is a three-dimensional gauge
field. Action (\ref{CSnab}) arises, for instance, from the
topological term in the four-dimensional Yang-Mills theory, \be
S_{CS}=\int
dx^4\mathrm{Tr\,}\epsilon^{\mu\nu\rho\sigma}F_{\mu\nu}F_{\rho\sigma}
\label{CSnab},\ \ee where the Wick rotation $t\equiv i\tau$ was
performed, and the integral in the r.h.s. is taken over the infinity
sphere, the Chern-Simons field being the projection of the
Yang-Mills field on the sphere. The Chern-Simons action $S_{CS}$
being independent of the metric approves referring it to as a
\emph{topological theory} \cite{CS,GuadMarMin1}.

Although is not included in the standard classical Yang-Mills
action, the topological term might be created by non-perturbative
corrections \cite{Rub}. On the other hand, the value of the
classical Yang-Mills action in the Euclidian space, when being
finite\footnote{For this to take place the Yang-Mills field must be
locally the pure gauge in the infinity sphere, the Chern-Simons
action giving then the number of covers of an $SU(2)$ subgroup of the gauge group by the infinite sphere \cite{BPST,Rub}.}, is restricted from below by the
value of $S_{CS}$ evaluated for a Yang-Mills field from the same
topological class. As a result, self-dual fields, which satisfy the
first-order equation
$F_{\mu\nu}=\epsilon_{\mu\nu\rho\sigma}F^{\rho\sigma}$ so that
$S_{YM}$ equals $S_{CS}$, are particular solutions of the Euclidian
Yang-Mills equations \cite{BPST}. The corresponding solutions in the
Minkowski space are referred to as Yang-Mills \emph{instantons}, the
interest to these solutions being inspired by Polyakov result
\cite{Pol}, which consisted in demonstrating the solutions of a
similar kind being responsible for the confinement in a toy model
(lattice two-dimensional electrodynamics).

The classical equations of motion for the Chern-Simons action \be
\epsilon^{ijk}\big(\p_jA_k+A_jA_k\big)=\epsilon^{ijk}F_{jk}=0 \ee
require for vanishing of the field tensor $F_{jk}$. However, the
corresponding potential is not necessarily a pure gauge. Particular
examples of the non-trivial classical Chern-Simons potentials are
given by projecting the already mentioned instantonic solutions of
the Yang-Mills equations \cite{BPST} on the infinity sphere.

\bigskip

One can also consider the Chern-Simons action in Minkowski signature
(there is no factor if $i$ in (\ref{CSnab}) then). As usual, the two
variants of the theory are related by the Wick rotation \cite{PesSch}
$t\rightarrow it$, or, equivalently, $A_0\rightarrow iA_0$. We start
from considering the more intuitively clear Euclidian version in
sec.\ref{sec:CSex},\ref{sec:lnum}, and \ref{sec:CSfram}, switching
to the Minkowski version in sec.\ref{sec:Konts}, since the statement
discussed there refers to the light-cone gauge, which can be
selected in the Minkowski signature (see \cite{LabNew} and
sec.\ref{sec:abgauge}).

\paragraph{Wilson lines and loops}
A Wilson line is an observable introduced in a field gauge theory
\cite{PesSch}. The Wilson line in an abelian gauge theory by
definition equals \be
W^{ab}\big[A(x),\gamma\big]\equiv\exp\int_{\gamma}dx^{\mu}A_{\mu}\left(x^{\mu}\right).\label{Wab}\ee
This quantity gives the phase increment of a wave function as the
particle subjected to the field $A$  passes the line $\gamma$.

The Wilson line in a non-abelian gauge theory is defined with help
of a notion of the path exponential. The path exponential of a
non-abelian gauge field $\vec A(t)$ over a curve $\gamma$, a
parameter $t$ is selected on the curve, is an operator that by
definition satisfies \be W(t+\delta t)=W(t)\big(1+\vec A(t)\cdot\vec
n(t)\delta t \big), \label{WAmult}\ee with $\vec n(t)$ being the
tangent vector to the curve. One writes \be
W[A(x),\gamma]\equiv\mathrm{Pexp}\left(\int_{0[\gamma]}^t\vec
A(s)\cdot\vec n(s)ds\right)\equiv
\mathrm{Pexp}\left(\int_{\gamma}\vec A\cdot dl\right).
\label{Wnab}\ee This quantity represents the finite gauge
transformation responsible for mixing of the matter fields in a
multiplet as they pass the line $\gamma$.

A Wilson line for the closed contour $\gamma$ is called a Wilson
loop. The trace of this quantity over the gauge group is a gauge
invariant and independent of the reference point on the contour
quantity. The quantity can be observed intermediately in the
phenomena like Aharonov-Bohm effect.

\paragraph{Wilson averages.}

In quantum field theory, one introduces the notion of the Wilson
average as well. This quantity is defined in perturbation theory as
an average of the formal series for the path exponential, \be \ckl
W_{\gamma}(A)\brr\equiv\sum_{k=0}^{\infty}g^k\int_{\tilde{x}}^{\tilde{\tilde{x}}}dx_k^{i_1}
\int_{\tilde{x}}^{x_1}dx_k^{i_2}\ldots\int_{\tilde{x}}^{x_{k-1}}dx^{i_1}
\ckl A_{i_k}(x_k)\ldots
A_{i_2}(x_2)A_{i_1}(x_1)\brr,\label{Wnabperp} \nn\ee where all
integrals are taken over the arcs of contour $\gamma$.

\paragraph{Wilson averages in Chern-Simons theory.}

A remarkable property of the Chern-Simons theory is that it can be
reduced to an abelian theory by the proper gauge fixing (see
\cite{GuadMarMin1} and sec.\ref{sec:Konts}). As a result, the
Chern-Simons Wilson average is in fact equal to the Wilson line
evaluated on the proper solution of the classical equations of
motion in the corresponding gauge. Because the classical
Chern-Simons equations are the curvature vanishing equations, the
Wilson loop contains then a contour integral of a closed one-form,
being thus unaffected by the smooth deformations of the contour and
giving a topological invariant.

In case of an everywhere regular Chern-Simons field, which satisfies
the classical equations of motion in the \emph{entire} space, all
Wilson loops will take the same and trivial value. Unlike that, the
Chern-Simons field having a line-like singularity gives results in
appearing of topologically distinct Wilson loops, the contours
variously intertwined with the line of the field singularity. The
value of the Wilson loop depends then only on the topological class
of the contour.

The singular locus of the Chern-Simons field can be accounted for by
putting the corresponding delta-function in the r.h.s. of classical
equations of motion, or, equivalently, by adding the source term to
the Chern-Simons action. Roughly speaking, such a term in turn can be
obtained by inserting the second Wilson line under the average sign,
joining then the exponent to the action as an interaction term of
the Chern-Simons field with a line-like field source. Hence, a
correlator of the two Wilson averages is in this sense similar to a
one Wilson loop evaluated over a classical Chern-Simons field,
generated by the source placed along the other loop. This is in fact
the idea of perturbative evaluation of the Wilson averages in an
abelian gauge \cite{LabNew, LabPer, DSS:Konz, MorSm} (see
sec.\ref{sec:CSfram} for details).

\subsubsection{Simplest examples of Chern-Simons fields and Wilson averages\label{sec:CSex}}

To better illustrate what kind of theory is under discussion, we
write down explicitly the Chern-Simons actions for the simplest
gauge groups, considering them from the standpoint of some physical
and geometrical analogies. A more broad and detailed discussion of the Chern-Simons theory from the physical standpoint can be found in \cite{CSphys}.

\paragraph{Gauge group $U(1)$.}
In this simplest case, the Chern-Simons field is an abelian gauge
field, which may be considered as a static magnetic field satisfying
the Maxwell equations \be \mathrm{rot}\vec H(\vec x)=4\pi \vec
j(\vec x),\ \mathrm{div}\vec H=0.\label{Hdiv} \ee The first equation
can be obtained by variation of the effective action \be
S_{CS}^{ab}[A_0=H_z,\ A_x=H_x,\ A_y=H_y]=\frac{\kappa}{4\pi}\int
d^3x\epsilon^{ijk}A_i\p_jA_k\stackrel{\kappa=1}{=\!=}\frac{1}{4\pi}\int
dV\ \vec{H}\cdot \mathrm{rot} \vec H+\int dV\ \vec{H}\cdot \vec
j,\ee which coincides with the abelian Chern-Simons action, the
magnetic \emph{field} standing for the Chern-Simons
\emph{potential}. The second equation may be considered then as the gauge
fixing condition $\p_kA_k=0$, a gauge transformation of the
Chern-Simons field $A_k\rightarrow A_k+\p_kf$ corresponding to
adding a rotor-free magnetic field, which satisfies the homogeneous
Maxwell equations.


An abelian Wilson loop is then an exponentiated circulation of the
Chern-Simons field over a closed circuit \be
W(\gamma)\left[A_0=H_z,\ A_x=H_x,\ A_y=H_y\right]\equiv
\exp\left(\oint_{\gamma}\vec dl\vec H\right)\label{Wlab}. \ee When
evaluated on a Maxwell equations solution, this quantity is equal to
the circulation of a circuit magnetic field over the circuit. This
quantity is not well defined for an infinitely thin circuit.
However, if there are two intertwined circuits, the quantity
contains a well defined contribution, which is the circulation of
the one circuits magnetic field over the other circuit. Due to the
integral form of the corresponding Maxwell equation, such a
cross term is proportional to the linking number of the
contours.

\paragraph{Gauge group $SU(2)$.}
The simplest case of a non-abelian Chern-Simons theory contains the
triple of gauge fields $(\vec A_1,\vec A_2,\vec A_3)$ entering the
action \be S=\int d^3x\left(\sum_{a=1}^3\vec
A_a\cdot\mathrm{rot}\vec A+(\vec A_1,\vec A_2, \vec A_3)\right),
\label{CSSU2}\ee where a dot stands for scalar product and the
parentheses stand for a mixed product of the vectors, $(\vec a,\vec
b,\vec c)\equiv \vec a\cdot [\vec b\times \vec c]=\vec b\cdot [\vec
c\times \vec a]=\vec c\cdot [\vec a\times \vec b]$. The same action
can be presented in the form The same action can be presented in the
more standard form \be S=\int
d^3x\mathrm{Tr}\left\{\epsilon^{ijk}\left(\hat A_i\p_j \hat
A_k+\frac{2i}{3}\hat A_i\hat A_j\hat
A_k\right)\right\},\label{CSSU2g} \ee with $\hat A$ being an
anti-hermitian matrix, \be
W=\left(\begin{array}{cc}iA_3&iA_1+A_2\\iA_1-A_2&-iA_3\end{array}\right)=i\sum_{a=1}^3A^ab_a,
\ee which is expanded over the Pauli matrices \be
b_1=\left(\begin{array}{cc}0&1\\1&0\end{array}\right),\ \
b_2=\left(\begin{array}{cc}0&-i\\i&0\end{array}\right),\ \
b_3=\left(\begin{array}{cc}1&0\\0&-1\end{array}\right),\nn\\
b_ab_b=\frac{1}{2}[b_a,b_b]=\epsilon_{abc}b_c. \ee The action being
presented as (\ref{CSSU2g}), it is easier to demonstrate its
invariance under a transformation generated by an arbitrary unitary
matrix with the unit determinant and generally with the coordinate
dependent entries, \be A_k\rightarrow
\Omega^{-1}A_k\Omega+\Omega^{-1}\p_k\Omega,\ \
\Omega=\left(\begin{array}{cc}a(x,y,z)&b(x,y,z)\\-\bar{b}(x,y,z)&\bar{a}(x,y,z)\end{array}\right)\in
SU(2),\label{SU2trans} \ee where the bar stands for the complex
conjugate. Transformation (\ref{SU2trans}) is called an $SU(2)$
\emph{gauge transforation} of the theory, which is thus by
definition an $SU(2)$ \emph{gauge theory}.

\paragraph{Classical equations the $SU(2)$ theory}
Variation of action (\ref{CSSU2}) in the fields $\vec A_1$, $\vec
A_2$, $\vec A_3$ gives rise to the classical equations of motion,
\be \mathrm{rot}\vec A_1=[\vec A_2,\vec A_3],\ \mathrm{rot}\vec
A_2=[\vec A_3,\vec A_1],\ \ \mathrm{rot}\vec A_3=[\vec A_1,\vec
A_2], \label{cleq}\ee respectively. Alternatively, one can vary the
action in form (\ref{CSSU2}) w.r.t. the matrix-valued field $\hat
A$, obtaining the same equations, this time presented as the
curvature vanishing equations, \be \p_i\hat A_j-\p_j\hat A_i+[\hat
A_i,\hat A_j]=0. \label{0curv}\ee

\paragraph{Solutions of the classical equations through the scalar potential.}
The classical equations of motion being the curvature-vanishing
equations, a solution is locally expressed as \be
A_{\mu}=\Omega^{-1}\p_{\mu}\Omega,\label{locsol} \ee where
$\Omega\in SU(2)$ is a gauge group element, namely, a $2\times 2$
special ($\det\Omega=1$) unitary
($\Omega\Omega^{\dagger}=\mathds{1}$) matrix. One can
straightforwardly verify that (\ref{locsol}) satisfies
(\ref{0curv}), using that $\p_{\mu}(\Omega\Omega^{-1})=0$ to express
$\p_{\mu}\Omega^{-1}$, namely, \be
\p_{\mu}A_{\nu}-\p_{\nu}A_{\mu}=\p_{\nu}\Omega^{-1}\p_{\mu}\Omega+\Omega^{-1}\p^2_{\nu\mu}\Omega-(\mu\leftrightarrow\nu)=
\Omega^{-1}\p_{\nu}\Omega\cdot\Omega^{-1}\p_{\mu}\Omega-(\mu\leftrightarrow\nu)=-[A_{\mu},A_{\nu}]
\ee If the scalar potential $\Omega$ is everywhere regular,
then the Chern-Simons field $A_{\mu}$ is everywhere vanishing up to
gauge transformation (\ref{SU2trans}) generated by $\Omega$. Unlike
that, $\Omega$ having at least one special point can not be used as
a matrix of a gauge transformation in the entire space, and the
corresponding solution for $A$ can be highly non-trivial.

\paragraph{``Abelian'' solution with a line-like source}
In the context of the knot theory, the solutions of the classical
equations with a line-like singularity are especially interesting (see discussion in sec.\ref{sec:CSWav}). The simplest of
such solutions is obtained from the above presented solution by
ignoring $z$ dependence of the scalar potential and by
setting the parameter $a$ to be zero, \be
\tilde{\tilde{\Omega}}\left(x,y,z=0,a=0\right)=\Omega\left(x,y,z=0,a=0\right)=\frac{1}{r}
\left(\begin{array}{cc}0&ix+y\\ix-y&0\end{array}\right). \ee The
corresponding components of the field $A$, \be
\tilde{\tilde{A}}_x=A_x\left(x,y,z=0,a=0\right)=-\frac{y}{r^2}\left(\begin{array}{cc}i&0\\0&-i\end{array}\right),\
\
\tilde{\tilde{A}}_y=A_y\left(x,y,z=0,a=0\right)=\frac{x}{r^2}\left(\begin{array}{cc}i&0\\0&-i\end{array}\right),\\
\tilde{\tilde{A}}_z=0, \nn\ee are regular everywhere but the line
$x=y=0$. In fact, this case reduces to the abelian case, both the
non-vanishing field components being proportional to one and the
same matrix everywhere in the space. However, it is already
instructive to examine the various Wilson lines in this case.

\paragraph{Wilson loops for the ``abelian'' solution}
The pase increment of a minimally coupled to the field (with the
charge $g$) particle passing an infinitely small segment of the
circle $x^2+y^2=1$ is given in this case by the diagonal matrix \be
\omega\equiv
A_x\frac{dx}{d\varphi}+A_y\frac{dy}{d\varphi}=-ydA_x+xA_y=\frac{1}{r^2+a^2}
\left(\begin{array}{cc}i&0\\0&-i\end{array}\right). \ee All the
matrices on the particle path commuting, the path exponential in
expression for Wilson line (\ref{WAmult}) reducing to the plain
matrix exponential, which yields \be W\big[\gamma(\psi)\big]=
\left(\begin{array}{cc}e^{ig\psi}&0\\0&e^{-ig\psi}\end{array}\right).
\ee In turn, the Wilson loop \be \Tr
W\big[\gamma(2\pi)\big]=2\cos(2\pi g) \ee is a gauge invariant
quantity that measures the phase increment of the particle passed
along the closed contour. It can be shown that the phase shift
corresponding to a closed contour is given by a non-unity matrix
only for the contours winded over the line $x=y=0$ of $\Omega$
singularity.

The presented solution in fact an example of the \emph{abelian} Chern-Simons field, since the non-vanishing components of the vector potential commute at each point, as well as operators of the various Wilson lines.

\paragraph{``Non-abelian'' solution with a line-like source}
Now we present an example of an ``essentially non-abelian'' classical Chern-Simons field (see Appendix \ref{App:SU2sol} for how the corresponding expression may be obtained and presented visually) Namely, one should take the scalar potential
\be
\Omega(x,y,z)=\Omega(x,y)=\frac{1}{\sqrt{\left(x^2+y^2\right)\left(x^2+y^2+a^2\right)}}
\left(\begin{array}{cc}x^2+y^2&a(ix+y)\\a(ix-y)&x^2+y^2\end{array}\right),
\ee
which gives rise to the vector potential
\be
\arraycolsep=0.5mm
A_x=\Omega^{-1}\p_x\Omega=\frac{a}{\left(x^2+y^2\right)\left(x^2+y^2+a^2\right)}
\left(\begin{array}{cc}-iay&i\left(-x^2+y^2\right)-2xy\\i\left(-x^2+y^2\right)&iay\end{array}\right),\nn\\
A_y=\Omega^{-1}\p_y\Omega=\frac{a}{\left(x^2+y^2\right)\left(x^2+y^2+a^2\right)}
\left(\begin{array}{cc}iax&-ixy+\left(x^2-y^2\right)\\-ixy-\left(x^2-y^2\right)&-iax\end{array}\right),\nn\\
A_z=\Omega^{-1}\p_z\Omega=0.\label{SU2cl}
\ee

Note, that different components of the field $A$ at the same point
do not commute, e.g., \be
\frac{1}{2}[A_y,A_x]=\frac{a^2}{\left(x^2+y^2\right)\left(x^2+y^2+a^2\right)^2}
\left(\begin{array}{cc}x^2+y^2&a(iy-x)\\a(iy+x)&-i\left(x^2-y^2\right)\end{array}\right),
\ee so that it is not just the antisymmetric derivative
$\p_xA_y-\p_yA_x$, but the non-abelian field tensor $F$, which
vanishes.

\paragraph{Wilson loops for the ``non-abelian'' solution.}
This time an infinitely small phase increment along the same line is
given by the matrix \be \omega\equiv
A_x\frac{dx}{d\varphi}+A_y\frac{dy}{d\varphi}=-ydA_x+xA_y=\frac{a}{x^2+y^2+a^2}
\left(\begin{array}{cc}ia&x-iy\\-x-iy&-ia\end{array}\right)=
\left(\begin{array}{cc}iA& Be^{-i\phi}\\-B
e^{i\phi}&-iA\end{array}\right),\\\nn A=\frac{a^2}{\rho^2+a^2},\ \
B=\frac{a\rho}{\rho^2+a^2},\ \  x=\rho\cos\phi,\ y=\rho\sin\phi. \ee
Because the phase shifts in the different points of the path do not
commute, the Wilson line, which by definition equals \be
\Pexp\int_{0[x=\rho\cos\phi,y=\rho\sin\phi]}^{\alpha}\omega
d\phi\equiv \lim_{\substack{N\rightarrow\infty\\q^N=\exp(i\alpha)}}
\prod_{i=0}^N\left\{\left(\begin{array}{cc}1&0\\0&1\end{array}\right)+
\frac{\phi}{N}\left(\begin{array}{cc}iA&Bq^{-1}\\-Bq&-iA\end{array}\right)\right\},
\ee no longer reduces to the plain matrix exponential.

\subsubsection{Gauging out of the cubic term in the non-abelian Chern-Simons theory.\label{sec:abgauge}}
An essential property of action (\ref{CSnab}) is that the cubic term
vanishes in certain gauges, which are called \emph{abelian} gauges.
This property is essential for our discussion, because it is in fact
a \emph{Gaussian} average, the properties of which reproduces the
discussed integral presentation for the knot invariants.

Any gauge where the three matrices $A_x$, $A_y$, $A_z$ (in euclidian
signature) or $A_0$, $A_x$, $A_y$ (in Minkowski signature) are
linearly dependent is an abelian gauge. One may verify that this
indeed may be achieved by a gauge transformation; moreover, the
quantum correction do not create the cubic term either
\cite{LabPer}. In Minkowski metric, the most common abelian gauges
are the temporial gauge $A_0=0$ \cite{MorRos,MorSm,LabNew} and a
light-cone gauge $\vec n\cdot\vec A=0$ with $\vec n\cdot\vec n=0$.
The light-cone gauge with $\vec n=(1,1,0)$ can be reformulated as a
\emph{holomorphic} gauge \cite{LabPer}, by passing to the euclidian
signature via the Wick rotation $A_0\rightarrow -iA_0$, and by
setting \be A_t=A_y,\ \ \ A_z=A_x-iA_0\ \ \ A_{\bar z}=A_x+iA_0.\ee
The gauge-fixing condition then reads \be A_{\bar
z}=0.\label{holg}\ee The Chern-Simons theory in the holomorphic
gauge is the most explored the moment. For instance, the main
statement about relation of the knot invariants to the perturbative
Chern-Simons theory, which we formulate and discuss in
sec.\ref{sec:Konts}, refers to this gauge \cite{LabPer, LabNew}.
Unlike that, non much is known about the Chern-Simons theory in the
temporial gauge, there being a lot more questions that answers
\cite{MorRos,MorSm,LabNew}. To examine this gauge is a highly
intriguing problem since the $\mathcal{R}$-matrix representation for
the knot polynomials \cite{ReshTur,MorSm}, which is the main subject
of the present text and which we discuss in details in
sec.\ref{sec:Rmat}, is conjecturally related to the perturbative
expansion of the Chern-Simons Wilson average in the holomorphic
gauge \cite{MorRos, MorSm}.

\subsection{Linking number as a contribution to abelian Wilson average\label{sec:lnum}}
The first motivation for the QFT interpretation of the Vassiliev
invariants is the integral formula for the simplest of this of this
invariants, which is the linking number of two closed curves
$\mathcal{C}_1$ and $\mathcal{C}_2$ \cite{ChumDuzMost}, \be
\mathcal{L}(\mathcal{C}_1,\mathcal{C}_2)=\oint_{\mathcal{C}_1}dy^{i}\oint_{\mathcal{C}_2}dx^{i}
\cfrac{\epsilon_{ijk}\left(x^{k}-y^{k}\right)}{|\vec x-\vec
y|^3}.\label{linkint} \ee The integration kernel may be considered as a
Green function of the abelian Chern-Simons theory  (see
sec.\ref{sec:Gav} for details), the action being \be
S^{abelian}_{CS}=\frac{\kappa}{4\pi}\int d^3x
\epsilon^{kij}A_{k}\p_{i}A_{j}.\label{CSab} \ee The corresponding
integral arises then as a second order contribution to abelian
Wilson average (\ref{Wab}). More precisely, (\ref{Wab}) contains the
double integral of the Green function, which is divergent.
Postponing the interpretation of this phenomenon for the
sec.\ref{sec:CSfram}, we just notice now that presenting in case of
a \emph{link} crossing term \be \ckl W_{\mathrm{cross}}^{(2)}({{\cal
C}_1}\otimes{{\cal C}_2})\brr =\ckl\oint_{{\cal C}_2}
dx^{i}\oint_{{\cal C}_1} dy^{j}A_{i}(x)A_{j}(y)\brr=\oint_{{\cal
C}_2} dx^{i}\oint_{{\cal C}_1} dy^{j}\ckl
A_{i}(x)A_{j}(y)\brr=\nn\\= \oint_{{\cal C}_2} dx^{i}\oint_{{\cal
C}_1} dy^{j} G_{ij}(z)\label{WA2cr} \ee is well-defined, giving just
linking number (\ref{linkint}).

\subsection{Properties of the knot invariants as general properties of the ordered exponential and of the Gaussian average.\label{sec:WAgen}}
As already mentioned in the introduction, an interpretation of more
general knot invariants in terms of a non-abelian QFT is essentially
motivated by the observation that the Vassiliev invariants possess
the integral representation and can be assembled into a generating
function that has the properties of the gaussian average of the
ordered exponential \cite{LabPer, LabNew}. In the  section, we
recall the relevant properties of the ordered exponential and of the
gaussian average, briefly mentioning the corresponding properties of
the knot invariants.

\subsubsection{Relevant properties of the ordered exponential\label{sec:Pexp}}
\paragraph{Composition property as the defining property of the ordered exponential.}
First of all, the ordered exponential by definition possess the
composition property \be
\boxed{\mathrm{Pexp}{\int_a^bdtF(t)}=\mathrm{Pexp}{\int_a^cdtF(t)}\,\mathrm{Pexp}{\int_c^bdtF(t)}}\label{Pexp:comp}
\ee  One can demonstrate explicitly that this property holds at any
order of the perturbative expansion, \be
\mathrm{Pexp}{\int_a^bdtF(t)}=\mathds{1}+\int_a^bdtF(t)+\frac{1}{2}\int\int_{a\le
s<t\le b}dtdsF(t)F(s)+\frac{1}{2}\int\int_{a\le t<s\le
b}dtdsF(s)F(t)+\ldots. \ee Splitting the last explicitly written out
summand as \be
\int_a^bdt\int_a^tdsF(s)F(t)=\int_a^cdt\int_a^tdsF(s)F(t)+\int_b^cdt\int_c^tdsF(t)F(s)+\int_b^cdt\int_a^cdsF(t)F(s),
\ee and taking into account that
\be\int_b^cdt\int_a^cdsF(t)F(s)=\left(\int_b^cdt
F(t)\right)\left(\int_a^cdsF(s)\right),\ee one obtains the result to
agree with the r.h.s. of (\ref{Pexp:comp}), \be
\mathrm{Pexp}{\int_a^cdtF(t)}\mathrm{Pexp}{\int_c^bdtF(t)}=\nn\\=\Big(\mathds{1}+\int_a^bdtF(t)+
\int_a^bdt\int_a^tdsF(s)F(t)+\ldots\Big)\Big(\mathds{1}+\int_c^bdtF(t)+\int_a^bdt\int_a^tdsF(s)F(t)+\ldots\Big),\ee
up to second order.

\paragraph{Tensor product presentation.}
Property (\ref{Pexp:comp}) enables one to identically rewrite a path
exponential in form of a tensor contraction, which is similar to the
one entering the definition of knot invariants of our interest (see
sec.\ref{sec:stmod}). Indeed, \be
\mathrm{Tr\,Pexp}{\oint_{\gamma}tF(t)}=
\mathrm{Tr}\left\{\mathrm{Pexp}{\int_a^bdtF(t)}\mathrm{Pexp}{\int_b^cdtF(t)}\mathrm{Pexp}{\int_c^ddtF(t)}\mathrm{Pexp}{\int_d^adtF(t)}\right\}
=S^{ij}_{kl}\bar{S}^{kl}_{ji}, \label{Pexp:tens}\ee where we
introduced the operators \be
S=\mathrm{Pexp}{\int_a^bdtF(t)}\otimes\mathrm{Pexp}{\int_c^ddtF(t)},\
\ \
\bar{S}=\mathrm{Pexp}{\int_b^cdtF(t)}\otimes\mathrm{Pexp}{\int_d^adtF(t)},
\ee and used the identity \be L^i_jM^j_kP^k_lQ^l_i=\big(L\otimes
P\big)^{ik}_{jl}\big(M\otimes Q\big)^{jl}_{ki}. \ee

\subsubsection{Relevant properties of the Gaussian average\label{sec:Gav}}
\paragraph{Definition of the Gaussian average}
The next property of the generating function for the Vassiliev
invariants is it having a structure of a \emph{sum over pairings},
same to a Gaussian average. Namely, there is the formula of
Gaussian integration \cite{PesSch} \be \ckl
F_iF_j\brr\equiv\frac{\int
\prod_{i=1}^NdF_i\,F_kF_l\exp(-\frac{1}{2}\sum_{i,j=1}^NK_{ij}F^kF^l)}{\int
\prod_{i=1}^NdF_i\,\exp(-\frac{1}{2}\sum_{i,j=1}^NK_{ij}F^kF^l)}=K^{-1}_{ij},
\ee which is straightforward to verify for a finite $N$.
In analogy with the finite-dimensional case, the Gaussian average of the product of two operators $F(t)$ and
$F(s)$ w.r.t. a quadratic action
\be S\left[F\right]=\int_0^T ds\int_0^T dt \mathcal{K}(t,s)F(t)F(s)\ee
 (the most common case is $\mathcal{K}(t,s)=\ddot\delta(t-s)$) by definition equals
\be
\ckl F(t)F(s)\brr_S=G(t,s)\label{Gav:Green}
\ee
where the r.h.s. contains the Green
function, which by definition satisfies
\be
\mathcal{K}(t,u)=\int_0^T ds G(t,s)\mathcal{K}(s,u),
\ee
being in the sense the inverse of the kinetic operator $dtds\mathcal{K}(t,s)$.

Any other correlator by definition either vanishes, if containing an odd number of operators,

\be \ckl
\prod_{i=1}^{2k-1}F\left(t_i\right)\brr\equiv0,\label{Gav:odv}
\ee or is expressed via the pairwise correlators, if containing an even number of operators,

\be \ckl \prod_{i=1}^{2k-1}F\left(t_i\right)\brr\equiv\sum_{\sigma}\prod_{i=1}^{2k-1}\ckl
F\left(t_i\right)F\left(t_{\sigma(i)}\right)\label{Gav:Wick}\brr \ee The r.h.s.
of (\ref{Gav:Wick}) contains \emph{the sum over all pairings}
$\sigma$ of the numbers from $1$ to $2k$, and the equality is
referred to as \emph{Wick theorem}, which also can be derived by a
straightforward computation in case of plain (not functional)
Gaussian integral.

\paragraph{Peculiar cases when the average factorizes}
If one takes the Gaussian average of the both parts of composition
property (\ref{Pexp:comp}), the r.h.s. does \textbf{not} decompose
into the product of the two averages generally. However, the
decomposition takes place in the particular case when the operators
$F(t)$ for $a<t<b$ do not correlate with the operators $F(s)$ for
$b<t<c$. Moreover, due to the Wick theorem, it suffices to require
the pairwise correlators of fields from different regions to vanish,
\be \ckl F(t)F(s)\brr=0\mbox{ for } a<t<c<s<b\Rightarrow
\ckl\mathrm{Pexp}{\int_a^bdtF(t)}\brr=\ckl\mathrm{Pexp}{\int_a^cdtF(t)}\brr\ckl\mathrm{Pexp}{\int_c^bdtF(t)}\brr\label{Gav:fac}
\ee The simplest and rather common case where property
(\ref{Gav:fac}) holds is the case \be \ckl F(t)F(s)\brr=\delta(t-s),
\ee which corresponds to averaging with the weight \be
\exp(-S)=\exp\left(-\int dtF^2(t)\right). \ee More generally,
(\ref{Gav:fac}) holds for two distant regions if the correlator
decreases fast enough, \be \ckl F(t)F(t+b-a)\brr\ll\ckl
F(t)F(t+\epsilon)\brr,\ \ \epsilon\ll b-a, \ee e.g., for \be \ckl
F(t)F(s)\brr=\frac{1}{(t-s)^2+T^2},\ \ T\ge b-a. \ee

Property (\ref{Gav:fac}) of the Gaussian average match the
factorization property of the Vassiliev invariants and HOMFLY
polynomials for the disjoint union of links (see the next paragraph)
\cite{Pras}. Moreover, the combinatorial representation for the
Kontsevich integral \cite{ChumDuzMost, DSS:Konz} can be derived with
help of this property, if one takes into account identity
(\ref{Pexp:tens}) for the path exponential as well.

\paragraph{Expansion of the averaged trace over the traces of the algebra generators products.}
As soon as $F(t)$ is an operator, the ordered exponential of $F(t)$
is an operator as well. Generally, an averaged trace of the operator
is \textbf{not} equal to the trace of the averaged operator. In
particular, property (\ref{Gav:fac}) do not hold for the traces of
the corresponding operators. However, if the operator $F(t)$ takes
values in a Lie algebra, one may expand all the operators over
algebra generators $T_a$, \be \ckl F(t)F(s)\brr=0\mbox{ for }
a<t<c<s<b\Rightarrow\
\ckl\mathrm{Tr\,}\mathrm{Pexp}{\left(\int_a^bdt\sum_aF^a(t)T^a\right)}\brr
=\nn\\=\ckl\mathrm{Tr\,}\left\{\mathrm{Pexp}{\left(\int_a^cdt\sum_a
F^a(t)T_a\right)}\mathrm{Pexp}{\left(\int_c^bdtF^a(t)T^a\right)}\right\}\brr=\nn\\=
\ckl\mathrm{Tr\,}\left\{\sum_a\left(\mathrm{Pexp}{\int_a^cdt
F(t)}\right)^{\!\!a}\!\!T_a\sum_b\left(\mathrm{Pexp}{\int_c^bdtF(t)}\right)^{\!\!b}\!\!T_b\right\}\brr=\nn\\=
\sum_{a,b}\ckl\left(\mathrm{Pexp}{\int_a^cdt
F(t)}\right)^{\!\!a\phantom{b}\!\!}\brr\ckl\left(\mathrm{Pexp}{\int_c^bdtF(t)}\right)^{\!\!b}\brr\mathrm{Tr\,}T_aT_b.\label{Gav:grfac}\ee
The product in (\ref{Gav:grfac}) is substituted now with the sum,
each term being a product of the \emph{group factor}
$\mathrm{Tr\,}T_aT_b$ and of the \emph{coordinate factor}, which is
given by the corresponding component of the path exponential, and
for which decomposition (\ref{Gav:fac}) still takes place.

A similar structure of the Kontsevich integral \cite{ChumDuzMost} motivates comparing it with an averaged trace of the ordered exponential \cite{LabPer,LabNew,DSS:Konz}.

\subsubsection{Green functions giving rise to contour independent integrals\label{sec:contind}}
The above observations on the generating function for the Vassiliev
invariants in the integral representation can be summarized into a
claim, that the generating function has the structure of a Gaussian
average of a path exponential. One more property of the integral
representation, in turn, enables one to relate the Vassiliev
invariants with the Lagrangian Chern-Simons theory. The property
consists in the contour independence of the corresponding integrals.
For instance, these integrals can be presented as integrals of a
holomorphic functions in a complex plane.

On the other hand, if a Green function is a closed two form
($\p_{\mu}G_{\mu\nu}(x-y)=0$), or a holomorphic two-form
($\p_zG(z-w)=0$), then the integrals of Gaussian averages entering
the perturbative series for the average of the path exponential are
independent of integration contour, i.e., are topologically
invariant. Hence, if one interprets the integral kernel of the Vassiliev invariants as the Green functions if pairwise correlators in a quantum field theory, the gradient vanishing or holomorphic condition should arise as the classical equations of motion in the theory. Example of a proper theory is given by an abelian Chern-Simons theory with the action (\ref{CSab}) \cite{LabPer, LabNew}. Modulo the subtleties discussed in Appendix \ref{App:CSGreen}, the Green function of the abelian Chern-Simons theory is
\be
G_{ij}(x-y)=4\pi\epsilon_{ijk}\p^k\frac{1}{|\vec x-\vec
y|},\label{Grab} \ee
if the Lorenz gauge where $\p_kA_k=0$ is selected.

\subsection{Knot invariants as link invariants: framing of knot from the Chern-Simons theory standpoint\label{sec:CSfram}}
Now we address to the question of the divergent contributions in
perturbative expansion for the Wilson average, which we have already
encountered with in sec.\ref{sec:lnum}.

\subsubsection{Second order framing contribution}
Second order term in perturbative expansion for the Wilson average
contains, apart from well defined cross term (\ref{WA2cr}), the
diagonal terms with the both integrals being taking over the same
link component. Moreover, in case of a knot, not a link, the answer
includes such diagonal term only.  Yet, substituting of Green
function to (\ref{Grab}) to second order term (\ref{WA2cr}) in
expansion for the Wilson average, one indeed obtains a divergent
integral for the coinciding integration contours
$\mathcal{C}_1=\mathcal{C}_2$.

The singularity we came across with is conventionally resolved by
shifting the second integration contour relative to the first (and
the original) one \cite{FrohKing}, \be \oint_{\cal C}dy^{i}\
\longrightarrow\ \oint_{{\cal C}^{\prime}}dy^{i}.\label{frreg} \ee
The corresponding second order contribution to the Wilson average
takes then form of cross terms (\ref{WA2cr}), with the original and
shifted contours as the two integration contours. The resulting
integral yields then the linking number of these contours.

Arising of a new parameter as a result of the regularization
procedure is a common point in QFT. E.g., in case of UV
regularization of a perturbative QFT, this new variable is
associated with the considered energy scale. In turn, the maximum
(or somehow averaged) distance $\varepsilon$ between the contours
${\cal C}$ and ${\cal C}^{\prime}$ can be considered as the
regularization parameter in the problem under discussion. As usual,
the answer does not have a definite limit as $\varepsilon$ tends to
$0$; instead, all possible values of this limit are parameterized by
a new independent variable, which in the case is the linking number
of the contours ${\cal C}$ and ${\cal C}^{\prime}$. A possible
interpretation of the observed phenomenon is that CS Wilson average
is related not to a closed \emph{contour}, but to a \emph{closed}
ribbon. The quantity thus depends not only on the shape of contour,
but on the number of the ribbon intertwinings as well. On the other
hand, the knot invariants we consider are also invariants of the
ribbon knots in fact, what can be shown from the standpoint of
completely independent definition of these invariants (see
\cite{Pras} and sec.\ref{sec:Rfram}). Such a correspondence is one
of the main inspiration sources for looking at the knot invariants
as on the QFT observables.

\subsubsection{Higher orders framing contributions}
At a first glance, it seems that taking into account higher orders
in the expansion for the Wilson average requires for introducing
more and more new contours, $\mathcal{C}^{\prime\prime}$,
$\mathcal{C}^{\prime\prime\prime}$, $\ldots$. Actually, the
situation is different. As follows from the Wick theorem (see
\cite{PesSch} and sec.\ref{sec:Gav}), a term in the perturbation
series for the Wilson average is expressed via the second order
term. In particular, the fourth order term reads \be
\ckl\left(\oint_{\mathcal{C}}dx^{i}A_{i}(x)\right)^4\brr=\ckl
\oint_{\mathcal{C}}dx^{i}\oint_{\mathcal{C}}dy^{j}\oint_{\mathcal{C}}du^{k}\oint_{\mathcal{C}}dv^{l}
A_{i}(x)A_{j}(y)A_{k}(u)A_{l}(v)\brr=\nn\\=
\ckl\oint_{\mathcal{C}}dx^{i}\oint_{\mathcal{C}}dy^{j}
A_{i}(x)A_{j}(y)\brr\ckl\oint_{\mathcal{C}}du^{k}\oint_{\mathcal{C}}dv^{l}A_{k}(u)A_{l}(v)\brr+\nn\\+
\ckl\oint_{\mathcal{C}}dx^{i}\oint_{\mathcal{C}}du^{k}
A_{i}(x)A_{k}(u)\brr\ckl\oint_{\mathcal{C}}dy^{j}\oint_{\mathcal{C}}dv^{l}A_{j}(y)A_{l}(v)\brr+\nn\\+
\ckl\oint_{\mathcal{C}}dx^{i}\oint_{\mathcal{C}}dv^{l}
A_{i}(x)A_{l}(v)\brr\ckl\oint_{\mathcal{C}}dy^{j}\oint_{\mathcal{C}}du^{k}A_{j}(y)A_{k}(u)\brr=\nn\\
= \oint_{\mathcal{C}}dx^{i}\oint_{\mathcal{C}}dy^{j}
G_{ij}(x-y)\oint_{\mathcal{C}}du^{k}\oint_{\mathcal{C}}dv^{l}G_{kl}(u-v)+\nn\\+
\oint_{\mathcal{C}}dx^{i}\oint_{\mathcal{C}}du^{k}
G_{ik}(x-u)\oint_{\mathcal{C}}dy^{j}\oint_{\mathcal{C}}dv^{l}G_{jl}(y-v)+\nn\\+
\oint_{\mathcal{C}}dx^{i}\oint_{\mathcal{C}}dv^{l}
G_{il}(x-v)\oint_{\mathcal{C}}dy^{j}\oint_{\mathcal{C}}du^{k}G_{jk}(y-u)=
3\left(\oint_{\mathcal{C}}dx^{i}\oint_{\mathcal{C}}dy^{j}
G_{ij}(x-y)\right)^2. \ee Hence, it is enough to substitute exactly
half of the contours appearing in each pairing by the contour
$\mathcal{C}^{\prime}$. After that all terms of the perturbative
expansion assemble into the exponential of the linking number.
Indeed, all the odd order terms vanish, and each term of order $2k$
yields the sum over $(2k-1)!!$ pairings, each pairing yielding the
contribution $n^k$. In somewhat symbolic notations, that reads \be
\ckl \exp(W)\brr=\sum_{k=0}^{\infty}\hbar^k\frac{\ckl W^k \brr
}{k!}=\sum_{k=0}^{\infty}(2k-1)!!\frac{h^kn^k}{(2k)!}=\sum_{k=0}^{\infty}\frac{n^k}{2^kk!}=\exp\left(\frac{nh}{2}\right).
\label{abfr}\ee

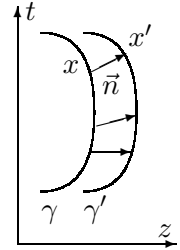
\begin{wrapfigure}{r}{80pt}
\begin{picture}(80,120)(0,-10)
\put(15,-10){\vector(0,1){90}}\put(15,-10){\vector(1,0){60}}
\qbezier(24,70)(44,70)(44,40)
\qbezier(44,40)(44,10)(24,10)
\qbezier(40,70)(60,70)(60,40)
\qbezier(60,40)(60,10)(40,10)
\put(43,25){\vector(1,0){15}}
\put(45,35){\vector(4,1){15}}
\put(42,55){\vector(2,1){14}}
\put(18,75){$\small{t}$}\put(68,-7){$\small{z}$}
\put(32,55){$\small{x}$}\put(57,65){$\small{x^{\prime}}$}
\put(47,47){$\small{\vec n}$}\put(24,0){$\small{\gamma}$}\put(40,0){$\small{\gamma^{\prime}}$}
\end{picture}
\caption{Framing of the knot from the non-abelian gauge theory
standpoint.\label{fig:fram}}
\end{wrapfigure}

\subsubsection{Framing in non-abelian theory.\label{sec:nafr}}
A similar approach is applied to the non-abelian theory, although
its realization is more involved then. First of all, the Wilson line
is now given by path ordered exponential (\ref{Wnab}), so that each
integral in perturbative expansion (\ref{Wnabperp}) is taken not
over the entire curve but from a selected origin point to the point
associated with the inner integration variable. Hence, a shifted
contour should be now introduced together with the matching rule of
its points to the points of the original contour
(fig.\ref{fig:fram}). Equivalently, one should associate each point
with of the original contour with a vector pointing out to the
corresponding point of the shifted contour. The resulting
construction is referred to as \emph{framing} of a knot. Although
evaluating of the framing contribution is a non-abelian theory is
not as simple as the above presented calculation for the abelian
case, the calculation can be carried out explicitly \cite{LabPer}.
The result reads that the framing dependence separates of the rest
perturbation series as a (\ref{abfr})-like factor. Hence, if the
integration contour contains several connection component, or is
artificially split into several parts, only the contribution to the
Wilson average with the integration variables running over the
different parts are non-trivial, while those with the variables
running over the same part, after being regularized, contribute to
the framing factor. By this reason, we concentrate on the well
defined cross contribution in the remaining of the section,
neglecting the singular contributions.

In sec.\ref{sec:Konts}, we explicitly evaluate a non-abelian analog
of the linking number, which has the same form as the framing
factor.

\subsection{Kontsevich integral for the Vassiliev invariants as the perturbative expansion for the Chern-Simons Wilson average in the holomorphic gauge\label{sec:Konts}}
In the present section, we finally address to the main part of our
discussion on relations between the knot invariants and the
perturbatively formulated TQFT. The up to day view on the
correspondence of the knot polynomials to the perturbative
Chern-Simons theory relies on the following two statements:
\begin{itemize}
\item{Kontsevich integral provides an integral representation for the HOMFLY polynomial.}
\item{Kontsevich integral may be term-wise related to the perturbative expansion for the Chern-Simons Wilson average in the holomorphic gauge.}
\end{itemize}
After providing some comments on the first statement, we discuss the
second statement in details in the remaining part of the section.

\subsubsection{HOMFLY polynomial as a generating function for the Vassiliev invariants\label{sec:HOMVass}}

As we already mentioned, an essential part of interpretation of the
knot polynomials in the context of the perturbative TQFT is the
integral presentation for the certain knot invariants. These
invariants are known in knot theory as \emph{Vassiliev invariants}
\cite{ChumDuzMost}, the corresponding presentation for them is
referred to as \emph{Kontsevich integral} \cite{Konz}.

The Kontsevich integral for a separate Vassiliev invariant already
has a rather specific structure, which we discuss below. But is much
more important for our purposes that the integrals for
\emph{various} Vassiliev invariants are assembled in a generating
function, $\hbar$ being the formal parameter. On the one hand, this
generating function possesses the properties of the perturbative
expansion for the Wilson average, as we discuss in what follows. On
the other hand, one may verify the HOMFLY polynomial, when expanded
in the parameter $\hbar$ related to the formal variables entering
the polynomial as $q=e^{2\pi\hbar}$, $A=e^{2\pi N\hbar}$, to
reproduce the same generating function. These facts can be
considered as a \emph{perturbative} correspondence between the
HOMFLY polynomials and the Chern-Simons Wilson average.

Moreover, the Kontsevich integral as entire series is known
decompose into the tensor contraction of the certain elementary
constituents \cite{ChumDuzMost, LabNew, DSS:Konz}, thus providing
one more state model representation for the HOMFLY polynomial (see
sec.\ref{sec:stmod}), which is of the same kind as the
$\mathcal{R}$-representation \cite{ReshTur, MorSm}, which we discuss
in details in sec.\ref{sec:Rmat}, and as the $WZW$-representation
\cite{Kaul,Inds1,Inds2,Inds3,KaulLeq,GMMM} developed on the base of \cite{Witt}. Examining how
these representations are related to each other hence illuminates
the interference of the associated with them QFT ideas
\cite{LabNew,MorSm,DSS:Konz,DSS:KZ}.

\subsubsection{Structure of the Kontsevich integral\label{sec:KonstStruct}}
In this section, we formulate the properties of the Kontsevich
integral essential for comparing it with the perturbative expansion
for the Chern-Simons Wilson average.

\begin{itemize}
\item{The entire Kontsevich integral is an infinite
 series of the integrals of the increasing (even) multiplicity.}
\item{Each multiple integral is a $t$-\emph{ordered} integral, i.e., the integration variables satisfy $t_1\le t_2\le\ldots \le t_k$.}
\item{Each multiple integral is multiplied on its own \emph{group factor}, which is a certain contraction of a Lie algebra generators.}
\item{All the integrals with a multiplicity $2g$, together with their group factors, are enumerated by all \emph{parings} of the $2g$ points.}
\item{A kernel of each multiple integral is a product of \emph{Green functions} of a certain differential equation.}
\item{The entire series can be presented as a tensor contraction of certain ``elementary constituents''. }
\end{itemize}
For example, the series can start from $1$, proceeding with the 6
double integrals of the form \be
\int_{a}^{b}dt\underbrace{\frac{\dot{z_1}(t)-\dot{z_2}(t)}{z_1(t)-z_2(t)}}_{<12>}+
\int_{a}^{b}dt\underbrace{\frac{\dot{z_1}(t)-\dot{z_3}(t)}{z_1(t)-z_3(t)}}_{<13>}+\ldots,
\label{Konts2}\ee  which are followed by the fourfold integrals
having the structure \be \footnotesize{\begin{array}{r}
\sum_{a,b}\Tr\int_{a}^{b}dt\int_{a}^tds\underbrace{
2\underbrace{\frac{\dot{z_1}(t)-\dot{z_2}(t)}{z_1(t)-z_2(t)}\cdot
\frac{\dot{z_1}(s)-\dot{z_2}(s)}{z_1(s)-z_2(s)}T_aT_bT_aT_b}_{<12><12>}}_{<1212>}+
\underbrace{\frac{\dot{z_1}(t)-\dot{z_2}(t)}{z_1(t)-z_2(t)}\cdot
\frac{\dot{z_1}(s)-\dot{z_3}(s)}{z_1(s)-z_3(s)}\Big(\underbrace{T_aT_bT_aT_b}_{(12)(13)}+\underbrace{T_aT_bT_bT_a}_{<13><12>}\Big)}_{<1123>}+\nn\\+
\underbrace{\underbrace{\frac{\dot{z_1}(t)-\dot{z_2}(t)}{z_1(t)-z_2(t)}\cdot
\frac{\dot{z_3}(s)-\dot{z_4}(s)}{z_3(s)-z_4(s)}T_aT_aT_bT_b}_{<12><34>}+
\underbrace{\frac{\dot{z_1}(t)-\dot{z_2}(t)}{z_1(t)-z_2(t)}\cdot
\frac{\dot{z_3}(s)-\dot{z_4}(s)}{z_3(s)-z_4(s)}T_aT_aT_bT_b}_{<13><24>}+
\underbrace{\frac{\dot{z_1}(t)-\dot{z_2}(t)}{z_1(t)-z_2(t)}\cdot
\frac{\dot{z_3}(s)-\dot{z_4}(s)}{z_3(s)-z_4(s)}T_aT_aT_bT_b}_{<14><23>}}_{<1234>}
+ \ldots,\end{array}\label{Konts4}}
\ee and so on. The structure of the series matches then the
expansion of the Gaussian average \be
\mathrm{P}\prod_{k=1}^4\left(1+\int
dzA^a(z_k)T_a+A^a(z_k)A^b(z_k)T_aT_b+\ldots \right),\label{WA4}\ee
double integrals (\ref{Konts2}) corresponding to the second order
terms, fourfold integrals (\ref{Konts4}) corresponding to the fourth
order terms, etc. We label each term (\ref{Konts2}) and
(\ref{Konts4}) by the corresponding average (e.g., $<1234>$ stands
for $<A(z_1)A(z_2)A_(z_3)A_(z_4)>$) and by the corresponding pairing
(e.g., $<12><34>$ stands for $<A(z_1)A(z_2)><A_(z_3)A_(z_4)>$). The
P sign in (\ref{WA4}) means that all the multiple integrals in the
expansion are ordered w.r.t. subscripts of the integration
variables, e.g., \be\mathrm{P}\int\int
dz_1dz_2\equiv\int_a^bdz_2\int_a^{z_2}dz_1.\ee One of the subtleties
arising here is that the expansion of (\ref{WA4}) reproduces
(\ref{Konts2},\ref{Konts4}), as well as the higher order terms in
the Kontsevich integral, only up to the singular pairings like
$<11><22>$ coming from the correlator $<1122>$. Yet ignoring of
these terms has a certain sense in the framework of the
\emph{framing procedure} \cite{ChumDuzMost, LabNew}, which is the
standard way of regularizing the Wilson average discussed in
sec.\ref{sec:CSfram}.

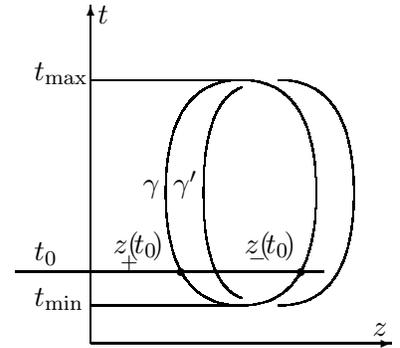
\begin{wrapfigure}{r}{145pt}
\begin{picture}(100,120)(-45,0)
\unitlength=0.5mm
\put(-10,0){\vector(0,1){90}}\put(-10,0){\vector(1,0){80}}
\put(-10,10){\line(1,0){40}}\put(-10,70){\line(1,0){40}}
\put(-30,19){\line(1,0){82}}\put(14,19){\circle*{2}}\put(46,19){\circle*{2}}
\qbezier(10,40)(10,70)(30,70)\qbezier(30,70)(50,70)(50,40)\qbezier(50,40)(50,10)(30,10)\qbezier(30,10)(10,10)(10,40)
\qbezier(40,70)(60,70)(60,40)\qbezier(60,40)(60,10)(40,10)
\qbezier(20,40)(20,64)(30,68)
\qbezier(20,40)(20,16)(30,12)
\put(-25,10){$\small{t_{\mathrm{min}}}$}\put(-25,70){$\small{t_{\mathrm{max}}}$}
\put(4,40){$\small{\gamma}$}\put(12,40){$\small{\gamma^{\prime}}$}
\put(-8,85){$\small{t}$}\put(65,2){$\small{z}$}
\put(-25,22){$\small{t}_0$}
\put(-4,24){$\small{z(\!t_0\!)}$}
\put(-2,20){$\scriptstyle{+}$}
\put(31,24){$\small{z(\!t_0\!)}$}
\put(32,21){$\scriptstyle{-}$}
\end{picture}
\caption{\label{fig:hollink}
The original ($\gamma$) and shifted ($\gamma^{\prime}$) integration contours in case of unknot with one minimum and one maximal points.}
\end{wrapfigure}

\subsubsection{Kernels of Vassiliev invariants as propagators of the complexificated Chern-Simons theory\label{sec:KontsKer}}
As a next step towards the Chern-Simons theory, one may observe that
all the kernels (we mean the denominators, relating the numerators
to the measure $\dot zdt=dz$) of integrals
(\ref{Konts2},\ref{Konts4}) are products of the Green functions of
the equation, which is the holomorphic condition (see Appendix
\ref{App:CSGreen} for details) \be \p_{\bar
z}\frac{\delta(t)}{z}=\delta(t)\delta(z)\delta(\bar z). \ee

The holomorphic condition, in turn, may arise as the classical
equations of motion
in the theory with the lagrangian \be \mathcal{L}=A_0(t,z,\bar
z)\p_{\bar z}A_{\bar z}(t,z,\bar z)-A_{\bar z}(t,z,\bar z)\p_{\bar
z}A_0(t,z,\bar z).\label{holac} \ee This the form, which the
Chern-Simons lagrangian takes in the holomorphic gauge
\cite{FrohKing, LabPer, LabNew}, as we discuss in
sec.\ref{sec:abgauge}.

The Lagrangian being quadratic, components of the Green function \be
G_{00}=G_{zz}=0,\ \ \ G_{0z}=-G_{z0}=\frac{\delta(t)}{z} \ee may be
considered as the Gaussian averages \be \brl A_zA_z\brr\equiv
G_{zz}=0,\ \ \brl A_0A_0\brr\equiv G_{00}=0,\ \ \brl
A_0A_z\brr=-\brl A_zA_0\brr\equiv G_{0z}. \label{holGreen}\ee

\subsubsection{Second order contribution to the Wilson average in the holomorphic gauge\label{sec:KontsWA}}
To complete our discussion on the relation of the Kontsevich
integral to the Chern-Simons Wilson average, we provide an
illustration of how evaluating of a perturbative contribution to the
Chern-Simons Wilson average can be reduced to evaluating integrals
of type (\ref{Konts2}) in the simplest case.

We evaluate explicitly the first non-vanishing contribution to the
Wilson average for the contour $\gamma$ placed as in
fig.\ref{fig:hollink}. Referring to sec.\ref{sec:CSfram}, we
regularize the divergent contributions to the average by introducing
the shifted contour $\gamma^{\prime}$, each point $x^{\prime}$
corresponding the point $x$ of the original contour, and moving one
point in each pairwise correlator to the shifted contour. For
instance, the second order contribution to the Wilson average then
equals \be \oint_{\gamma} dx^{\mu} \oint_{\gamma^{\prime}}
dx^{\prime\nu}\ckl A_{\mu}\left(t,z, \bar{z}\right)A_{\nu}\left(t,z,
\bar{z}\right)\brr\ee Then, using definition (\ref{Gav:Green}) of
the Gaussian average and explicit expression for the Green function
in the holomorphic gauge (\ref{holGreen}), the rewrite the
contribution as \be \oint_{\gamma} dz \oint_{\gamma^{\prime}}
dt^{\prime}\ckl
A_t(t,z,\bar{z})A_z(t^{\prime},z^{\prime},\bar{z}^{\prime})\brr+\oint_{\gamma}
dt \oint_{\gamma^{\prime}} dz^{\prime}\ckl
A_z(t,z,\bar{z})A_t(t^{\prime},z^{\prime},\bar{z}^{\prime})\brr=\nn\\=
\oint_{\gamma} \frac{dz}{z-z^{\prime}}+\oint_{\gamma^{\prime}}
\frac{dz^{\prime}}{z^{\prime}-z}.\ee The next step is in passing to
the integration over the selected axes. In the simplest case of the
contour having two critical points (as each contour in
fig.\ref{fig:hollink}), one should split the contour $\gamma$ into
two pieces, $\gamma_+$, $\gamma_-$, given by the explicit functions
$z=z_+(t)$ and $z=z_-(t)$, respectively, and one should split the
contour $\gamma^{\prime}$ in a similar way. The integral over $t$ is
composed then of four summands, \be
\int_0^1dt\frac{\dot{z}_+(t)-\dot{z^{\prime}}_+(t)}{z_+(t)-z^{\prime}_+(t)}+
\int_0^1dt\frac{\dot{z}_+(t)-\dot{z^{\prime}}_-(t)}{z_+(t)-z^{\prime}_-(t)}+
\int_1^0dt\frac{\dot{z}_-(t)-\dot{z^{\prime}}_-(t)}{z_-(t)-z^{\prime}_+(t)}+
\int_1^0dt\frac{\dot{z}_-(t)-\dot{z^{\prime}}_-(t)}{z_-(t)-z^{\prime}_-(t)}.
\ee To proceed with, we use contour independence of the original
integral to place the contours $\gamma$ and $\gamma^{\prime}$
parallel everywhere but the region, corresponding, e.g., to the
segments of contours pieces $\gamma_+^{\prime}$ and
$\gamma^{\prime}_-$ with $t_1\le t\le t_2$, where the contours
intertwine. The integral acquires the value only in the layer
$t_1\le t\le t_2$, because \be
\dot{z}_+(t)=\dot{z}_-(t)=\dot{z}_-^{\prime}(t)=\dot{z}_-^{\prime}(t),\
\ \ 0\le t\le t_1,\ \ t_2\le t\le 1, \ee so that the integrand
identically vanishes in all the remaining space. To continue, we
notice that the Green function in the holomorphic gauge is a
decreasing function of the argument absolute value, and we use the
contour independence once again to move the intertwined segments
$\gamma_+$ and $\gamma^{\prime}_+$ apart from the parallel segments
$\gamma_-$, $\gamma^{\prime}_-$ in the layer $t_1\le t\le t_2$ on
the infinitely increasing distance, \be
\left|z_+(t)-z^{\prime}_-(t)\right|\sim\left|z_-(t)-z^{\prime}_+(t)\right|\rightarrow\infty.
\ee The value of the integral must be independent of this distance,
hence, the terms with two variables running over the different
connection components vanish in this limit. Finally, we notice that
the non-intertwined pieces of contours can be placed vertically so
that they do not contribute to the integral as well, \be
\dot{z}_-(t)=\dot{z}^{\prime}_-(t)=0,\ \ \ t_1\le t\le t_2. \ee
Finally, for the integration contours placed as in
fig.\ref{fig:hollink}, the integral takes form \be
\mathcal{I}=\int_{t_1}^{t_2}dt\frac{\dot{z}_+(t)-\dot{z^{\prime}}_+(t)}{z_+(t)-z^{\prime}_+(t)},
\ee For instance, this example illustrates the Kontsevich integral
decomposing into the sum of the ``elementary contributions'' in the
considered simplest case. While these contributions are merely
summed up in the lowest perturbation order, the higher orders
reproduce their tensor contraction. This property of the Kontsevich
integral, well known in the Vassiliev invariants theory
\cite{ChumDuzMost}, can be derived as a property of the Chern-Simons
Wilson average \cite{LabNew}, with help of the properties of the
path exponential and of the Gaussian average discussed in
sec.\ref{sec:Pexp} and sec.\ref{sec:Gav}, respectively (if the
abelian gauge is selected, see sec.\ref{sec:abgauge}, and the
framing procedure is performed, see sec.\ref{sec:CSfram}).

We outline the decomposition of the Wilson average in the
holomorphic gauge into the elementary constituents of the Kontsevich
integral in Appendix \ref{App:eldec}, and we demonstrate the
explicit evaluation of the lowest order perturbative contributions
to the basic kinds of these constituents in Appendix \ref{App:elcalc}.

\section{Knot polynomial as an observable in a state model\label{sec:stmod}}
\subsection{Formulating the approach}
All the knot invariants enumerated in sec.\ref{sec:knmath} can be
introduced with help of the construction presented in L.Kauffman
paper \cite{Kauff}.

\begin{itemize}
\item{
A  knot or a link, a direction on each connection component
selected, corresponds to a directed four-valent planar graph with
two  kinds of vertices (fig.\ref{fig:invcr}), which is obtained by
projecting a knot on a plane, arrangement of lines in the
self-crossing w.r.t. the projection plane is kept. This graph is
called \emph{a knot (or link) diagram} (fig.\ref{fig:ltref}).}
\item{The two kinds of vertices on the knot diagram correspond to two kinds of four-script operators,
$S^{ij}_{kl}$ and $\tilde{S}^{ij}_{kl}$, respectively, superscripts
for incoming edges, subscripts for the out going once, and the left
pair for the upper line. Components of the operators commute, as
reads the first line of table \ref{tab:op} reads.}
\item{An entire knot diagram corresponds then to a contraction of
the operators along the edges (fig.\ref{fig:ltref}).}
\end{itemize}

\subsection{Definition of average}
As we discuss below, a knot invariant can not be constructed just as
an operator contraction. One more item is put then in the
construction. Namely, a knot polynomial is going to be an
\emph{average} of the operator product. One may keep in mind that
the operators now depend on additional parameters, averaging over
the parameters yielding a knot invariant. The case is indeed like
that in certain variants of the construction, for instance, in the
version we discuss in details in sec.\ref{sec:Rmat}. However, the
original variant \cite{Kauff} introduces the average just as a
formal operation, namely,
\begin{itemize}
\item{
An \emph{average} of the operator contraction is introduced as a
scalar quantity related to the contraction, the properties in
tab.\ref{tab:av} being satisfied.}
\end{itemize}

\be
\begin{array}{|rcl|c|}
\hline\\[-5mm]&&&\\
\ckl \hat{\mathcal{O}}_1(\lambda \hat{\mathcal{O}}_2+\mu
\hat{\mathcal{O}}_3)\hat{\mathcal{O}}_4\brr&=&
\lambda\ckl\hat{\mathcal{O}}_1\hat{\mathcal{O}}_2\hat{\mathcal{O}}_4\brr
+\mu\ckl
\hat{\mathcal{O}}_1\hat{\mathcal{O}}_3\hat{\mathcal{O}}_4\brr
&\mbox{Poly-linearity}\\[2mm]\hline\\[-8mm]&&&\\
\ckl
\Tr\hat{\mathcal{O}}_1\hat{\mathcal{O}}_2\hat{\mathcal{O}}_3\brr&=&\ckl
\Tr\hat{\mathcal{O}}_3\hat{\mathcal{O}}_1\hat{\mathcal{O}}_2\brr
&\begin{array}{cc}\mbox{Symmetry under}\\
\mbox{cyclic permutations}\end{array}\\[2mm]\hline\\[-8mm]&&&\\
\ckl\hat{\mathcal{O}}_1\Tr \hat{\mathcal{O}}_2\brr&=&\ckl
\hat{\mathcal{O}}_1\brr \Big<\Tr\mathds{1}\Big>\Tr
\hat{\mathcal{O}}_2
&\mbox{Decoupling of full contraction}\\[2mm]
\hline
\end{array}
\label{tab:av}\ee

\subsection{Operator identities}
Similarly, instead introducing the operators of the crossing points
explicitly, one requires that
\begin{itemize}
\item{The operators satisfy the identities listed from the second to the fourth lines in tab.\ref{tab:op} together with the corresponding deformations of
knot diagrams.}
\end{itemize}

\be
\begin{array}{|c|rcl|c|}
\hline\\[-5mm]&&&&\\
S^{ij}_{kl}S^{ab}_{cd}=S^{ab}_{cd}S^{ij}_{kl}&\multicolumn{3}{c|}{}&
\begin{array}{c}\mbox{Distant}\\\mbox{commutation}\end{array}\\[2mm]
\hline S^{ij}_{\alpha\beta}S^{\alpha
k}_{l\gamma}S^{\beta\gamma}_{mn}=S^{jk}_{\gamma\alpha}S^{\beta
i}_{n\beta}S^{\gamma\alpha}_{lm}
&\begin{picture}(40,30)(-7,7)\qbezier(0,23)(12,14)(29,2)
\qbezier(0,0)(5,3)(15,9)\qbezier(20,12)(25,15)(30,18)
\qbezier(5,-8)(5,-2)(5,0)\qbezier(5,6)(5,9)(5,16)\qbezier(5,22)(5,26)(5,28)
\put(27,16){\vector(3,2){1}} \put(27,3){\vector(3,-2){1}}
\put(5,12){\vector(0,1){1}}
\end{picture}
&=&
\begin{picture}(40,30)(-32,-13)\qbezier(0,-23)(-12,-14)(-29,-2)
\qbezier(0,0)(-5,-3)(-15,-9)\qbezier(-20,-12)(-25,-15)(-30,-18)
\qbezier(-5,8)(-5,2)(-5,0)\qbezier(-5,-6)(-5,-9)(-5,-16)\qbezier(-5,-22)(-5,-26)(-5,-28)
\put(-2,-1){\vector(3,2){1}} \put(-3,-21){\vector(3,-2){1}}
\put(-5,-10){\vector(0,1){1}}
\end{picture}&\mbox{RII}\\[8mm]
\hline \tilde
S^{ij}_{\alpha\beta}S^{\alpha\beta}_{kl}=S^{ij}_{\alpha\beta}\tilde{S}^{\alpha\beta}_{kl}=\delta^i_k\delta^j_l&
\begin{picture}(30,30)(-15,8)
\qbezier(-8,-8)(-5,-5)(-3,-3)
\qbezier(-3,23)(-5,25)(-8,28)
\qbezier(3,3)(10,10)(3,17)
\qbezier(8,-8)(5,-5)(0,0)
\qbezier(0,20)(5,25)(8,28)
\qbezier(0,0)(-10,10)(0,20)
\put(6,26){\vector(1,1){1}}\put(-6,26){\vector(-1,1){1}}
\put(-5,-5){\vector(1,1){1}}\put(5,-5){\vector(-1,1){1}}
\end{picture}&=&
\begin{picture}(30,30)(-10,-2)
\qbezier(-5,-15)(10,0)(-5,15)
\qbezier(15,-15)(0,0)(15,15)
\put(13,13){\vector(1,1){1}}\put(-3,13){\vector(-1,1){1}}
\put(-2,-12){\vector(1,1){1}}\put(12,-12){\vector(-1,1){1}}
\end{picture}
&\mbox{RIII}\\[8mm]
\hline \ckl S^{i\alpha}_{\alpha
j}\mathcal{O}^j_i\brr=\ckl\delta^{\alpha}_{\alpha}\brr\ckl\mathcal{O}^i_i\brr
&\begin{picture}(30,30)(-15,15)
\qbezier(-3,23)(-5,25)(-8,28)
\qbezier(-5,15)(5,25)(8,28) 
\qbezier(5,5)(10,10)(5,15)\qbezier(-5,5)(-10,10)(-5,15)\qbezier(-5,5)(0,0)(5,5)
\put(6,26){\vector(1,1){1}} \put(-5,25){\vector(1,-1){1}}
\end{picture}
&=&
\begin{picture}(30,10)(-15,-3)\qbezier(-15,0)(0,0)(15,0)\put(0,0){\vector(1,0){1}}\end{picture}&\mbox{RI}\\[8mm]
\hline\\[-8mm]&&&&\\
\begin{array}{rcl}
\left(S^{ij}_{\alpha\beta}-Aq\delta^{i}_{\alpha}\delta^{j}_{\beta}\right)
\left(S^{\alpha\beta}_{kl}-Aq^{-1}\delta^{\alpha}_{k}\delta^{\beta}_{l}\right)&=&0\\[2mm]
\Leftrightarrow\ \ \
A^{-1}S^{ij}_{kl}-A\tilde{S}^{ij}_{kl}&=&\left(q-q^{-1}\right)\delta^i_k\delta^j_l
\end{array}
& \multicolumn{3}{c|}{\begin{picture}(30,30)(-10,-2)
\qbezier(-5,-15)(5,0)(15,15)
\qbezier(15,-15)(11,-9)(7,-3)\qbezier(3,3)(-1,9)(-5,15)
\put(14,13){\vector(2,3){1}}\put(-4,13){\vector(-2,3){1}}
\end{picture}
-\begin{picture}(30,30)(-10,-2)
\qbezier(-5,-15)(-1,-9)(3,-3)\qbezier(7,3)(11,9)(15,15)
\qbezier(15,-15)(5,0)(-5,15)
\put(14,13){\vector(2,3){1}}\put(-4,13){\vector(-2,3){1}}
\end{picture}
= z\begin{picture}(30,30)(-10,-2)
\qbezier(-5,-15)(10,0)(-5,15)
\qbezier(15,-15)(0,0)(15,15)
\put(13,13){\vector(1,1){1}}\put(-3,13){\vector(-1,1){1}}
\put(-2,-12){\vector(1,1){1}}\put(12,-12){\vector(-1,1){1}}
\end{picture}}&\begin{array}{c}\mbox{Skein}\\\mbox{relations}\end{array}
\\[8mm]
\hline
\end{array}
\label{tab:op}\ee \begin{wrapfigure}{r}{120pt}
\begin{picture}(100,100)(-60,0)
\unitlength=0.5mm
\qbezier(-7,47)(-5,45)(-3,43)
\put(20,60){\vector(1,-1){1}}
\put(-30,50){\vector(-1,-1){1}}
\put(20,20){\vector(-1,-1){1}}
\qbezier(-10,30)(-20,20)(-30,30)\qbezier(-30,30)(-40,40)(-30,50)\qbezier(-30,50)(-20,60)(-13,53)
\qbezier(3,37)(10,30)(20,40)\qbezier(20,40)(30,50)(20,60)\qbezier(20,60)(10,70)(0,60)
\qbezier(0,60)(-20,40)(-13,33)\qbezier(-7,27)(10,10)(20,20)\qbezier(20,20)(30,30)(23,37)\qbezier(17,43)(10,50)(-10,30)
{\linethickness{0.4mm}\qbezier(3,57)(0,60)(-3,63)\qbezier(-19,40)(-15,40)(-11,40)\qbezier(-8,38)(-5,35)(-2,32)}
\put(-41,40){$i$}\put(-19,44){$b$}\put(0,29){$a$}\put(-3,47){$c$}
\put(10,48){$l$}\put(10,68){$k$}\put(10,7){$j$}
\end{picture}
\caption{$S^{ij}_{ab}S^{ak}_{lc} S^{bc}_{mi}S^{ml}_{kj}$
\label{fig:ltref}}
\end{wrapfigure}
According to the Reidemeister theorem, two closed curves can be
continuously transformed one into the other if and only if the
corresponding knot diagrams  are related by a sequence of three
elementary transformations listed in tab.\ref{tab:op} (lines 2-4),
which are referred to as \emph{Reidemeister moves}. Hence, the
corresponding constraints imposed on the operators guarantee
coinciding of averages associated with different diagrams
representing the same knot.

As it is reflected in the table, only two out of the three
Reidemeister moves (lines 2-3 of tab.\ref{tab:op}) can be imposed on
the operators themselves. The remaining condition (line 4 of
tab.\ref{tab:op}) turns out to be inconsistent with former two, when
being treated as an operator identity, and it can hold only under
the average sign \cite{Pras}.

\bigskip

\begin{center}
*\ *\ *
\end{center}
The constraints on the operators and on the averages enumerated
above turn out to be very restrictive, yet insufficient to calculate
the knot invariant. One should couple these constraints with some
additional data, which varies for different versions of the
construction.

\subsection{Jones polynomial for the trefoil knot as an average in the state model}
Outlined the general construction, we proceed now with a more
detailed presentation, relying on the explicit calculation of the
knot polynomial for a particular knot diagram with help of several
variants of the above construction.

\subsubsection{Ward identities\label{sec:stskein}}
The first way to compete the construction enables to determine plain
(uncolored) HOMFLY polynomials, as well as Alexander and Jones
polynomials, which are obtained from the HOMFLY polynomial one by
setting $A=1$ and $A=q^2$, respectively, and, with some
complications, for the plain Kauffman polynomial \cite{Kauff}.
Following this way, one should additionally require that
\begin{itemize}
\item{The operators satisfy the eigenvalue equation (line 5 of tab.\ref{tab:op}).}
\end{itemize}
Taking into account the identity in line 2 of tab.\ref{tab:op}, one
can rewrite the eigenvalue equation as a linear relation between the
direct and inverse crossing operators (also given in line 5 of
tab.\ref{tab:op}), which can be presented graphically (second column
of line 5 of tab.\ref{tab:op}).  The relations between the averages
that follow from this form of the equation are known as \emph{skein
relations} for knot polynomials \cite{Pras}. The resulting system of
constraints, together with the properties of the average listed in
table \ref{tab:av}, then enables to evaluate a knot polynomial for
arbitrary knot diagram. E.g., for knot diagram in
fig.\ref{fig:ltref}, which is a diagram of the trefoil knot, one has
\be \ckl
S^{\boxed{\scriptstyle{ij}}}_{ab}S^{a\boxed{\scriptstyle{k}}}_{\boxed{\scriptstyle{l}}c}
S^{bc}_{\boxed{\scriptstyle{mi}}}S^{ml}_{kj}\brr= \ckl
S^{\boxed{\scriptstyle{jk}}}_{bc}S^{c\boxed{\scriptstyle{i}}}_{\boxed{\scriptstyle{i}}a}
S^{ab}_{\boxed{\scriptstyle{\underline{lm}}}}S^{\underline{ml}}_{kj}\brr=
z\ckl S^{\underline{jk}}_{bc}S^{ci}_{ia}S^{ab}_{\underline{kj}}\brr+
\ckl S^{jk}_{kc}S^{ci}_{ij}\brr=\nn\\
=z^2\ckl S^{c\underline{i}}_{\underline{i}a}S^{ab}_{bc}\brr+
z\ckl\delta^b_b\brr\ckl S^{a\underline{i}}_{\underline{i}a}\brr+\ckl
S^{jk}_{kc}S^{c\underline{i}}_{\underline{i}j}\brr =z^2q\ckl
S^{a\underline{b}}_{\underline{b}a}\brr+ zq\big<\delta^b_b\big>\ckl
\delta^a_a\brr+q\ckl S^{jk}_{kj}S^{ci}_{ij}\brr= \nn\\=z^2q^2\ckl
\delta^a_a\brr+ zq\ckl\delta^a_a\brr^2+q^2\ckl
\delta^a_a\brr.\label{ltref1} \ee The above exercise is the standard
calculation of a HOMFLY polynomial with help of the skein relations,
which is usually represented graphically \cite{Pras}, but we
intentionally keep the operator language, which is the main subject
of the present text.

\bigskip

\begin{center}
*\ *\ *
\end{center}
According to the corresponding theorem \cite{Pras}, an average
corresponding to an arbitrary knot diagram can be calculated this
way. Moreover, one obtains one and the same quantity for all ways of
evaluating the average for a given knot diagram, as well as for all
diagrams representing the same knot\footnote{However, we omit here
certain subtleties; see sec.\ref{sec:Rfram} for details.}. This
approach thus provides a tool for calculating  HOMFLY polynomials
of, in principle, arbitrary knots, and this tool turns out to be
highly effective; for instance, the numerous HOMFLY polynomial
presented in \cite{katlas,indknot,indlink} were calculated in this
way.

However, the announced in \cite{Kauff} \emph{state model} for the
knot polynomial is not fully presented yet. Apart from that, this
approach can not be straightforwardly extended to the case of
\emph{colored} HOMFLY polynomials, which cause much more interest
today. These two points make one to search for a more explicit
variants of the construction.

\subsubsection{Explicit form of the crossing
operators\label{sec:stKauff}} An alternative way is to find an
explicit expression for the crossing operator $S$, which would
provide the operator satisfying the topological invariance
constraints. There is a proper expression of a very simple form in
the particular case $A=q^2$ (when the average, generally being equal
to the HOMFLY polynomial, reduces to the Jones polynomial), namely,
\be
S^{ij}_{kl}=q^{-1}\delta^i_l\delta^j_k-q^{-2}\epsilon^{ij}\epsilon_{kl},\label{Kauffmat}
\ee and \be \ckl\mathrm{Tr}\mathds{1}\brr\equiv
q+q^{-1}.\label{Kauftr} \ee Expression (\ref{Kauffmat}) is referred
to as \emph{Kauffman matrix} \cite{KauffTB}. Form of the operator
(\ref{Kauffmat}) guarantees satisfying of the constraints in lines
1-3 and 5 of tab.\ref{tab:op} (although the last one in not used
explicitly in this variant of the approach). Unlike that, the
constraint in line 4 of tab.\ref{tab:op}, does \emph{not} holds for
(\ref{Kauffmat}), being valid only under the trace sign (where an
arbitrary operator $\mathcal{O}^l_i$ can be inserted as well),
\be \ckl S^{ij}_{jl}\brr
=a\ckl\delta^i_l\brr\ckl\delta^j_j\brr-b\ckl\epsilon^{ij}\epsilon_{jl}\brr
=\left(a\ckl\delta^j_j\brr-b\right)\Tr\ckl\mathds{1}\brr=
\ckl\delta^{i}_{l}\brr\ \ \Rightarrow\ \ \
\Tr\ckl\mathds{1}\brr=\frac{1+b}{a},\label{Kauffmat} \ee we denote
the coefficients in (\ref{Kauffmat}) by $a\equiv q^{-1}$ and
$b\equiv -q^{-2}$ to better demonstrate the structure of the answer.
We see that, generally, the average is invariant under the R1
transformation only up to a factor, which equals the unity provided
that the trace of the averaged unity operator takes a particular
value, namely \be \Tr\ckl\mathds{1}\brr=q+q^{-1}.\label{unav} \ee
Since expression (\ref{Kauffmat}) reduces a general full contraction
of the operators to a polynomial in $\Tr\ckl\mathds{1}\brr$, it is
necessary and sufficient to fix the value of this quantity to
calculate an arbitrary average. In particular, on can take thus
value to be (\ref{unav}), so that the RI invariants holds.

The operators and the average sign being defined by (\ref{Kauffmat})
and (\ref{unav}), respectively, the Jones polynomial for knot
diagram \ref{fig:ltref} is calculated as follows: \be &\ckl
S^{ij}_{ab}S^{a
k}_{lc}S^{bc}_{mi}S^{ml}_{kj}\brr=a^4\ckl(\delta^i_b\delta^j_a)(\delta^a_c
\delta^k_l)(\delta^b_i\delta^c_m)(\delta^m_j\delta^l_k)\brr+b^4\ckl(\epsilon^{a
k}\epsilon_{lc})(\epsilon^{bc}\epsilon_{mi})(\epsilon^{ml}\epsilon_{kj})(\epsilon^{ij}\epsilon_{ab})\brr+\nn\\&
+a^3b\Big<(\delta^i_b\delta^j_a)(\delta^a_c
\delta^k_l)(\delta^b_i\delta^c_m)(\epsilon^{ml}\epsilon_{kj})+
(\delta^i_b\delta^j_a)(\delta^a_c
\delta^k_l)(\epsilon^{bc}\epsilon_{mi})(\delta^m_l\delta^j_k)+\nn\\&+
(\delta^i_b\delta^j_a)(\epsilon^{a
k}\epsilon_{lc})(\delta^b_i\delta^c_m)(\delta^m_l\delta^j_k)+
(\epsilon^{ij}\epsilon_{ab})(\delta^a_c
\delta^k_l)(\delta^b_i\delta^c_m)(\delta^m_j\delta^l_k)\Big>+\nn\\&
+ab^3\Big<(\delta^i_b\delta^j_a)(\epsilon^{a
k}\epsilon_{lc})(\epsilon^{bc}\epsilon_{mi})(\epsilon^{ml}\epsilon_{kj})+
(\epsilon^{ij}\epsilon_{ab})(\delta^a_c
\delta^k_l)(\epsilon^{bc}\epsilon_{mi})(\epsilon^{ml}\epsilon_{kj})+\nn\\&+
(\epsilon^{ij}\epsilon_{ab})(\epsilon^{a
k}\epsilon_{lc})(\delta^b_i\delta^c_m)(\epsilon^{ml}\epsilon_{kj})+
(\epsilon^{ij}\epsilon_{ab})(\epsilon^{a
k}\epsilon_{lc})(\epsilon^{bc}\epsilon_{mi})(\delta^m_j\delta^l_k)\Big>+\nn\\&
+a^2b^2\Big<(\delta^i_b\delta^j_a)(\delta^a_c
\delta^k_l)(\epsilon^{bc}\epsilon_{mi})(\epsilon^{ml}\epsilon_{kj})+
(\delta^i_b\delta^j_a)(\epsilon^{a
k}\epsilon_{lc})(\delta^b_i\delta^c_m)(\epsilon^{ml}\epsilon_{kj})+
(\delta^i_b\delta^j_a)(\epsilon^{a
k}\epsilon_{lc})(\epsilon^{bc}\epsilon_{mi})(\delta^m_j\delta^l_k)+\nn\\&+
(\epsilon^{ij}\epsilon_{ab})(\delta^a_c
\delta^k_l)(\delta^b_i\delta^c_m)(\epsilon^{ml}\epsilon_{kj})+
(\epsilon^{ij}\epsilon_{ab})(\delta^a_c
\delta^k_l)(\epsilon^{bc}\epsilon_{mi})(\delta^m_j\delta^l_k)+
(\epsilon^{ij}\epsilon_{ab})(\epsilon^{a k}\epsilon_{lc})(\delta^b_i\delta^c_m)(\delta^m_j\delta^l_k)\Big>+\nn\\
&=a^4\Tr\ckl\mathds{1}\brr^3+4a^3b\Tr\ckl\mathds{1}\brr^2+a^2b^2\left(2\Tr\ckl\mathds{1}\brr^3+4\ckl
\Tr\mathds{1}\brr\right)+
4ab^3\ckl \Tr\mathds{1}\brr^2+b^4\ckl \Tr\mathds{1}\brr\nn\\
&=\left(q+q^{-1}\right)\left(q^{-2}+q^{-6}-q^{-8}\right)=q^{-1}+q^{-3}+q^{-5}-q^{-9}.\label{ltref2}\ee
The answer coincides with (\ref{ltref1}) for $A=q^2$, as it should.

\subsubsection{Turn-over operators instead of average\label{sec:sttov}}
One more way to complete the construction is to get read of
averaging procedure at all, thus obtaining a fully explicit
presentation of the knot polynomial. This is indeed possible, at
least in many particular cases, provided that one inserts certain
additional operators in some edges of the knot diagram. The
corresponding procedure relies on a notion of \emph{cycle}, which is
by definition a closed path on a directed graph that can be passed
along the directions on edges. We call a cycle a \emph{simple cycle}
if each edge and each vertex is passed no more that once. The rule
reads then:
\begin{itemize}
\item{Each simple cycle on the knot diagram must contain exactly one additional operator.}
\end{itemize}
We recall that a simple cycle on a directed graph is a closed
directed path passing through no edge or vertex for twice (or for
more times). E.g., for the above considered knot diagram the proper
operator contraction takes form \be \ckl
S^{ij}_{ab}S^{ak}_{lc}S^{bc}_{mi}S^{ml}_{kj}\brr\equiv
S^{ij}_{ab}S^{\alpha k}_{lc}S^{\beta c}_{mi}S^{ml}_{\kappa
j}\mathcal{M}^a_{\alpha}\mathcal{M}^b_{\beta}\mathcal{M}^{\kappa}_k.
\label{ltref3}\ee In the case relevant to the Jones polynomial, the
new operator $\mathcal{M}$ takes the explicit form \be
\mathcal{M}=\left(\begin{array}{cc}q&0\\0&q^{-1}\end{array}\right).
\ee Each averaged unity in the calculation of the previous section
is substituted then with the trace of the operator $\mathcal{M}$,
\be \ckl \delta_a^a\brr\ \ \longrightarrow\ \ \ \mathcal{M}^a_a, \ee
the result literally reproducing (\ref{ltref2}).

\subsubsection{Explicit definition of average
\label{sec:stbr}}
The next variant of the approach is already very close to the
variant of our main interest. It relies on the explicit definition
of the averaging procedure. Namely, one should first substitute the
crossing operators with their projections $S_r$ on certain
subspaces, defined with help of the corresponding projectors
$\mathcal{P}_r$ as \be S_r\equiv\mathcal{P}_rS\mathcal{P}_r, \ee and
then sum the contractions over these subspaces with certain weights
$w_r$, \be \ckl S\ldots S\brr\equiv_r\sum w_r\ckl S_r\ldots S_r\brr,
\ee To carry out this procedure explicitly, it is useful first to
rewrite the contraction of the crossing operators as the trace of a
matrix product. For instance, contraction (\ref{fig:ltref}) can be
rewritten as, \be
S^{\underline{sj}}_{\boxed{\scriptstyle{ab}}}\delta^{\underline{t}}_{\boxed{\scriptstyle{\kappa}}}\cdot
S^{\boxed{\scriptstyle{\alpha
k}}}_{\overline{lc}}\delta^{\boxed{\scriptstyle{\beta}}}_{\overline{u}}\cdot
S^{\overline{uc}}_{\underline{\underline{mi}}}\delta^{\overline{l}}_{\underline{\underline{v}}}\cdot
S^{\underline{\underline{mv}}}_{\underline{tj}}\delta^{\underline{\underline{i}}}_{\underline{s}}
=S^{(2)\beta\alpha
k}_{ucl}S^{(1)ucl}_{imv}S^{(2)imv}_{sjt}S^{(1)sjt}_{ba\kappa}=\Tr\left\{S^{(2)}S^{(1)}S^{(2)}S^{(1)}\right\}\ee
where each of the matrices $S^{(i)I}_L$, $i=1,2$, has the two
multi-indices corresponding two the triples of the sub- and
superscripts as \be S^{(1)I}_L\equiv
S^{(1)ijk}_{\underline{lm}n}\equiv
S^{ij}_{\underline{ml}}\delta^k_n,\ \ S^{(2)I}_L\equiv
S^{(2)ijk}_{l\underline{mn}}\equiv
\delta^i_lS^{jk}_{\underline{nm}}. \ee We emphasize that reducing of
the tensor contraction to the trace of the matrix product requires
for introducing the two operators instead of one. The completed
operation can be presented graphically, as redrawing initial knot
diagram (fig.\ref{fig:ltref}) as of a braid closure. In the case relevant to the Jones
polynomial, explicit expressions (\ref{Kauffmat}) for the operators
$S$ yield the corresponding expressions for the operators $S^{(1)}$
and $S^{(2)}$, \be
S^{(1)ijk}_{lmn}=q^{-1}\delta^i_l\delta^j_m\delta^k_n-q^{-2}\epsilon^{ij}\epsilon_{ml}\delta^k_n,\
\ \ \
S^{(2)ijk}_{lmn}=q^{-1}\delta^i_l\delta^j_m\delta^k_n-q^{-2}\delta^i_l\epsilon^{jk}\epsilon_{nm},
\ee the matrices of the operators being, respectively, \be
q^2S^{(1)}=\begin{array}{||cccccccc||r}
\multicolumn{1}{c}{111}&112&121&122&211&212&221&\multicolumn{1}{c}{222}&ijk/lnm\\[2mm]
q&&&&&&&&111\\&q&&&&&&&112\\&&q+1&&-1&&&&121\\&&&q+1&&-1&&&122\\
&&-1&&q+1&&&&211\\&&&-1&&q+1&&&212\\&&&&&&q&&221\\&&&&&&&q&222
\end{array},
\ee and \be q^2S^{(2)}=\begin{array}{||cccccccc||r}
\multicolumn{1}{c}{111}&112&121&122&211&212&221&\multicolumn{1}{c}{222}&ijk/lnm\\[2mm]
q&&&&&&&&111\\&q+1&-1&&&&&&112\\&-1&q+1&&&&&&121\\&&&q&&&&&122\\
&&&&q&&&&211\\&&&&&q+1&-1&&212\\&&&&&-1&q+1&&221\\&&&&&&&q&222
\end{array}.
\ee The averaging procedure is defined then with help of an
additional operator $Q$, as a sum over the $Q$ eigenvalues of the
matrix products projected on the related to each eigenvalue
subspaces, with the eigenvalues as the weights, \be
\ckl\Tr\left\{S^{(2)}S^{(1)}S^{(2)}S^{(1)}\right\}\brr\equiv
\sum_{\substack{\lambda=q^3,q,\\q^{-1},q^{-3}}}\lambda\,
\Tr\left\{S_{\lambda}^{(2)}S_{\lambda}^{(1)}S_{\lambda}^{(2)}S_{\lambda}^{(1)}\right\},
\ \ S^{(i)}\equiv\mathcal{P}_{\lambda}S^{(i)}\mathcal{P}_{\lambda},\
\ i=1,2.\label{ltref4} \ee Of course, the thus defined averaging
procedure is equivalent to just inserting the operator $Q$ under the
trace sign, \be
\ckl\Tr\left\{S^{(2)}S^{(1)}S^{(2)}S^{(1)}\right\}\brr\equiv
\Tr\left\{\mathcal{Q}S^{(2)}S^{(1)}S^{(2)}S^{(1)}\right\}.\label{ltref4v1}\ee,
However, the basic quantities in some variants of the approach, in
particular, in the variant we discuss in details in
sec.\ref{sec:braids}, are the $Q$ eigenvalues and the projected
matrices $\mathcal{S}_{\lambda}$, rather than the operators
themselves. Hence, it is expression (\ref{ltref4}), and not
(\ref{ltref4v1}), which is taken as a definition of the average
then.

In the selected basis, the operator $\mathcal{Q}$ and the
corresponding projectors are given by the diagonal matrices \be
\begin{array}{ccrcccccccl}
&&111&112&121&122&211&212&221&222\\
\mathcal{Q}&=&\mathrm{diag}\Big(q^3&q&q&q&q^{-1}&q^{-1}&q^{-1}&q^{-3}\Big)\\
\mathcal{P}_1&=&\mathrm{diag}\Big(1&0&0&0&0&0&0&0\Big)\\
\mathcal{P}_2&=&\mathrm{diag}\Big(0&1&1&1&0&0&0&0\Big)\\
\mathcal{P}_3&=&\mathrm{diag}\Big(0&0&0&0&1&1&1&0\Big)\\
\mathcal{P}_4&=&\mathrm{diag}\Big(0&0&0&0&0&0&0&1\Big)\\
\end{array}.
\ee One can explicitly verify that the operator $\mathcal{Q}$ is
related to the operators $\mathcal{M}$ from the previous section as
\be
\mathcal{Q}^I_L\equiv\mathcal{Q}^{ijk}_{lmn}=\mathcal{M}^i_l\mathcal{M}^j_m\mathcal{M}^k_n,
\ee so that the trace in (\ref{ltref4}) coincides with contraction
(\ref{ltref3}).

\subsubsection{Character expansion\label{sec:stchar}}
The subspaces and weights entering the definition of the average can
not be selected just arbitrarily, because they should provide the
invariance of the resulting quantity under RI (see
sec.\ref{sec:braids} for details). However, there are many
equivalent ways to present the subspaces and weights, which
correspond to rewriting (\ref{ltref4}) various bases. In particular,
the approach variant we are mostly interested in relies on the
choice of the subspaces and of the corresponding weights, which
leads to presenting the average as the linear combination \be
\ckl\Tr\left\{S^{(2)}S^{(1)}S^{(2)}S^{(1)}\right\}\brr\equiv
\sum_{r=s,a}[\mathrm{dim}]_q^r\Tr\left\{S^{(2)}_rS^{(1)}_rS^{(2)}_rS^{(1)}_r\right\}
=\Tr\left\{\mathcal{Q}S^{(2)}S^{(1)}S^{(2)}S^{(1)}\right\},\label{ltref5}\\
\ S^{(i)}_r\equiv \mathcal{P}_rS^{(i)}\mathcal{P}_r,\ \ r=s,f,\
i=1,2, \nn\ee where the quantities $[\mathrm{dim}]_q^f$  and
$[\mathrm{dim}]_q^s$ are \emph{quantum dimensions} \cite{KlimSch} of
the $SU(2)$ representations obtained by the group acting on vectors
and rank three permutation tensors, respectively \cite{JonesBr}. We
postpone the precise definitions of these quantities for the next
section, just presenting here the explicit form of the corresponding
decomposition for the above example. Expression (\ref{ltref5}) can
be derived from expression (\ref{ltref4v1}), with help of the fact
that both the matrices $S^{(1)}$ and $S^{(2)}$ are block-diagonal in
some distinguished basis, \be S^{(i)}_{8\times 8}
=\left(\begin{array}{ccc} S^{(i)s}_{4\times 4}\\&S^{(i)a}_{2\times
2}\\&&S^{(i)a}_{2\times 2}
\end{array}\right),\ \ \ i=1,2,
\ee the matrix $\mathcal{Q}$  being just diagonal in the same basis
and having the explicit form \be
\mathcal{Q}=\mathrm{diag}\Big(\underbrace{q^3\ q\ q^{-1}\ q^{-3}}_s\
\underbrace{q\ q}_a\ \underbrace{q^{-1}\ q^{-1}}_a\Big). \ee Trace
(\ref{ltref4v1}) decomposes then into sum (\ref{ltref5}), with
traces of the $\mathcal{Q}$ operator over the corresponding
subspaces as the weights,
\be[\mathrm{dim}]_q^s=q^3+q+q^{-1}+q^{-3},\ \ \ \
[\mathrm{dim}]_q^f=q+q^{-1}.\label{qdims}\ee For $q=1$ these weights
reproduce the dimensions   $[\mathrm{dim}]_q^f(q=1)=2$ and
$[\mathrm{dim}]_q^s(q=1)=4$ of the fundamental $SU(2)$
representation and of the space of the rank three permutation tensors
on it, respectively.

The explicit form of the constituent blocks is \be
S^{(1)}_s=S^{(1)}_a=q^{-1}\mathds{1}_{4\times 4},\ \
S^{(1)}_a=\left(\begin{array}{cc}q^{-1}\\&-q^{-3}\end{array}\right),\
\ \ \
S^{(2)}_a=\left(\begin{array}{cccc}-\frac{1}{q^3+q^5}&\frac{\sqrt{1+q^2+q^4}}{q^2+q^4}
\\\frac{\sqrt{1+q^2+q^4}}{q^2+q^4}&\frac{q}{1+q^2}\end{array}\right),\label{Sblocks}
\ee We do not give the precise definition of the corresponding
subspaces here (see sec.\ref{sec:Rbr} for it).

\bigskip

Substituting explicit formulas (\ref{Sblocks}) and (\ref{qdims}) for
blocks and quantum dimensions in expression for the average
(\ref{ltref5}), and using the identities $\Tr A^2=\sum A_{ij}A_{ji}$
and \be
\left(\frac{-q^{-1}}{q^3+q^5}\right)^2+\left(\frac{-q}{q^3(1+q^2)}\right)^2-2q^{-1}(-q^{-3})\frac{1+q^2+q^4}{\left(q^2+q^4\right)^2}=-q^{-8},
\ee one finally obtains the value of the average \be
\ckl\Tr\left\{S^{(2)}S^{(1)}S^{(2)}S^{(1)}\right\}\brr=q^{-4}[\mathrm{dim}]_q^s-q^{-8}[\mathrm{dim}]_q^f
=q^{-4}(q^3+q+q^{-1}+q^{-3})-q^{-8}(q+q^{-1})=\nn\\
=q^{-1}+q^{-3}+q^{-5}-q^{-9}, \ee which reproduces the Jones
polynomial for the trefoil knot once again.

\bigskip

Unfortunately, such simple expressions for the crossing operators
and averages of the unity (or for the edge operators) are
unavailable already in case of HOMFLY polynomials, nothing to say
about the colored knot polynomials. A generalization of the above
outlined approach is yet possible, and we describe it in the next
section.

\bigskip

The last variant of the approach turns out to be very effective as a
tool of calculating HOMFLY polynomials, both plain and colored. For
instance, the computational technology developed in \cite{MMM1,MMM2,MMM3,IMMM1,IMMM2,IMMM3,AMMM1,AMMM2,AMMM3,AM1,AM2}
relies just on this variant. What is even more important, expression
(\ref{ltref5}) gives rise to a new interesting quantity, which is
obtained by substituting the quantum dimensions $[\mathrm{dim}]_q^f$
and $[\mathrm{dim}]_q^s$ of the $SU(2)$ representations by
\emph{characters} of these representations, $\chi_f=t_1$ and
$\chi_s=\frac{1}{2}\left(t_1^2+t_2\right)$, respectively.
Substituting the special values $t_1=q+q^{-1}$ and $t_2=q^2+q^{-2}$
for the formal variables, obtains the quantum dimensions again. The
new quantity is referred to as \emph{extended} HOMFLY polynomial
\cite{MMM1}. Such quantity is no longer a \emph{knot} invariant,
being a \emph{braid} invariant instead (see sec.\ref{sec:braids} for
details). The extended HOMFLY polynomial is expected to arise
naturally in the context of the matrix models \cite{AlMMM} and integrable
hierarchies \cite{MMM1}, where a character decomposition for an
average is widely used technic.

A drawback of the above representation for the average is that it
relies on representing a knot as the closure of a braid. However,
there is a vast amount of knots with the braid representations being
much more involved that other representations of the same knots (the
most common example is given by the \emph{twist} knots
\cite{katlas}). Such knots can be effectively treated with by means
of different variants of the operator contraction approach
\cite{Kaul,Inds1,Inds2,Inds3,KaulLeq,GMMM}. Apart form that, such an interesting knot invariant as
a \emph{Khovanov} polynomial \cite{Khov,BarNat} is also introduced with
help of the same operators related to the crossings on the knot
diagram, but the \emph{homologies} of the operators matter in this
case. Expressions for averages like (\ref{ltref4},\ref{ltref5}) turn
out to be insufficient then, explicit expressions for the very
operators like (\ref{Kauffmat}) being needed instead.

\section{Knot polynomial as an averaged trace of a braid group element\label{sec:braids}}

\be
\small{\begin{array}{|c|c|c|c|} \hline\mbox{Braid group}&\mbox{Symmetric group}&\mbox{Hecke algebra}&\mbox{Colored Hecke algebra}\\
\hline
b_ib_{i+1}b_i=b_{i+1}b_ib_{i+1}&\sigma_i\sigma_{i+1}\sigma_i=\sigma_{i+1}\sigma_i\sigma_{i+1}&
h_ih_{i+1}h_i=h_{i+1}h_ih_{i+1}&H_iH_{i+1}H_i=H_{i+1}H_iH_{i+1}\\[2mm]
b_ib_j=b_jb_i,\
\scriptstyle{|i-j|\ne1}&\sigma_i\sigma_j=\sigma_j\sigma_i,\
\scriptstyle{|i-j|\ne1}&h_ih_j=h_jh_i,\
\scriptstyle{|i-j|\ne1}&H_iH_j=H_jH_i,\ \scriptstyle{|i-j|\ne1}\\[4mm]
&\sigma_i^2=\mathds{1}&h_i^2=\mathds{1}+(q-q^{-1})h_i=0&H_m=\sum_{k=0}^{m-1}a_kH_i^k\\
&\Updownarrow&\Updownarrow&\Updownarrow\\
&(\sigma_i-1)(\sigma_i+1)=0&(h_i-q)(h_i+q^{-1})=0&
\prod_{k=1}^m(H_k-\lambda_k)=0\\[1mm]\hline
\end{array}}\label{tab:symmbr}\ee

In this section, we discuss one more approach to the knot polynomials. Namely, the knot polynomials that are obtained in the framework of the state model approach \cite{Kauff} can be alternatively presented with help of matrix representations of the braid group \cite{JonesBr}. This is not a coincidence, but rather a demonstration of a deep relation between the two approaches. This relation becomes clear from the standpoint of the $\mathcal{R}$-matrix approach, which we address to in sec.\ref{sec:Rmat}, and which is in fact a version of the both approaches at the same time. Hence, the present section, together with the preceding one, should be considered as two preliminary parts of the subject presented in the following section.

\subsection{Properties of braids and braid closures\label{sec:brgen}}
The below discussed representation for knot polynomials relies on representing a knot as a braid closure. Hence, we start from recalling the basic properties of such representation. Throughout the section, we put attention of a relation between the braid group and the permutation group, this relation implying that the discuss knot invariants may naturally arise as the observable quantities in any physical model, which possess the permutation symmetry.

\subsubsection{The braid group an extension of the permutation group\label{sec:brsymm}}
In the present section, we recall the definitions of the braid and permutation groups, and briefly review their properties essential for the following presentation. We also discuss a permutation group extension known as \emph{Hecke algebra}, on which the below procedure of constructing a knot invariant essentially relies on.
\be
\begin{array}{rclrclrcl}
\multicolumn{9}{c}{\mbox{Equivalence transformations of braids}}\\[1mm]
\multicolumn{3}{c}{\mbox{I}}&\multicolumn{3}{c}{\mbox{II}}&\multicolumn{3}{c}{\mbox{III}}\\
\begin{picture}(48,84)(-12,-36)
\qbezier(0,48)(0,42)(0,36)
\qbezier(0,36)(3,33)(6,30)\qbezier(6,30)(9,27)(12,24)
\qbezier(12,24)(12,18)(12,12)
\qbezier(12,12)(15,9)(18,6)\qbezier(18,6)(21,3)(24,0)
\qbezier(24,0)(24,-6)(24,-12) \qbezier(24,-12)(24,-18)(24,-24)
\qbezier(24,-24)(24,-30)(24,-36)
\qbezier(12,48)(12,42)(12,36)
\qbezier(12,36)(9,33)(8,32)\qbezier(4,28)(3,27)(0,24)
\qbezier(0,24)(0,18)(0,12) \qbezier(0,12)(0,6)(0,0)
\qbezier(0,0)(0,-6)(0,-12)
\qbezier(0,-12)(3,-15)(6,-18)\qbezier(6,-18)(9,-21)(12,-24)
\qbezier(12,-24)(12,-30)(12,-36)
\qbezier(24,48)(24,42)(24,36) \qbezier(24,36)(24,30)(24,24)
\qbezier(24,24)(24,18)(24,12)
\qbezier(24,12)(21,9)(20,8)\qbezier(16,4)(15,3)(12,0)
\qbezier(12,0)(12,-6)(12,-12)
\qbezier(12,-12)(9,-15)(8,-16)\qbezier(4,-20)(3,-21)(0,-24)
\qbezier(0,-24)(0,-30)(0,-36)
\qbezier(0,39)(0,42)(1,45)\qbezier(0,39)(0,42)(-1,45)
\qbezier(12,39)(12,42)(13,45)\qbezier(12,39)(12,42)(11,45)
\qbezier(24,39)(24,42)(25,45)\qbezier(24,39)(24,42)(23,45)
\qbezier(0,-33)(0,-30)(1,-27)\qbezier(0,-33)(0,-30)(-1,-27)
\qbezier(12,-33)(12,-30)(13,-27)\qbezier(12,-33)(12,-30)(11,-27)
\qbezier(24,-33)(24,-30)(25,-27)\qbezier(24,-33)(24,-30)(23,-27)
\end{picture}
&\begin{picture}(12,12)(0,-36)\put(0,0){=}\end{picture}&
\begin{picture}(48,84)(-12,-36)
\qbezier(0,48)(0,42)(0,36) \qbezier(0,36)(0,30)(0,24)
\qbezier(0,24)(0,18)(0,12)
\qbezier(0,12)(3,9)(6,6)\qbezier(6,6)(9,3)(12,0)
\qbezier(12,0)(12,-6)(12,-12)
\qbezier(12,-12)(15,-15)(18,-18)\qbezier(18,-18)(21,-21)(24,-24)
\qbezier(24,-24)(24,-30)(24,-36)
\qbezier(12,48)(12,42)(12,36)
\qbezier(12,36)(15,33)(18,30)\qbezier(18,30)(21,27)(24,24)
\qbezier(24,24)(24,18)(24,12) \qbezier(24,12)(24,6)(24,0)
\qbezier(24,0)(24,-6)(24,-12)
\qbezier(24,-12)(21,-15)(20,-16)\qbezier(16,-20)(15,-21)(12,-24)
\qbezier(12,-24)(12,-30)(12,-36)
\qbezier(24,48)(24,42)(24,36)
\qbezier(24,36)(21,33)(20,32)\qbezier(16,28)(15,27)(12,24)
\qbezier(12,24)(12,18)(12,12)
\qbezier(12,12)(9,9)(8,8)\qbezier(4,4)(3,3)(0,0)
\qbezier(0,0)(0,-6)(0,-12) \qbezier(0,-12)(0,-18)(0,-24)
\qbezier(0,-24)(0,30)(0,-36)
\qbezier(0,39)(0,42)(1,45)\qbezier(0,39)(0,42)(-1,45)
\qbezier(12,39)(12,42)(13,45)\qbezier(12,39)(12,42)(11,45)
\qbezier(24,39)(24,42)(25,45)\qbezier(24,39)(24,42)(23,45)
\qbezier(0,-33)(0,-30)(1,-27)\qbezier(0,-33)(0,-30)(-1,-27)
\qbezier(12,-33)(12,-30)(13,-27)\qbezier(12,-33)(12,-30)(11,-27)
\qbezier(24,-33)(24,-30)(25,-27)\qbezier(24,-33)(24,-30)(23,-27)
\end{picture}&
\begin{picture}(60,72)(-12,-36)
\qbezier(0,36)(0,30)(0,24)
\qbezier(0,24)(3,21)(6,18)\qbezier(6,18)(9,15)(12,12)
\qbezier(12,12)(12,6)(12,0) \qbezier(12,0)(12,-6)(12,-12)
\qbezier(12,-12)(12,-18)(12,-24)
\qbezier(12,36)(12,30)(12,24)
\qbezier(12,24)(9,21)(8,20)\qbezier(4,16)(3,15)(0,12)
\qbezier(0,12)(0,6)(0,0) \qbezier(0,0)(0,-6)(0,-12)
\qbezier(0,-12)(0,-18)(0,-24)
\qbezier(24,36)(24,30)(24,24) \qbezier(24,24)(24,18)(24,12)
\qbezier(24,12)(24,6)(24,0)
\qbezier(24,0)(27,-3)(30,-6)\qbezier(30,-6)(33,-9)(36,-12)
\qbezier(36,-12)(36,-18)(36,-24)
\qbezier(36,36)(36,30)(36,24) \qbezier(36,24)(36,18)(36,12)
\qbezier(36,12)(36,6)(36,0)
\qbezier(36,0)(33,-3)(32,-4)\qbezier(28,-8)(27,-9)(24,-12)
\qbezier(24,-12)(24,-18)(24,-24)
\qbezier(0,27)(0,30)(1,33)\qbezier(0,27)(0,30)(-1,33)
\qbezier(12,27)(12,30)(13,33)\qbezier(12,27)(12,30)(11,33)
\qbezier(24,27)(24,30)(25,33)\qbezier(24,27)(24,30)(23,33)
\qbezier(36,27)(36,30)(37,33)\qbezier(36,27)(36,30)(35,33)
\qbezier(0,-21)(0,-18)(1,-15)\qbezier(0,-21)(0,-18)(-1,-15)
\qbezier(12,-21)(12,-18)(13,-15)\qbezier(12,-21)(12,-18)(11,-15)
\qbezier(24,-21)(24,-18)(25,-15)\qbezier(24,-21)(24,-18)(23,-15)
\qbezier(36,-21)(36,-18)(37,-15)\qbezier(36,-21)(36,-18)(35,-15)
\end{picture}
&\begin{picture}(12,12)(0,-36)\put(0,0){=}\end{picture}&
\begin{picture}(60,72)(-12,-36)
\qbezier(24,36)(24,30)(24,24)
\qbezier(24,24)(27,21)(30,18)\qbezier(30,18)(33,15)(36,12)
\qbezier(36,12)(36,6)(36,0) \qbezier(36,0)(36,-6)(36,-12)
\qbezier(36,-12)(36,-18)(36,-24)
\qbezier(36,36)(36,30)(36,24)
\qbezier(36,24)(33,21)(32,20)\qbezier(28,16)(27,15)(24,12)
\qbezier(24,12)(24,6)(24,0) \qbezier(24,0)(24,-6)(24,-12)
\qbezier(24,-12)(24,-18)(24,-24)
\qbezier(0,36)(0,30)(0,24) \qbezier(0,24)(0,18)(0,12)
\qbezier(0,12)(0,6)(0,0)
\qbezier(0,0)(3,-3)(6,-6)\qbezier(6,-6)(9,-9)(12,-12)
\qbezier(12,-12)(12,-18)(12,-24)
\qbezier(12,36)(12,30)(12,24) \qbezier(12,24)(12,18)(12,12)
\qbezier(12,12)(12,6)(12,0)
\qbezier(12,0)(9,-3)(8,-4)\qbezier(4,-8)(3,-9)(0,-12)
\qbezier(0,-12)(0,-18)(0,-24)
\qbezier(0,27)(0,30)(1,33)\qbezier(0,27)(0,30)(-1,33)
\qbezier(12,27)(12,30)(13,33)\qbezier(12,27)(12,30)(11,33)
\qbezier(24,27)(24,30)(25,33)\qbezier(24,27)(24,30)(23,33)
\qbezier(36,27)(36,30)(37,33)\qbezier(36,27)(36,30)(35,33)
\qbezier(0,-21)(0,-18)(1,-15)\qbezier(0,-21)(0,-18)(-1,-15)
\qbezier(12,-21)(12,-18)(13,-15)\qbezier(12,-21)(12,-18)(11,-15)
\qbezier(24,-21)(24,-18)(25,-15)\qbezier(24,-21)(24,-18)(23,-15)
\qbezier(36,-21)(36,-18)(37,-15)\qbezier(36,-21)(36,-18)(35,-15)
\end{picture}&
\begin{picture}(36,72)(-12,-36)
\qbezier(0,36)(0,30)(0,24)
\qbezier(0,24)(3,21)(6,18)\qbezier(6,18)(9,15)(12,12)
\qbezier(12,12)(12,6)(12,0)
\qbezier(12,0)(9,-3)(6,-6)\qbezier(6,-6)(3,-9)(0,-12)
\qbezier(0,-12)(0,-18)(0,-24)
\qbezier(12,36)(12,30)(12,24)
\qbezier(12,24)(9,21)(8,20)\qbezier(4,16)(3,15)(0,12)
\qbezier(0,12)(0,6)(0,0)
\qbezier(0,0)(3,-3)(4,-4)\qbezier(8,-8)(9,-9)(12,-12)
\qbezier(12,-12)(12,-18)(12,-24)
\qbezier(0,27)(0,30)(1,33)\qbezier(0,27)(0,30)(-1,33)
\qbezier(12,27)(12,30)(13,33)\qbezier(12,27)(12,30)(11,33)
\qbezier(0,-21)(0,-18)(1,-15)\qbezier(0,-21)(0,-18)(-1,-15)
\qbezier(12,-21)(12,-18)(13,-15)\qbezier(12,-21)(12,-18)(11,-15)
\end{picture}
&\begin{picture}(12,12)(0,-36)\put(0,0){=}\end{picture}&
\begin{picture}(36,72)(-12,-36)
\qbezier(0,36)(0,0)(0,-24)\qbezier(12,36)(12,0)(12,-24)
\qbezier(0,27)(0,30)(1,33)\qbezier(0,27)(0,30)(-1,33)
\qbezier(12,27)(12,30)(13,33)\qbezier(12,27)(12,30)(11,33)
\qbezier(0,-21)(0,-18)(1,-15)\qbezier(0,-21)(0,-18)(-1,-15)
\qbezier(12,-21)(12,-18)(13,-15)\qbezier(12,-21)(12,-18)(11,-15)
\end{picture}\\
b_1b_2b_1&=&b_2b_1b_2& b_1b_3&=&b_3b_1& b_1\tilde{b_1}&=&1
\end{array}
\label{fig:Bfix}\ee

\paragraph{Generators of the permutation group.} By definition, a generator $\sigma_i$ of the permutation group of $n+1$ elements corresponds to a transposition of the elements $i$
and $i+1$ in the sequence. Taking product of two the permutations is merely doing them
successively. The group unity $\mathds{1}$ corresponds to the trivial permutation (all the elements remain on their positions).

\paragraph{Generators of the braid group.} A generator $b_i$ of the $n+1$-strand braid group  corresponds to a intertwining of the strands $i$ and $i+1$ in a braid section. Taking a product of two braids is, by definition, plating one after the other on the same strands. The unity in the $n+1$-strand braid group is the trivial $n+1$-strand braid, which consists of $n+1$ unintertwined strands.

\paragraph{Symmetric group constraints.} Two products of the
permutation group generators are equivalent if and only if they
realize the same permutation, e.g., \be
\sigma_1\sigma_1xy=\sigma_1yx=\mathds{1}xy,\ \ \
\sigma_1\sigma_3\,xyzt=\sigma_3\sigma_1\,xyzt=yxtz,\ \ \mbox{and}\ \
\sigma_1\sigma_2\sigma_1\,xyz=\sigma_2\sigma_1\sigma_2\,xyz=zyx.\ee
In fact, all relations between the permutation group generators are exhausted by the above examples. Namely, the products of the group generators are in one to one correspondence with the permutations provided that the group generators satisfy \cite{Fulton}:
\be
\sigma_i\sigma_{i+1}\sigma_i=\sigma_{i+1}\sigma_i\sigma_{i+1},\ \ i=1,\ldots,n-1,\label{perm1}\\
\sigma_i\sigma_j=\sigma_j\sigma_i,\ \ i,j=1,\ldots,n,\ |i-j|\ne 1,\label{perm2}\\
\sigma_i^2=\mathds{1},\ \ \ \ i=1,\ldots,n.\label{perm3}\ee

\paragraph{Braid group constraints.}
In turn, the braid group generators satisfy their own constraints,
accordingly with that two braids, differing when projected on a
plane, may be \emph{isotopic}, i.e., being continuously transformed into each other in the
three-dimensional space. According to the Artin theorem \cite{Pras},
such a transformation may be presented as a sequence of the elementary transformations shown in fig.\ref{fig:Bfix}. While transformation (\ref{fig:Bfix}-III) just determines the
inverse of a braid group generators, transformations (\ref{fig:Bfix}-I,II) give rise the following identities for the braid group generators:
\be
b_ib_{i+1}b_i=b_{i+1}b_ib_{i+1},\ \ i=1,\ldots,n-1,\label{br1}\\
b_ib_j=b_jb_i,\ \ i,j=1,\ldots,n.\ |i-j|\ne 1,\label{br2}
\ee

\paragraph{Relating a permutation to a braid.}
Constraints (\ref{br1},\ref{br2}) coincide with constraints
(\ref{perm1},\ref{perm2}) for the permutation group generators.
Hence, one may relate a permutation to a braid, intertwining strands
$i$ and $i+1$ when the elements $i$ and $i+1$ are permuted, isotopic
braids being related to the same permutation due to
(\ref{br1},\ref{br2}). However, the inverse is generally wrong,
since there is no analog of constraint (\ref{perm3}) for the braid
group. In particular, any two-strand braid corresponds either to the
permutation $xy$ or to the permutation $yx$, when containing even or
odd number of crossings, respectively, while any two-stand braids
with the different numbers of crossings are not isotopic.

A one to one correspondence may be established between $n+1$-strand braids and replacements of $n+1$ indistinguishable points in the plane, a generator $b_i$ of the braid group corresponding to moving the points $i$ and $i+1$ \emph{continuously} on the positions of each other, in the selected, e.g., clockwise, direction. Although the final arrangement of points coincides with the initial one, such operation my have a non-trivial effect, e.g., when acting on a function with algebraic or logarithmic singularities in the given points \cite{Shab}.

\paragraph{Relating a ``quantum permutation'' to a braid.}
The continuous transposition operators described above may still satisfy some relations, apart from braid group relations (\ref{br1},\ref{br2}), in each particular case. The simplest non-trivial and important in many respects  example is given by \emph{Hecke algebra} \cite{KauffTB}.
By definition, generators $h_i$ with $i=\overline{1,n}$ of the Hecke algebra satisfy the constraints
\be
h_ih_{i+1}h_i=h_{i+1}h_ih_{i+1},\ \ i=1,\ldots,n-1,\label{heck1}\\
h_ih_j=h_jh_i,\ \ i,j=1,\ldots,n.\ |i-j|\ne 1,\label{heck2}\\
h_i^2=(q-q^{-1})h_i+1,\ \ \ q^n\ne
1,\ \ n\in\mathds{Z},\label{heck3} \ee where $q$ is a new formal
variable, and thorough the present text we suppose $q$ being not a root of unity
(since the properties of the Hecke algebra strictly differ in the
opposite case \cite{KlimSch}).

A Hecke algebra may be considered as a deformation of the permutation group, with the elements being enumerated, a transposition respecting the order numbers, and, apart from the permutations, their formal linear combinations being considered, namely,
\be
h_i\{\underbrace{\ldots }_{i-1}x_kx_l\underbrace{\ldots }_{n-i}\}=\{\ldots x_lx_k\ldots\}, \  l>k,\nn\\
h_i\{\ldots x_kx_l\ldots\}=(q-q^{-1})\{\ldots x_kx_l\ldots\}+\{\ldots x_lx_k\ldots\},\ \ l<k.
\ee
From this stand point, an element of a Hecke algebra is sometimes referred to as a \emph{quantum} or $q$-\emph{permutation} \cite{KauffTB}.

\paragraph{Extensions of the permutation group with the polynomial constraints.}
Hecke algebra constraint (\ref{heck3}) can be considered as a
particular case of a more general constraint, having the form \be
H_m=\sum_{k=0}^{m-1}a_kH_i^k,\label{colheck3} \ee where generators
$H_i$ the new algebra are supposed to satisfy
(\ref{heck1},\ref{heck2}) as well. Moreover, all three relations
(\ref{perm3}), (\ref{heck3}), and (\ref{colheck3}) can be considered
as the spectral equations for the algebra generators, \be
(\sigma_i-\mathds{1})(\sigma_i+\mathds{1})=0,\ \
(h_i-q\mathds{1})(h_i+q^{-1}\mathds{1})=0,\ \ \mbox{and}\ \
\prod_{k=1}^m(H_k-\lambda_k\mathds{1})=0,\label{permevals} \ee
respectively. The braid group representations with a general form of
constraint (\ref{colheck3}) give rise to the \emph{colored} HOMFLY
polynomials. Yet, we do not go into further details here, this point
is much more natural to discuss in from the standpoint of the
$\mathcal{R}$-matrix approach, we address to in section
\ref{sec:Rmat}.

\subsubsection{Braids closures and knots\label{sec:brcl}}

\be
\begin{array}{rclrclrcl}
\multicolumn{9}{c}{\mbox{Equivalence transformations of braids closures.}}\\[1mm]
\multicolumn{3}{c}{\mbox{I}}&\multicolumn{3}{c}{\mbox{II}}&\multicolumn{3}{c}{\mbox{II}^{\prime}}\\
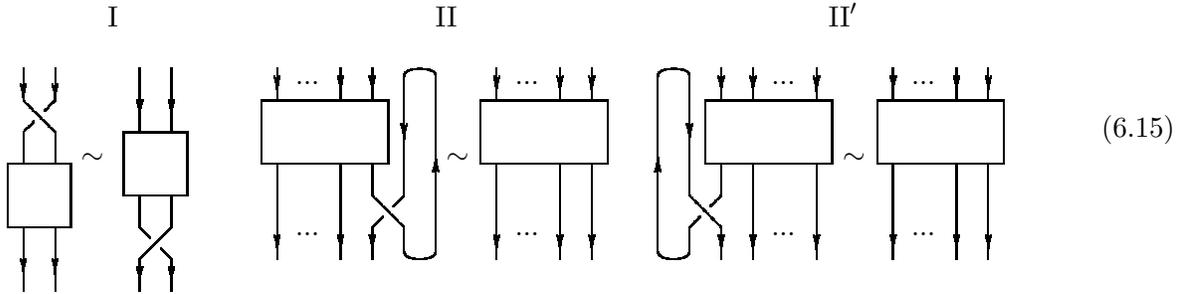
\begin{picture}(18,96)(0,-36)
\qbezier(0,48)(0,42)(0,36)
\qbezier(0,36)(3,33)(6,30)\qbezier(6,30)(9,27)(12,24)
\qbezier(12,24)(12,18)(12,12) \qbezier(12,-12)(12,-18)(12,-24)
\qbezier(12,-24)(12,-30)(12,-36)
\qbezier(12,48)(12,42)(12,36)
\qbezier(12,36)(9,33)(8,32)\qbezier(4,28)(3,27)(0,24)
\qbezier(0,24)(0,18)(0,12) \qbezier(0,-12)(0,-18)(0,-24)
\qbezier(0,-24)(0,-30)(0,-36)
\qbezier(-6,12)(0,12)(18,12)\qbezier(-6,-12)(0,-12)(18,-12)
\qbezier(-6,12)(-6,0)(-6,-12)\qbezier(18,12)(18,0)(18,-12)
\qbezier(0,36)(0,39)(1,42)\qbezier(0,36)(0,39)(-1,42)
\qbezier(12,36)(12,39)(13,42)\qbezier(12,36)(12,39)(11,42)
\qbezier(0,-30)(0,-27)(1,-24)\qbezier(0,-30)(0,-27)(-1,-24)
\qbezier(12,-30)(12,-27)(13,-24)\qbezier(12,-30)(12,-27)(11,-24)
\end{picture}
&\begin{picture}(6,48) \put(-6,48){$\sim$}
\end{picture}&
\begin{picture}(24,96)(0,-48)
\qbezier(0,-48)(0,-42)(0,-36)
\qbezier(0,-36)(3,-33)(6,-30)\qbezier(6,-30)(9,-27)(12,-24)
\qbezier(12,-24)(12,-18)(12,-12) \qbezier(12,12)(12,18)(12,24)
\qbezier(12,24)(12,30)(12,36)
\qbezier(12,-48)(12,-42)(12,-36)
\qbezier(12,-36)(9,-33)(8,-32)\qbezier(4,-28)(3,-27)(0,-24)
\qbezier(0,-24)(0,-18)(0,-12) \qbezier(0,12)(0,18)(0,24)
\qbezier(0,24)(0,30)(0,36)
\qbezier(-6,-12)(0,-12)(18,-12)\qbezier(-6,12)(0,12)(18,12)
\qbezier(-6,-12)(-6,0)(-6,12)\qbezier(18,-12)(18,0)(18,12)
\qbezier(0,18)(0,21)(1,24)\qbezier(0,18)(0,21)(-1,24)
\qbezier(12,18)(12,21)(13,24)\qbezier(12,18)(12,21)(11,24)
\qbezier(0,-45)(0,-42)(1,-39)\qbezier(0,-45)(0,-42)(-1,-39)
\qbezier(12,-45)(12,-42)(13,-39)\qbezier(12,-45)(12,-42)(11,-39)
\end{picture}
&\begin{picture}(72,84)(-18,-48) \qbezier(0,36)(0,30)(0,24)
\qbezier(0,0)(0,-6)(0,-12) \qbezier(0,-12)(0,-18)(0,-24)
\qbezier(0,-24)(0,-30)(0,-36)
\qbezier(24,36)(24,30)(24,24) \qbezier(24,0)(24,-6)(24,-12)
\qbezier(24,-12)(24,-18)(24,-24) \qbezier(24,-24)(24,-30)(24,-36)
\qbezier(36,36)(36,30)(36,24) \qbezier(36,0)(36,-6)(36,-12)
\qbezier(36,-12)(39,-15)(42,-18)\qbezier(42,-18)(45,-21)(48,-24)
\qbezier(48,-24)(48,-30)(48,-33)
\qbezier(48,33)(48,18)(48,0) \qbezier(48,0)(48,-6)(48,-12)
\qbezier(48,-12)(45,-15)(44,-16)\qbezier(40,-20)(39,-21)(36,-24)
\qbezier(36,-24)(36,-30)(36,-36)
\qbezier(-6,24)(0,24)(42,24)\qbezier(-6,0)(0,0)(42,0)
\qbezier(-6,24)(-6,12)(-6,0)\qbezier(42,24)(42,12)(42,0)
\qbezier(60,33)(60,0)(60,-33)
\qbezier(60,33)(60,36)(54,36)\qbezier(54,36)(48,36)(48,33)
\qbezier(60,-33)(60,-36)(54,-36)\qbezier(54,-36)(48,-36)(48,-33)
\put(7,30){.}\put(10,30){.}\put(13,30){.}
\put(7,-27){.}\put(10,-27){.}\put(13,-27){.}
\qbezier(0,27)(0,30)(1,33)\qbezier(0,27)(0,30)(-1,33)
\qbezier(24,27)(24,30)(25,33)\qbezier(24,27)(24,30)(23,33)
\qbezier(36,27)(36,30)(37,33)\qbezier(36,27)(36,30)(35,33)
\qbezier(0,-33)(0,-30)(1,-27)\qbezier(0,-33)(0,-30)(-1,-27)
\qbezier(24,-33)(24,-30)(25,-27)\qbezier(24,-33)(24,-30)(23,-27)
\qbezier(36,-33)(36,-30)(37,-27)\qbezier(36,-33)(36,-30)(35,-27)
\qbezier(48,9)(48,12)(49,15)\qbezier(48,9)(48,12)(47,15)
\qbezier(60,3)(60,0)(61,-3)\qbezier(60,3)(60,0)(59,-3)
\end{picture}&\begin{picture}(6,60)(0,0)\put(0,48){$\sim$}\end{picture}&
\begin{picture}(66,84)(-3,-48)
\qbezier(0,36)(0,30)(0,24) \qbezier(0,0)(0,-6)(0,-12)
\qbezier(0,-12)(0,-18)(0,-24) \qbezier(0,-24)(0,-30)(0,-36)
\qbezier(24,36)(24,30)(24,24) \qbezier(24,0)(24,-6)(24,-12)
\qbezier(24,-12)(24,-18)(24,-24) \qbezier(24,-24)(24,-30)(24,-36)
\qbezier(36,36)(36,30)(36,24) \qbezier(36,0)(36,-6)(36,-12)
\qbezier(36,-12)(36,-18)(36,-24)
\qbezier(36,-24)(36,-30)(36,-36)
\qbezier(-6,24)(0,24)(42,24)\qbezier(-6,0)(0,0)(42,0)
\qbezier(-6,24)(-6,12)(-6,0)\qbezier(42,24)(42,12)(42,0)
\put(7,30){.}\put(10,30){.}\put(13,30){.}
\put(7,-27){.}\put(10,-27){.}\put(13,-27){.}
\qbezier(0,27)(0,30)(1,33)\qbezier(0,27)(0,30)(-1,33)
\qbezier(24,27)(24,30)(25,33)\qbezier(24,27)(24,30)(23,33)
\qbezier(36,27)(36,30)(37,33)\qbezier(36,27)(36,30)(35,33)
\qbezier(0,-33)(0,-30)(1,-27)\qbezier(0,-33)(0,-30)(-1,-27)
\qbezier(24,-33)(24,-30)(25,-27)\qbezier(24,-33)(24,-30)(23,-27)
\qbezier(36,-33)(36,-30)(37,-27)\qbezier(36,-33)(36,-30)(35,-27)
\end{picture}
&
\begin{picture}(48,84)(-48,-48)
\qbezier(0,36)(0,30)(0,24) \qbezier(0,0)(0,-6)(0,-12)
\qbezier(0,-12)(0,-18)(0,-24) \qbezier(0,-24)(0,-30)(0,-36)
\qbezier(-24,36)(-24,30)(-24,24) \qbezier(-24,0)(-24,-6)(-24,-12)
\qbezier(-24,-12)(-24,-18)(-24,-24)
\qbezier(-24,-24)(-24,-30)(-24,-36)
\qbezier(-36,36)(-36,30)(-36,24) \qbezier(-36,0)(-36,-6)(-36,-12)
\qbezier(-36,-12)(-39,-15)(-40,-16)\qbezier(-44,-20)(-45,-21)(-48,-24)
\qbezier(-48,-24)(-48,-30)(-48,-33)
\qbezier(-48,33)(-48,18)(-48,0) \qbezier(-48,0)(-48,-6)(-48,-12)
\qbezier(-48,-12)(-45,-15)(-42,-18)\qbezier(-42,-18)(-39,-21)(-36,-24)
\qbezier(-36,-24)(-36,-30)(-36,-36)
\qbezier(6,24)(0,24)(-42,24)\qbezier(6,0)(0,0)(-42,0)
\qbezier(6,24)(6,12)(6,0)\qbezier(-42,24)(-42,12)(-42,0)
\qbezier(-60,33)(-60,0)(-60,-33)
\qbezier(-60,33)(-60,36)(-54,36)\qbezier(-54,36)(-48,36)(-48,33)
\qbezier(-60,-33)(-60,-36)(-54,-36)\qbezier(-54,-36)(-48,-36)(-48,-33)
\put(-11,30){.}\put(-14,30){.}\put(-17,30){.}
\put(-11,-27){.}\put(-14,-27){.}\put(-17,-27){.}
\qbezier(0,27)(0,30)(-1,33)\qbezier(0,27)(0,30)(1,33)
\qbezier(-24,27)(-24,30)(-25,33)\qbezier(-24,27)(-24,30)(-23,33)
\qbezier(-36,27)(-36,30)(-37,33)\qbezier(-36,27)(-36,30)(-35,33)
\qbezier(0,-33)(0,-30)(-1,-27)\qbezier(0,-33)(0,-30)(1,-27)
\qbezier(-24,-33)(-24,-30)(-25,-27)\qbezier(-24,-33)(-24,-30)(-23,-27)
\qbezier(-36,-33)(-36,-30)(-37,-27)\qbezier(-36,-33)(-36,-30)(-35,-27)
\qbezier(-48,9)(-48,12)(-49,15)\qbezier(-48,9)(-48,12)(-47,15)
\qbezier(-60,3)(-60,0)(-61,-3)\qbezier(-60,3)(-60,0)(-59,-3)
\end{picture}&
\begin{picture}(6,60)\put(0,48){$\sim$}\end{picture}&
\begin{picture}(42,84)(-3,-48)
\qbezier(0,36)(0,30)(0,24) \qbezier(0,0)(0,-6)(0,-12)
\qbezier(0,-12)(0,-18)(0,-24) \qbezier(0,-24)(0,-30)(0,-36)
\qbezier(24,36)(24,30)(24,24) \qbezier(24,0)(24,-6)(24,-12)
\qbezier(24,-12)(24,-18)(24,-24) \qbezier(24,-24)(24,-30)(24,-36)
\qbezier(36,36)(36,30)(36,24) \qbezier(36,0)(36,-6)(36,-12)
\qbezier(36,-12)(36,-18)(36,-24)
\qbezier(36,-24)(36,-30)(36,-36)
\qbezier(-6,24)(0,24)(42,24)\qbezier(-6,0)(0,0)(42,0)
\qbezier(-6,24)(-6,12)(-6,0)\qbezier(42,24)(42,12)(42,0)
\put(7,30){.}\put(10,30){.}\put(13,30){.}
\put(7,-27){.}\put(10,-27){.}\put(13,-27){.}
\qbezier(0,27)(0,30)(1,33)\qbezier(0,27)(0,30)(-1,33)
\qbezier(24,27)(24,30)(25,33)\qbezier(24,27)(24,30)(23,33)
\qbezier(36,27)(36,30)(37,33)\qbezier(36,27)(36,30)(35,33)
\qbezier(0,-33)(0,-30)(1,-27)\qbezier(0,-33)(0,-30)(-1,-27)
\qbezier(24,-33)(24,-30)(25,-27)\qbezier(24,-33)(24,-30)(23,-27)
\qbezier(36,-33)(36,-30)(37,-27)\qbezier(36,-33)(36,-30)(35,-27)
\end{picture}
\end{array}
\label{fig:Bcl} \ee

So far we have considered the braids and their equivalence
transformations. However, a knot is the closure of a braid
\cite{Pras}. Hence, to construct a knot invariant, one should treat
braids closures instead. Their equivalence transformations include
those of the braids, as well as two more transformations. The first
one is pulling a sequence of crossings $b_1b_2\ldots b_k$ from the
beginning of the braid to its end through the closure; the reversed
sequence, $b_k\ldots b_2b_1$, appears then at the braids end,
(\ref{fig:Bcl}-I). The transformation is called the first Markov
transform.

The second transformation consists in contracting a loop in
fig.\ref{fig:Bcl}-II or that in fig.\ref{fig:Bcl}-II$^{\prime}$, and is referred
to the second Markov transformation.

According to the Markov theorem \cite{Pras}, any two isotopic (related by a continuous deformation in the three-dimensional space) braid closures are transformed into
each other by a sequence elementary transformations, which include group
multiplication law (\ref{fig:Bfix}-III), Artin transformations
(\ref{fig:Bfix}-I,II), and Markov transformations,
(\ref{fig:Bcl}-I,II). In particular, transformation
(\ref{fig:Bcl}-II$^{\prime}$) can be performed this way, and hence should
not be included in the list.

\

There are some other knot
representations of a similar kind, as the braid representation of a knot \cite{katlas}. In particular, a
knot can be related to a braid, the strands being directed
differently (several ones ``upwards'', the remaining ones ``downwards''). A ``closure'' of such a braid includes an auxiliary
element, \cite{MMM3}. A conventional notation for Pretzel knots
\cite{katlas},\cite{GMMMSl},\cite{MMSl2} is this type one. All such
braid-like representations can be used to represent a knot invariant in a similar way, as we discuss in what follows.

\subsubsection{Unlinked knots and multiplication property\label{sec:fact}}

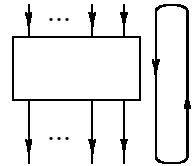
\begin{wrapfigure}{r}{60pt}
\begin{picture}(66,48)(-6,-12)
\qbezier(0,36)(0,30)(0,24) \qbezier(0,0)(0,-12)(0,-24)
\qbezier(24,36)(24,30)(24,24) \qbezier(24,0)(24,-12)(24,-24)
\qbezier(36,36)(36,30)(36,24) \qbezier(36,0)(36,-12)(36,-24)
\qbezier(48,33)(48,18)(48,0) \qbezier(48,0)(48,-12)(48,-21)
\qbezier(-6,24)(0,24)(42,24)\qbezier(-6,0)(0,0)(42,0)
\qbezier(-6,24)(-6,12)(-6,0)\qbezier(42,24)(42,12)(42,0)
\qbezier(60,33)(60,0)(60,-21)
\qbezier(60,33)(60,36)(54,36)\qbezier(54,36)(48,36)(48,33)
\qbezier(60,-21)(60,-24)(54,-24)\qbezier(54,-24)(48,-24)(48,-21)
\put(7,30){.}\put(10,30){.}\put(13,30){.}
\put(7,-15){.}\put(10,-15){.}\put(13,-15){.}
\qbezier(0,27)(0,30)(1,33)\qbezier(0,27)(0,30)(-1,33)
\qbezier(24,27)(24,30)(25,33)\qbezier(24,27)(24,30)(23,33)
\qbezier(36,27)(36,30)(37,33)\qbezier(36,27)(36,30)(35,33)
\qbezier(0,-21)(0,-18)(1,-15)\qbezier(0,-21)(0,-18)(-1,-15)
\qbezier(24,-21)(24,-18)(25,-15)\qbezier(24,-21)(24,-18)(23,-15)
\qbezier(36,-21)(36,-18)(37,-15)\qbezier(36,-21)(36,-18)(35,-15)
\qbezier(48,9)(48,12)(49,15)\qbezier(48,9)(48,12)(47,15)
\qbezier(60,3)(60,0)(61,-3)\qbezier(60,3)(60,0)(59,-3)
\end{picture}
\caption{A braid closure with a ``free'' strand.\label{fig:Bmult}}
\end{wrapfigure}

One degenerated case was left thus far beyond the scope of our
consideration. Namely, if a strand of braid is never intertwined
with the other ones (it can be either the first or the last strand in a
braid section, see fig.\ref{fig:Bmult}), then the closure of the
strand, which is just a circle, is separated from the closure of the
braid. More generally, the first $k$ strands may never cross the
$m-k$ last ones, or the braid may be reduced to one of the type by
the equivalence transformations; the closure is a disjoint union of
two or more knots or links. From the formal stand point, an
invariant of a disjoint union a quantity, independent of the
invariants of the components. However, many link invariants, among
them Alexander, Jones, Kauffman, and HOMFLY polynomials, by
definition possess a \emph{multiplication property} \cite{Pras}.
Namely,
\begin{itemize}
\item{
An invariant of a disjoint union of links is equal to the
product of the invariants of the components.}
\end{itemize}
In fact, it suffices to require for factorization of the invariant related to braid with last strand unlinked, the general property following from this fact \cite{JonesBr}.

The multiplication property for the braid in fig.\ref{fig:Bmult} completes the list of the constraints fig.\ref{fig:Bfix},\ref{fig:Bcl} on a quantity related to a braid, in order the one to be a knot invariant.


\subsection{Algorithm of constructing a knot inavriant\label{sec:brinv}}

Completed a general discussion on constructing a knot invariant as a braid group invariant, we proceed with presenting the precise algorithm. First,
\begin{step}
A knot is represented as the closure of a braid.\label{st:clos}
\end{step}
Then, one should construct a representation of a braid group, i.e., to
\begin{step}
Associate each of braid group generator $b_i$ with a matrix $B_i$, a
product of generators corresponding to the product of the
matrices.\label{st:brrep}
\end{step}
The next step relies on the correspondence discussed above and
summarized in tab.\ref{tab:symmbr}. Namely, a representation of the
permutation group, as well as a representation of the Hecke algebra
is at the same time a representation of the braid group with the
same number of generators \cite{Jones} (we recall that we suppose
the parameter $q$ entering the Hecke algebra constraints to be not a
root of unity). Moreover, the finite-dimensional irreducible
representation of the Hecke algebra and of the permutation group are
known then be in a one to one correspondence, the latter ones being
obtained from the former ones for $q=1$. In turn, dimensions of all
the permutation group irreducible representations are well known in
representation theory \cite{Fulton, LL3}, what enables explicitly
making the following ansatz:
\begin{step}
Dimension of the matrices is the dimension of an irreducible representation of the permutation group.\label{st:dim}
\end{step}
E.g., for the one-strand braid they are $1\times 1$ since the only
(irreducible) representation of permutation group one generator is the
one-dimensional representation. As we demonstrate in
sec.\ref{sec:2str}, matrices for the two-strand braid are of the
same size, since both (irreducible) representations of the permutation
group with two elements are one-dimensional. In turn, the permutation
group with three elements has one- and two-dimensional irreducible
representations, hence the three-strand matrices can be either one-,
or two-dimensional (see sec.\ref{sec:3str}), etc.

Explicit matrix expressions for Hecke algebra generators are in fact also known in representation theory (they are given, e.g., in \cite{JonesBr}), However, we find it more illustrative not just present the corresponding expressions, but to derive them from the first principles in the particular cases.
Namely, we are going to
\begin{step}
Write down the matrices $B_i$ of the proper size with undefined matrix elements and
impose the braid group constraints (\ref{br1},\ref{br2}) on them.
(\ref{br2})\label{st:breq}
\end{step}
Using a freedom in selected the basis, we suppose the matrix $B_1$ to be diagonal.
We will then
\begin{step} Solve the obtained
system of equations w.r.t. to the elements of the matrices $B_i$,
$i>1$, expressing them via the eigenvalues of the matrix
$B_1$.\label{st:brsol}
\end{step}
After that, a matrix representing an arbitrary braid can be calculated explicitly by taking the proper product of the matrices $B_i$ representing the braid group generators. This matrix depends on $\dim B_i$ of formal parameters $\lambda_j$, which are the $B_1$ eigenvalues.

The above steps provide a braid group representation, i.e., a braid is related to a matrix invariant under transformations (fig.\ref{fig:Bfix}) of the braid. To pass to a knot invariant,  one must construct a quantity that is invariant under braid closure transformations (fig.\ref{fig:Bcl},\ref{fig:Bmult}) as well. To guarantee the invariance under transformation (fig.\ref{fig:Bcl}-I) one can just
\begin{step}
Take the trace of the matrix $B$ related to the braid. \label{st:brtr}
\end{step}
To achieve the invariance under remaining transformations (fig.\ref{fig:Bcl}-II,III) and (fig.\ref{fig:Bmult}) is much more difficult. As we discussed in sec.\ref{sec:stmod}, ``extra'' conditions can be taken into account by taking a kind of ``\emph{average}'' (see sec.\ref{sec:stmod}).
If we continue following from the first principles, in the case to take the average implies to
\begin{step}
Write a linear combination of the traces $\Tr B$ of matrices from different braid group representations with undefined coefficients,\label{st:brch}
\end{step}
then
\begin{step}
impose (fig.\ref{fig:Bcl}-II,III) and (fig.\ref{fig:Bmult}) on the obtained expressions,\label{st:eqs}
\end{step}
and, finally,
\begin{step}
Solve the resulting equations w.r.t. the coefficients, expressing them via the $B_1$ eigenvalues.\label{st:sol}
\end{step}
We interrupt the general description at this point, sending a reader
to \cite{JonesBr} for the sequel, and proceed with carrying out the
formulated steps explicitly in the particular cases.



\subsection{Explicit calculating of the knot invariant in particular cases}
\subsubsection{Two-strand braids\label{sec:2str}}

\paragraph{Step 1.}
A two strand braid has the form $b_1^n$, where $b_1$ is the only
braid group generator. The closures of various two-strand braids
(fig.\ref{fig:B2n}) represent the so named series $T^{2,n}$ of torus
knots (for $n$ odd) and links (for $n$ even). In particular,
$T^{2,0}$ is a pair of unlinked unknots, and $T^{2,1}$ is the once
intertwined unknot (fig.\ref{fig:tcirc}-I), $T^{2,2}$ is the Hopf
link, $T^{2,3}$ is a trefoil knot (fig.\ref{fig:tref}), $T^{2,4}$ is
the Solomon link, etc. (see figures at \cite{katlas}).

\paragraph{Step 2.}
There are $2!=2$ permutations of two elements, $xy$ and $yx$. The
space of their formal linear combinations is the two-dimensional, a
basis can be chosen as \be X_S=\frac{1}{2}(xy+yx),\ \ \
X_A=\frac{1}{2}(xy-yx).\label{x2} \ee One refers to the formal
expressions $X_S$ and $X_A$ as to one-dimensional representations of
the permutation group $S_2$, implying that are conserved, up to
factor, subjected to the permutation group generator $b_1$ that
permutes $x$ and $y$, \be b_1X_S=X_S,\ \ b_1X_A=-X_A.\label{X2ev}
\ee

\paragraph{Step 3.}
According to the ansatz of \textbf{Step\,\ref{st:dim}}, one should
take two matrices $1\times1$, denote them $\lambda$ and $\mu$. No
constraints are imposed on the single generator of the two-strand
braid group, hence \textbf{Steps\,\ref{st:breq}-\ref{st:brsol}} are
omitted, and we go to

\paragraph{Step 6.}
According to the ansatz of \textbf{Step\,\ref{st:brrep}} a
two-strand braid is associated with a matrix product \be
B\left(b_1^n\right)=\left(h^I_1\right)^n,\ \ \ I=S\ \ \mbox{or}\ \
A. \ee where $n$ is the number of crossings, positive or negative
depending if the crossings are of direct or inverse orientation,
respectively (see fig.\ref{fig:invcr}).

\

Finally, the indeed non-trivial step, already in the case concerned,
is

\paragraph{Step 7.}
Following the ansatz of \textbf{Step~\ref{st:brch}}, we write the
invariant of a braid in the form \be
\lambda^n\chi_S+\mu^n\chi_A,\label{2str} \ee where $n$ is the number
of crossings, positive or negative depending on their orientation
(see \ref{fig:Bfix}-II), the eigenvalues  $\lambda$ and $\mu$ are
considered as formal variables, while  $\chi_S$ and $\chi_A$ are
(\emph{weight}) coefficients to define.

To obtain a knot (or link), not just a braid invariant, one has to
impose some extra conditions on (\ref{2str}). E.g., one may observe
that the closure of the two parallel strands reproduces a pair of
unlinked unknots, while the closure of the two once intersected
strands reproduces the single unknot with a contractible loop. The
value of invariant of the unknot is not defined at the moment, and
we denote it just $\chi$; due to the multiplication property discussed in sec.\ref{sec:fact}, the
invariant of a pair of unlinked unknots equals $\chi^2$. The
resulting constraints read: \be
\begin{array}{rclccrcl}
\mathds{1}^{(2)}&\sim&\mathds{1}^{(1)}\otimes \mathds{1}^{(1)}&\Rightarrow&\chi^2&=&\chi_S+\chi_A,\\
h^{(2)}_1&\sim&\mathds{1}^{(1)}&\Rightarrow&\chi&=&\lambda\chi_S+\mu\chi_A,\label{in2}
\end{array}\ee
System (\ref{in2}) enables one to express the coefficients of
irreducible representations in terms of the eigenvalues of the
corresponding operators and value of the invariant for the unknot:
\be \chi_S=\cfrac{(\mu\chi-1)\chi}{\mu-\lambda},\ \ \
\chi_A=\cfrac{(\lambda\chi-1)\chi}{\lambda-\mu}.\label{ch2} \ee

\

The resulting expression for invariant of the knot represented as
the closure of a two-strand braid with $n$ crossings oriented the
upper one in \ref{fig:Bfix}-II, or with $-n$ crossings oriented as
the lower one, is \be
H(\mathfrak{B}^{(2)};\lambda,\mu)=\lambda^n\cfrac{(\mu\chi-1)\chi}{\mu-\lambda}+
\mu^n\cfrac{(\lambda\chi-1)\chi}{\lambda-\mu}.\label{H2str}, \ee
where $\lambda$ and $\mu$ can be substituted by arbitrary numbers or
just left as formal parameters.

\subsubsection{Three-strand braids\label{sec:3str}}
\paragraph{Step 1.}
A three-strand braid has the form
$b_1^{a_1}b_2^{b_1}b_1^{a_2}b_2^{b_2}\ldots$, where $b_1$ and $b_2$
are the braid group generators, while $a_1,\ a_2,\ \ldots$ and
$b_1,\ b_2,\ \ldots$ are integer numbers, positive or negative. Not
all braids with various $a$ and $b$ are inequivalent; many of them
are isotopic, and hence must be equivalent as braid group elements.
The corresponding formal expressions are brought to each other with
help of the braid group relation \be
b_1b_2b_1=b_2b_1b_2,\label{3br1}\ee which is a particular case of
(\ref{br1}) for $m=3$. The smallest values of $a$ and $b$ correspond
to unknots, there unlinked ones for the trivial braid $\mathds{1}$
(all $a$ and $b$ equal zero), two unlinked ones (one once
intertwined) for the braids $b_1^{\pm 1}$ and $b_2^{\pm 1}$, and one
twice intertwined for the braids $b_1^{s_1}b_2^{s_2}\sim
b_2^{s_2}b_1^{s_1}$, where $s_1$ and $s_2$ equal $1$ or $-1$
independently of each other. The first non-trivial knot is obtained
for $a_1=a_2=1$, and $b_1=b_2=1$; this is the trefoil knot with an
additional contractible loop on a wire (fig.\ref{fig:ltref}).
Setting all $a$ and $b$ equal one ($2n$ of them non-zero), one
obtains a braid of the form $(b_1b_2)^n$, whose close is a torus
knot (for $n$ not multiplies $3$) or link (otherwise) out of the
series named $T^{3,n}$, which starts form the twice intertwined
unknot ($n=1$), the trefoil knot with a contractible loop ($n=2$),
and the Borromean rings link ($n=3$, $L6a4$ or $6_2^3$ in
\cite{katlas}). The simplest so called twist knots are represented
by three-stand braids, namely, the figure-eight, or two half-twist
knot ($4_1$ in \cite{katlas}) is the closure of the braid
$b_1(b_2)^{-1}b_1(b_2)^{-1}$, and the next, three half-twist knot
($5_2$ in \cite{katlas}) is the closure of the braid
$b_1^3b_2b_1^{-1}b_2$. An infinitely many topologically different
non-torus and non-twist knots are among the closures of three-strand
braids, see \cite{katlas}.

\paragraph{Step 2.}
There are $3!=6$ permutations of six elements,
$\left\{xyz,\,xzy,\,yxz,\,yzx,\,zxy,\,zyx\right\}$. In analogy with
the permutation group of two elements, two linear combinations are
conserved up to a factor by all permutations, \be
X_S=xyz+yxz+xzy+yzx+zxy+zyx\equiv(xyz),\ \ \ b_1X=b_2X=X,\label{x3}
\ee and \be X_A\equiv xyz-yxz-xzy+yzx+zxy-zyx\equiv[xyz].\ \ \
b_1X=b_2X=-X,\label{x111} \ee Hereafter, we use parentheses for a
full symmetrization, i.e., for the sum over all permutations, and we
use square brackets for a full antisymmetrization, i.e., for the sum
over all even permutations minus the sum over all odd permutations.

The linear space spanned by the permutations of three elements has
the dimension $3!=6$, a generic vector being of the
form\footnote{subscripts of the coefficients reflect the sequence of
permutations that yields a given element, i.e., $a_{121}$ is the
coefficient of $zyx=b_1b_2b_1xyz$.} \be
X=axyz+a_1yxz+a_2xzy+a_{12}yzx+a_{21}zxy+a_{121}zyx.\label{3sym} \ee
The cases, (\ref{x3}) with $a=a_1$, $a_2=a_{12}$, $a_{21}=a_{121}$,
and (\ref{x111}) with $a=a_2$, $a_1=a_{21}$, $a_{12}=a_{121}$,
correspond to the two common eigenvectors of the group generators,
and thus to the two one-dimensional eigenspaces of the permutation
group. Apart from that, the generators have a common
four-dimensional eigenspace (complementary to the subspace spanned
by the two common eigenvectors), given by the system of constraints:
\be
\left\{\begin{array}{rcl}a+a_1+a_2+a_{12}+a_{21}+a_{121}&=&0,\\
a-a_1-a_2+a_{12}+a_{21}-a_{121}&=&0
\end{array}\right.
\Leftrightarrow
\left\{\begin{array}{rcl}a+a_{12}+a_{21}&=&0,\\
a_1+a_2+a_{121}&=&0
\end{array}\right.,\label{AS21}
\ee i.e. coefficients of the permutations mutually related by a
cyclic shift must sum up to zero.
It is straightforward to verify that this property is covariant
under the transpositions, both of the first two and of the last two
elements. Hence, (\ref{AS21}) indeed defines a subspace of
(\ref{3sym}) invariant under the action of the permutation group.

The four-dimensional space given by (\ref{AS21}) is by definition a
representation of the permutation group, but not an irreducible one.
It turns out to decompose into two two-dimensional irreducible
representations. More precisely, it includes the two-dimensional
subspace \be X_{AS}=a(xy)z+b(yz)x+c(zx)y\equiv
axyz+ayxz+byzx+bzyx+czxy+cxzy, \nn\\ a+b+c=0.\label{x21} \ee each
vector of which generates a two-dimensional irreducible
representation of the permutation group. Indeed, $X_{AS}$ is, by
construction, an eigenvector of the first generator: \be
b_1X_{AS}=X_{AS};\label{21ev} \ee in addition, it satisfies the
identity \be
X_{AS}+b_2X_{AS}+b_1b_2X_{AS}=(a+b+c)(xyz)=0.\label{21ldep} \ee To
verify it, let us notice, that (\ref{21ldep}) contains $3\cdot 6=18$
summands and includes each of 6 monomials exactly 3 times, with
different coefficients, as they arise from different brackets in
(\ref{x21}), e.g., \be
(a+b+c)xyz&=&(axzy+b_2 cxzy+b_1b_2byzx),\nn\\
(a+b+c)yxz&=&(ayxz+b_2 byzx+b_1b_2cxzy), \ee and similarly for the
other monomials. Relations (\ref{21ev}) and (\ref{21ldep}) guarantee
that the representation in question is two-dimensional, since any
$X_{AS}$ from subspace (\ref{x21}) and the corresponding $b_2X_{AS}$
are turned by the group generators into linear combinations of each
other, \be
\begin{array}{|c||c|c|}
\hline
&X&b_2X\\
\hhline{|=||=|=|}
b_1&X&-X-b_2X\\
\hline
b_2&b_2X&X\\
\hline
\end{array}
\label{21rep}\ee

In fact, it is not necessary to start exactly from (\ref{x21}) to
construct a two-dimensional irreducible representation of the
permutation group. Subspace (\ref{AS21}) is a direct sum of two
two-dimensional subspaces of the first generators eigenvectors, with
the eigenvalues $1$ and $-1$, respectively. On the other hand, same
subspace (\ref{AS21}) is a similar direct sum w.r.t. the second
generator. A plane containing simultaneously two first generators
eigenvectors and two second generators eigenvectors  (the ones with
distinct eigenvalues, otherwise there would be more common
eigenvectors, while their is non) is a two-dimensional common
eigenspace of the generators. One has to parameterize all these
planes in one or another way. In the what follows, we checked and
used, that each first generators eigenvector with the eigenvalue $1$
from subspace (\ref{AS21}) generates, together with its image under
the second generator, a two-dimensional common eigenspace of the
generators; hence, these eigenspaces are in one to one
correspondence with the first generators eigenvectors. One can use
the first generators eigenvectors with the eigenvalue $-1$ equally
well. The latter ones also lie in subspace (\ref{AS21}) and have the
form \be X_{SA}=a[xy]z+b[yz]x+c[zx]y\equiv
axyz-ayxz+byzx-bzyx+czxy-cxzy, \ \ a+b+c=0, \ee and satisfy the
identities, similar to (\ref{21ev},\ref{21ldep}): \be
b_1X_{SA}=-X_{SA}, \ee and \be
X_{SA}-b_2X_{SA}+b_1b_2X_{SA}=(a+b+c)(xyz)=0, \ee the last one being
verified with help of the equalities \be
(a+b+c)xyz&=&(axzy-b_2 (-cxzy)+b_1b_2byzx),\nn\\
-(a+b+c)yxz&=&(-ayxz-b_2 byzx-b_1b_2cxzy), \ee etc. One could also
start from eigenvectors of the second generator instead. Of course,
in all cases one will obtain the same set of two-dimensional planes,
just differently parameterized. Each of these planes is a common
eigenspace of the two permutation group generators, thus being an
invariant subspace (in other words, a space of a representation) of
the entire permutation group. The group generators acting on such a
subspace can be presented as $2\times 2$ matrices. In particular, if
one selects a basis as $\{X_{AS},\ b_2X_{AS}\}$, for any $X_{AS}$
from (\ref{AS21}), the matrices are read from table (\ref{21rep}),
\be b_1=\left(\begin{array}{rr}1&-1\\0&-1\end{array}\right),\ \ \
b_2=\left(\begin{array}{cc}0&1\\1&0\end{array}\right),\ \
X\equiv\left(\begin{array}{c}1\\0\end{array}\right),\ \
b_2X\equiv\left(\begin{array}{c}0\\1\end{array}\right)\label{sgm}
\ee
For practical purposes, a basis of eigenvectors of $b_1$ (or $b_2$) is more convenient: 
\be b_1=\left(\begin{array}{cc}1&0\\0&-1\end{array}\right),\ \ \
b_2=\left(\begin{array}{cc}-\frac{1}{2}&\frac{\sqrt{3}}{2}\\\frac{\sqrt{3}}{2}&\frac{1}{2}\end{array}\right),\
\ \ X\equiv\left(\begin{array}{c}1\\0\end{array}\right),\ \
\frac{1}{\sqrt{3}}\big(X+2b_2X\big)\equiv\left(\begin{array}{c}0\\1\end{array}\right).\label{Ssgm}
\ee where the normalization factor in front of the second
eigenvector is chosen so that the matrix of $b_2$ is permutation.
Independence of these matrices of $a$, $b$, and $c$, entering
expression (\ref{x21}) for $X$, means that all the representations
with various $a$, $b$, and $c$ are isomorphic.

\

In explicit calculations, it is convenient to take a certain
$X_{AS}$ from (\ref{x21}), the conventional choice is \be a=-b=1\
\mbox{in}\ (\ref{x21})&\Rightarrow&X=(xy)z-(zx)y,\ \
X+2b_2X=
2[yz]x-[xy]z-[zx]y,\\
a=b=1,\ c=-2\ \mbox{in}\
(\ref{x21})&\Rightarrow&X=(xy)z+(zx)y-2(yz)x,\ \
X+2b_2X=
3[xy]z-3[zx]y.\nn \ee

\

Finally, we have constructed three distinct representations of the
permutation group with there elements, two one-dimensional ones,
(\ref{x3}), (\ref{x111}), and a two-dimensional one, generated by
any vector of form (\ref{x21}). In fact, there is an entire
two-dimensional space of such a vectors, given by (\ref{x21}). Each
of these vectors generates a two-dimensional representation,
altogether filling a four-dimensional space, and, together with the
two one-dimensional representations, covering the entire
six-dimensional space spanned by the $3!=6$ permutations of six
elements.

\paragraph{Step 3.}

We have obtained three distinct irreducible representations of the
permutation group with three elements, one-dimensional ones
(\ref{x3}) and (\ref{x111}, and a two-dimensional one (\ref{x21}); a
known theorem claims that there no other ones  \cite{Fulton}. The
ansatz of \textbf{Step \ref{st:dim}} dictates then to take three
pairs of matrices,  $h^{SS}_i$ and $h^{AA}_i$ of the size $1\times
1$, and $h^{AS}_i$ of the size $2\times 2$, with $i=1,2$. The next
step is to determine the explicit form of the matrices from the
braid group relations.

\paragraph{Step 4.}
Matrices $h_1$ and $h_2$ must satisfy group relations (\ref{3br1}).
For the one-dimensional representations, it gives merely \be
h_1^{SS}=h_2^{SS}=\alpha,\ \ \ h^{AA}_1=h^{AA}_2=\delta, \ee while
for the two dimensional representation a non-trivial equations is
obtained, \be
&h_1^{SA}h_2^{SA}h_1^{SA}=h_2^{SA}h_1^{SA}h_2^{SA},\nn\\\nn\\
&\left(
\begin{array}{cc}
\beta&0\\
0&\gamma
\end{array}\right)\left(
\begin{array}{cc}
a&b\\
c&d
\end{array}\right)
\left(
\begin{array}{cc}
\beta&0\\
0&\gamma
\end{array}\right)=
\left(
\begin{array}{cc}
a&b\\
c&d
\end{array}\right)
\left(
\begin{array}{cc}
\beta&0\\
0&\gamma
\end{array}\right)
\left(
\begin{array}{cc}
a&b\\
c&d
\end{array}\right),\label{3stYB}
\ee
\paragraph{Step 5.}
Equality (\ref{3stYB}) is satisfied if $h_2=h_1$. Apart from that,
there is the non-trivial solution for $h_2$: \be
a=\frac{\gamma^2}{\gamma-\beta},\ \
bc=\frac{\beta\gamma(\beta\gamma-\beta^2-\gamma^2)}{(\beta-\gamma)^2},\
\ d=\frac{\beta^2}{\beta-\gamma}.\label{YB3s} \ee

Since (\ref{YB3s}) gives the value only of the product $bc$, there
remains an arbitrariness, which, however, will not affect on the
invariants, where only the traces of matrix products enter. Indeed,
the matrices $h^{SA}_1$ and $h^{AS}_2$ satisfy group relation
(\ref{3br1}) and the eigenvalue equation \be
(h^{SA}_1-\beta)(h^{SA}_1-\gamma)=0,\ \
(h^{SA}_2-\beta)(h^{SA}_2-\gamma)=0.\label{ev21} \ee As already
said, the matrices lie in a representation of the Hecke algebra,
whose dimension is the same to that of the corresponding permutation
group \cite{JonesBr}, i.e., equals $(2+1)!=6$ for the case. Hence, a
product of the matrices, each one equals $h^{SA}_1$ or $h^{SA}_2$ or
is inverse to one of them, reduces, with help of eigenvalue equation
(\ref{ev21}) and relation (\ref{3br1}), to a linear combination of
$6$ basis elements, which can be chosen, e.g., as \be \mathrm{Id},\
\ h_1,\ \ h_2,\ \ h_1h_2,\ \ h_2h_1,\ \ \ h_2h_1h_2. \ee This is
easy to verify in each particular case, e.g. (the omitted
superscript $SA$ is assumed), \be
\underbrace{h_2h_1h_2}_{=h_1h_2h_1}h_1^{-1}\underbrace{h_2h_1h_2}_{=h_1h_2h_1}=
h_1\underbrace{h_2h_1h_2}_{h_1h_2h_1}h_1
=h_1^2h_2h_1^2=\big((\beta+\gamma)h_1-\beta\gamma\big)h_2\big((\beta+\gamma)h_1-\beta\gamma\big)=\nn\\=
\beta^2\gamma^2h_2-(\beta+\gamma)\beta\gamma\big(h_1h_2+h_2h_1\big)+(\beta+\gamma)^2h_1h_2h_1.
\nn\ee A trace of a braid group element, which enters the knot
invariant definitions, therefore, expands over the basis traces as
well; the latter ones are independent of $b$, \be
\mathrm{Tr}\,\mathds{1}=2,\ \
\mathrm{Tr}\,h_1=\mathrm{Tr}\,h_2=\beta+\gamma, \ \
\mathrm{Tr}\,h_1h_2=\mathrm{Tr}\,h_2h_1=a\beta+d\gamma,\ \
\mathrm{Tr}\,h_1h_2h_1=a\beta^2+d\gamma^2,\ \ \nn\ee and, hence, so
does the knot invariant. When discussing the matrices themselves, we
will fix this arbitrariness so that the matrices turn permutation;
the result then is

\be
h^{SA}_1=\left(\begin{array}{cc}\beta&0\\0&\gamma\end{array}\right),\ \ \left(h^{SA}_1\right)^{-1}=\left(\begin{array}{cc}\beta^{-1}&0\\0&\gamma^{-1}\end{array}\right)\nn\\\nn\\
h^{SA}_2=\left(\begin{array}{cc}\frac{\gamma^2}{\gamma-\beta}&
\frac{\sqrt{\beta\gamma(\beta\gamma-\beta^2-\gamma^2)}}{\beta-\gamma}\\\\
\frac{\sqrt{\beta\gamma(\beta\gamma-\beta^2-\gamma^2)}}{\beta-\gamma}&\frac{\beta^2}{\gamma-\beta}\end{array}\right),\nn\\\nn\\
\left(h^{SA}_2\right)^{-1}=\left(\begin{array}{cc}\frac{\beta}{\gamma(\gamma-\beta)}&
\frac{1}{\beta-\gamma}\sqrt{\frac{(\beta\gamma-\beta^2-\gamma^2)}{\beta\gamma}}\\\\
\frac{1}{\beta-\gamma}\sqrt{\frac{(\beta\gamma-\beta^2-\gamma^2)}{\beta\gamma}}&
\frac{\gamma}{\beta(\gamma-\beta)}\end{array}\right). \label{B21}\ee

\paragraph{Step 6.} Following the ansatz of \textbf{Step}~\ref{st:brrep}, we write the invariant of a three-strand braid in the form \be
h^I\left(b_1^{a_1}b_2^{b_1}\ldots b_1^{a_k}b_2^{b_k}\right)=
\prod_{i=1}^k\left(h_1^I\right)^{a_i}\left(h_2^I\right)^{b_i},\ \ \
I=SS,\ AS,\ AA. \ee

\paragraph{Step 7.}

At this step, we have obtained an invariant of a braid group
conjugation class. The ansatz of \textbf{Step~\ref{st:brch}}
dictates then to search for invariant of a knot presented as the
closure of a three-strand braid $b_1^{a_1}b_2^{b_1}\ldots
b_1^{a_k}b_2^{b_k}$ in the form \be \sum_{I=S,A,SA} \chi_I\
\mathrm{Tr}\prod_k\left(h_1^I\right)^{a_k}\left(h_2^I\right)^{b_k}=\chi_{SS}\alpha^n+\chi_{SA}
\mathrm{Tr}\left\{\left(h^{SA}_1\right)^{a_1}\left(h^{SA}_2\right)^{b_1}\ldots\right\}+\chi_{AA}\delta^n,
\ee and we proceed with determining the there standing weight
coefficients from matching the invariant values for braids with
different numbers of strands representing same knots. For
three-strand braids, this step is even less trivial compared to that
for two-stand braid, so we split it into two substeps.

\paragraph{Step 7-a: co-product for operators and weight coefficients.}
This step is based on the multiplication property (see sec.\ref{sec:fact}), which implies that \be
\left(b_1^{(3)}\right)^n \sim\left(b_1^{(2)}\right)^n\otimes
\mathds{1}^{(1)}\ \Rightarrow\ \lambda^n\chi_S\chi+\mu^n\chi_A\chi
=\alpha^n\chi_{SS}+(\beta^n+\gamma^n)\chi_{SA}+\delta^n\chi_{AA}\label{un3st}
\ee for an arbitrary $n$. Relation (\ref{un3st}) can be considered as a
system of infinitely many homogeneous linear equations. The only
solution is \be
\chi_S\chi=\chi_A\chi=\chi_{SS}=\chi_{SA}=\chi_{AA}=0, \ee unless
the equations are linearly dependent. A principle minor occupying in
the lines with $n=0,\ldots,5$ is a Wandermonde determinant, \be
\det\left\|\begin{array}{cccccc}1&1&1&1&1&1\\
\lambda&\mu&\alpha&\beta&\gamma&\delta\\
\lambda^2&\mu^2&\alpha^2&\beta^2&\gamma^2&\delta^2\\
\lambda^3&\mu^3&\alpha^3&\beta^3&\gamma^3&\delta^3\\
\lambda^4&\mu^4&\alpha^4&\beta^4&\gamma^4&\delta^4\\
\lambda^5&\mu^5&\alpha^5&\beta^5&\gamma^5&\delta^5
\end{array}\right\|=\begin{array}{r}\\\\(\alpha-\beta)(\alpha-\gamma)(\alpha-\delta)(\beta-\gamma)(\beta-\delta)(\gamma-\delta)\cdot\\
\cdot(\lambda-\alpha)(\lambda-\beta)(\lambda-\gamma)(\lambda-\delta)\cdot\\\cdot(\mu-\alpha)(\mu-\beta)(\mu-\gamma)(\mu-\delta);
\end{array}
\label{Wand}\ee any other one differs just by a factor of type
$\lambda^{m_1}\mu^{m_2}\alpha^{m_3}\beta^{m_4}\gamma^{m_5}\delta^{m_6}$.
Analysis of smaller sized minors shows, that rank of the matrix in
the l.h.s. of (\ref{Wand}), which is the number of linearly
independent solutions of (\ref{un3st}), is equal to the number of
distinct eigenvalues among $\lambda$, $\mu$, $\alpha$, $\beta$,
$\gamma$, $\delta$, the coefficient before each one becoming one of
the relations generating the space of solutions of (\ref{un3st}).

There is a distinguished solution, which has a profound group theory
sense: \be \lambda=\alpha=\beta,\ \
\mu=\gamma=\delta\nn\\
\chi\chi_S=\chi_{SS}+\chi_{SA},\ \
\chi\chi_A=\chi_{SA}+\chi_{AA}.\label{co3st} \ee Relations
(\ref{co3st}) are produced by a coproduct structure on the braid
group \cite{ReshTur}.

Note, that (\ref{un3st}) is satisfied as long as (\ref{co3st}) does;
in particular, the coefficients are thus far completely independent
of eigenvalues. The last step consists in relating them two.

\paragraph{Step 7-b.}

It remaining step relies on transformation (\ref{fig:Bcl}-II). As
already mentioned, it is the corresponding constraint on the
coefficients and eigenvalues that turn a braid invariant to a knot
invariant. There is an infinite set on constraints of the form \be
\left(b_1^{(3)}\right)^{n-1}b_2^{(3)}
\sim\left(b_1^{(2)}\right)^{n-1}\ \Rightarrow\
\lambda^{n-1}\chi_S+\mu^{n-1}\chi_A=
\lambda^n\chi_{SS}+\cfrac{\lambda^{n+1}-\mu^{n+1}}{\lambda-\mu}\chi_{SA}+\mu^n\chi_{AA},\label{R1st3}
\ee for all integer $n$. Any three of these equations are linearly
dependent since \be \lambda\left\|
\begin{array}{c}
\lambda^n\\
\lambda^m\\
\lambda^k
\end{array}
\right\|- \mu\left\|
\begin{array}{c}
\mu^n\\
\mu^m\\
\mu^k
\end{array}
\right\|= \left\|
\begin{array}{c}
\lambda^{n+1}-\mu^{n+1}\\
\lambda^{m+1}-\mu^{m+1}\\
\lambda^{k+1}-\mu^{k+1}
\end{array}
\right\|. \ee Hence, one can determine all three three-strand
coefficients, selecting any two of constrains (\ref{R1st3}), e.g.
with $n=1,2$ and completing them by any one of (\ref{co3st}), e.g.
with $n=0$: \be
\begin{array}{rclccrcrcrcrcr}
\mathds{1}^{(3)}
&\sim&\mathds{1}^{(2)}\otimes \mathds{1}^{(1)}&\Rightarrow&\chi\chi_S&+&\chi\chi_A&=&\chi_{SS}&+&2\chi_{SA}&+&\chi_{AA},\\
b_2^{(3)}
&\sim&\mathds{1}^{(2)}&\Rightarrow&\chi_S&+&\chi_A&=&\lambda\chi_{SS}&+&(\lambda+\mu)\chi_{SA}&+&\mu\chi_{AA},\\
b_1^{(3)}b_2^{(3)}
&\sim&h_1^{(2)}&\Rightarrow&\lambda\chi_S&+&\mu\chi_A&=&
\lambda^2\chi_{SS}&+&\lambda\mu\chi_{SA}&+&\mu^2\chi_{AA},
\end{array}\ee
wherefrom, taking into account (\ref{ch2}), one gets \be
\chi_{SS}=\frac{\chi(\mu\chi-1)(\lambda^2\chi+\mu-\lambda)}{(\mu-\lambda)\lambda\mu},\
\
\chi_{AA}=\frac{\chi(\lambda\chi-1)(\mu^2\chi+\lambda-\mu)}{(\lambda-\mu)\lambda\mu},
\label{ch3}\ee and \be
\chi_{SA}=\frac{\chi(\mu\chi-1)(\lambda\chi-1)}{\lambda\mu}. \nn\ee
It can be checked that selecting any other three linearly
independent constrains from of (\ref{R1st3}) and (\ref{co3st})
provides the same answer.

\paragraph{The result.}
Finally, the invariant of the knot presented as the closure of a
three-strand braid is expressed via the two-strand eigenvalues as
\be H\left(b_1^{a_1}b_2^{b_1}\ldots b_1^{a_k}b_2^{b_k}\right)=
\lambda^n\frac{\chi(\mu\chi-1)(\lambda^2\chi+\mu-\lambda)}{(\mu-\lambda)\lambda\mu}
+\mu^n\frac{\chi(\lambda\chi-1)(\mu^2\chi+\lambda-\mu)}{(\lambda-\mu)\lambda\mu}+\nn\\
+\mathrm{Tr}\prod_k\left(h^{SA}_1\right)^{a_k}\left(h^{SA}_2\right)^{b_k}
\frac{\chi(\mu\chi-1)(\lambda\chi-1)}{\lambda\mu},\label{H3str} \ee
where $h_1^{SA}$ and $h_2^{SA}$, as well as the inverse matrices,
are given by (\ref{B21}) with $\beta=\lambda$ and $\gamma=\mu$. Let
us emphasize, that $\lambda$ and $\mu$ enter (\ref{H3str}) just as
formal variables; substituting for them \emph{arbitrary} numbers,
or, equivalently, taking the coefficient of their \emph{any} powers
provides a numeric knot invariant.

\subsubsection{Four-strand braids\label{sec:4str}}
\paragraph{Step 1.} A four-strand braid has the form
\be
\mathfrak{h^{(4)}}=b_1^{a_1}\sigma^{b_1}_2\sigma^{c_1}_3b_1^{a_1}\sigma^{b_1}_2\sigma^{c_1}_3
\ldots b_k^{a_1}\sigma^{b_1}_k\sigma^{c_1}_k.\label{br4} \ee The
particular case of $h^{(4)}=\left(b_1b_2b_3\right)^k$ corresponds to
the torus knots (even $k$) and links (odd $k$) of the so named
series $T^{4,k}$. The four and five halve-twist knots ($6_1$ and
$7_2$ in \cite{katlas}, respectively), as well as all knots with no
more then $7$ crossings can be presented as the closures of
four-strand braids \cite{katlas}.

\paragraph{Step 2.}
As usual, we begin from listing the irreducible representations of
the permutation group, this time acting on four elements. Similarly to
the previous cases, there are two one-dimensional representations,
fully permutation one, \be X_{SSS}=(xyzt),\ \ \ \
b_1X_{SSS}=b_2X_{SSS}=b_3X_{SSS}=X_{SSS},\label{x4} \ee where all
the permutations of $xyzt$ enter with the same coefficient, and
fully antisymmetric one, \be X_{AAA}=[xyzt],\ \ \ \ \ \
b_1X_{AAA}=b_2X_{AAA}=b_3X_{AAA}=-X_{AAA},\label{x1111} \ee where
coefficients of all the permutations are equal up to a sign, being
plus for even permutations, and minus for odd permutations (we
recall that in all formulas parentheses and the square brackets
stand for stand, correspondingly, for the fully symmetric and
antisymmetric combinations constructed from the embarked elements).
Apart from the one-dimensional representations, the permutation group
with four elements has several more complicated ones. These
representations are constructed even less trivial than a two
dimensional representation of the permutation group with three
elements. For this reason, we sketch briefly the common
representation theory approach to the problem, before listing the
representations in question explicitly.



An important here theorem states that the irreducible
representations of a permutation group with $k$ elements are in one-to
one correspondence with the partitions of $k$ \be
k=k_1+k_2+\ldots+k_m,\ \ \ k_1\ge k_2\ge\ldots \ge k_m,\ \ \ k_1,\
k_2,\ \ldots,\ k_m\ \in\ \mathds{N}\label{partk} \ee Moreover, an
irreducible representation corresponding to a given partition of $k$
is constructed within the approach explicitly, as a formal linear
combination of permutations. We start from revising the cases of two
and three elements on this language. A partition of $k$ defined as
(\ref{partk}) we denote by $[k_1k_2\ldots k_m]$.

There two partitions of $2$, $2$ and $1+1$ (we write $[2]$ and
$[11]$ for them). They correspond to two irreducible representations
(\ref{x2}) of the permutation group with two elements, where the
elements are distributed over the round brackets as $2$, $X_S\equiv
X_2=(xy)$, and as $1+1$,\ $X_A\equiv X_{11}=a(x)(y)+b(y)(x)=axy+byx$
with $a+b=0$. Similarly, three elements are distributed over the
three round brackets in there irreducible representations
(\ref{x3},\ref{x21},\ref{x111}) of the permutation group according to
one of three partitions of $3$, $X_S\equiv X_3$, $X_{SA}\equiv
X_{21}$, $X_A\equiv X_{111}$.

In turn, all the irreducible representations of the permutation
group with four elements the are enumerated by partitions of 4
\cite{Fulton, LL3}. The representation corresponding to a partition
$[k_1\ge k_2\ge k_3\ge k_4]$  ($k_1+k_2+k_3+k_4=4$) is constructed
in the two following steps. First, one writes $xyzt$, putting the
parentheses around the first $k_1$, next $k_2$, next $k_3$, and last
$k_4$ elements; the parentheses (symmetrization sign) are separated
by a group product, e.g., \be (xy)(zt)\equiv
(xy+yx)(zt+tz)=xyzt+yxzt+xytz+yxtz. \ee Then, one lists all
permutations of $xyzt$ that remained inequivalent provided the
parentheses standing as described. These permutations are assembled
then to a linear combination, with the coefficient satisfying a
certain system of linear equations. We do not describe here neither
a general form of this system, nor rules for constructing it,
restricting ourselves by listing it explicitly in the particular
cases. Roughly speaking, a linear combination associated with a
given partition is to be, first, an eigenspace of the entire
permutation group, second, linearly independent of the linear
combinations for the previous partitions (in the inverse
lexicographic order). Such a system is in the most cases excessively
defined, and we also verify, for the cases considered, that each
solution of this system gives rise to an irreducible representation
of the permutation group. However, it can be shown, that all the
representations corresponding to a given partition are isomorphic
\cite{Fulton}. One can equivalently start from putting the square
brackets (antisymmetrization sign) in accordance with each
partition, and then impose some other (constructed in a similar
manner as for expressions with the parentheses) systems of linear
equations on the coefficients of the corresponding linear
combinations. The representation with parentheses for a certain
partition will be then isomorphic to the representation with the
square brackets for the dual partition (in the case $[4]$ is dual to
$[1111]$, $[31]$ is dual to $[211]$, and $[22]$ is dual to itself).
We use the easiest way of these two in each case, in one case
demonstrating their equivalence.

The simplest representations correspond to the partitions $[4]$ and
$[1111]$; they are already listed fully symmetric and antisymmetric
representations correspondingly, $X_4\equiv X_{SSS}$ and
$X_{[1111]}\equiv X_{AAA}$. Next comes the partition $[31]$. The
corresponding irreducible representation of the permutation group with
four elements is a straightforward analog of $[21]$-type
representation of the permutation group with three elements: \be
X_{31}=a(xyz)t+b(yzt)x+c(ztx)y+d(txy)z,\ \ \ a+b+c+d=0\label{x31}
\ee The general form of the linear combination is dictated by the
anzatz for a $[31]$-type representation, while the constraint on the
coefficients arises as a condition of linear independence of
$X_{31}$ and $X_{4}$ (which is obtained for $a=b=c=d$), being
formulated in a permutation (i.e., invariant under entire permutation
group) form. The constructed linear combination is fully permutation
under the permutations of the first three elements, i.e., \be
b_1X_{31}=b_2X_{31}=X_{31},\label{31ev} \ee and also satisfies the
identity \be
(\mathds{1}+b_3+b_2b_3+b_1b_2b_3)X_{31}=(a+b+c+d)(xyzt)=0.\label{31ldep}
\ee Identity (\ref{31ldep}) is similar to identity (\ref{21ldep})
for $21$-type representation of a permutation group. Each permutation
of $xyzt$ appears in the resulting expression for the four times,
picked up from one of the four summands in (\ref{x31}) by one of the
four group elements entering (\ref{31ldep}), e.g., \be
axyzt+db_3\,xytz+cb_2b_3\,xzty+bb_1b_2b_3\,ztyx=(a+b+c+d)xyzt, \ee
and similarly for other monomials. The constructed representation
turns out to be three-dimensional, with a basis can be chosen as
$\{X,\ b_3X,\ b_2b_3X\}$. To verify that, it suffices to check that
the action of the permutation group generators on the basis elements
expands, in account for (\ref{31ev},\ref{31ldep}), over the same
basis; the corresponding expressions are summarized in the table:
\be
\begin{array}{|c||c|c|c|}
\hline
&X&b_3X&b_2b_3X\\
\hhline{|=||=|=|=|}
b_1&X&b_3X&-X-b_3X-b_2b_3X\\
\hline
b_2&X&b_2b_3X&b_3X\\
\hline
b_3&b_3X&X&b_2b_3b_2X=b_2b_3X\\
\hline
\end{array}
\label{31rep}\ee Note that \emph{any} linear combination of type
(\ref{x31}), there is a three-dimensional space of them, gives rise
to a three-dimensional representation of the permutation group; all
these representations are isomorphic, since the action of the group
generators on the basis vectors is given in all cases by the same
table (\ref{31rep}). According the above mentioned rule, the
representation for the transposed partition, which is for the case
is $[211]$, is obtained by substituting all parentheses with the
square brackets, and demanding (in a permutation manner) that all the
obtained expression is linearly independent of $X_{1111}$ (which
corresponds to $a=-b=c=-d=1$). \be
X_{211}=a[xyz]t+b[yzt]x+c[ztx]y+d[txy]z,\ \ \ a-b+c-d=0\label{x211}
\ee Similarly to $X_{31}$, their are the identities valid for this
representation \be
b_1X_{211}=b_2X_{211}=-X_{211},\label{211ldef}\\
(\mathds{1}-b_3+b_2b_3-b_1b_2b_3)X_{211}=(a+b+c+d)(xyzt)=0,\nn \ee
the latter one being verified for each permutation, e.g., for
$xyzt$, as \be
(a+b+c+d)xyzt=axyzt-b_3(-dxytz)+b_2b_3cxzty-b_1b_2b_3(-byztx). \ee
As a result, a $[211]$-type representation of the permutation group is
of the same dimension three, as a $[31]$-type one. Basisses can
be chosen similarly in both cases; in case of $[211]$, the group
generators act on the basis vectors as: \be
\begin{array}{|c||c|c|c|}
\hline
&X&b_3X&b_2b_3X\\
\hhline{|=||=|=|=|}
b_1&-X&b_3X&-X+b_3X-b_2b_3X\\
\hline
b_2&-X&b_2b_3X&b_3X\\
\hline
b_3&b_3X&X&b_2b_3b_2X=-b_2b_3X\\
\hline
\end{array}
\label{211rep} \ee The above table confirm that the presented
$X_{211}$ indeed gives rise to a three-dimensional representation of
the permutation group. Again, (\ref{x211}) sets an entire
three-dimensional set of isomorphic $[211]$-type representations,
each one can be alternatively defined via action (\ref{211rep}) of
the group generators on the basic elements.

One more partition remains, namely $[22]$. The corresponding
irreducible representation of the permutation group has a slightly
more involved structure, than the above listed ones. A general
linear combination fitting the $[22]$-type anzatz includes six
summands: \be
X_{22}=a_1(xy)(zt)+a_2(zt)(xy)+a_3(xzr)(yt)+a_4(yt)(xz)+a_5(yz)(xt)+a_6(xt)(yz).\label{x22}
\ee The very ansatz implies that \be
X_{22}=b_1X_{22}=b_3X_{22}.\label{x22ev} \ee A linear space spanned
by all permutations of $4$ elements has the dimension $4!=24$. We
have already established, that there are two one-dimensional
representations ($[4]$- and $[1111]$-type ones), and two
tree-dimensional spaces of three-dimensional representations
($[31]$- and $[211]$-type ones). According the mentioned theorem
\cite{Fulton}, each vector of the remained subspace belongs to in a
$[22]$-type representation. Hence, the dimension of this subspace,
which is $4!-2\cdot1\cdot 1-2\cdot3\cdot 3=4$, equals the dimension
$[22]$-type representation multiplied by its multiplicity, i.e., by
the number of linearly independent (\ref{x22})-type expressions
allowed by the corresponding constraints. Actually, the
representation in question has the dimension two and the
multiplicity two, as we demonstrate in the below. Four constraints
on six coefficients are imposed in accordance with that. These
constraints can be presented in one of two equivalent forms: \be
\left\{
\begin{array}{ccccccc}
a_1&+&a_2&+&a_6&=&0,\\
a_1&+&a_3&+&a_5&=&0,\\
a_2&+&a_3&+&a_4&=&0,\\
a_4&+&a_5&+&a_6&=&0
\end{array}
\right.\ \ \Leftrightarrow\ \ \left\{
\begin{array}{ccccccc}
a_3&+&a_4&+&a_5&=&0,\\
a_2&+&a_4&+&a_6&=&0,\\
a_1&+&a_5&+&a_6&=&0,\\
a_1&+&a_2&+&a_3&=&0
\end{array}
\right.\label{22const}\ee The second equations are obtained from the
first ones by a termwise subtraction from the equality \be
a_1+a_2+a_3+a_4+a_5+a_6=0, \ee which, in turn, arises as a termwise
sum (being canceled by $2$) of all the equations, either the first
ones, or the second ones. Each of the first equations collects the
coefficients of terms in (\ref{x22}) that have in the second
parentheses  $x$, $y$, $z$, or $t$ in the second parenthesis. The
second equations are organized in a similar manner w.r.t. to the
first parenthesis. Such combinations of the coefficients arise in
the expressions of the form \be
a_1xyzt+a_2b_2xzyt+a_6b_1b_2yzxt&=&(a_1+a_2+a_6)xyzt,\nn\\
a_1xyzt+a_2b_2xzyt+a_3b_3b_2xtyz&=&(a_1+a_2+a_3)xyzt, \ee and in the
similar ones for other permutations. Hence, a combination of form
(\ref{x22}) constrained by (\ref{22const}) satisfies, in addition to
(\ref{x22ev}), the identities \be
\begin{array}{l}
(\mathds{1}+b_2+b_1b_2)X_{22}=\\[1mm](a_1+a_2+a_6)(xyz)t+(a_1+a_3+a_5)(txy)z+(a_2+a_3+a_4)(ztx)y+(a_4+a_5+a_6)(yzt)x
\end{array}\\
=0,\nn\\
\begin{array}{l}
(\mathds{1}+b_2+b_3b_2)X_{22}=\nn\\[1mm](a_3+a_4+a_5)t(xyz)+(a_2+a_4+a_6)z(txy)+(a_1+a_5+a_6)y(ztx)+(a_1+a_2+a_3)x(yzt)
\end{array}\\
=0.\nn \ee Dimension of the examined representation equals two, what
is checked by acting by the group generators on the basic elements,
which can be chosen as $\{X,\ b_2X\}$: \be
\begin{array}{|l||l|l|}
\hline
&X&b_2X\\
\hhline{|=||=|=|}
b_1&X&-X-b_2X\\
\hline
b_2&b_2X&X\\
\hline
b_3&X&-X-b_2X\\
\hline
\end{array}
\ee

\

To summarize, we have constructed formal linear combinations of
permutations of four elements, corresponding to two one-dimensional
representations ($[4]$- and $[1111]$-type ones), two
three-dimensional spaces of vectors, each one giving rise to a
three-dimensional representation of the permutation group (a $[31]$-
and $[211]$-type one, correspondingly), and a two dimensional space
of vectors, each one giving rise to a two-dimensional representation
of the permutation group (a $[22]$-type one). The described spaces do
not intersect by construction and form altogether a linear space of
the dimension $2\cdot 1\cdot 1+2\cdot 3\cdot 3 +1\cdot 2\cdot
2=24=4!$. Hence, any linear combination of the $4!$ permutations of
four elements is expanded over vectors from the spaces of the listed
irreducible representations.

\paragraph{Step 3.}
We have constructed five distinct irreducible representations of the
permutation group with four elements, one-dimensional ones (\ref{x4})
and (\ref{x1111}), two-dimensional one (\ref{x22}), and
three-dimensional ones (\ref{x31}) and (\ref{x211}). It is known
that there are no other, inequivalent ones \cite{Fulton}.
\textbf{Step \ref{st:dim}} then tells one to take five triples
matrices of the corresponding sizes. After the general scheme of
constructing the irreducible representation was discussed, it is
natural to label these matrices with the partitions. There are
$1\times 1$ matrices $h^4_i$ and $h^{1111}_i$, three-dimensional
ones $h^{31}_i$ and $h^{211}_i$, and two-dimensional ones
$h^{22}_i$, with $i=1,2,3$. We omit the superscripts unless the size
of matrices is essential.

\paragraph{Step 4.} The next step is to impose on the matrices the group relations, which for the case take the form
\be b_1b_2b_1=b_2b_1b_2,\ \ b_3b_2b_3=b_2b_3b_2,\ \
b_1b_3=b_3b_1.\label{bgr4} \ee
\paragraph{Step 5.} One must then solve the constraints w.r.t. to the matrix elements, expressing
them all via a necessary number of free parameters. Similarly to the
three-strand case, ones of the matrices eigenvalues can be taken as
such parameters. We approach to the problem at several steps.
\paragraph{Step 5-a: coinciding of eigenvalues from the II Artin constraint.}

One conclusion can be done for an arbitrary braid. Namely, the
matrices $h^Y_i$, a partition $Y$ given, a position $i$ in a braid
section vary, have the same eigenvalues. This follows from the II
Artin constraint, which can be brought to the form \be
(h_i-\lambda)h_{i+1}h_i=h_{i+1}h_i(h_{i+1}-\lambda), \ee Taking the
determinants from both sides, one obtains that the characteristic
polynomials of the matrices coincide, \be
\det(h_i-\lambda)=\det(h_{i+1}-\lambda),\label{eveq} \ee what is
equivalent to the coincidence of all the eigenvalues up to a
permutation; it is natural then to enumerate the basis vectors
$e^{(k)}$ so that $\lambda^{(k)}_i=\lambda^{(k)}_{i+1}$.

\paragraph{Step 5-b: commuting of non-adjacent operators.}
Other steps in solving (\ref{bgr4}) are not so straightforward. The
next in simplicity one concerns the third the constraints, which
reflects the commutation of the non-adjacent crossings. For
one-dimensional matrices the property is held trivially. The
corresponding two-dimensional matrices must be both diagonal, their
eigenvalues coincide, as already established, \be
h^{22}_1=\left(\begin{array}{cc}\lambda_{22,1}&\\&\lambda_{22,2}\end{array}\right)\
\ \Rightarrow\ \
h^{22}_3=\left(\begin{array}{cc}\lambda_{22,1}\\&\lambda_{22,2}\end{array}\right),\
\ \mbox{or}\ \
h^{22}_3=\left(\begin{array}{cc}\lambda_{22,1}\\&\lambda_{22,2}\end{array}\right).
\label{22B3}\ee The same might be true for the three-dimensional
matrices. A case is more involved if two of three eigenvalues
coincide; the corresponding matrices then commute provided that they
have a block structure: \be
h^{31}_1=\left(\begin{array}{ccc}\lambda_{31,1}&\\&\lambda_{31,1}\\&&\lambda_{31,2}\end{array}\right)\
\ \Rightarrow\ \
h^{31}_3=\left(\begin{array}{ccc}a&b\\b&c\\&&\lambda_{31,2}\end{array}\right),\label{bl31}
\ee and similarly for the other three-dimensional representations
\be
h^{211}_1=\left(\begin{array}{ccc}\lambda_{211,1}&\\&\lambda_{211,1}\\&&\lambda_{211,2}\end{array}\right)\
\ \Rightarrow\ \
h^{211}_3=\left(\begin{array}{ccc}\tilde{a}&\tilde{b}\\\tilde{b}&\tilde{c}\\&&\lambda_{211,2}\end{array}\right).\label{bl211}
\ee We examine this possibility in the next paragraph, postponing
the question of whether this is the case until \textbf{Step
\ref{st:brch}}.

\paragraph{Step 5-b: form of blocks from the II Artin constraint.}
We return now to the first two of constraints (\ref{bgr4}), this
time using them to express the non-diagonal matrix elements via the
eigenvalues; for three-dimensional matrices we take ans\"atze
(\ref{bl31},\ \ref{bl211}).

The first of equations (\ref{bgr4}) reduces to (\ref{3stYB}), both
for the two-dimensional matrices, and for the two-by blocks in the
three-dimensional matrices. The only difference is that $\beta$ and
$\gamma$ are substituted by the corresponding eigenvalues. Since
(\ref{3stYB}) has the only (modulo the above mentioned subtleties)
solution, one just writes \be
&h^{22}_2=h^{SA}_2(\beta=\lambda_{22,1},\ \gamma=\lambda_{22,2}),\nn\\
&h^{31}_2=\left(\begin{array}{ccc}\lambda_{31,1}\\&h^{SA}_1(\lambda_{31,1},\lambda_{31,2})\end{array}\right),\
\
h^{211}_2=\left(\begin{array}{ccc}\lambda_{211,1}\\&h^{SA}_2(\lambda_{211,1},\lambda_{211,2})\end{array}\right).
\label{22B2}\ee The second of equations (\ref{bgr4}) makes one to
select the first possibility for $h_3^{22}$ in (\ref{22B3}). For
each of the three-dimensional matrices, a separate non-trivial
matrix equation is obtained. There are no non-trivial solutions for
arbitrary $2\times 2$ blocks entering $h^{31}_2$ and $h^{211}_2$.
For the two by two blocks as in (\ref{22B2}), one obtains that the
matrix elements in (\ref{bl31}) and (\ref{bl211}) are \be
a=\frac{\lambda_2^3}{\lambda_1^2-\lambda_1\lambda_2+\lambda_2^2},\
c=\frac{\lambda_1^3}{\lambda_1^2-\lambda_1\lambda_2+\lambda_2^2},\
b^2=\frac{(\lambda_1-\lambda_2)\sqrt{-\lambda_1
\lambda_2(\lambda_1^2+\lambda_2^2)}}{\lambda_1^2-\lambda_1\lambda_2+\lambda_2^2},
\ee where $\lambda_1$ and $\lambda_2$ should be substituted by the
eigenvalues of the corresponding matrices.

\paragraph{Step 6.}
As \textbf{Step \ref{st:brrep}} prescribes, an invariant of a
four-strand braid has the form \be
h^I\left(b_1^{a_1}\sigma^{b_1}_2\sigma^{c_1}_3 \ldots
b_1^{a_k}\sigma^{b_k}_2\sigma^{c_k}_3\right)=
\prod_{i=1}^k\left(h^I_1\right)^{a_i}\left(h^I_2\right)^{b_i}\left(h^I_3\right)^{c_i}.
\ee

\paragraph{Step 7.}
It remains to specify the coefficients in the linear combination \be
H\left(\mathfrak{B}^{(4)}\right)=\sum_{\substack{I=4,311,22\\211,1111}}\chi_I\mathrm{Tr}h^I\left(\mathfrak{B}^{(4)}\right)
\ee that yields a knot invariant. Again, we split this step into two
ones.
\paragraph{Step 7-a: co-product for operators and weight coefficients.}

If the last two strands of a four-strand braid enter no crossings,
then the closure of the braid consists of pairwise unlinked two
unknots and the knot or link that is the closure of the braid placed
in the first two strands. Due to the multiplication property (see sec.\ref{sec:fact}), the invariant of
the former one must decompose into the product of the three latter
ones. In the sense specified in sec.\ref{sec:fact},  we write \be
\left(\sigma^{(4)}_1\right)^n\sim\left(\sigma^{(2)}_1\right)^n\otimes
\mathds{1}^{(1)}\otimes \mathds{1}^{(1)}, \ee with the corresponding
constraints on knot invariants being \be
\lambda^n_4\chi_4+\left(\lambda^n_{31,1}+\lambda^n_{31,2}+\lambda^n_{31,3}\right)\chi_{31}
+\left(\lambda^n_{22,1}+\lambda^n_{22,2}\right)\chi_{22}+\nn\\+
\left(\lambda^n_{211,1}+\lambda^n_{211,2}+\lambda^n_{211,3}\right)\chi_{211}
+\lambda^n_{1111}\chi_{1111}=\lambda^n\chi_2\chi^2_1+\mu^n\chi_{11}\chi^2_1.
\ee
Similarly to that in the three-strand case, a homogenous linear system of constraints on $\chi_2\chi^2_1$, $\chi_{11}\chi^2_1$, and $\chi_{Y,i}\equiv\chi_Y$ (with $Y$ 
running over the partitions of $4$, and $i$ running from $1$ to the
number of corresponding eigenvalues) was obtained. It is
straightforward to verify that a principle minor of this system is
proportional to the Wandermond determinant composed of the
eigenvalues. Hence, non-trivial solution for characters are their
for some of the eigenvalues coincide. In particular, the system is
satisfied if \be
\chi_2\chi_1^2=\left(\chi_3+\chi_{21}\right)\chi_1=\chi_4+2\chi_{31}+\chi_{22}+\chi_{211},\nn\\
\chi_{11}\chi_1^2=\left(\chi_{21}+\chi_{111}\right)\chi_1=\chi_{31}+\chi_{22}+2\chi_{211}+\chi_{1111}.\label{coch}
\ee and \be\lambda_4=\lambda,\ \
\lambda_{31,1}=\lambda_{31,2}=\lambda, \lambda_{31,3}=\mu,\ \
\lambda_{22,1}=\lambda,\ \lambda_{22,2}=\mu,\nn\\
\lambda_{211,1}=\lambda,\ \ \lambda_{211,2}=\lambda_{211,3}=\mu,
\lambda_{1111}=\mu.\label{coev} \ee This is the solution, which
enters in the definition of the knot invariant of the interest. The
case is that rules (\ref{coch}) and (\ref{coev}) reflect a
co-algebra structure on the braid group, similarly to (\ref{co3st})
in case of three-strand braids. Although any other solution might
give rise to \emph{a} knot invariant, only the named one is
systematically studied. The reason is, probably, in that the
additional structure enables to re-define the knot invariant in an
explicit and concise manner \cite{Tur,ReshTur,MorSm}, not just as a
solution of infinitely many equations. We will partially approve the
made choice in sec.\ref{sec:Rbr}, by arriving at the same answer by
at the first glance completely different method.


\paragraph{Step 7-b: weight coefficients via eigenvalues.}

In remains to express the weight coefficients via eigenvalues. Just
as in the previous cases, some of the necessary relations follow
from the second of the equivalence transformations, specific for
braid closures, (\ref{fig:Bcl}-I,II). To determine the five
characters, one needs overall five independent equations, e.g., \be
\mathds{1}^{(4)} \sim \mathds{1}^{(3)}\otimes\mathds{1}\
\Rightarrow\
\chi_4+\chi_{31}+\chi_{22}+\chi_{211}+\chi_{1111}=\label{evch40}\\=\chi_1\chi_3+\chi_1\chi_{21}+\chi_1\chi_{111},\nn\ee
\be h_3^{(4)}\sim \mathds{1}^{(3)} \ \Rightarrow\
\lambda\chi_4+(2\lambda+\mu)\chi_{31}+(\lambda+\mu)\chi_{22}+(\lambda+2\mu)\chi_{211}+\mu\chi_{1111}=\\=
\lambda\chi_3+(\lambda+\mu)\chi_{21}+\mu\chi_{111}, \nn\ee \be
h_2^{(4)}h_3^{(4)} \sim h_2^{(3)}\ \Rightarrow\
\lambda^2\chi_4+(\lambda^2+\lambda\mu)\chi_{31}+\lambda\mu\chi_{22}+(\lambda\mu+\mu^2)\chi_{211}+\mu^2\chi_{1111}=\\
=\lambda^2\chi_3+\lambda\mu\chi_{21}+\mu^2\chi_{111}, \nn\ee \be
h_1^{(4)}h_2^{(4)}h_3^{(4)} \sim h_1^{(3)}h_2^{(3)}\ \Rightarrow\
\lambda^3\chi_4+\mu\lambda^2\chi_{31}+\mu^2\lambda\chi_{211}+\mu^3\chi_{1111}=
\\=\lambda^2\chi_3+\lambda\mu\chi_{21}+\mu^2\chi_{111},\nn\ee
\be
h_1^{(4)}h_2^{(4)}h_1^{(4)}h_2^{(4)}h_1^{(4)}h_2^{(4)}h_3^{(4)}\sim h_1^{(3)}h_2^{(3)}h_2^{(3)}h_1^{(3)}h_2^{(3)}\ \Rightarrow\ \nn\\
\lambda^7\chi_4-\lambda^4\mu^3\chi_{31}-\lambda^3\mu^3(\lambda+\mu)\chi_{22}-\lambda^3\mu^4\chi_{211}+\mu^7\chi_{1111}=
\label{evch44}\\=\lambda^6\chi_3-2\lambda^3\mu^3\chi_{21}+\mu^6\chi_{111}.
\nn\ee
Note, that to produce five linearly independent equations, one has to take at least one braid with a strand entering no crossings,
and at least one three-stand braid not reducible to a two-strand one. 
The solution reads, \be
\chi_4&=&\frac{\chi_1(\mu\chi_1-1)(\mu^2\chi_1+\lambda-\mu)(\mu^3\chi_1+\lambda^2-\lambda\mu+\mu^2)}
{(\lambda-\mu)^2(\lambda^2+\mu^2)(\lambda^2+\mu^2-\lambda\mu)},\\
\chi_{31}&=&-\frac{\chi_1(\lambda\chi_1-1)(\mu\chi_1-1)(\mu^2\chi_1+\lambda-\mu)}
{(\lambda-\mu)^2(\lambda^2+\mu^2)},\nn\\
\chi_{22}&=&\frac{\lambda\mu\chi^2_1(\mu\chi_1-1)(\lambda\chi_1-1)}{(\lambda-\mu)^2(\lambda^2+\mu^2-\lambda\mu)},\nn\\
\chi_{211}&=&-\frac{\chi_1(\lambda\chi_1-1)(\mu\chi_1-1)(\lambda^2\chi_1+\mu-\lambda)}{(\lambda-\mu)^2(\lambda^2+\mu^2)},\nn\\
\chi_{1111}&=&\frac{\chi_1(\lambda\chi_1-1)(\lambda^2\chi_1+\mu-\lambda)(\lambda^3\chi_1+\lambda^2-\lambda\mu+\mu^2)}
{(\lambda-\mu)^2(\lambda^2+\mu^2)(\lambda^2+\mu^2-\lambda\mu)}.
\label{ch4}\ee
\paragraph{The result.}
Putting everything together, we obtain that the invariant of the
knot presented as the closure of a four-strand braid is computed as
\be H\left(b_1^{a_1}\sigma^{b_1}_2\sigma^{c_1}_3 \ldots
b_1^{a_k}\sigma^{b_k}_2\sigma^{c_k}_3\right)=
\sum_{Y\vdash4}\chi_Y\mathrm{Tr}\prod_{i=1}^k\left(h^I_1\right)^{a_i}\left(h^Y_2\right)^{b_i}\left(h^Y_3\right)^{c_i},
\ee where $Y\vdash4$ means that $Y$ runs over partitions of $4$
($Y=[4],\ [31],\ [22],\ [211],\ [1111]$), the matrices $h^Y$ are \be
&h^4_1=h^4_2=\lambda,\ \ \,h^{1111}_1=h^{1111}_2=\mu,\nn\\[2mm]
&h^{22}_2=h^{SA}_2(\lambda,\mu)=\left(\begin{array}{cc}\frac{\mu^2}{\mu-\lambda}&
\frac{\sqrt{\lambda\mu(\lambda\mu-\lambda^2-\mu^2)}}{\lambda-\mu}\\\\
\frac{\sqrt{\lambda\mu(\lambda\mu-\lambda^2-\mu^2)}}{\lambda-\mu}&\frac{\lambda^2}{\mu-\lambda}\end{array}\right),\nn\\[2mm]
&h^{31}_2=\left(\begin{array}{ccc}\lambda\\&h^{SA}_1(\lambda,\mu)\end{array}\right),\
\
h^{211}_2=\left(\begin{array}{ccc}\mu\\&h^{SA}_2(\lambda,\mu)\end{array}\right).
\ee and the weight coefficients $\chi_I$ are given by (\ref{ch4}).
\subsection{Weight coefficients as $SU(N)$ characters\label{sec:char}}
The obtained expressions for the weight coefficients have a
remarkable property, which reveals a deeper underlining structure.
Namely, the crossing matrices arise as a deformation of the
permutation group generators. If one substitutes now the two-strand
eigenvalues in (\ref{ch2},\ref{ch3},\ref{ch4}) by the eigenvalues of
the rang two permutation groups generator on the symmetric and
antisymmetric representations (\ref{x2}), respectively, i.e.,
$\lambda=1$ and $\mu=-1$, the weight coefficients turn into the
quotients of the gamma function resembling expressions,\footnote{We
label all the coefficients by partitions this times; in particular,
$\chi_2\equiv\chi_S$, $\chi_{11}\equiv\chi_A$,
$\chi_{SS}\equiv\chi_3$, $\chi_{SA}\equiv\chi_{21}$, and
$\chi_{AA}\equiv\chi_{111}$.} \be \chi_2=\frac{\chi(\chi+1)}{2},\ \
\chi_{11}=\frac{\chi(\chi-1)}{2}\label{dm2} \ee \be
\chi_3=\frac{\chi(\chi+1)(\chi+2)}{6},\ \
\chi_{21}=\frac{\chi(\chi+1)(\chi-1)}{3},\ \
\chi_{111}=\frac{\chi(\chi-1)(\chi-2)}{6}\label{dm3} \ee \be
\chi_4=\frac{\chi(\chi+1)(\chi+2)(\chi+3)}{24},\ \ \chi_{31}=\frac{\chi(\chi+1)(\chi+2)(\chi-1)}{8},\ \ \chi_{22}=\frac{\chi^2_1(\chi+1)(\chi-1)}{3},\label{dm4}\\
\chi_{211}=\frac{\chi(\chi+1)(\chi-1)(\chi+2)}{8},\ \
\chi_{1111}=\frac{\chi(\chi+1)(\chi+2)(\chi+3)}{24}.\nn \ee
In particular, for $\chi$ being equal to any integer $N$, a
coefficient $\chi_Y$ for a Young diagram $Y$ equals the dimension
of $su(N)$ representation corresponding to the Young diagram $Y$; as
well as permutation group irreducible representations, $su(N)$
irreducible representations are in one-to-one correspondence with
the Young diagrams \cite{Fulton, LL3}.

The observed property admits a generalization.
As already mentioned in sec.\ref{sec:brsymm},  the permutation group
admires a deformation called Hecke algebra, or $q$-permutation group.
As we will see in sec.\ref{sec:Rbr}, the generator of the
$q$-permutation group acts on its two irreducible representation with
the eigenvalues $\lambda=q$ and $\mu=-q^{-1}$, with $q$ begin a
formal parameter entering group relations. Substituting the so
deformed eigenvalues in (\ref{ch2},\ref{ch3},\ref{ch4}), one obtains
that (\ref{dm2}-\ref{dm4}) is substituted by \be
\chi_2=\frac{[N][N+1]}{[2]},\ \
\chi_{11}=\frac{[N][N-1]}{[2]}\label{qdm2} \ee \be
\chi_3=\frac{[N][N+1][N+2]}{[2][3]},\ \
\chi_{21}=\frac{[N][N+1][N-1]}{[3]},\ \
\chi_{111}=\frac{[N][N-1][N-2]}{[2][3]},\label{qdm3} \ee \be
\chi_4=\frac{[N][N+1][N+2][N+3]}{[2][3][4]},\ \ \chi_{31}=\frac{[N][N+1][N+2][N-1]}{[2][4]},\ \ \chi_{22}=\frac{[N]^2[N+1][N-1]}{[3]},\nn\\
\chi_{211}=\frac{[N][N+1][N-1][N-2]}{[2][4]},\ \
\chi_{1111}=\frac{[N][N+1][N+2][N+3]}{[2][3][4]}.\label{qdm4}\nn \ee
with the standard notation
$[N]\equiv\frac{q^N-q^{-N}}{q-q^{-1}}=q^{N-1}+q^{N-3}+\ldots+q^{-N-1}$
used. For $q=1$, $[N]=N$, and (\ref{qdm2}-\ref{qdm4}) reduce to
(\ref{dm2}-\ref{dm4}) with $\chi_1=N$. Quantities
$(\ref{qdm2}-\ref{qdm4})$ are referred to as \emph{quantum
dimensions} of \emph{quantum group} $U_q(SU(N))$ irreducible
representations \cite{ReshTur}. They can be considered traces of the
exponentiated  quadratic Casimir operator over the corresponding
$SU(N)$ representations \cite{MorSm}.

\section{Constructing a knot polynomial from $\mathcal{R}$-matrices \label{sec:Rmat}}

In this section, we finally address to the approach we are especially interested in. As we already mentioned, this approach can be developed either on the ground of the state model approach (see sec.\ref{sec:stmod}), or on the ground of the braid group approach (see sec.\ref{sec:braids}). Below we outline the construction, which we discuss in details throughout the section.

In complete analogy with state model approach the with braid group approach, we first pass from the
knot diagram to a \emph{cut} knot diagram, and relate it to an
operator product, which is invariant under Reidemeister moves II
(fig.\ref{fig:RIII}) and III (fig.\ref{fig:RII}) and depending on
some additional parameters. Afterwards, we define an ``average''
over these parameters, which will be invariant under the
Reidemeister move I (fig.\ref{fig:RI}) as well. This procedure has
different versions, already outlined in sec.\ref{sec:stmod}. We
mostly concentrate here on the turn-over operators approach
introduced in sec.\ref{sec:sttov}, since it being very poorly
presented in literature. We also give a notion of the more common
approach, or rather the approaches, which are based on selecting a
certain direction on the projection plane \cite{KirResh,MorSm,Kaul,Inds1,Inds2,Inds3,Inds6,Inds10,KaulLeq,GuJock,MMM4,MMM5,GMMM,FrohKing,LabPer,LabNew,DSS:Konz,DSS:KZ}.

\begin{wrapfigure}{r}{65pt}
\begin{picture}(36,24)(0,-15)
\qbezier(0,24)(6,18)(12,12)\qbezier(12,12)(18,6)(24,0)
\qbezier(24,24)(18,18)(16,16)\qbezier(8,8)(6,6)(0,0)
\qbezier(0,0)(2,2)(2,5)\qbezier(0,0)(2,2)(4,2)
\qbezier(24,0)(22,2)(22,5)\qbezier(24,0)(22,2)(20,2)
\put(3,24){$i$}\put(18,24){$j$}\put(0,6){$l$}\put(22,6){$k$}
\put(5,-24){$R^{ij}_{kl}$}
\end{picture}
\begin{picture}(24,24)(0,-15)
\qbezier(0,24)(6,18)(8,16)\qbezier(16,8)(18,6)(24,0)
\qbezier(24,24)(18,18)(12,12)\qbezier(12,12)(6,6)(0,0)
\qbezier(0,0)(2,2)(2,5)\qbezier(0,0)(2,2)(4,2)
\qbezier(24,0)(22,2)(22,5)\qbezier(24,0)(22,2)(20,2)
\put(5,24){$j$}\put(20,24){$i$}\put(-2,6){$k$}\put(22,6){$l$}
\put(5,-24){$\tilde{R}^{ij}_{kl}$}
\end{picture}
\label{fig:invcr} \caption{Direct and inverse crossings.}
\end{wrapfigure}
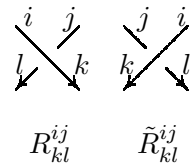

\subsection{The notion of $\mathcal{R}$-matrix\label{sec:Rdef}}
In sec.\ref{sec:braids} we considered the Hecke algebra elements, discussing there how a symmetry generalizing the permutation symmetry can be realized with help of these elements. In fact, Hecke algebra elements are particular cases of the \emph{quantum} $\mathcal{R}$-\emph{matrices} \cite{Pras,InvScatt,Baxt,KlimSch}. Namely, an operator satisfying permutation group constraints (\ref{perm1},\ref{perm2}), generally \emph{not} satisfying (\ref{perm3}) is by definition $\mathcal{R}$-matrix. An eigenvalue equation of a general form (\ref{colheck3}) holds for an $\mathcal{R}$-matrix instead of (\ref{perm3}).

As we discussed above, symmetric group constraints (\ref{perm1}),\ref{perm2}) can be associated with the transformations of a knot diagram in lines 1 and 2 of tab.\ref{tab:op}, respectively.  These transformations, it turn, give rise to the operator equations the right column of the table. A four script operator satisfying these constraints is thus another representation of an $\mathcal{R}$-matrix. Equation (\ref{perm3},\ref{tab:op} --- line 2), represented in any of the two equivalent forms, is referred to as \emph{the Yang-Baxter equation} \cite{Pras}.

The $\mathcal{R}$-matrices the most naturally arise in the context of \emph{quantum groups} \cite{KlimSch}. However, this subject is beyond the scope of our presentation. Here we just outline the several properties of a quantum $\mathcal{R}$-matrix, which are essential for using it as a tool for constructing the knot polynomials. Namely,
\begin{center}
A quantum $\mathcal{R}$-matrix
\begin{itemize}
\item{Is explicitly constructed for each representation of each Lie group}
\item{Depends on a formal variable $q$ (which is referred to as a \emph{quantum} parameter)}
\item{Satisfies the Yang-Baxter equation, which is associated with the III Reidemeister transformation of the knot planar diagram.}
\end{itemize}
\end{center}
We also present the explicit form of the simplest $\mathcal{R}$-matrix and verify that it indeed satisfies the Yang-Baxter equation in sec.\ref{sec:Rexpl}.

\subsection{A contraction of $\mathcal{R}$-matrices as a cut knot diagram invariant\label{sec:Rinv}}
\subsubsection{The general part of the state model approach adopted to the $\mathcal{R}$-matrix approach}
\begin{wrapfigure}{r}{135pt}
\begin{picture}(60,20)(-10,-20)
\put(0,0){\circle*{2}}\put(30,0){\circle*{2}}\put(25,0){\vector(1,0){1}}
\qbezier(0,0)(15,0)(30,0)\qbezier(15,-5)(15,0)(15,5)
\put(-5,-12){$\small{A}$}\put(25,-12){$\small{B}$}
\end{picture}
\begin{picture}(10,10)(10,-20)\put(0,0){\vector(1,0){20}}\end{picture}
\begin{picture}(60,60)(-10,-20)
\put(10,0){\circle*{2}}\put(20,0){\circle*{2}}\put(0,0){\vector(-1,0){1}}\put(30,0){\vector(1,0){1}}
\qbezier(0,0)(5,0)(10,0)\qbezier(20,0)(25,0)(30,0)
\put(5,-12){$\small{A}$}\put(15,-12){$\small{B}$}
\end{picture}
\caption{Cutting of the edge on a knot diagram.\label{fig:cutedge}}
\end{wrapfigure}
We start from recalling the first steps of the constructing a knot
polynomial in the state model approach (see sec.\ref{sec:stmod}),
this time formulating them in a way especially convenient for the
following presentation.

First,
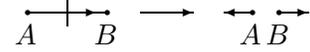
\begin{step}
A knot is related to a \emph{knot diagram}.\label{st:diag}
\end{step}
The definition of a knot diagram is given in sec.\ref{sec:stmod}.
Now we pass from a knot diagram to a \emph{cut} knot diagram.
Namely, we
\begin{step}
Cut some edges on the knot diagram so that the cut diagram is
related to a collection of self-crossing free curves. \label{st:cut}
\end{step}
To cut an internal edge directed from vertex $A$ to vertex $B$ means
substitute the edge with the pair of edge incoming the vertex $A$
and the edge outgoing the vertex $B$. (see fig.\ref{fig:cutedge}).

The proper cutting of a knot diagram can be carried out as follows.
The first cut is made on an arbitrary edge. Then, one follows in the
direction selected on the edges, at each crossing selecting the edge
corresponding to the proceeding of an already passed edge on the
original curve (i.e., edge $k$ proceeds edge $i$, and edge $l$
proceeds edge $j$ in fig.\ref{fig:invcr}). Following the edges in a
knot diagram this way, one stops before encountering with the
already passed edge for the first time. The next cut is done on the
edge where one stopped. The procedure is then repeated, the made cut
as a new starting point. In case of a knot, one terminates at the
first cut by the end. In case of a link, the procedure is carried
out for each connection component separately. An example of the
resulting cut diagram is presented in fig.\ref{fig:tref}.
After the procedure is completed,
\begin{step}
Each incoming edge on the cut diagram acquires its own
number.\label{st:num}
\end{step}
The same number can be related to the corresponding segment of the
original curve after being cut. One obtains then a collection of
enumerated directed curve segments, corresponding to unclosed
polygons composed of edges of the cut knot diagram, the crossing on
the diagram corresponding now to intersections of these segments.
Hence, a crossing is related now to a pair of numbers associated
with the corresponding curve segments. The following two steps then
are to
\begin{step}
Associate the polygon on the cut diagram attached with incoming
edge $\alpha$ to a vector space $\mathcal{L}_{\alpha}$,\label{st:vec}
\end{step}
and to
\begin{step}
Associate a direct crossing (see fig.\ref{fig:invcr}) of polygons
$\alpha$ and $\beta$ with a linear operator $S^{(\alpha,\beta)}$
acting in the space $\mathcal{L}_{\alpha}\otimes
\mathcal{L}_{\beta}$.\label{st:op}
\end{step}
Written down in components, this operator has four tensor indices,
$S^{(\alpha,\beta)ij}_{kl}$, the superscripts running in the space
of origin, $i\in \mathcal{L}_{\alpha}$, $j\in \mathcal{L}_q$ and
being related to the incoming strands, the subscripts running in the
space of image, $k\in \mathcal{L}_{\alpha}$, $l\in \mathcal{L}_q$
and being related to the outgoing strands; the first script in each
pair being related to the edge corresponding to the upper curve
segment in the planar projection, fig.\ref{fig:invcr}. We reflect
this fact by labeling each edge by an index running in the
corresponding $\mathcal{L}$. Alternatively, one may write the
operator with one superscript and one subscript,
$S^{(\alpha,\beta)I}_J$, both running in the space
$\mathcal{L}_{\alpha}\otimes \mathcal{L}_{\beta}$.

The general part of the construction is completed by assuming that
\begin{step}
Associate an inverse crossings (see fig.\ref{fig:invcr}) of polygons
$\alpha$ and $\beta$ with the operator $\tilde{S}^{(\alpha,\beta)}$
inverse to operator $S^{(\alpha,\beta)}$ on the space
$\mathcal{L}_{\alpha}\otimes \mathcal{L}_q$.\label{st:invop}
\end{step}
Other words, the operators $S^{(\alpha,\beta)}$ and
$\tilde{S}^{(\alpha,\beta)}$ by definition satisfy the identities in
line 3 of tab.\ref{tab:op}.

Once each edge of the cut knot diagram is labeled by a script,
\begin{step}
The cut knot diagram is related to a certain contraction of (yet
undefined) operators $S^{(\alpha,\beta)}$ and
$\tilde{S}^{(\alpha,\beta)}$, the contractions coinciding
identically for a pair of diagrams related a sequence of the second
Reidemeister moves (line 3 in tab.\ref{tab:av}).\label{st:contr}
\end{step}

\subsubsection{The $\mathcal{R}$-matrices as crossing operators.} The last step completes the general part of the
construction, we now turning to the specific $R$-matrix part of it.
Namely, one takes the following ansatz,
\begin{step}
A space $\mathcal{L}_{\alpha}$ is the space representation
$Q_{\alpha}$ of a Lie group $G$,\label{st:rep}
\end{step}
and
\begin{step}
The operator $S^{(\alpha,\beta)}$ is the corresponding
\emph{quantum}
$\mathcal{R}_{Q_{\alpha},Q_{\beta},G}(q)$\emph{-matrix}.\label{st:Rm}
\end{step}
The general definition of the quantum $\mathcal{R}$-matrix is given
above in sec.\ref{sec:Rdef} (rigorously speaking, $G$ is not a Lie
group but the corresponding \emph{quantum group}, but we ignore the
difference here modulo the remark in sec.\ref{sec:Rdef}). Here we
emphasize once again that the \emph{explicit}, although very
complicated expressions for arbitrary quantum $\mathcal{R}$-matrices
are available in representation theory \cite{KlimSch}.
Substituting one of the corresponding expressions for the crossing
operators in the previously obtained operator contraction, one
obtains that
\begin{step}
A cut knot diagram with $n$ incoming and $n$ outgoing edges
corresponds to an $(n,n)$ type tensor depending on the Lie group
$G$, its representation $Q$, and on the formal parameter $q$.\label{st:cuttens}
\end{step}
Moreover, \emph{by definition} of the $\mathcal{R}$-matrix,\label{st:cutinv}
\begin{step}
The same tensors correspond to any pair of cut knot diagrams related
by a sequence of the second (line 3 of tab.\ref{tab:op}) and third
(line 2 of tab.\ref{tab:op}) Reidemeister moves.
\end{step}
In the following, we concentrate on the particular case of the
representations $Q$ being the \textbf{fundamental} representation of
the $SU(N)$ group, keeping $N$ as a free parameter. The obtained
knot invariants are (uncolored) HOMFLY polynomials (see
sec.\ref{sec:knmath} and \ref{sec:stmod}) in the case, after the
substitution $A=q^N$ and analytic continuation to arbitrary complex
$A$ being done. Although the same knot polynomial can be obtained
equally in many other approaches (e.g., in the once discussed in
\ref{sec:stskein} and \ref{sec:braids}), we chose this simplest case
as an illustration to the $\mathcal{R}$-matrix approach, which, as
already mentioned, enables constructing the colored HOMFLY
polynomials for knots \cite{IMMM1,IMMM2,IMMM3,AMMM2,AM1,IndsMut,GuJock} and multi-colored HOMFLY polynomials for links
(taking higher representations of $SU(N)$ as the representation $Q$)
\cite{Inds6,Inds7,Inds10,AM1,MMM4,MMM5,GMMM}, as well as Kauffman
polynomials \cite{Kauff}, plain \cite{MorSm,Sal} and colored \cite{ChenChen} (considering the $SO(N)$ group instead of $SU(N)$), and even more general knot invariants \cite{colE8} (considering the exceptional groups; very few explicit answers is yet available for these cases).

\subsection{Explicit verifying the topological invariance constraints for the simplest $\mathcal{R}$-matrix \label{sec:Rexpl}}
In the present section, we complete presentation of the $\mathcal{R}$-matrix approach to the knot invariants by writing out explicitly
the form of these operators in the simplest case (the
obtained knot invariant is an (uncolored) HOMFLY polynomial in this case)., and
by verifying for them the constraints providing a topological
invariance of the contraction. According to the Reidemeister theorem
\cite{Pras}, three constraints are enough to impose; they are
illustrated in (\ref{fig:RII}),(\ref{fig:RIII}), and (\ref{fig:RI})
and referred to as Reidemeister moves. Each of these constraints is
associated with a singular transformation of a planar diagram
corresponding to non-singular transformation of the knot. The
theorem states that any two diagrams of the (topologically) same
knots can be transformed into each other by a combination of three
Reidemeister moves. Hence, any quantity that satisfy the three
corresponding constraints would be a knot invariant.

\subsubsection{Yang-Baxter equation}
The next constraint corresponds to the third Reidemeister move (in
the next coming evaluation, these are delta-symbols, not the
$R$-matrices, stand in all crossings; the crossings where
$R$-matrices still stand are labeled by dots): \be
\begin{picture}(100,90)(-30,10)
\put(0,0){\line(0,-1){15}}\put(0,0){\line(0,1){15}}
\put(0,15){\line(0,1){10}}\put(0,35){\line(0,1){10}}
\put(-5,-3){\line(-5,-3){10}}\put(5,3){\line(5,3){15}}
\put(-5,-3){\line(-5,-3){10}}\put(5,3){\line(5,3){15}}
\put(30,18){\line(5,3){10}}
\put(0,30){\line(-5,3){15}}\put(0,30){\line(5,-3){25}}
\put(25,15){\line(5,-3){15}}
\put(2,-20){$l$}\put(-12,-15){$n$} \put(2,40){$i$}\put(-12,40){$j$}
\put(37,10){$m$}\put(37,25){$k$}
\put(-10,10){$a$}\put(12,26){$b$}\put(12,-4){$c$}
\qbezier(0,12)(0,15)(1,18)\qbezier(0,12)(0,15)(-1,18)
\qbezier(10,6)(12,7)(13,9)\qbezier(10,6)(12,7)(14,7)
\qbezier(15,21)(13,22)(12,24)\qbezier(15,21)(13,22)(11,22)
\put(0,0){\circle*{2}}\put(0,30){\circle*{2}}\put(25,15){\circle*{2}}
\end{picture}&=&\begin{picture}(100,90)(-20,20)
\put(40,40){\line(0,1){15}}\put(40,40){\line(0,-1){15}}
\put(40,25){\line(0,-1){10}}\put(40,5){\line(0,-1){10}}
\put(45,43){\line(5,3){10}}\put(35,37){\line(-5,-3){15}}\put(10,22){\line(-5,-3){10}}
\put(40,10){\line(5,-3){15}}\put(40,10){\line(-5,3){25}}\put(15,25){\line(-5,3){15}}
\put(38,57){$i$}\put(52,52){$k$} \put(38,-15){$l$}\put(52,-10){$m$}
\put(3,35){$j$}\put(3,10){$n$}
\put(45,22){$a$}\put(22,35){$c$}\put(22,8){$b$}
\qbezier(40,22)(40,25)(41,29)\qbezier(40,22)(40,25)(39,29)
\qbezier(25,31)(27,32)(28,34)\qbezier(25,31)(27,32)(29,32)
\qbezier(25,19)(23,20)(22,22)\qbezier(25,19)(23,20)(21,20)
\put(40,40){\circle*{2}}\put(40,10){\circle*{2}}\put(15,25){\circle*{2}}
\end{picture}\nn\\[1cm]
R^{ji}_{ba}R^{bk}_{cm}R^{ac}_{nl}&=&R^{ik}_{ac}R^{jc}_{bn}R^{ba}_{ml}
\label{fig:RIII}\ee This constraint is called a Yang-Baxter
equation, and its solutions are studied in the integrable models
theory under the name of $R$-matrices \cite{Baxt}. In particular, a
one-parametric family of solutions is constructed explicitly for a
finite-dimensional representation of a regular Lie group
\cite{Jimbo}. Here we restrict ourselves by verifying the
Yang-Baxter equation for the simplest of the solution, which
corresponds to the fundamental representation of the group $su(N)$.
Non-zero elements of the examined solution are: \be R_{ii}^{ii}=q;\
\ R^{ij}_{ij}=1,\ \ i\ne j;\ \ \ R^{ij}_{ji}=q-q^{-1},\ \ 1\le
i<j\le N,\label{slnf} \ee where $N$ is an integer number, and $q$ is
a formal parameter of the family. This solution gives rise to
(uncolored) HOMFLY polynomials (see examples below).

\

Let us verify that (\ref{slnf}) satisfies the Yang-Baxter equation
explicitly. Since we write the solution in a certain basis, the
notations are \emph{not} covariant any longer; in particular,
\emph{no sum} after repeated indices is assumed default. Solution
(\ref{slnf}) can be presented as \be
R^{ij}_{kl}=a_{ij}\delta_i^j\delta_k^l+b_{ij}\delta_i^j\delta_k^l,\
\ a_{ij}=1+(q-1)\delta_{ij},\ \
b_{ij}=(q-q^{-1})\theta_{ij}\label{Rfund} \ee with
$\theta_{ij}\equiv 1$ for $i<j$, and $\theta_{ij}\equiv 0$
otherwise. Verifying of the YB equation can be then carried out
graphically. Four out of eight pairs of diagrams merely coincide:

\vspace{-2cm} \be
\begin{array}{rclrcl}
\begin{picture}(100,90)(-30,10)
\put(0,0){\line(0,-1){15}}\put(0,0){\line(0,1){15}}
\put(0,15){\line(0,1){10}}\put(0,35){\line(0,1){10}}
\put(-5,-3){\line(-5,-3){10}}\put(5,3){\line(5,3){15}}
\put(30,18){\line(5,3){10}}
\put(0,30){\line(-5,3){15}}\put(0,30){\line(5,-3){25}}
\put(25,15){\line(5,-3){15}}
\put(2,-20){$l$}\put(-12,-15){$n$} \put(2,40){$i$}\put(-12,40){$j$}
\put(37,10){$m$}\put(37,25){$k$}
\qbezier(0,12)(0,15)(1,18)\qbezier(0,12)(0,15)(-1,18)
\qbezier(10,6)(12,7)(13,9)\qbezier(10,6)(12,7)(14,7)
\qbezier(15,21)(13,22)(12,24)\qbezier(15,21)(13,22)(11,22)
\end{picture}&=&\begin{picture}(100,90)(-20,20)
\put(40,40){\line(0,1){15}}\put(40,40){\line(0,-1){15}}
\put(40,25){\line(0,-1){10}}\put(40,5){\line(0,-1){10}}
\put(45,43){\line(5,3){10}}\put(35,37){\line(-5,-3){15}}\put(10,22){\line(-5,-3){10}}
\put(40,10){\line(5,-3){15}}\put(40,10){\line(-5,3){25}}
\put(15,25){\line(-5,3){15}}
\put(38,57){$i$}\put(52,52){$k$} \put(38,-15){$l$}\put(52,-10){$m$}
\put(3,35){$j$}\put(3,10){$n$}
\qbezier(40,22)(40,25)(41,29)\qbezier(40,22)(40,25)(39,29)
\qbezier(25,31)(27,32)(28,34)\qbezier(25,31)(27,32)(29,32)
\qbezier(25,19)(23,20)(22,22)\qbezier(25,19)(23,20)(21,20)
\end{picture}&
\begin{picture}(100,90)(-30,10)
\qbezier(-10,-6)(0,0)(0,15)\qbezier(35,21)(25,15)(35,9)\qbezier(-10,36)(0,30)(0,15)
\qbezier(0,-15)(0,0)(15,9)\qbezier(0,45)(0,30)(10,24)\qbezier(10,24)(25,15)(15,9)
\put(2,-20){$l$}\put(-12,-15){$n$} \put(2,40){$i$}\put(-12,40){$j$}
\put(37,10){$m$}\put(37,25){$k$}
\qbezier(0,12)(0,15)(1,18)\qbezier(0,12)(0,15)(-1,18)
\qbezier(19,13)(18,15)(19,19)\qbezier(19,13)(20,15)(16,18)
\qbezier(31,12)(30,15)(32,17)\qbezier(31,12)(30,15)(29,18)
\end{picture}&=&
\begin{picture}(100,90)(-20,20)
\qbezier(50,46)(40,40)(40,25)\qbezier(5,19)(15,25)(5,31)\qbezier(50,4)(40,10)(40,25)
\qbezier(40,55)(40,40)(25,31)\qbezier(40,-5)(40,10)(30,16)\qbezier(30,16)(15,25)(25,31)
\put(38,57){$i$}\put(52,52){$k$} \put(38,-15){$l$}\put(52,-10){$m$}
\put(3,35){$j$}\put(3,10){$n$}
\qbezier(40,22)(40,25)(41,29)\qbezier(40,22)(40,25)(39,29)
\qbezier(22,23)(20,25)(23,28)\qbezier(22,23)(20,25)(21,29)
\qbezier(9,22)(11,24)(11,27)\qbezier(9,22)(10,25)(8,27)
\end{picture}\\[1cm]
a_{ji}a_{jk}a_{ik}&=&a_{kj}a_{ik}a_{ji}&b_{ji}^2b_{ik}&=&b_{ij}b_{ik}^2
\end{array}
\ee

\vspace{-2cm}

\be
\begin{array}{rclrcl}
\begin{picture}(100,90)(-30,10)
\put(0,0){\line(0,-1){15}}\put(0,0){\line(0,1){15}}
\put(-5,-3){\line(-5,-3){10}}\put(5,3){\line(5,3){15}}\put(30,18){\line(5,3){10}}
\put(25,15){\line(5,-3){15}}
\qbezier(-10,36)(0,30)(0,15)
\qbezier(0,45)(0,30)(25,15)
\put(2,-20){$l$}\put(-12,-15){$n$} \put(2,40){$i$}\put(-12,40){$j$}
\put(37,10){$m$}\put(37,25){$k$}
\qbezier(0,12)(0,15)(1,18)\qbezier(0,12)(0,15)(-1,18)
\qbezier(10,6)(12,7)(13,9)\qbezier(10,6)(12,7)(14,7)
\qbezier(15,22)(13,23)(13,25)\qbezier(16,21)(12,23)(11,23)
\end{picture}&=&\begin{picture}(100,90)(-20,20)
\put(40,40){\line(0,1){15}}\put(40,40){\line(0,-1){15}}
\put(45,43){\line(5,3){10}}\put(35,37){\line(-5,-3){15}}\put(10,22){\line(-5,-3){10}}
\put(15,25){\line(-5,3){15}}
\qbezier(50,4)(40,10)(40,25)
\qbezier(40,-5)(40,10)(15,25)
\put(38,57){$i$}\put(52,52){$k$} \put(38,-15){$l$}\put(52,-10){$m$}
\put(3,35){$j$}\put(3,10){$n$}
\qbezier(40,22)(40,25)(41,29)\qbezier(40,22)(40,25)(39,29)
\qbezier(25,31)(27,32)(28,34)\qbezier(25,31)(27,32)(29,32)
\qbezier(25,18)(23,20)(22,22)\qbezier(25,18)(23,20)(21,20)
\end{picture}
&\begin{picture}(100,90)(-30,10)
\put(0,15){\line(0,1){10}}\put(0,35){\line(0,1){10}}
\put(30,18){\line(5,3){10}}
\put(0,30){\line(-5,3){15}}\put(0,30){\line(5,-3){25}}
\put(25,15){\line(5,-3){15}}
\qbezier(-10,-6)(0,0)(0,15)
\qbezier(0,-15)(0,0)(20,12)
\put(2,-20){$l$}\put(-12,-15){$n$} \put(2,40){$i$}\put(-12,40){$j$}
\put(37,10){$m$}\put(37,25){$k$}
\qbezier(0,12)(0,15)(1,18)\qbezier(0,12)(0,15)(-1,18)
\qbezier(10,5)(12,7)(13,9)\qbezier(10,5)(13,7)(15,7)
\qbezier(15,21)(13,22)(12,24)\qbezier(15,21)(13,22)(11,22)
\end{picture}&=&\begin{picture}(100,90)(-20,20)
\put(40,25){\line(0,-1){10}}\put(40,5){\line(0,-1){10}}
\put(10,22){\line(-5,-3){10}}
\put(40,10){\line(5,-3){15}}\put(40,10){\line(-5,3){25}}
\put(15,25){\line(-5,3){15}} \qbezier(50,46)(40,40)(40,25)
\qbezier(40,55)(40,40)(20,28)
\put(38,57){$i$}\put(52,52){$k$} \put(38,-15){$l$}\put(52,-10){$m$}
\put(3,35){$j$}\put(3,10){$n$}
\qbezier(40,22)(40,25)(41,29)\qbezier(40,22)(40,25)(39,29)
\qbezier(25,31)(27,32)(28,35)\qbezier(25,31)(27,32)(29,33)
\qbezier(25,19)(23,20)(22,22)\qbezier(25,19)(23,20)(21,20)
\end{picture}
\\[1cm]
a_{jk}a_{ik}b_{ji}&=&a_{jk}a_{ik}b_{ji}&a_{ji}a_{jk}b_{ik}&=&a_{ji}a_{jk}b_{ik}
\end{array}\ee

One more pair coincides unless $i=j<k$ or $j<k=i$:

\vspace{-2cm}

\be
\begin{picture}(100,90)(-30,10)
\put(0,0){\line(0,-1){15}}\put(0,0){\line(0,1){15}}
\put(0,15){\line(0,1){10}}\put(0,35){\line(0,1){10}}
\put(-5,-3){\line(-5,-3){10}}
\put(-15,39){\line(5,-3){20}}
\qbezier(35,21)(25,15)(35,9)
\qbezier(5,27)(25,15)(5,3) \put(2,-20){$l$}\put(-12,-15){$n$}
\put(2,40){$i$}\put(-12,40){$j$} \put(37,10){$m$}\put(37,25){$k$}
\qbezier(0,12)(0,15)(1,18)\qbezier(0,12)(0,15)(-1,18)
\qbezier(14,12)(15,15)(16,18)\qbezier(14,12)(15,15)(13,18)
\qbezier(31,12)(30,15)(32,17)\qbezier(31,12)(30,15)(29,18)
\end{picture}&=&
\begin{picture}(100,90)(-20,20)
\put(40,40){\line(0,1){15}}\put(40,40){\line(0,-1){15}}
\put(40,25){\line(0,-1){10}}\put(40,5){\line(0,-1){10}}
\put(45,43){\line(5,3){10}}
\put(35,13){\line(5,-3){20}}
\qbezier(5,19)(15,25)(5,31)
\qbezier(35,13)(15,25)(35,37) \put(38,57){$i$}\put(52,52){$k$}
\put(38,-15){$l$}\put(52,-10){$m$} \put(3,35){$j$}\put(3,10){$n$}
\qbezier(40,22)(40,25)(41,29)\qbezier(40,22)(40,25)(39,29)
\qbezier(25,23)(24,25)(27,28)\qbezier(25,23)(24,25)(24,28)
\qbezier(9,22)(11,24)(11,27)\qbezier(9,22)(10,25)(8,27)
\end{picture}\mbox{          if not }i=j<k\mbox{ or }j<k=i\label{fig:YB1}\\[1cm]
a_{ij}a_{ji}b_{jk}&=&a_{ik}a_{ki}b_{kj}\nn \ee and the three
remaining ones groups to coincide \vspace{-2cm} \be
\begin{array}{rcrclc}
\begin{picture}(100,90)(-30,10)
\put(0,0){\line(0,-1){15}}\put(0,0){\line(0,1){15}}
\put(-5,-3){\line(-5,-3){10}}
\qbezier(-10,36)(0,30)(0,15)\qbezier(35,21)(25,15)(35,9)
\qbezier(0,45)(0,30)(10,24)\qbezier(10,24)(25,15)(5,3)
\put(2,-20){$l$}\put(-12,-15){$n$} \put(2,40){$i$}\put(-12,40){$j$}
\put(37,10){$m$}\put(37,25){$k$}
\qbezier(0,12)(0,15)(1,18)\qbezier(0,12)(0,15)(-1,18)
\qbezier(16,13)(17,15)(17,19)\qbezier(16,13)(18,15)(14,18)
\qbezier(31,12)(30,15)(32,17)\qbezier(31,12)(30,15)(29,18)
\end{picture}&+&\begin{picture}(100,90)(-30,10)
\put(0,15){\line(0,1){10}}\put(0,35){\line(0,1){10}}
\put(-15,39){\line(5,-3){20}}
\qbezier(-10,-6)(0,0)(0,15)
\qbezier(35,21)(25,15)(35,9)
\qbezier(0,-15)(0,0)(15,9)
\qbezier(5,27)(25,15)(15,9) \put(2,-20){$l$}\put(-12,-15){$n$}
\put(2,40){$i$}\put(-12,40){$j$} \put(37,10){$m$}\put(37,25){$k$}
\qbezier(0,12)(0,15)(1,18)\qbezier(0,12)(0,15)(-1,18)
\qbezier(18,12)(18,15)(19,17)\qbezier(18,12)(18,15)(16,17)
\qbezier(31,12)(30,15)(32,17)\qbezier(31,12)(30,15)(29,18)
\end{picture}&=&\begin{picture}(100,90)(-20,20)
\put(10,22){\line(-5,-3){10}}
\put(15,25){\line(-5,3){15}}
\qbezier(50,46)(40,40)(40,25)\qbezier(50,4)(40,10)(40,25)
\qbezier(40,55)(40,40)(20,28)\qbezier(40,-5)(40,10)(15,25)
\put(38,57){$i$}\put(52,52){$k$} \put(38,-15){$l$}\put(52,-10){$m$}
\put(3,35){$j$}\put(3,10){$n$}
\qbezier(40,22)(40,25)(41,29)\qbezier(40,22)(40,25)(39,29)
\qbezier(25,31)(27,32)(28,35)\qbezier(25,31)(27,32)(29,33)
\qbezier(25,18)(23,20)(22,22)\qbezier(25,18)(23,20)(21,20)
\end{picture}&\mbox{if not }i=j<k\label{fig:YB2}\\[1cm]
a_{ij}b_{ji}b_{ik}&+&a_{ji}b_{ij}b_{jk}&=&a_{ji}b_{ik}b_{jk}
\end{array}
\ee and \vspace{-2cm} \be
\begin{array}{rclclc}
\begin{picture}(100,90)(-30,10)
\put(30,18){\line(5,3){10}}
\put(25,15){\line(5,-3){15}}
\qbezier(-10,-6)(0,0)(0,15)\qbezier(-10,36)(0,30)(0,15)
\qbezier(0,-15)(0,0)(20,12)\qbezier(0,45)(0,30)(25,15)
\put(2,-20){$l$}\put(-12,-15){$n$} \put(2,40){$i$}\put(-12,40){$j$}
\put(37,10){$m$}\put(37,25){$k$}
\qbezier(0,12)(0,15)(1,18)\qbezier(0,12)(0,15)(-1,18)
\qbezier(10,5)(12,7)(13,9)\qbezier(10,5)(13,7)(15,7)
\qbezier(15,22)(13,23)(13,25)\qbezier(16,21)(12,23)(11,23)
\end{picture}
&=&\begin{picture}(100,90)(-20,20)
\put(40,25){\line(0,-1){10}}\put(40,5){\line(0,-1){10}}
\put(35,13){\line(5,-3){20}}
\qbezier(50,46)(40,40)(40,25)
\qbezier(5,19)(15,25)(5,31)
\qbezier(40,55)(40,40)(25,31)
\qbezier(35,13)(15,25)(25,31) \put(38,57){$i$}\put(52,52){$k$}
\put(38,-15){$l$}\put(52,-10){$m$} \put(3,35){$j$}\put(3,10){$n$}
\qbezier(40,22)(40,25)(41,29)\qbezier(40,22)(40,25)(39,29)
\qbezier(22,24)(21,25)(23,28)\qbezier(22,24)(21,25)(21,29)
\qbezier(9,22)(11,24)(11,27)\qbezier(9,22)(10,25)(8,27)
\end{picture}&+&\begin{picture}(100,90)(-20,20)
\put(40,40){\line(0,1){15}}\put(40,40){\line(0,-1){15}}
\put(45,43){\line(5,3){10}}
\qbezier(50,4)(40,10)(40,25)\qbezier(5,19)(15,25)(5,31)
\qbezier(40,-5)(40,10)(30,16)\qbezier(30,16)(15,25)(35,37)
\put(38,57){$i$}\put(52,52){$k$} \put(38,-15){$l$}\put(52,-10){$m$}
\put(3,35){$j$}\put(3,10){$n$}
\qbezier(40,22)(40,25)(41,29)\qbezier(40,22)(40,25)(39,29)
\qbezier(24,22)(23,25)(26,28)\qbezier(24,22)(23,25)(24,29)
\qbezier(9,22)(11,24)(11,27)\qbezier(9,22)(10,25)(8,27)
\end{picture}&\mbox{if not }j<i=k\label{fig:YB3}\\[1cm]
a_{ik}b_{ji}b_{jk}&=&a_{ik}b_{ji}b_{ik}&+&a_{ik}b_{jk}b_{ki}\nn
\end{array}
\ee

due to the identities
$\theta_{ji}\theta_{ik}+\theta_{ij}\theta_{jk}=\theta_{ik}\theta_{jk}$
and
$\theta_{ji}\theta_{ik}+\theta_{jk}\theta_{ki}=\theta_{ji}\theta_{jk}$,
which hold unless $i=j<k$, or $j<i=k$, correspondingly. The cases
$i=j<k$ and $j<k=i$ need for a separate treatment; one has to take
into account that not all diagrams (\ref{fig:YB1}-\ref{fig:YB3}) are
independent for some of $i$, $j$, $k$ coinciding. For $i=j<k$, the
diagrams (\ref{fig:YB1}) and (\ref{fig:YB2}) contribute to the same
component
$\arraycolsep=0mm\footnotesize{\Big(\begin{array}{ccc}i&i&k\\l&n&m\end{array}\Big)}$;
adding the equalities termwise, one gets correct in the case in case
identity, \be
\begin{array}{rclcl}
&\boxed{i=j<k}\\[-2cm]
\begin{picture}(100,90)(-30,10)
\put(0,0){\line(0,-1){15}}\put(0,0){\line(0,1){15}}
\put(0,15){\line(0,1){10}}\put(0,35){\line(0,1){10}}
\put(-5,-3){\line(-5,-3){10}}
\put(-15,39){\line(5,-3){20}}
\qbezier(35,21)(25,15)(35,9)
\qbezier(5,27)(25,15)(5,3) \put(2,-20){$l$}\put(-12,-15){$n$}
\put(2,40){$i$}\put(-12,40){$i$} \put(37,10){$m$}\put(37,25){$k$}
\qbezier(0,12)(0,15)(1,18)\qbezier(0,12)(0,15)(-1,18)
\qbezier(14,12)(15,15)(16,18)\qbezier(14,12)(15,15)(13,18)
\qbezier(31,12)(30,15)(32,17)\qbezier(31,12)(30,15)(29,18)
\end{picture}&=&
\begin{picture}(100,90)(-20,20)
\put(40,40){\line(0,1){15}}\put(40,40){\line(0,-1){15}}
\put(40,25){\line(0,-1){10}}\put(40,5){\line(0,-1){10}}
\put(45,43){\line(5,3){10}}
\put(35,13){\line(5,-3){20}}
\qbezier(5,19)(15,25)(5,31)
\qbezier(35,13)(15,25)(35,37) \put(38,57){$i$}\put(52,52){$k$}
\put(38,-15){$l$}\put(52,-10){$m$} \put(3,35){$i$}\put(3,10){$n$}
\qbezier(40,22)(40,25)(41,29)\qbezier(40,22)(40,25)(39,29)
\qbezier(25,23)(24,25)(27,28)\qbezier(25,23)(24,25)(24,28)
\qbezier(9,22)(11,24)(11,27)\qbezier(9,22)(10,25)(8,27)
\end{picture}&+&\begin{picture}(100,90)(-20,20)
\put(10,22){\line(-5,-3){10}}
\put(15,25){\line(-5,3){15}}
\qbezier(50,46)(40,40)(40,25)\qbezier(50,4)(40,10)(40,25)
\qbezier(40,55)(40,40)(20,28)\qbezier(40,-5)(40,10)(15,25)
\put(38,57){$i$}\put(52,52){$k$} \put(38,-15){$l$}\put(52,-10){$m$}
\put(3,35){$i$}\put(3,10){$n$}
\qbezier(40,22)(40,25)(41,29)\qbezier(40,22)(40,25)(39,29)
\qbezier(25,31)(27,32)(28,35)\qbezier(25,31)(27,32)(29,33)
\qbezier(25,18)(23,20)(22,22)\qbezier(25,18)(23,20)(21,20)
\end{picture}\\[1cm]
a_{ii}^2b_{ik}&=&a_{ik}a_{ki}b_{ik}&+&a_{ii}b_{ik}^2
\end{array}
\ee\vspace{1cm} due to the identity $q^2=1+q(q-q^{-1})$. Similarly,
for $i=j<k$, equality takes place for the sums of diagrams
(\ref{fig:YB2}) and (\ref{fig:YB3}) contributing to the component
$\arraycolsep=0mm\footnotesize{\Big(\begin{array}{ccc}i&j&i\\l&n&m\end{array}\Big)}$,
\be
\begin{array}{rcrcl}
&\boxed{j<i=k}\\[-2cm]
\begin{picture}(100,90)(-30,10)
\put(0,0){\line(0,-1){15}}\put(0,0){\line(0,1){15}}
\put(0,15){\line(0,1){10}}\put(0,35){\line(0,1){10}}
\put(-5,-3){\line(-5,-3){10}}
\put(-15,39){\line(5,-3){20}}
\qbezier(35,21)(25,15)(35,9)
\qbezier(5,27)(25,15)(5,3) \put(2,-20){$l$}\put(-12,-15){$n$}
\put(2,40){$i$}\put(-12,40){$j$} \put(37,10){$m$}\put(37,25){$i$}
\qbezier(0,12)(0,15)(1,18)\qbezier(0,12)(0,15)(-1,18)
\qbezier(14,12)(15,15)(16,18)\qbezier(14,12)(15,15)(13,18)
\qbezier(31,12)(30,15)(32,17)\qbezier(31,12)(30,15)(29,18)
\end{picture}&+&
\begin{picture}(100,90)(-30,10)
\put(30,18){\line(5,3){10}}
\put(25,15){\line(5,-3){15}}
\qbezier(-10,-6)(0,0)(0,15)\qbezier(-10,36)(0,30)(0,15)
\qbezier(0,-15)(0,0)(20,12)\qbezier(0,45)(0,30)(25,15)
\put(2,-20){$l$}\put(-12,-15){$n$} \put(2,40){$i$}\put(-12,40){$j$}
\put(37,10){$m$}\put(37,25){$k$}
\qbezier(0,12)(0,15)(1,18)\qbezier(0,12)(0,15)(-1,18)
\qbezier(10,5)(12,7)(13,9)\qbezier(10,5)(13,7)(15,7)
\qbezier(15,22)(13,23)(13,25)\qbezier(16,21)(12,23)(11,23)
\end{picture}
&=&
\begin{picture}(100,90)(-20,20)
\put(40,40){\line(0,1){15}}\put(40,40){\line(0,-1){15}}
\put(40,25){\line(0,-1){10}}\put(40,5){\line(0,-1){10}}
\put(45,43){\line(5,3){10}}
\put(35,13){\line(5,-3){20}}
\qbezier(5,19)(15,25)(5,31)
\qbezier(35,13)(15,25)(35,37) \put(38,57){$i$}\put(52,52){$i$}
\put(38,-15){$l$}\put(52,-10){$m$} \put(3,35){$j$}\put(3,10){$n$}
\qbezier(40,22)(40,25)(41,29)\qbezier(40,22)(40,25)(39,29)
\qbezier(25,23)(24,25)(27,28)\qbezier(25,23)(24,25)(24,28)
\qbezier(9,22)(11,24)(11,27)\qbezier(9,22)(10,25)(8,27)
\end{picture}\\[1cm]
a_{ji}a_{ij}b_{ji}&+&a_{ii}b_{ji}^2&=&b_{ji}a_{ii}
\end{array}
\ee\vspace{1cm}



\subsubsection{Inverse crossings\label{sec:RII}}
The next (in fact, the most simple out of the three ones) constraint
reflects the invariance under the second Reidemeister move,
(\ref{fig:RII}). \be
\begin{array}{rclrcl}
\begin{picture}(60,60)(-30,-5)
\qbezier(-12,30)(3,0)(-12,-30)
\qbezier(-7,10)(-9,0)(-7,-10)\qbezier(0,30)(-3,23)(-5,16)\qbezier(0,-30)(-3,-23)(-5,-16)
\qbezier(-3,22)(-2,25)(0,26)\qbezier(-3,22)(-2,25)(-3,27)
\qbezier(-9,23)(-10,26)(-12,27)\qbezier(-9,23)(-10,26)(-9,28)
\qbezier(-10,-26)(-9,-23)(-7,-22)\qbezier(-10,-26)(-9,-23)(-10,-21)
\qbezier(-1,-28)(-2,-25)(-4,-24)\qbezier(-1,-28)(-2,-25)(-1,-23)
\put(1,-29){$l$}\put(1,22){$j$}\put(-17,21){$i$}\put(-18,-29){$k$}
\put(-15,0){$b$}\put(-2,0){$a$}
\end{picture}&=&
\begin{picture}(60,60)(-30,-5)
\qbezier(12,30)(-3,0)(12,-30) \qbezier(-12,30)(3,0)(-12,-30)
\qbezier(-9,23)(-10,26)(-12,27)\qbezier(-9,23)(-10,26)(-9,28)
\qbezier(-10,-26)(-9,-23)(-7,-22)\qbezier(-10,-26)(-9,-23)(-10,-21)
\qbezier(9,23)(10,26)(12,27)\qbezier(9,23)(10,26)(9,28)
\qbezier(10,-26)(9,-23)(7,-22)\qbezier(10,-26)(9,-23)(10,-21)
\put(12,-29){$l$}\put(12,22){$j$}\put(-17,21){$i$}\put(-18,-29){$k$}
\end{picture},
\\[1cm]
\sum_{a,b}R^{ij}_{ab}\tilde{R}^{ab}_{kl}&=&\delta^i_k\delta^j_l.\label{fig:RII}
\end{array}\ee

The condition gives rise to the constraints, which relates the
operators corresponding to the crossings of the mutually inverse
orientations. Because (\ref{fig:RII}) is a system of $N^2$ linear
equations on $N^2$ variables (which can be verified to
non-degenerate), matrix elements of the inverse crossing operators
are determined therefrom explicitly and unambiguously. The most
simple is to write for them the ansatz, similar to expression
(\ref{Rfund}), \be
\tilde{R}^{ij}_{kl}=\tilde{a}_{ij}\delta^i_k\delta^j_l+\tilde{b}_{ij}\delta^i_l\delta^j_k.\label{invRanz}
\ee Equations (\ref{fig:RII}) then take the form \be
\sum_{p,q}\tilde{R}^{ij}_{pq}R^{pq}_{kl}=
\big(\tilde{a}_{ij}a_{ij}+\tilde{b}_{ij}b_{ji}\big)\delta^i_k\delta^j_l+
\big(\tilde{a}_{ij}b_{ij}+\tilde{b}_{ij}a_{ji}\big)\delta^i_l\delta^j_k=\delta^i_k\delta^j_l.
\label{RIIcomps}\ee With $a_{ij}$ and $b_{ij}$ from (\ref{Rfund}),
there are three distinct non-trivial cases, which give (note that
$\tilde{b}_{ii}=0$ by definition) \be
\begin{array}{rcccl}
i=k\le j=l&\Rightarrow&\tilde{a}_{ij}a_{ij}=1&\Rightarrow&\tilde{a}_{ii}=q^{-1},\ \ \tilde{a}_{ij}=1,\  i>j\\
i=k>j=l&\Rightarrow&\tilde{a}_{ij}a_{ij}+\tilde{b}_{ij}b_{ji}=1&\Rightarrow&\tilde{b}_{ij}=0,\ \ i>j,\\
i=l<j=k&\Rightarrow&\tilde{a}_{ij}b_{ij}+\tilde{b}_{ij}a_{ji}=0&\Rightarrow&\tilde{b}_{ij}=q^{-1}-q.\label{invRcompts}\\
\end{array}
\label{invRsol}\ee Yang-Baxter equation (\ref{fig:RIII}) holds for
the obtained solution automatically, since the matrix elements for
the inverse crossing operators appeared to be related with that for
the direct ones by a plain change of parameter,
$\tilde{a}_{ij}(q)=a_{ij}(q^{-1})$ and
$\tilde{b}_{ij}(q)=b_{ij}(q^{-1})$, while (\ref{Rfund}) satisfies
(\ref{fig:RIII}) for an \emph{arbitrary} value of $q$.

Thereby, if one puts operators (\ref{Rfund}) in the direct
crossings, one should must put the operators (recall that
$\theta_{ij}=1$ for $i<j$, and $\theta_{ij}=0$ for $i\ge j$) \be
R^{ij}_{kl}=\tilde{a}_{ij}\delta_i^j\delta_k^l+\tilde{b}_{ij}\delta_i^j\delta_k^l,\
\ \tilde{a}_{ij}=1+(q^{-1}-1)\delta_{ij},\ \
\tilde{b}_{ij}=(q^{-1}-q)\theta_{ij}\label{invRfund} \ee in the
inverse crossings.

\subsection{First Reidemeister move and turn-over operators}
\subsubsection{RI invariance as a condition on the $R$-matrix contraction}
One more transformation of the knot, which is smooth though looking
singular at a planar projection, is contracting of a loop: \be
\begin{picture}(60,60)(-30,-5)
\put(0,0){\oval(24,24)[r]} \put(0,0){\oval(24,24)[lb]}
\qbezier(0,12)(-7,10)(-6,-3) \put(-12,0){\line(0,1){24}}
\put(-6,-15){\line(0,-1){15}} \qbezier(12,0)(11,-5)(12,-7)
\qbezier(12,0)(12,-4)(8,-6)
\qbezier(-12,12)(-12,15)(-11,18)\qbezier(-12,12)(-12,15)(-13,18)
\qbezier(-6,-28)(-6,-25)(-5,-22)\qbezier(-6,-28)(-6,-25)(-7,-22)
\put(-18,16){$i$}
\put(-15,-30){$k$} \put(-5,16){$a$}\put(-2,-22){$b$}
\end{picture}&=&
\begin{picture}(60,60)(-30,-5)
\put(-12,20){\line(0,-1){50}}
\qbezier(-12,-8)(-12,-5)(-11,-2)\qbezier(-12,-8)(-12,-5)(-13,-2)
\put(-10,16){$i$}\put(-10,-30){$k$}
\end{picture}\nn\\[1cm]
\sum_{a,b}\mathcal{M}^b_aR^{ia}_{bk}&=&\delta^i_k\label{fig:RI} \ee
The above rules relate the left and right figures with the partial
contraction of the $R$-matrix $\sum_{a,b}R^{ia}_{bk}$ and with the
unity operator $\delta^i_k$, respectively. It is straightforward to
verify that these expressions are not equal; the $R$-matrix do not
possess the desired property.

\subsubsection{Turn-over operators}

To make a contraction of the $R$-matrices invariant under
transformation (\ref{fig:RI}) as well, one introduces one more
element in the construction
\cite{ReshTur},\cite{KirResh},\cite{MorSm}. Namely, the scripts of
$R$-matrix are contracted now with help of a new operator
$\mathcal{M}$, which we call \emph{a turn-over operator}. Equality
(\ref{fig:RI}) is a definition of the operator $\mathcal{M}$, and it
enables one to determine elements of the turn-over operator
explicitly.

\paragraph{Explicit expression for the turn-over operators\label{sec:Mop}}

We will use the explicit expression (\ref{Rfund}) for elements of
the $R$-matrix in a selected basis, and we suppose that the operator
$\mathcal{M}$ is diagonal in the same basis,
\be\mathcal{M}_j^i=m_i\delta_j^i;\ee as we will verify, such a
solution exists, and it is unique, as follows from the dimension
counting. Let us start from the particular values of $N$ in
(\ref{Rfund}). For $N=2$, constraints (\ref{fig:RI}) take the
explicit form \be \left\{\begin{array}{lcl}
m_1R^{11}_{11}+m_2R^{12}_{21}&=&1,\\
m_2R^{22}_{22}&=&1
\end{array}\right.
\ \Rightarrow\ \ \left\{\begin{array}{lcl}
qm_1+(q-q^{-1})m_2&=&1,\\
qm_2&=&1
\end{array}\right.,\label{Meq2}
\ee and have the solution \be m_1=q^{-3},\ \ m_2=q^{-1}\label{Msol2}
\ee For $N=3$, one has the system \be \left\{\begin{array}{lcl}
m_1R^{11}_{11}+m_2R^{12}_{21}+m_3R^{13}_{31}&=&1,\\
m_2R^{22}_{22}+m_3R^{23}_{32}&=&1,\\
m_3R^{33}_{33}&=&1
\end{array}\right.
\ \Rightarrow\ \ \left\{\begin{array}{lcl}
qm_1+(q-q^{-1})\left(m_2+m_3\right)&=&1,\\
qm_2+(q-q^{-1})m_3&=&1,\\
qm_3&=&1,
\end{array}\right.
\label{Meq3}\ee the solution being \be m_1=q^{-5},\ \ m_2=q^{-3},\ \
m_3=q^{-1}.\label{Msol3} \ee Now it is easy to write down both the
constraints and their solution for generic $N$. Equations
(\ref{fig:RI}) can be rewritten as \be
m_iR^{ii}_{ii}+\sum_{j=i+1}^Nm_jR^{ij}_{ji}=1\ \ \Rightarrow\ \
qm_i+(q-q^{-1})\sum_{j=i+1}^Nm_j=1,\label{meq} \ee wherefrom one
expresses explicitly the non-zero elements of the turn-over
operator, \be \boxed{m_i=q^{2i-2N-1},\ \ \
i=1,\ldots,N.}\label{Msol} \ee A contraction of the $R$-matrix in
the other two scripts, in turn, corresponds contracting loop
(\ref{fig:loops}-II), and should be carried out with help of another
operator, $\mathcal{M}^{\prime}$, which is determined by the system
of constraints \be
\sum_{a,b}\mathcal{M}^{\prime b}_aR^{aj}_{lb}&=&\delta^j_l.\ee

\be
\begin{picture}(60,60)(0,-36)
\qbezier(0,24)(6,18)(12,12)\qbezier(12,12)(18,6)(24,0)
\qbezier(24,24)(18,18)(16,16)\qbezier(8,8)(6,6)(0,0)
\qbezier(24,0)(36,-12)(36,12)\qbezier(24,24)(36,36)(36,12)
\qbezier(0,0)(2,2)(2,5)\qbezier(0,0)(2,2)(4,2)
\qbezier(24,0)(22,2)(22,5)\qbezier(24,0)(22,2)(20,2)
\put(0,-18){$\mathcal{M}R$}\put(12,-33){I}
\end{picture}
\begin{picture}(60,60)(-12,-36)
\qbezier(0,24)(6,18)(12,12)\qbezier(12,12)(18,6)(24,0)
\qbezier(24,24)(18,18)(16,16)\qbezier(8,8)(6,6)(0,0)
\qbezier(0,0)(-12,-12)(-12,12)\qbezier(0,24)(-12,36)(-12,12)
\qbezier(0,0)(2,2)(2,5)\qbezier(0,0)(2,2)(4,2)
\qbezier(24,0)(22,2)(22,5)\qbezier(24,0)(22,2)(20,2)
\put(-4,-18){$\mathcal{M}^{\prime}R$}\put(6,-33){II}
\end{picture}
\begin{picture}(60,60)(0,-36)
\qbezier(0,24)(6,18)(8,16)\qbezier(16,8)(18,6)(24,0)
\qbezier(24,24)(18,18)(12,12)\qbezier(12,12)(6,6)(0,0)
\qbezier(24,0)(36,-12)(36,12)\qbezier(24,24)(36,36)(36,12)
\qbezier(0,0)(2,2)(2,5)\qbezier(0,0)(2,2)(4,2)
\qbezier(24,0)(22,2)(22,5)\qbezier(24,0)(22,2)(20,2)
\put(0,-18){$\tilde{\mathcal{M}}\tilde{R}$}\put(6,-33){III}
\end{picture}
\begin{picture}(60,60)(-12,-36)
\qbezier(0,24)(6,18)(8,16)\qbezier(16,8)(18,6)(24,0)
\qbezier(24,24)(18,18)(12,12)\qbezier(12,12)(6,6)(0,0)
\qbezier(0,0)(-12,-12)(-12,12)\qbezier(0,24)(-12,36)(-12,12)
\qbezier(0,0)(2,2)(2,5)\qbezier(0,0)(2,2)(4,2)
\qbezier(24,0)(22,2)(22,5)\qbezier(24,0)(22,2)(20,2)
\put(-4,-18){$\tilde{\mathcal{M}}^{\prime}\tilde{R}$}\put(3,-33){IV}
\end{picture}
\label{fig:loops} \ee E.g., the system for $N=2$ takes the explicit
form \be \left\{\begin{array}{rcl}
m^{\prime}_1R^{11}_{11}&=&1,\\
m^{\prime}_1R^{12}_{21}+m^{\prime}_2R^{22}_{22}&=&1,
\end{array}\right.
\ \Rightarrow\ \ \left\{\begin{array}{rcl}
qm^{\prime}_1&=&1,\\
(q-q^{-1})m^{\prime}_1+qm^{\prime}_2&=&1,
\end{array}\right.
\ee and it has the solution \be m^{\prime}_1=q^{-1},\ \
m^{\prime}_2=q^{-3}.\label{Mtsol2} \ee For $N=3$, one has \be
\left\{\begin{array}{rcl}
m^{\prime}_1R^{11}_{11}&=&1,\\
m^{\prime}_1R^{12}_{21}+m^{\prime}_2R^{22}_{22}&=&1,\\
m^{\prime}_1R^{13}_{31}+m^{\prime}_2R^{23}_{32}+m^{\prime}_3R^{33}_{33}&=&1,
\end{array}\right.
\ \Rightarrow\ \ \left\{\begin{array}{rcl}
qm^{\prime}_1&=&1,\\
(q-q^{-1})m^{\prime}_1+qm^{\prime}_2&=&1,\\
(q-q^{-1})\left(m^{\prime}_1+m^{\prime}_2\right)+qm^{\prime}_3&=&1,\\
\end{array}\right.
\ee what gives \be m^{\prime}_1=q^{-1},\ \ m^{\prime}_2=q^{-3},\ \
m^{\prime}_3=q^{-5}.\label{Mtsol3} \ee Finally, the constraints on
the non-zero matrix elements and their solutions for generic $N$
are, respectively, \be
\sum_{j=1}^{i-1}m^{\prime}_jR^{ji}_{ij}+m^{\prime}_iR^{ii}_{ii}=1\ \
\Rightarrow\ \
(q-q^{-1})\sum_{j=1}^{i-1}m^{\prime}_j+qm^{\prime}_i=1,\label{mteq}
\ee and \be \boxed{m^{\prime}_j=q^{-2i+1},\ \ \
i=1,\ldots,N.}\label{Mtsol} \ee

\paragraph{Loops attached to the inverse vertices.}

To complete the construction under discussion, One also should
introduce the operators $\tilde{\mathcal{M}}$ and
$\tilde{\mathcal{M}}^{\prime}$ to be inserted in contractions of the
\emph{inverse} $R$-matrix in one and in the other pairs of scripts,
and associated with contractions of the loops in
fig.\ref{fig:loops}-III and -IV, respectively. The matrix elements
of the operators must satisfy the system of constraints, which,
according to (\ref{invRfund}), are obtained from (\ref{meq}) and
(\ref{mteq}), correspondingly, by changing $q$ for $q^{-1}$.
Therefore, the same change of parameters relates the unique
solutions of the systems, \be \boxed{\tilde{m}_i=q^{2N-2i+1},\ \ \
i=1,\ldots,N,}\label{Misol} \ee and \be
\boxed{\tilde{m}^{\prime}_i=q^{2i-1},\ \ \
i=1,\ldots,N.}\label{Mitsol} \ee Hence, there are in fact only two
independent operators, \be \mathcal{M}=\tilde{\mathcal{M}}^{-1}\ \
\mbox{and}\ \ \ \mathcal{M}^{\prime}=\tilde{\mathcal{M}}^{\prime-1}.
\ee


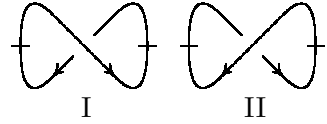
\begin{wrapfigure}{r}{140pt}
\begin{picture}(60,36)(-12,-12)
\qbezier(0,24)(6,18)(12,12)\qbezier(12,12)(18,6)(24,0)
\qbezier(24,24)(18,18)(16,16)\qbezier(8,8)(6,6)(0,0)
\qbezier(0,0)(-12,-12)(-12,12)\qbezier(0,24)(-12,36)(-12,12)
\qbezier(24,0)(36,-12)(36,12)\qbezier(24,24)(36,36)(36,12)
\qbezier(0,0)(2,2)(2,5)\qbezier(0,0)(2,2)(4,2)
\qbezier(24,0)(22,2)(22,5)\qbezier(24,0)(22,2)(20,2)
\qbezier(-15,12)(-12,12)(-9,12)\qbezier(33,12)(36,12)(39,12)
\put(11,-15){I}
\end{picture}
\begin{picture}(60,36)(-12,-12)
\qbezier(0,24)(6,18)(8,16)\qbezier(16,8)(18,6)(24,0)
\qbezier(24,24)(18,18)(12,12)\qbezier(12,12)(6,6)(0,0)
\qbezier(0,0)(-12,-12)(-12,12)\qbezier(0,24)(-12,36)(-12,12)
\qbezier(24,0)(36,-12)(36,12)\qbezier(24,24)(36,36)(36,12)
\qbezier(0,0)(2,2)(2,5)\qbezier(0,0)(2,2)(4,2)
\qbezier(24,0)(22,2)(22,5)\qbezier(24,0)(22,2)(20,2)
\qbezier(-15,12)(-12,12)(-9,12)\qbezier(33,12)(36,12)(39,12)
\put(9,-15){II}
\end{picture}
\caption{Diagrams of the twisted circles, both are equivalent to the
unknot.\label{fig:tcirc}}
\end{wrapfigure}

\subsubsection{Topological normalization of the turn-over operators and framing of a knot\label{sec:Rfram}}
\emph{A priory}, there are four distinct turn-over operators
associated with contractions of the four distinct loops in
fig.\ref{fig:loops}, two including the direct crossing, the other
two the inverse one. However, we will see that in a more general
case a turn-over operator can be inserted in an arbitrary edge that
belongs to a cycle on a knot diagram. Hence, there should be just
two distinct operators, inserted in the counter-clock and clock-wise
cycles, respectively. Yet, we obtained four explicit distinct
expressions (\ref{Msol},\ref{Mtsol},\ref{Misol},\ref{Mitsol}) for
the operators associated with the corresponding loops. This problem
is resolved if one rescales the turn-over operators in some proper
way, bringing them in a so called \emph{topological} normalization.
The four turn-over operators equal pairwise in this normalization.

Rigorously speaking, equation (\ref{fig:RI}), which follows from RI
invariance, determines the turn-over operators uniquely, not up to a
factor. Rescalling of the operators, hence, is possible only
together with deformation of the equation itself. The operators $R-$
and $\mathcal{M}-$ satisfying the resulting deformed equations give
rise to so called \emph{framed} knot invariants, which generalize
the notion of plain knot invariants. The latter ones still can be
obtained as contractions of the rescaled $R-$ and $\mathcal{M}-$
operators, if one properly rescales the entire contraction at the
last step of the computation.

\paragraph{Twisted circles as full contractions of $R$-matrices\label{sec:tcirc}}


As the first demonstration of a problem with inserting the correct
turn-over operator, consider an invariant of the twisted circle
(fig.\ref{fig:tcirc}-I). Following the algorithm of
sec.\ref{sec:Rinv}, one makes two cuts as in the diagram. There is
only one crossing, and the corresponding invariant equals the
contraction of the $R$-matrix with two $\mathcal{M}$-matrices
inserted in the two pairs of indices, \be
\sum_{i,j,a,b}R^{ij}_{ab}\mathcal{M}^i_b\mathcal{M}^{\prime j}_{a}=
\sum_{i,j}R^{ij}_{ji}m_im^{\prime}_j=\sum_{i=1}^Nm_i=\sum_{j=1}^Nm^{\prime}_j=\sum_{i=1}^Nq^{-2i+1}=
q^{-N}\frac{q^N-q^{-N}}{q-q^{-1}}\equiv q^{-N}[N].\label{tcircp}
\ee In turn, a similar diagram with the crossing substituted with
the inverse one (fig.\ref{fig:tcirc}-II) gives the value of the
invariant \be
\sum_{i,j,a,b}\tilde{R}^{ij}_{ab}\mathcal{M}^i_b\mathcal{M}^{\prime
j}_{a}=
\sum_{i,j}\tilde{R}^{ij}_{ji}\tilde{m}_i\tilde{m}^{\prime}_j=\sum_{i=1}^N\tilde{m}_i=\sum_{j=1}^N\tilde{m}^{\prime}_j
=\sum_{i=1}^Nq^{2i-1}=q^{N}[N].\label{tcircm} \ee On the other hand,
(\ref{tcircp}) must coincide with (\ref{tcircm}) for a properly
defined knot invariant, since both diagrams in fig.\ref{fig:tcirc}
are projections of the curves that can be continuously transformed
into a plain circle, i.e. to the unknot. One can cure this problem
by putting one more element into the $R$-matrix construction.

\begin{wrapfigure}{r}{60pt}
\begin{picture}(45,45)
\put(16,15){\circle{30}} \qbezier(-3,15)(0,15)(3,15)
\end{picture}
\caption{A plain circle (the unknot) requires for inserting a
turn-over operator.\label{fig:circ}}
\end{wrapfigure}
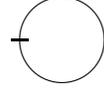

\paragraph{Topological normalization}
The problem is considered even clearly as one considers a plain circle.
Following the program of sec.\ref{sec:Rinv}, one should treat its
planar projection as a diagram with no vertices and with one cut
(fig.\ref{fig:circ}). Rigorously speaking, the above rules do not
determine the corresponding value of the invariant at all. In
analogy with (\ref{fig:RI}), one may associate with this diagram the
trace of a turn-over operator. One of the four turn-over operators
$\mathcal{M}$, $\mathcal{M}^{\prime}$, $\mathcal{\tilde{M}}$,
$\mathcal{\tilde{M}}^{\prime}$ can be taken equally well in the
case. Unfortunately, their traces differ, as we already verified
above, \be
\sum_{i=1}^N\mathcal{M}_i^i=\sum\limits_{i=1}^N\mathcal{M}^{\prime
i}_i&=q^{-N}[N],\ \ \
\sum\limits_{i=1}^N\tilde{\mathcal{M}}_i^i=\sum\limits_{i=1}^N\tilde{\mathcal{M}}^{\prime
i}_i&=q^N[N].\label{Mtr} \ee

\paragraph{Invariants of framed knots in $R$-matrix construction}
A clue to the resolution is given by the following observation. The
expressions (\ref{tcircp}) and (\ref{tcircm}) are obtained from the
knot diagrams with the different values of a \emph{relative}
invariant, a \emph{writhe number}, which by definition equals the
number of direct crossings minus the number of inverse crossings
(see fig.\ref{fig:invcr}). A writhe number remains unchanged under
the second and third Reidemeister moves, (\ref{fig:RII}) and
(\ref{fig:RIII}), respectively, but increases or decreases by one by
the first Reidemeister move, (\ref{fig:RI}). Answers (\ref{tcircp})
and (\ref{tcircm}) coincide provided that one multiplies them by
factors $q^{wN}$ with the corresponding writhe numbers $w=1$ and
$w=-1$ for figs.\ref{fig:tcirc}-I and II, respectively. This
observation, however, does not solve the problem with plain circle,
for which $w=0$. This problem can be avoided by rescaling of the
crossing operators as \be \mathcal{M}\rightarrow \mathfrak{M}\equiv
q^{-N}\mathcal{M},\ \ \ \mbox{and}\ \
\mathcal{M}^{\prime}\rightarrow \mathfrak{M}^{\prime}\equiv
q^{-N}\mathcal{M}^{\prime},\label{Mresc} \ee The non-vanishing
matrix elements then become \be \mathfrak{m}_i=q^{2i-N-1},\ \
\mathfrak{m}^{\prime}_i=q^{N-2i+1},\ \
\tilde{\mathfrak{m}}_j=q^{N-2i+1},\ \
\tilde{\mathfrak{m}}^{\prime}_i=q^{2i-N-1},\ \ \
i=1,\ldots,N.\label{mresc} \ee All the four traces in (\ref{Mtr})
equal then $[N]$; (\ref{tcircp}) then matches (\ref{tcircm}) and the
answer for the plain circle after each expression is multiplied on
the factor of $q^{-wN}$ with its own $w$ (the factor is inverse to
the previously suggested ones). Motivated by these considerations,
one may introduce then two more rules,
\begin{step}
Rescale the turn-over operators so that traces of all the four
operators equal,\label{st:Mtr}
\end{step}
and
\begin{step}
Multiply the result of \textbf{Steps \ref{st:diag}-\ref{st:cutinv}} by
$q^{-wN}$, with $w$ being the writhe number of knot diagram involved
in the evaluation.\label{st:fr}
\end{step}
If one follows the above rules, both diagrams in
fig.\ref{fig:tcirc}, as well that in fig.\ref{fig:circ}, yield one
and the same value $[N]$ of the invariant for the unknot. However,
one has to check whether the suggested rescaling does not break
already imposed constraints
(\ref{fig:RIII},\ref{fig:RII},\ref{fig:RI}). Equalities
(\ref{fig:RIII}) and (\ref{fig:RII}) which do not involve turn-over
operators, indeed, remain unaffected. Unlike them, condition
(\ref{fig:RI}) gains an extra factor of $q^{N}$ for loops in
fig.\ref{fig:loops}-I,II, and $q^{-N}$ for loops in
fig.\ref{fig:loops}-III,IV. Deformed this way, relation
(\ref{fig:RI}) is called a \emph{q-Reidemeister-I move}. Since a
loop in (\ref{fig:RI}) can be contracted in the three-dimensional
space, a supposed deformation of the constraint is inept, unless one
endows a knot with an additional structure. The needed structure can
be represented graphically, by substituting a knot by a knotted
\emph{ribbon}. The first Reidemeister move is not an equivalence
transformation of such knots, since it causes a twist of the ribbon.
Formally speaking, a knot, or each component of a link should be
associated now with an integer number, which equals a number of
the ribbon intertwinings and is changed by one under the first
Reidemeister move. The obtained object is called a \emph{framed}
knot or link \cite{Pras}. We have demonstrated in the elementary
example that a knot invariant of the studied type is, in fact, an
invariant not just of a knot, but of a \emph{framed}
knot\footnote{In fact, framing of a knot consists in substituting a
curve with a ribbon, or, equivalently, in introducing a normal
vector bundle on it. The power of the framing factor equals then
the number on ribbon intertwinings \cite{Pras}.} \cite{ReshTur}.
Somewhat surprisingly, attempt of interpreting the same invariant as
a QFT observable lead one to the same conclusion \cite{GuadMarMin1} (see
sec.\ref{sec:CSfram} for details).

Although required in the definition of the knot invariant of the
studied type, choice of framing does not affect the answer, provided
that one follows \textbf{Step \ref{st:fr}} (the corresponding factor
is called a \emph{framing factor}). The independence of the answer on the writhe number used in the calculation knot diagram  becomes
even more explicit if the multiplying of the entire answer on a
framing factor is substituted by the rescaling of each crossing
operator as $R\rightarrow q^{-N}R$, and each inverse crossing
operator as $\tilde{R}\rightarrow q^{N}\tilde{R}$.  Both constraints
(\ref{fig:RIII}), (\ref{fig:RII}) are preserved by such a change,
and (\ref{fig:RI}) holds again without any extra factors. However,
the framing independence  property breaks both for knots in
topologically non-trivial (other than $R_3$ or $S_3$) spaces
\cite{Pras} and for some other (yet conjectured) generalizations of
the studied quantities \cite{Dan}.

\subsubsection{Turn-over operators and loop version of the RII move.}
\be
\begin{picture}(60,60)(-30,-5)
\qbezier(-12,30)(3,0)(-12,-30)
\qbezier(-7,10)(-9,0)(-7,-10)\qbezier(0,30)(-3,23)(-5,16)\qbezier(0,-30)(-3,-23)(-5,-16)
\qbezier(-3,-22)(-2,-25)(0,-26)\qbezier(-3,-22)(-2,-25)(-3,-27)
\qbezier(-9,23)(-10,26)(-12,27)\qbezier(-9,23)(-10,26)(-9,28)
\qbezier(-10,-26)(-9,-23)(-7,-22)\qbezier(-10,-26)(-9,-23)(-10,-21)
\qbezier(-1,28)(-2,25)(-4,24)\qbezier(-1,28)(-2,25)(-1,23)
\put(1,-29){$l$}\put(1,22){$j$}\put(-17,21){$i$}\put(-18,-29){$k$}
\put(-15,0){$b$}\put(-2,0){$a$}
\end{picture}&=&
\begin{picture}(60,60)(-30,-5)
\qbezier(12,30)(-3,0)(12,-30) \qbezier(-12,30)(3,0)(-12,-30)
\qbezier(-9,23)(-10,26)(-12,27)\qbezier(-9,23)(-10,26)(-9,28)
\qbezier(-10,-26)(-9,-23)(-7,-22)\qbezier(-10,-26)(-9,-23)(-10,-21)
\qbezier(9,-23)(10,-26)(12,-27)\qbezier(9,-23)(10,-26)(9,-28)
\qbezier(10,26)(9,23)(7,22)\qbezier(10,26)(9,23)(10,21)
\put(13,-29){$l$}\put(12,22){$j$}\put(-17,21){$i$}\put(-18,-29){$k$}
\end{picture}\nn\\[1cm]
\mathcal{R}^{ia}_{bl}\tilde{\mathcal{R}}^{cj}_{ka}\mathcal{M}^a_c&=&\delta^i_k\mathcal{M}^j_l.\label{fig:vRII}
\ee

Attempt of constructing a topologically invariant contraction of the
$R$-matrices comes across with another problem as well. Namely, the
Reidemeister theorem \cite{Pras} is related to \emph{undirected}
knot diagrams. The elementary equivalence transformation of
\emph{directed} graphs, in turn, include several versions of
(\ref{fig:RIII}) and (\ref{fig:RIII}) with variously directed edges
(compare, e.g., fig.\ref{fig:RII} and \ref{fig:vRII}).

The Reidemeister \cite{Pras} claims that two knot diagrams are
equivalent if and only if they are related by a combination of
Reidemeister moves (\ref{fig:RIII}), (\ref{fig:RII}) and
(\ref{fig:RI}). However, $R$-matrix contractions are invariant only
under versions of the transformations with certain directions of
edges, as in (\ref{fig:RII}) and (\ref{fig:RIII}). In particular,
they are \emph{not} invariant under transformation (\ref{fig:vRII}),
as one can verify straightforwardly.

This problem is closely related to the problem of RI invariance, since the both transformations delete a \emph{cycle}, i.e., a closed directed curve on a knot diagram. Similarly to the former case, transformation
(\ref{fig:vRII}) can be associated with the corresponding equality,
which also includes a turn-over operator.

A difference between the directed graphs in the l.h.s. of
(\ref{fig:RII}) and (\ref{fig:vRII}) is that the latter one contains
a \emph{cycle} (one can pass a closed path following the direction
of arrows on the diagram), while the former one does not. One can
then extract from here the empiric rule to
\begin{step}
Insert exactly one turn-over operator in one cycle, using the
operators $\mathcal{M}$ and $\mathcal{M}^{\prime}$ for the
counter-clock and clock-wise oriented cycles,
respectively.\label{st:tov}
\end{step}
For instances, possible placements of the cuts for the two different diagrams of the trefoil knot are shown in fig.\ref{fig:tref}-I and fig.\ref{fig:ltref}. However, the above rule sets a new problem.

\subsubsection{Commutation of the turn-over operators with the $R$-matrices}

Rule (\textbf{Step} \ref{st:tov}) does not specify where exactly one
should insert the turn-over operators. Moreover, the construction
would be self-consistent only if \emph{all} ways to insert the
turn-over operators in accordance with the rule gave the same
result. It is not straightforward to understand whether such
independence takes place generally. On the one hand, the turn-over
operator does \emph{not} commute with the $R$-matrix, what follows
just from the explicit expressions for the operators in the
considered particular case, \be
\sum_a\mathcal{\mathcal{M}}^a_kR^{il}_{al}\ne\sum_a\mathcal{\mathcal{M}}^i_aR^{aj}_{kl}\
\ \Leftarrow\ \ m_kR^{ij}_{kl}\ne m_iR^{ij}_{kl}.\label{McommN} \ee
One the other hand, one generally can \emph{not} just move a
turn-over operator from one edge to another since this would violate
rule \textbf{Step} \ref{st:tov}.

\paragraph{The considered particular ($su_n$ fundamental) case.}

One may observe here that the $R$-matrix commutes with a \emph{pair}
of the turn-over operators instead,
\be\sum_{a,b}\mathcal{M}^i_a\mathcal{M}^j_bR^{ab}_{kl}=\sum_{a,b}\mathcal{M}^a_k\mathcal{M}^b_lR^{ij}_{kl}.\label{RMM}
\ee The above equality is straightforward to verify for the here
considered ($su_n$ fundamental) $R$-matrices and the corresponding
$\mathcal{M}$-matrices, which satisfy \be
m_im_jR^{ij}_{kl}=m_km_lR^{ij}_{kl},\label{McommT}
\ee wherefrom comes (\ref{RMM}) in this particular case.

\paragraph{General case.}
Commutation relation (\ref{RMM}) holds for generic $R$- and
$\mathcal{M}$-operators as well, as a corollary of certain group
theory facts \cite{MorSm}. To sketch the prove, let us rewrite
(\ref{RMM}) in the script-free form, \be
\mathcal{M}\otimes\mathcal{M}\cdot
R=R\cdot\mathcal{M}\otimes\mathcal{M},\label{RMMtens} \ee where both
the operators $\mathcal{M}\otimes\mathcal{M}$ and $R$ are maps of
the space $V\otimes V$ (recall that a vector space $V$ of a Lie
algebra representation is attached to each edge on the cut diagram,
see sec.\ref{sec:Rinv}). Equality (\ref{RMMtens}) then follows of
the well-known $R$-matrix property \cite{Baxt}. Namely, the
$R$-matrix commutes with each algebra operator when acting on the
tensor product of the representations. The operator $\mathcal{M}$,
in turn, can be presented a formal series in the algebra operators,
only pairwise commuting operators entering the series \cite{MorSm}.
For an ordinary Lie algebra, these generators act on the tensor
product of the representations merely as $V\otimes V\rightarrow
TV\otimes V+V\otimes TV$. However, in the construction that we
consider $T$ are in fact generators of a \emph{quantum} Lie algebra;
they satisfy somehow deformed commutation relations and act on a
tensor product of the representations yielding a more generic linear
combination of tensor monomials, $\Delta(T)\Big(V\otimes
V\Big)\equiv\sum T_{a_1}\ldots T_{a_i}V\otimes T_{b_1}\ldots
T_{b_j}V$, which, for instance, is not necessary permutation w.r.t.
a permutation of the tensor factors (the rule specifying this
combination for each algebra operators is called a \emph{co-product}
rule). The commutative subalgebra is yet the same for the
corresponding ``classical'' and quantum Lie algebras, operators $H$
of the subalgebra still acting on the tensor product just as
$HV\otimes V+V\otimes HV$. Hence, the turn-over operator, being the
exponential of such an operator, acts on the same space as
$\mathcal{M}V\otimes\mathcal{M}V$, i.e., just with the operator
entering (\ref{RMMtens}). This proves the equality.

\be
\begin{array}{rcl}
\begin{picture}(24,30)(0,10)
\qbezier(0,24)(6,18)(12,12)\qbezier(12,12)(18,6)(24,0)
\qbezier(24,24)(18,18)(16,16)\qbezier(8,8)(6,6)(0,0)
\qbezier(0,0)(2,2)(2,5)\qbezier(0,0)(2,2)(4,2)
\qbezier(24,0)(22,2)(22,5)\qbezier(24,0)(22,2)(20,2)
\qbezier(0,18)(3,21)(6,24)
\qbezier(24,18)(21,21)(18,24)
\end{picture}
&\longrightarrow&
\begin{picture}(42,30)(0,10)
\qbezier(0,24)(6,18)(12,12)\qbezier(12,12)(18,6)(24,0)
\qbezier(24,24)(18,18)(16,16)\qbezier(8,8)(6,6)(0,0)
\qbezier(0,0)(2,2)(2,5)\qbezier(0,0)(2,2)(4,2)
\qbezier(24,0)(22,2)(22,5)\qbezier(24,0)(22,2)(20,2)
\qbezier(15,3)(18,6)(21,9)
\qbezier(9,3)(6,6)(3,9)
\end{picture}
\end{array}
\label{fig:cuts}\\[2mm]\nn\ee

Move of the turn-over operators (fig.\ref{fig:cuts}) corresponding
to commutation relation (\ref{fig:cuts}) might imply that one
operator is removed from a loop, being substituted by another one.
Were any two allowed by \textbf{Step} \ref{st:tov} placements of the
turn-over operators related by a sequence of (\ref{fig:loops})-type
moves, (\ref{RMM}) would ensure equality of the operator
contractions for all such placements. However, a more careful
treatment is needed before one can make a general statement.

\subsubsection{Conventional non-covariant approach: the extremum point operators}
A conventional approach to the problem of the turn-over operators
\cite{KirResh},\cite{MorSm} is to refuse from the ``covariance'' of
the construction. Namely, \textbf{Steps} \ref{st:Mtr}- \ref{st:tov}
are substituted by the following
\begin{step}
A direction on the projection plane is selected, and an operator
(which we call an \emph{extremum point} operator) is inserted in
each extremum point, where a tangent to the knot diagram vector is
orthogonal to the direction.\label{st:extp}
\end{step}
Hence, there are four operators, $Q_+$, $Q_-$, $\tilde{Q}_+$,
$\tilde{Q}_-$ for the four distinct extremum points, two minimum and
two maximum, each pair differing by the direction selected on the
curve. The pairwise products of the extremum point operators must be
equal to either to the unity operator,  \be
Q_+Q^{\prime}_-=Q^{\prime}_-Q_+=\mathds{1},\ \ \ Q_+^{
\prime}Q_-=Q_-Q_+^{ \prime}=\mathds{1}.\label{Extop}\ee or to a turn
over operators, there being two distinct ones the topological
normalization, \be Q_-Q_+=Q_+Q_-=\mathfrak{M},\ \ \
Q^{\prime}_+Q^{\prime}_-=Q^{\prime}_-Q^{\prime}_+=\tilde{\mathfrak{M}}.\label{MinMax}\ee
The corresponding theorem \cite{KirResh} claims that the conditions
(\ref{MinMax}) together with defining constraints (\ref{fig:RIII}),
(\ref{fig:RII}) and (\ref{fig:RI}) on the $R$- and $M$- matrices
ensure invariance of the operators contraction constructed with help
of \textbf{Steps} \ref{st:diag}-\ref{st:cutinv} and \ref{st:extp}
under all elementary equivalence transformations of directed graphs.
Moreover, we already mentioned above that the vector space where run
the scripts of the operators is a representation space of a Lie
group, or, more precisely, of its certain deformation referred to as
\emph{quantum group}; the obtained contraction is be invariant of
this groups transformations as well \cite{KirResh}.

Condition (\ref{MinMax}) in fact does not specify the extremum point
operators uniquely. In particular, one may additionally require that
the minimum or maximum point operator equals the unity operator,
the other one being then equal to the corresponding turn-over
operator.

\bigskip

Following another version of the non-covariant approach
\cite{Kaul,Inds1,Inds2,Inds3,Inds6,Inds10,KaulLeq,GuJock,MMM4,MMM5,GMMM}, one, instead of introducing the extremum point
operators, takes a ``\emph{matrix element}'', i.e., a certain
component of the tensor related to a cut knot diagram, or, more
precisely, a linear combinations of these components, the
coefficients provide the RI invariance. This approach can be shown
to be equivalent to the extremum point operators approach
\cite{KirResh}, but we are not to go in details the present text.

\subsubsection{Examples of using the covariant approach}

One can still operate in covariant terms of the turn-over operators
at least in certain particular cases.

To complete our discussion on the turn-over operators, let is
illustrate that the covariant "one operator for one loop" approach
works at least in particular cases. We consider explicitly three
examples of the knot, corresponding to the closure of a two-strand
braid, a multiply intertwined circle, and to a pair of
counter-oriented strands that can be separated in the
three-dimensional space. Each case has its own points to discuss.
It seems especially interesting to us that the topological
invariance constrains can be formulated, at least for the particular
case of the ($su_N$ fundamental) $R$- and $\mathcal{M}$-matrices
that we consider, intermediately for the operator products, not only
for the full contractions that yield the knot invariants.

\paragraph{Examples. I. Two-strand braids.}
We start with evaluating the invariant of a torus knot or link
$T^{2,n}$, presented as the closure of a two-strand braid with $n$
crossings. A main subtlety arises here is that one should insert the
\emph{different} turn-over operators for the counter-clock wise and
clock-wise closures. The obtained values of the invariant yet must
be the same.\footnote{The two described planar figures, although
representing the same braid closure, can \emph{not} be continuously
transformed one into the other in the projection plane. The mirror
reflection of an arc, which relates two such figures, should be
considered as one more equivalence transformation of a knot diagram,
the transformation being referred to as \emph{zero Reidemeister
move}. We did not consider this transformation yet, because the
tensor contractions associated with the two diagrams in the state
model approach, as well as the related to them the braid group
elements in another approach, coincide \emph{identically}, and the
same is true for averages of these quantities. However, the
situation turns different as soon one introduces the turn-over
operators.} Below we demonstrate this being indeed true for
turn-over operators (\ref{Msol},\ref{Mtsol}), although looking out
as a non-trivial fact.

A two-strand braid with $n$ crossings corresponds to an operator
(all scripts run from $1$ to $N$),
\begin{equation}
h^{ij}_{kl}\equiv
\mathcal{R}^{ij}_{j_2i_2}\mathcal{R}^{i_2j_2}_{j_3i_3}\ldots
\mathcal{R}^{i_nj_n}_{lk}.
\end{equation}
The matrix forms of the operators are ($R$ differs from
$\mathcal{R}$ by the permutation of the subscripts,
$\mathcal{R}^{ij}_{kl}\equiv R^{ij}_{lk}$, and matrix indices are a
multi-index $I=(ij)$ with $i,j$ running from $1$ to $N$ and $i<j$)
\begin{equation}
R=\begin{array}{||ccc|cccc||c}\multicolumn{1}{c}{\ldots}&ii&\ldots&\ldots&ij&ij&\multicolumn{1}{c}{\ldots}&\\\ddots&&&&&&&\vdots
\\&q&&&&&&ii\\&&\ddots&&&&&\vdots
\\\cline{1-8}&&&\ddots&&&&\vdots\\&&&&q-q^{-1}&1&&ij\\&&&&1&&&ji\\&&&&&&\ddots&\vdots\end{array},
\end{equation}
and
\begin{equation}
B=R^n=\begin{array}{||ccc|cccc||c}\multicolumn{1}{c}{\ldots}&ii&\ldots&\ldots&ij&ij&\multicolumn{1}{c}{\ldots}&\\\ddots&&&&&&&\vdots
\\&q&&&&&&ii\\&&\ddots&&&&&\vdots
\\\cline{1-8}&&&\ddots&&&&\vdots\\&&&&P_{n+1}&P_n&&ij\\&&&&P_n&P_{n-1}&&ji\\&&&&&&\ddots&\vdots\end{array},
\end{equation}
where $P_n=P_n(q)$ are determined from the recurrent relations
\begin{equation}
P_0=0,\ P_1=1,\ P_{n+1}=\left(q-q^{-1}\right)P_n+P_{n-1}\ \ \
\Rightarrow\ \ \
P_n=\sum_{k=0}^{n-1}(-1)^kq^{n-2k-1}=\frac{q^n+(-1)^{n+1}q^{-n}}{[2]}.\label{rec}
\end{equation}
All the non-zero matrix elements of $B$ have the form either
$h^{ij}_{ij}$ or $h^{ij}_{ji}$, i.e., the subscripts are a
permutation of the superscripts. Hence, the partial contraction
corresponding to closing of the second strand is diagonal in the
selected basis,
\begin{equation}
b^i_l\equiv
h^{ij}_{kl}\mathcal{M}^l_j=\sum_{l=1}^Nh^{ik}_{lk}m_k=\delta^i_l\sum_{k=1}^Nh^{ik}_{ik}m_k\equiv
b_i\delta^i_l.\label{pcont}
\end{equation}
Moreover, all diagonal elements turn out to be equal to each other,
\begin{equation}
\begin{array}{r}
b_i=h^{ii}_{ii}m_i+\sum_{j=1}^{i-1}h^{ij}_{ij}m_j+\sum_{j=i+1}^{N}h^{ij}_{ij}m_j
=q^nm_i+P_{n+1}\sum_{j=1}^{i-1}m_j+P_{n-1}\sum_{j=i+1}^{N}m_j=\\=
q^n\sum_{i=1}^Nm_i-P_n\left(q\sum_{j=1}^{i-1}m_j+q^{-1}\sum_{j=i+1}^{N}m_j\right)=q^n[N]-P_n[N-1].\label{diagpc}
\end{array}
\end{equation}
where we substituted $m_i=q^{N-2i+1}$ and used the relations
\begin{equation}
P_{n+1}=q^n-q^{-1}P_n\ \ \ \mbox{and}\ \ \ P_{n-1}=q^n-qP_n,
\end{equation}
following from (\ref{rec}). Substituting the explicit expression for
$P_N$ from (\ref{rec}) into (\ref{pcont}), one obtains that partial
contraction (\ref{pcont}) takes form
\begin{equation}
h^{ij}_{kl}\mathcal{M}^l_j=\delta^k_l\left(\frac{q^n[N+1]+(-1)^nq^{-n}[N-1]}{[2]}\right)\equiv\delta^k_l\mathcal{H}^{2,n}_r,
\end{equation}
where a factor of delta-symbol is the \emph{reduced} HOMFLY
polynomial for the $T^{2,n}$ knot or link, which is the closure of
the considered braid. Hence, both possible full contractions
give the \emph{unreduced} HOMFLY polynomial
of the same link,
\begin{equation}
h^{ij}_{kl}\mathcal{M}^l_j\mathcal{M}^k_i=h^{ij}_{kl}\mathcal{M}^l_j\mathcal{M}^{\prime
k}_i=\mathcal{M}^i_i\mathcal{H}^{2,n}_r=\mathcal{M}^{\prime
i}_i\mathcal{H}^{2,n}_r=[N]\mathcal{H}^{2,n}_r\equiv\mathcal{H}^{2,n}.
\end{equation}

\paragraph{II. Multiply intertwined circle.}
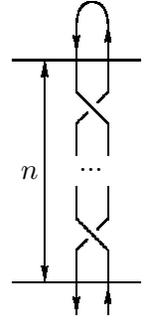
\begin{wrapfigure}{r}{60pt}
\begin{picture}(60,102)(-24,-36)
\qbezier(0,60)(0,54)(0,48) \qbezier(0,48)(0,42)(0,36)
\qbezier(0,36)(3,33)(6,30)\qbezier(6,30)(9,27)(12,24)
\qbezier(12,24)(12,18)(12,12)
\qbezier(0,0)(0,-6)(0,-12)
\qbezier(0,-12)(3,-15)(6,-18)\qbezier(6,-18)(9,-21)(12,-24)
\qbezier(12,-24)(12,-30)(12,-36) \qbezier(12,-36)(12,-42)(12,-48)
\qbezier(12,60)(12,54)(12,48) \qbezier(12,48)(12,42)(12,36)
\qbezier(12,36)(9,33)(8,32)\qbezier(4,28)(3,27)(0,24)
\qbezier(0,24)(0,18)(0,12)
\qbezier(12,0)(12,-6)(12,-12)
\qbezier(12,-12)(9,-15)(8,-16)\qbezier(4,-20)(3,-21)(0,-24)
\qbezier(0,-24)(0,-30)(0,-36) \qbezier(0,-36)(0,-42)(0,-48)
\qbezier(-24,48)(0,48)(24,48)\qbezier(-24,-36)(0,-36)(24,-36)
\qbezier(-12,48)(-12,0)(-12,-36)
\qbezier(0,60)(0,70)(6,70)\qbezier(6,70)(12,70)(12,60)
\put(-21,3){$n$}\put(1,6){.}\put(4,6){.}\put(7,6){.}
\qbezier(0,51)(0,54)(1,57)\qbezier(0,51)(0,54)(-1,57)
\qbezier(12,61)(12,58)(13,55)\qbezier(12,61)(12,58)(11,55)
\qbezier(0,-48)(0,-45)(1,-42)\qbezier(0,-48)(0,-45)(-1,-42)
\qbezier(12,-38)(12,-41)(13,-44)\qbezier(12,-38)(12,-41)(11,-44)
\qbezier(-12,48)(-12,45)(-11,42)\qbezier(-12,48)(-12,45)(-13,42)
\qbezier(-12,-36)(-12,-33)(-11,-30)\qbezier(-12,-36)(-12,-33)(-13,-30)
\end{picture}
\caption{The ``fake closure'' of the antiparallel braid.\label{fig:antpar}}
\end{wrapfigure}
The next example that we consider is a two-strand antiparallel
braid. According to the general rule, one should insert the
turn-over operators not only in the ``closure'', but in \emph{each}
section, since a section of an antiparallel braid is related to a
cycle (fig.\ref{fig:antpar}). We restrict ourselves by considering
the simplest case, when the strands of the braids are matched as in
fig.\ref{fig:antpar}, so that the braid can be unplaited in the
three-dimensional space. We demonstrate explicitly that the proper
partial contraction of the $R$- and $\emph{M}$- matrices is
proportional to the unity operator, as it should be.

An antiparallel two-strand braid one can associate with the
contraction of the crossing and turn-over operators
\begin{equation}
 \mathcal{R}^{ia}_{kl}\mathcal{M}^j_a\mathcal{R}^{nb}_{jm}\mathcal{M}^k_b\ldots\equiv\big(RM\big)^n,
\end{equation}
where for $R$ one should substitute the $sl_n$ solution
\begin{equation}
R^{ii}_{ii}=q,\ \ \ R^{ij}_{ij}=1\mbox{  for  }i\ne j,\ \ \
R^{ij}_{ji}=q-q^{-1}\mbox{  for  }i< j,
\end{equation}
and $M$ is determined form the RI invariance condition (are factor
of $q^{-N}$ is present since we use the so called topological
framing, which is a separate story),
\begin{equation}
R^{ia}_{bj}M^b_a=q^{-N}\delta^i_j\ \ \ \Rightarrow\ \ \
M^b_a=m_a\delta^b_a,\ \ m_a=q^{N-2a+1}.
\end{equation}
The matrix forms of the operators are
\begin{equation}
R=\left(\begin{array}{cc}S_{N\times
N}\\&\mathrm{Id}_{N(N-1)}\end{array}\right),\ \
S=\left(\begin{array}{cccc}q\\q-q^{-1}&q\\q-q^{-1}&q-q^{-1}&q\\\
\ldots&\ldots&\ldots&\ldots\end{array}\right),
 \end{equation}
and
$M=\mathrm{diag}(m_1,\ m_2,\ \ldots,\ m_N, 0, \ldots, 0)$.
The ``fake'' closure on the one end gives an $n$ times intertwined
line; the corresponding partial contraction might be
\begin{equation}
\begin{array}{r}
\mathcal{M}^k_l\mathcal{R}^{ia}_{kl}\mathcal{M}^j_a\mathcal{R}^{nb}_{jm}\mathcal{M}^k_b\ldots\equiv\mu\big(RM\big)^n
=q^{-Nn}(1,\ 1,\ \ldots,\ 1, 0, \ldots, 0)=q^{-Nn}\delta^p_q,
\end{array}
\end{equation}
with
$\mu=(m_1,\ m_2,\ \ldots,\ m_N,\ 0,\ \ldots,\ 0)$.
Hence, the RI invariance holds as an operator identity in this
particular case, if one inserts one turn-over operator for one
cycle.

\paragraph{III. Loop version of RII.}
As the last example, let us demonstrate that relation
(\ref{fig:vRII}) holds as an operator identity provided that one
inserts the turn-over operator in the loop,
\begin{equation}
\mathcal{R}^{ia}_{bl}\tilde{\mathcal{R}}^{cj}_{ka}\mathcal{M}^a_c=\delta^i_k\mathcal{M}^j_l.\label{lRII}
\end{equation}
Indeed, using explicit expressions (\ref{Rfund}) and (\ref{Msol})
for elements of the $\mathcal{R}$- and $\mathcal{M}$-matrices, one
obtains for the components of (\ref{lRII}) that do not vanish
identically,
\begin{equation}
\begin{array}{rl}
i=k=j=l,&\sum_{a=1}^N\mathcal{R}^{ja}_{aj}\tilde{\mathcal{R}}^{aj}_{ja}m_a=
\mathcal{R}^{jj}_{jj}\tilde{\mathcal{R}}^{jj}_{jj}m_j=q q^{-1}m_j=m_j,\\[1mm]
i=k\ne
l=j,&\sum_{a=1}^N\mathcal{R}^{ia}_{ai}\tilde{\mathcal{R}}^{aj}_{ja}m_b=
\mathcal{R}^{ij}_{ij}\tilde{\mathcal{R}}^{ij}_{ij}m_i=1\cdot1\cdot m_i=m_i,\\[1mm]
i=l\ne
k=j,&\sum_{a=1}^N\mathcal{R}^{ia}_{bj}\tilde{\mathcal{R}}^{bj}_{ia}m_b
=\mathcal{R}^{ii}_{ii}\tilde{\mathcal{R}}^{ij}_{ji}m_i+
\sum_{a=i+1}^{j-1}
\mathcal{R}^{ia}_{ai}\tilde{\mathcal{R}}^{aj}_{ja}m_a+\mathcal{R}^{ij}_{ji}\tilde{\mathcal{R}}^{jj}_{jj}m_j,\\
\end{array}
\end{equation}
i.e., in all cases one reproduces the r.h.s. of (\ref{lRII}), which
equals $\delta^i_k\delta^j_lm_j$.


\subsection{Explicit evaluation of the invariant for the trefoil knot\label{sec:tref}}

\begin{wrapfigure}{r}{215pt}
\begin{picture}(75,80)(-40,-40)
\qbezier(0,30)(-24,30)(-9,-5)\qbezier(0,10)(39,5)(27,-15)
\qbezier(-27,-15)(-15,-35)(9,-5)\qbezier(-16,7)(-36,0)(-27,-15)\qbezier(0,10)(-8,10)(-10,9)
\qbezier(14,10)(18,30)(0,30)\qbezier(9,-5)(12,0)(13,4)
\qbezier(2,-17)(18,-30)(27,-15)\qbezier(-4,-12)(-7,-9)(-9,-5)
\qbezier(-2,30)(0,30)(4,31)\qbezier(-2,30)(2,30)(4,28)
\qbezier(29,-11)(27,-15)(27,-18)\qbezier(29,-11)(27,-15)(25,-15)
\qbezier(-26,-17)(-27,-15)(-27,-12)\qbezier(-26,-17)(-27,-15)(-30,-13)
\put(-3,15){$a$}\put(-17,-6){$c$}\put(11,-10){$b$}\put(-13,-16){$d$}
\put(-2,33){$i$}
\put(-26,7){$k$}
\put(-40,-15){$j$}\put(32,-15){$l$}
\qbezier(-28,6)(-25,3)(-22,0) \qbezier(-12,-8)(-10,-6)(-6,-2)
\put(-1,-40){I}
\end{picture}
\begin{picture}(70,80)(-40,-40)
\qbezier(0,30)(24,30)(9,-5)\qbezier(0,10)(-39,5)(-27,-15)
\qbezier(27,-15)(15,-35)(-9,-5)\qbezier(16,7)(36,0)(27,-15)\qbezier(0,10)(8,10)(10,9)
\qbezier(-14,10)(-18,30)(0,30)\qbezier(-9,-5)(-12,0)(-13,4)
\qbezier(-2,-17)(-18,-30)(-27,-15)\qbezier(4,-12)(7,-9)(9,-5)
\qbezier(-2,30)(0,30)(4,31)\qbezier(-2,30)(2,30)(4,28)
\qbezier(29,-11)(27,-15)(27,-18)\qbezier(29,-11)(27,-15)(25,-15)
\qbezier(-26,-17)(-27,-15)(-27,-12)\qbezier(-26,-17)(-27,-15)(-30,-13)
\put(-3,-40){II}
\end{picture}
\begin{picture}(60,80)(-40,-40)
\qbezier(0,30)(-24,30)(-9,-5)\qbezier(0,10)(39,5)(27,-15)
\qbezier(-27,-15)(-15,-35)(9,-5)\qbezier(-16,7)(-36,0)(-27,-15)\qbezier(0,10)(-8,10)(-10,9)
\qbezier(14,10)(18,30)(0,30)\qbezier(9,-5)(12,0)(13,4)
\qbezier(2,-17)(18,-30)(27,-15)\qbezier(-4,-12)(-7,-9)(-9,-5)
\qbezier(2,30)(0,30)(-4,31)\qbezier(2,30)(-2,30)(-4,28)
\qbezier(25,-18)(27,-15)(27,-12)\qbezier(25,-18)(27,-15)(29,-15)
\qbezier(-28,-13)(-27,-15)(-27,-18)\qbezier(-28,-13)(-27,-15)(-24,-17)
\put(-6,-40){III}
\end{picture}
\caption{I. Cut diagram of the trefoil knot. II. Diagram of the
reflected knot I, \emph{inequivalent} to the original one. III.
Another projection of knot I.}\label{fig:tref}
\end{wrapfigure}
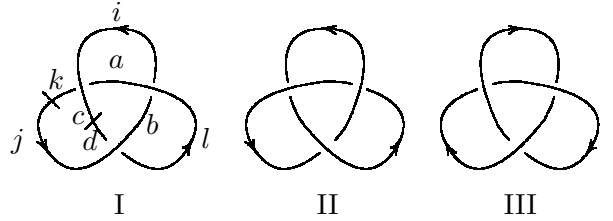

After all discussions, we are finally ready to complete an explicit
evaluation of the studied invariant for the simplest knot, which is
the trefoil knot. The $\mathcal{R}$ matrix construction dictates,
with all conclusions and assumptions of
sec.\ref{sec:Rinv}-\ref{sec:Rexpl} taken into account, to related
with the knot diagram in \ref{fig:tref} the expression \be
q^{-wN}\sum_{\substack{i,j,k,l,\\a,b,c,d}}R^{kc}_{ai}R^{ia}_{bl}R^{lb}_{dj}\mathfrak{M}^d_c\mathfrak{M}^{
j}_k=
q^{-3N}\sum_{\substack{i,k,l,\\a,b,c}}R^{kc}_{ai}R^{ia}_{bl}R^{lb}_{ck}\mathfrak{m}_c\mathfrak{m}_k.\label{trefcont}\ee
Form of solution (\ref{Rfund}) for $\mathcal{R}$-matrix implies that
scripts of the non-zero summands in (\ref{trefcont}) are related
according to one of the lines in the tabular \be
\begin{array}{|c|c|c|c|c|c|c|}
\hline
&\begin{array}{cc}k&c\\a&i\end{array}&\begin{array}{cc}i&a\\b&l\end{array}&
\begin{array}{cc}l&b\\c&k\end{array}&
\begin{array}{c}k\\c\end{array}&\begin{array}{c}c\\k\end{array}\\
\hline
k=a=l=c=i=b&\begin{array}{cc}i&i\\i&i\end{array}&\begin{array}{cc}i&i\\i&i\end{array}&
\begin{array}{cc}i&i\\i&i\end{array}&
\begin{array}{c}i\\i\end{array}&\begin{array}{c}i\\i\end{array}\\
\hline
k=a=l<c=i=b&\begin{array}{cc}k&i\\k&i\end{array}&\begin{array}{cc}i&k\\i&k\end{array}&
\begin{array}{cc}k&i\\i&k\end{array}&
\begin{array}{c}k\\i\end{array}&\begin{array}{c}i\\k\end{array}\\
\hline
k=a=b<c=i=l&\begin{array}{cc}k&i\\k&i\end{array}&\begin{array}{cc}i&k\\k&i\end{array}&
\begin{array}{cc}i&k\\i&k\end{array}&
\begin{array}{c}k\\i\end{array}&\begin{array}{c}i\\k\end{array}\\
\hline
k=i=b<c=a=l&\begin{array}{cc}k&l\\l&k\end{array}&\begin{array}{cc}k&l\\k&l\end{array}&
\begin{array}{cc}l&k\\l&k\end{array}&
\begin{array}{c}k\\l\end{array}&\begin{array}{c}l\\k\end{array}\\
\hline
k=a=b=c=i=l<k&\multicolumn{5}{|c|}{-\!\!\!-}\\
\hline
k=i=b=c=a=l<k&\multicolumn{5}{|c|}{-\!\!\!-}\\
\hline
k=i=l=c=a=b<k&\multicolumn{5}{|c|}{-\!\!\!-}\\
\hline
k=i=l<c=a=b&\begin{array}{cc}k&a\\a&k\end{array}&\begin{array}{cc}k&a\\a&k\end{array}&
\begin{array}{cc}k&a\\a&k\end{array}&
\begin{array}{c}k\\a\end{array}&\begin{array}{c}a\\k\end{array}\\
\hline
\end{array}.
\label{scrtref}\ee Assembling all listed in (\ref{scrtref})
contributions and substituting the values of the $R$-matrix elements
from (\ref{Rfund}), we get \be
H^{3_1}_N=q^{-3N}\Big\{\sum_{i=1}^NR^{ii}_{ii}R^{ii}_{ii}R^{ii}_{ii}\mathfrak{m}_im_i+
\sum_{\substack{i,k=1,\\i<k}}^N\left(3R^{ik}_{ik}R^{ik}_{ik}R^{ik}_{ki}+
R^{ki}_{ik}R^{ki}_{ik}R^{ki}_{ik}\right)\mathfrak{m}_i\mathfrak{m}_k\Big\}\equiv\nn\\\equiv
q^{-3N}\big\{q^3\alpha_N+\left(3(q-q^{-1})+(q-q^{-1})^3\right)\beta_N\big\}=q^{-3N}\big\{q^3(\alpha_N+\beta_N)-q^{-3}\beta_N\big\},
\ee where we introduced the notations \be
\alpha_N\equiv\sum_{i=1}^N(\mathfrak{m}_i)^2,\ \
\beta_N\equiv\sum_{\substack{i,k=1,\\i<k}}^N\mathfrak{m}_i\mathfrak{m}_k\
\ \Rightarrow\ \ \alpha_N+\beta_N=\sum_{\substack{i,k=1,\\i\le
k}}^N\mathfrak{m}_i\mathfrak{m}_k. \ee Calculating the sums
explicitly for $\mathfrak{m}$ from (\ref{Msol}), we obtain them to
be related with the already familiar from sec.\ref{sec:char}
quantities, \be \beta_N&=&\sum_{1\le i<j\le
N}q^{2i+2j-2N-2}=\frac{[N][N-1]}{[2]}=\chi_{11},\nn\\
\alpha_N+\beta_N&=&\sum_{1\le i\le j\le
N}q^{2i+2j-2N-2}=\frac{[N][N+1]}{[2]}=\chi_2.\label{mchars} \ee
Finally, we gets the answer \be
H^{3_1}_N(N,q)=q^{-3N}\left\{q^3\frac{[N][N+1]}{[2]}-q^{-3}\frac{[N][N-1]}{[2]}\right\}.\label{Hur31}
\ee The formula enables an analytic continuation to arbitrary values
of $N$, merely by substituting $q^N$ for a new independent variable
$A$. Divided by the value of the invariant for the unknot
$[N]\equiv\frac{A-A^{-1}}{q-q^{-1}}$, the resulting expression
yields the HOMFLY polynomial for the trefoil knot \cite{katlas}
(with $a=A^{-1}$ and $z=q-q^{-1}$), \be
\mathcal{H}^{3_1}_N(A,q)=\frac{A^{-3}}{[2]}\left\{q^3\frac{Aq-A^{-1}q^{-1}}{q-q^{-1}}-
q^{-3}\frac{Aq^{-1}-A^{-1}q}{q-q^{-1}}\right\}=
A^{-2}(q^2+q^{-2})-A^{-4}.\label{Hr31} \ee

\subsubsection{Mirror symmetry}
If one changes all crossings in fig.\ref{fig:tref}-I for the inverse
ones, one obtains the knot diagram in fig.\ref{fig:tref}-II.
Evaluation of the associated knot invariant repeats the above
calculation almost literally, with only difference that the
operators $R$ and $\mathfrak{M}$ are substituted by the
corresponding inverse operators $\tilde{R}$ and
$\tilde{\mathfrak{M}}$. The matrix elements of the latter ones are
given by (\ref{invRanz},\ref{invRcompts}), and (\ref{Misol}), respectively, and they are
obtained from the corresponding elements of $R$ and $\mathfrak{M}$
by substituting $q\rightarrow q^{-1}$. In addition, the writhe
number in the framing factor changes for the opposite one. Hence,
the answer for the reflected diagram is \be
H^{\tilde{3_1}}_N(N,q)=H^{3_1}_N(N,q^{-1})=q^{3N}\left\{q^{-3}\frac{[N][N+1]}{[2]}-q^3\frac{[N][N-1]}{[2]}\right\}.\label{mHur31}
\ee
Expressions (\ref{Hur31}) and (\ref{mHur31}) are \emph{different}, they are not even proportional to each other. This agrees with the inequivalence of knots represented by diagrams in figs.\ref{fig:tref}-I and II, which are named as \emph{left}- and \emph{right}-hand trefoils, respectively. The above reasoning applies to an arbitrary knot diagram, and one obtains that the HOMFLY polynomials of the knot $\mathcal{K}$ and its mirror image $\tilde{\mathcal{K}}$ are related as $H^{\tilde{\mathcal{K}}}(q)=H^{\mathcal{K}}(q^{-1})$. 

In contrast to the considered transformation, reversing the
orientation of a knot diagram (i.e., reflecting the directions of
arrows) does \emph{not} affect the answer at all, since all the
direct crossings remaind direct ones, and the inverse crossings
remain the inverse ones (fig.\ref{fig:tref}-III). The result is a
diagram of the same knot, just projected on the ``ceiling'' instead
of the ``floor''.

\subsection{From $\mathcal{R}$-matrices approach to the braid group approach\label{sec:Rbr}}
In this section, we discuss how the $\mathcal{R}$-matrix approach is
related to the braid group approach (see sec.\ref{sec:braids}). A
naive guess might be that the braid crossing operators $B$ are just
particular cases of the $\mathcal{R}$-matrix, written down in a
certain basis. The real case is very close to that, but is still
somewhat different. In the current section, we discuss the relation
between the $\mathcal{R}$-matrices and the matrices representing the braid group elements, which we considered in sec.\ref{sec:braids},
comparing them explicitly in the simplest examples.

As we already mentioned, the history went just the opposite way. The
$R$-matrices approach was originally formulated applied to braids
only \cite{Tur,ReshTur}. The method was then developed into a fine
working computational tool in the same terms, and was extended to
wider class of knot representations only recently
\cite{IMMM1,IMMM2,IMMM3,AMMM2,AM1,Inds6,Inds7,AM1,MMM4,MMM5,GMMM,
Inds10,IndsMut,GuJock}. Constructing a consistent $R$-matrix
formalism valid for \emph{arbitrary} knot diagrams is a separate
problem \cite{MorSm}, and it is still far from being solved
exhaustively.

\subsubsection{Reduction of four-indices operators to matrices. Twisted $R$-matrices}
\begin{wrapfigure}{r}{45pt}
\begin{picture}(50,80)(0,-45)
\qbezier(0,36)(0,30)(0,24)\qbezier(0,24)(6,18)(12,12)\qbezier(12,12)(12,6)(12,0)
\qbezier(0,0)(6,-6)(12,-12)\qbezier(12,-12)(12,-18)(12,-24)
\qbezier(12,36)(12,30)(12,24)\qbezier(12,24)(9,21)(8,20)\qbezier(4,16)(3,15)(0,12)\qbezier(0,12)(0,6)(0,0)
\qbezier(12,0)(9,-3)(8,-4)
\qbezier(4,-8)(3,-9)(0,-12)\qbezier(0,-12)(0,-18)(0,-24)
\qbezier(0,27)(0,30)(1,33)\qbezier(0,27)(0,30)(-1,33)
\qbezier(12,27)(12,30)(13,33)\qbezier(12,27)(12,30)(11,33)
\qbezier(0,2)(0,5)(1,8)\qbezier(0,2)(0,5)(-1,8)
\qbezier(12,2)(12,5)(13,8)\qbezier(12,2)(12,5)(11,8)
\qbezier(12,-22)(12,-19)(13,-16)\qbezier(12,-22)(12,-19)(11,-16)
\qbezier(0,-22)(0,-19)(1,-16)\qbezier(0,-22)(0,-19)(-1,-16)
\put(2,30){$i$}\put(14,30){$j$}
\put(2,2){$a$}\put(14,2){$b$} \put(2,-26){$k$}\put(14,-26){$l$}
\put(0,-45){$R^{ij}_{ba}R^{ab}_{lk}$}
\end{picture}
\caption{A pair of successive crossings.\label{fig:R2br}}
\end{wrapfigure}

A serial connection of crossings, as in fig.\ref{fig:R2br}, is
associated with a contraction of the $R$-matrices in a \emph{pair}
of indices, as we already have considered when we considered the second
Reidemeister move. Such a combination can be considered as a matrix
product. Namely, one can introduce a \emph{twisted} $R$-matrix,
which differs by a permutation of the subscripts \be
\mathcal{R}^{ij}_{kl}\equiv R^{ij}_{lk}. \ee In both matrices the
superscripts are placed at the incoming and the lowers ones at the
outgoing strands; however, while the two lest scripts of $R$ are
related to the upper line w.r.t. to the projection plane, while the
two left scripts of $\mathcal{R}$ are related to the left strand in
the braid. Contraction in fig.\ref{fig:R2br} is rewritten by this
trick as a matrix product \be \sum_{a,b}R^{ij}_{ba}R^{ab}_{lk}=
\sum_{a,b}\mathcal{R}^{ij}_{ab}\mathcal{R}^{ab}_{kl}\equiv\sum_J\mathcal{R}^I_J\mathcal{R}^J_K,\label{Rmul}
\ee with the multi-indices $I$, $J$, and $K$ standing for the pairs
of indices $(i,j)$, $(a,b)$, and $(k,l)$ correspondingly. The
constraint due to the second Reidemeister move is rewritten then as
\be \sum_J\mathcal{R}^I_J\tilde{\mathcal{R}}^J_K=\delta^I_K\ \
\Rightarrow\ \
\tilde{\mathcal{R}}^I_J=\left(\mathcal{R}^{-1}\right)^I_J, \ee i.e.
successive crossings of mutually inverse orientations correspond to
the mutually inverse matrices. Finally, the full contraction,
associated with the braid closure, would corresponds to taking the
trace, were not there the turn-over operators.
Before proceeding with writing out the explicit expression, we
discuss one more difference between regular and twisted
$R$-matrices.

\subsubsection{Eigenvectors of regular and twisted crossing operators in a two-strand braid\label{sec:Rev}}
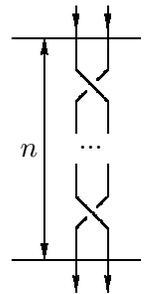
\begin{wrapfigure}{r}{48pt}
\begin{picture}(60,102)(-24,-36)
\qbezier(0,60)(0,54)(0,48) \qbezier(0,48)(0,42)(0,36)
\qbezier(0,36)(3,33)(6,30)\qbezier(6,30)(9,27)(12,24)
\qbezier(12,24)(12,18)(12,12)
\qbezier(0,0)(0,-6)(0,-12)
\qbezier(0,-12)(3,-15)(6,-18)\qbezier(6,-18)(9,-21)(12,-24)
\qbezier(12,-24)(12,-30)(12,-36) \qbezier(12,-36)(12,-42)(12,-48)
\qbezier(12,60)(12,54)(12,48) \qbezier(12,48)(12,42)(12,36)
\qbezier(12,36)(9,33)(8,32)\qbezier(4,28)(3,27)(0,24)
\qbezier(0,24)(0,18)(0,12)
\qbezier(12,0)(12,-6)(12,-12)
\qbezier(12,-12)(9,-15)(8,-16)\qbezier(4,-20)(3,-21)(0,-24)
\qbezier(0,-24)(0,-30)(0,-36) \qbezier(0,-36)(0,-42)(0,-48)
\qbezier(-24,48)(0,48)(24,48)\qbezier(-24,-36)(0,-36)(24,-36)
\qbezier(-12,48)(-12,0)(-12,-36)
\put(-21,3){$n$}\put(1,6){.}\put(4,6){.}\put(7,6){.}
\qbezier(0,51)(0,54)(1,57)\qbezier(0,51)(0,54)(-1,57)
\qbezier(12,51)(12,54)(13,57)\qbezier(12,51)(12,54)(11,57)
\qbezier(0,-48)(0,-45)(1,-42)\qbezier(0,-48)(0,-45)(-1,-42)
\qbezier(12,-48)(12,-45)(13,-42)\qbezier(12,-48)(12,-45)(11,-42)
\qbezier(-12,48)(-12,45)(-11,42)\qbezier(-12,48)(-12,45)(-13,42)
\qbezier(-12,-36)(-12,-33)(-11,-30)\qbezier(-12,-36)(-12,-33)(-13,-30)
\end{picture}
\caption{A two-strand braid with $n$ crossings.\label{fig:B2n}}
\end{wrapfigure}

Generally, $R$ (and hence $\mathcal{R}$) maps a tensor product of
vector spaces, associated with the incoming strands to that of the
ones associated with the outgoing strands. Since all the four spaces
are of the same dimension, either $R$ or $\mathcal{R}$ (but not they
both at the same time) can be looked at as an automorphism. In
particular, one can find the eigenvalues and eigenvectors of these
operators. As we will see in the next section, this will notably
simplify the computation in case of several operators $R$ contacted
as in fig.\ref{fig:B2n}.  From the practical point of view, these
are eigenvalues of $\mathcal{R}$ (not of $R$), which are needed, due
to the multiplication rule (\ref{Rmul}). Yet, we explore the
eigenvalue problem for both matrices to better demonstrate a
difference in their properties.

The very form of (\ref{Rfund}) supposes that the entire
$N^2$-dimensional space spanned by $\xi_i\eta_j$ ($i,j=1,\ldots,n$)
decomposes into a sum of $N$ one-dimensional eigenspaces spanned by
$\xi_i\eta_i$ ($i=1,\ldots,n$) and $\frac{N(N-1)}{2}$
two-dimensional spaces spanned by $\xi_i\eta_j$ and $\eta_j\xi_i$
($i,j=1,\ldots,n$ and $i<j$). The same statement is valid for
$\mathcal{R}$. Hence, one already has $N$ coinciding eigenvalues
$R^{ii}_{ii}=\mathcal{R}^{ii}_{ii}=q$ with the corresponding
eigenvectors, both of $R$ and $\mathcal{R}$: \be\lambda_i=q,\ \
X_i=\xi_i\eta_i,\ \ i=1,\ldots,N.\label{evecii}\ee It remains to
examine one of $\frac{N(N-1)}{2}$ identical two-dimensional spaces.
Making use of expression (\ref{Rfund}) for $i<j$, one gets the
eigenvalue problem for $R$ \be
\sum_{i,j}R^{ij}_{kl}\left(\alpha\xi_i\eta_j+\beta\xi_j\eta_i\right)=
\alpha\xi_k\eta_l+\big((q-q^{-1})\alpha+\beta\big)\xi_l\eta_k=\lambda(\alpha\xi_k\eta_l+\beta\xi_l\eta_k),\label{Rev}
\ee whence $\lambda=1$, $\alpha=0$, and the only (for a given pair
$k,l$) eigenvector is $\xi_k\eta_l$; there is also the adjoint
vector $\xi_k\eta_l-\xi_l\eta_k$, i.e., $R$ is a Jordan cell w.r.t.
to the stated eigenvalue problem. Unlike that, $\mathcal{R}$, whose
eigenvalue problem differs from (\ref{Rev}) by a permutation of
indices $k$ and $l$ in the last term of the equality, has two
distinct eigenvalues; they are the two roots of the characteristic
equation \be \lambda^2-(q-q^{-1})\lambda-1=0, \ee which, in turn is
obtained as consistency condition of the system \be
\lambda\beta&=&\alpha,\label{tRev}\\
\lambda\alpha&=&(q-q^{-1})\alpha+\beta, \ee on the eigenvectors
components. Hence, the eigenvalues and the corresponding
eigenvectors are \be
\lambda_+=q,&x^+_{ij}=q\xi_i\eta_j+\xi_j\eta_i,\label{evecijs}\\
\lambda_-=-1/q,&x^-_{ij}=\xi_i\eta_j-q\xi_j\eta_i.\label{evecija}
\ee The observed difference between $R$ and $\mathcal{R}$ is considered
especially well from the matrix form of their corresponding cells;
say, one has for $i,j$ running the values $1,2$
\be
R^{(12)}_{(12)}\equiv\left(\begin{array}{cc}R^{12}_{12}&R^{12}_{21}\\[1mm]R^{21}_{12}&R^{21}_{21}\end{array}\right)=
\left(\begin{array}{cc}1&q-q^{-1}\\0&1\end{array}\right),\nn\\\nn\\
\mathcal{R}^{(12)}_{(12)}\equiv\left(\begin{array}{cc}R^{12}_{21}&R^{12}_{12}\\[1mm]R^{21}_{21}&R^{21}_{12}\end{array}\right)=
\left(\begin{array}{cc}q-q^{-1}&1\\1&0\end{array}\right). \ee

Following the general rule, one obtains the eigenvalues and the
eigenvectors of the inverse $R$- and $\mathcal{R}$-matrices by the
substituting $q\rightarrow q^{-1}$. It is straightforward to check
that the eigenvectors of the straight and inverse operators
coincide.

\subsubsection{Turn-over operators in a two-stand braid and character decomposition}
As discussed above, an operator contraction yielding a knot
invariant contains not only the crossing operators, but also the
turn-over operators, their explicit form determined in
sec.\ref{sec:Mop}, for crossing operators of form (\ref{Rfund}).
Two turn-over operators are needed in the case, since a two strand
braid is obtained from its closure by making two cuts. Modulo
discussion of sec.\ref{sec:Mop}-\ref{sec:Rbr}, one should write \be
H_N^n(q)\equiv\sum_{\substack{a,b,\ldots,c,d\\i,j,p,q}}\mathcal{R}^{ij}_{ab}\ldots
\mathcal{R}^{cd}_{pq}\mathfrak{M}^p_i\mathfrak{M}^{q}_j
=\sum_{\substack{A,\ldots,C\\I,P}}\mathcal{R}^I_A\ldots\mathcal{R}^C_P(\mathfrak{M}\otimes\mathfrak{M})^P_I
=
\mathrm{Tr\,}\left\{\mathcal{R}^{n}\mathfrak{M}\otimes\mathfrak{M}\right\},\label{trTurn}
\ee with $n$ being the number of crossings in the braid, a positive
or negative integer number depending on their orientation (see
fig.\ref{fig:invcr}).

The tail of the obtained contraction,
$\mathfrak{M}^p_i\mathfrak{M}^{q}_j$, is permutation under a
permutation of $i$ and $j$. Hence, both tensor monomials in
expression (\ref{Rev}), $\xi_i\eta_j$ and $\xi_j\eta_i$, are
multiplied at the end of (\ref{trTurn}) on the same factor, thus
remaining the eigenvectors of the entire operator
$\mathcal{R}^{n}\mathfrak{M}\otimes\mathfrak{M}$. Moreover, since
eigenvectors (\ref{Rev}) form a basis, the trace can be evaluated as
a sum of eigenvalues of the standing under the trace sign product
corresponding to all eigenvectors
(\ref{evecii},\ref{evecijs},\ref{evecija}). Similarly to the
examples of sec.\ref{sec:tcirc}-\ref{sec:tref}, each eigenvalue is
multiplied on a matrix element of the squared turn-over operator,
\be H_N^n(q)=q^n\underbrace{\sum_{i=1}^Nm_im_i}_{X_i=\xi_i\eta_i}+
q^n\underbrace{\sum_{\substack{i,j=1\\i<j}}^Nm_im_j}_{x^+_{ij}=q\xi_i\eta_j+\xi_j\eta_i}+
\underbrace{(-q)^{-n}\sum_{\substack{i,j=1\\i<j}}^Nm_im_j}_{x^-_{ij}=\xi_i\eta_j-q\xi_j\eta_i}=
q^n\alpha_N+\left(q^{n}+(-q)^{-n}\right)\beta_N, \ee so that the
multiples of similar factors assemble into the quantities $\alpha_N$
and $\beta_N$, which we introduced and evaluated in
sec.\ref{sec:tref}. Substituting the corresponding expressions, one
gets
\be
H_N^n(q)=
q^n\frac{[N][N+1]}{[2]}+(-q)^{-n}\frac{[N][N-1]}{[2]}\label{RH2str}
\ee coincides with (\ref{H2str}) obtained in (\ref{sec:braids}) by the
braid group method. The one by one blocks, associated in
sec.\ref{sec:braids} with the symmetric and antisymmetric
representations of the permutation group, arose here as eigenvalues of
the $R$-matrix. This is not a coincidence; as was already mentioned
in sec.\ref{sec:brsymm}, the blocks can be obtained as the eigenvalues
of the Hecke algebra, which is an extension of the permutation group \cite{KauffTB}. A major difference of the $\mathcal{R}$-matrices
from the braid crossing operators is that the first one has
degenerate eigenvalues, $q$ (the ``permutation'' one) with the
multiplicity $\frac{N(N+1)}{2}$, $-q^{-1}$  (the ``antisymmetric''
one) with the multiplicity $\frac{N(N-1)}{2}$, for the $\mathcal{R}$-matrix of form (\ref{Rfund}). When one evaluates trace
(\ref{trTurn}), each degenerate eigenvalue is multiplied on the
trace of the squared turn-over operator over the corresponding
``symmetric'' and ``antisymmetric'' subspaces. As a result, one
obtains just the weight coefficients, which were determined in
(\ref{H2str}) as solutions of the topological invariance constraints
(although (\ref{RH2str}) reproduces just a particular case of
(\ref{H2str}) with $\lambda=q$, $\mu=-q^{-1}$, $\chi=[N]$). Although
we solved the topological invariance constraints to determine the
matrix elements of the turn-over operators, an explicit expression
for the very operators is available in group theory, at least in the
particular case of braid representations \cite{ReshTur},
\cite{MorSm}. It is clear from the form of this expression, that the
coefficients of the $R$-matrix eigenvalues $q$ and $q^{-1}$ are
nothing but the \emph{quantum dimensions} of the first symmetric and
antisymmetric $SU(N)$ representations, respectively. We came to the
same result twice, in \ref{H2str} and in \ref{RH2str}, with help of
straightforward computations.

\subsubsection{Common eigenspaces of twisted crossing operators in a three-stand braid}

\begin{wrapfigure}{r}{40pt}
\begin{picture}(40,60)(0,-30)
\qbezier(0,36)(0,30)(0,24)\qbezier(0,24)(6,18)(12,12)\qbezier(12,12)(12,6)(12,0)
\qbezier(12,0)(18,-6)(24,-12)\qbezier(24,-12)(24,-18)(24,-24)
\qbezier(12,36)(12,30)(12,24)\qbezier(12,24)(9,21)(8,20)\qbezier(4,16)(3,15)(0,12)\qbezier(0,12)(0,6)(0,0)\qbezier(0,0)(0,-12)(0,-24)
\qbezier(24,36)(24,18)(24,0)\qbezier(24,0)(21,-3)(20,-4)\qbezier(16,-8)(15,-9)(12,-12)\qbezier(12,-12)(12,-18)(12,-24)
\qbezier(0,27)(0,30)(1,33)\qbezier(0,27)(0,30)(-1,33)
\qbezier(12,27)(12,30)(13,33)\qbezier(12,27)(12,30)(11,33)
\qbezier(24,27)(24,30)(25,33)\qbezier(24,27)(24,30)(23,33)
\qbezier(0,2)(0,5)(1,8)\qbezier(0,2)(0,5)(-1,8)
\qbezier(12,2)(12,5)(13,8)\qbezier(12,2)(12,5)(11,8)
\qbezier(24,2)(24,5)(25,8)\qbezier(24,2)(24,5)(23,8)
\qbezier(0,-22)(0,-19)(1,-16)\qbezier(0,-22)(0,-19)(-1,-16)
\qbezier(12,-22)(12,-19)(13,-16)\qbezier(12,-22)(12,-19)(11,-16)
\qbezier(24,-22)(24,-19)(25,-16)\qbezier(24,-22)(24,-19)(23,-16)
\put(2,30){$i$}\put(14,30){$j$}\put(26,30){$k$}
\put(2,0){$a$}\put(14,0){$b$}\put(26,0){$c$}
\put(2,-26){$l$}\put(12,-26){$m$}\put(26,-26){$n$}
\end{picture}
\caption{A fragment of a three-strand braid.\label{fig:R3br}}
\end{wrapfigure}
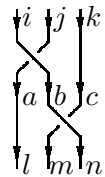
One can derive a generalization of (\ref{trTurn}) for an arbitrary
braid \cite{ReshTur, MorSm, MMM2}. An additional difficulty one
encounters with is a presence of several kinds of crossings, each
one bringing a contribution to (\ref{trTurn}) of its own form. E.g.,
there will be two distinct operators for two kinds of crossings in a
three-strand braid (fig.\ref{fig:R3br}). The operators do not
commute, and, hence, can not have a basis of common eigenvectors.
Nevertheless, they do have a number of common eigenvectors, while
the complimentary subspace decomposes into a sum of two-dimensional
common eigenspaces. We demonstrate this explicitly in what follows.

An braid version of the $R$-matrix approach provides an illustration the fact that twisted $R$-matrices form the group, which
extends the permutation group \cite{InvScatt}. Namely, solution to (\ref{Rfund})
for the operator $R^{ij}_{kl}$ can be considered as so referred to
$q$-permutation operator, whose action is defined on a tensor
product of vector spaces $V\otimes V$ as \be
R(q)\xi_i\eta_j=\left\{\begin{array}{rl}\xi_j\eta_i,&i>j\\q\xi_i\eta_i,&i=j,\\
(q+q^{-1})\xi_i\eta_j+\xi_j\eta_i,&i<j.\end{array}\right. \ee where
$\xi$ and $\eta$ are vectors of $V$, the subscripts run from $1$ to
$\mathrm{dim}\, V$. In particular, common eigenspaces of braid
$R$-matrices are constructed similarly to irreducible
representations of the permutation group (see sec.\ref{sec:brinv}). To
make this analogy explicit, we briefly review the corresponding formulas from
sec.\ref{sec:braids}, this time
presenting them in a form generalizable for $R$-matrices case.

First, let us notice that two irreducible representations (\ref{x2})
of rank two permutation group can be rewritten by introducing two
linear operators, which are called, correspondingly, symmetriser and
antisymmetrizer: \be (xy)=\frac{1}{2}\big(\mathds{1}+b_1\big)\equiv
Sxy,\ \ [xy]-\frac{1}{2}\big(\mathds{1}-b_1\big)\equiv
Axy,\label{SAdef} \ee and satisfy \be S^2=S,\ \ A^2=A,\ \ SA=AS=0,\
\ S+A=1,\label{SAprop} \ee thus being the orthogonal projectors.
Relations (\ref{SAprop}) follow straight from definitions
(\ref{SAdef}) and from the property $b_1^2=\mathds{1}$ of the
permutation group generator. The irreducible representations of the
permutation group can be determined just in terms of the introduced
operators. Since \be b_1S=S,\ \ b_1A=-A,\label{SAev} \ee any vectors
of type $SX$ and $AX$ are eigenvectors of $b_1$ with the eigenvalues
$1$ and $-1$ correspondingly. Similarly, one can introduce three
pairwise orthogonal projectors on the common eigenspaces of the
permutation group with three elements, \be
X_S=(xyz)=\frac{1}{6}\left(\mathds{1}+b_1+b_2+b_1b_2+b_2b_1+b_1b_2b_1\right)\equiv
SSxyz,
\label{SSdef}\\
X_{SA}=\frac{1}{2}\big(1+b_1)(a+bb_1b_2+cb_2b_1)xyz\equiv SAxyz,\nn\\
X_A=[xzy]=-\frac{1}{6}\left(\mathds{1}-b_1-b_2+b_1b_2+b_2b_1-b_1b_2b_1\right)\equiv
AAxyz, \nn \ee so that \be
AA\cdot SS=SS\cdot AA=SA\cdot SS=SS\cdot SA=SA\cdot AA=AA\cdot SA=0,\\
SS^2=SS,\ \ SA^2=SA,\ \ AA^2=AA.\nn \ee One can verify then, that
operators $SS_q$, $AA_q$, and $S_qSA_q$ give the projectors on three
distinct irreducible representations of the permutation group. In case
of the one-dimensional representations, one has to check that
$b_1SS=b_2SS=SS$ and $b_1AA=b_2AA=-AA$; this is done by substituting
for $SS$ and $AA$ their explicit expressions (\ref{SSdef}) and using
that $b_1^2=b_2^2=1$. For the two-dimensional representation, one
gets $b_1S\cdot AS=S\cdot AS$, and the check reduces then to
ensuring that the expressions $S_qSA_q$, $b_2S_qSA_q$, and
$b_1b_2S_qSA_q$ are linearly dependent, treated as formal
polynomials in group generators. This indeed follows form
(\ref{SSdef}) given that the squared group generators are the
unities: \be &\underbrace{\begin{array}{rcrrrrrrr}
S\cdot SA&=&2&+2b_1&-b_2&-b_1b_2&-b_2b_1&-b_1b_2b_1,\\
b_2S\cdot SA&=&-1&-b_1&+2b_2&-b_1b_2&+2b_2b_1&-b_1b_2b_1,\\
b_1b_2b_2SAS&=&-1&-b_1&-b_2&+2b_1b_2&-b_2b_1&+2b_1b_2b_1
\end{array}}_{}\nn\\&\Downarrow\nn\\
&(\mathds{1}+b_2+b_1b_2)SA=0.\label{21ldop} \ee Identity
(\ref{21ldop}) implies that any expression of the form $S\cdot SA
X$, where $X$ is a formal polynomial in permutations, generates a
two-dimensional representation of the permutation group with three
elements. Moreover, matrix expressions for group generators in this
representations, e.g., (\ref{Ssgm}), can derived from the operator
identities \be
\begin{array}{rclrcl}
b_1\cdot\mathds{1}&=&b_1,&
b_1\cdot\frac{\mathrm{1}+2b_2}{\sqrt{3}}&=&-\frac{\mathrm{1}+2b_2}{\sqrt{3}},\\
b_2\cdot\mathds{1}&=&-\frac{1}{2}\mathds{1}+\frac{\sqrt{3}}{2}\cdot\frac{\mathrm{1}+2b_2}{\sqrt{3}},&
b_2\cdot\frac{\mathrm{1}+2b_2}{\sqrt{3}}&=&\frac{\sqrt{3}}{2}\cdot\mathds{1}+\frac{1}{2}
\cdot\frac{\mathrm{1}+2b_2}{\sqrt{3}}.
\end{array}\ee

Similar formulas can be derived for permutation groups with more
elements. Then, there is a one-to-one correspondence of the
permutation group irreducible representations and common eigenspaces
of the braid $R$-matrices \cite{ReshTur, MMM2}. Moreover, the
partition-based approach discussed a little in sec.\ref{sec:4str}
can be extended to determine the common eigenspaces of the braid
R-matrices explicitly \cite{MMM2, AMMM1}, in principle, for
arbitrary braids. The only reservation should be done here.
Practically, the common eigenspaces are found not as irreducible
representations of $q$-permutation group, but as that of the quantum
group $U_q(SU(N)$, the two ones are related by the analog of
Schur-Weyl duality \cite{LL3},\cite{Fulton},\cite{qgroups}. An
output is an explicit expression for knot polynomials in terms of
eigenvalues of quantum $R$-matrices and quantum Racah coefficients
\cite{ReshTur},\cite{MMM2}. The first ones are known in full
generality, hence the entire problem is concentrated in evaluating
the second ones. While the case of $SU(2)$ group is a rather text-book
subject \cite{KlimSch}, a breakthrough beyond was made just recently
\cite{IMMM3,Inds9,GuJock}.
In the particular case of $R$-matrices we discuss here, which are
related to the fundamental representation of the quantum group
$SU(N)$ and give rise to the HOMFLY polynomials, all needed
ingredients are determined explicitly, and a concise and
computationally effective procedure is available \cite{AMMM3}. The
same tools enables to calculate colored HOMFLY polynomials as well,
by means of the cabling procedure \cite{AM1}. Finally, an attempt of
applying the approach to studies of superpolynomials was recently
performed \cite{AM2}, in the framework of elaborating a modified
Khovanov formalism \cite{DM3}.

As usual, we illustrate the approach discussed on the simplest
relevant example, that of braid $R$-matrices for a three-strand
braid. We start from recalling the even simpler case of the
two-strands. The generalizations of (\ref{SAdef}) by definition
satisfy \be S_q^2=S_q,\ \ A_q^2=A_q,\ \
S_qA_q=A_qS_q=0,\label{SAqdef} \ee and \be R_1S_q=qS_q,\ \
R_1A_q=-q^{-1}A_q.\label{SAqev} \ee The operators $S_q$ and $A_q$
are referred to as, correspondingly, $q$-symmertrizer and
$q$-antisymmertrizer; explicit expressions for them are \be
S_q&\equiv&\frac{1}{q[2]_q}\big(\mathds{1}+qR_1\big),\\
A_q&\equiv&-\frac{q}{[2]_q}\big(\mathds{1}-q^{-1}R_1\big), \ee and
properties (\ref{SAqdef}, \ref{SAqev}) are verified with help of
generalization of the identity $\sigma_1^2=1$ for the permutation
group generator: \be R_1^2-(q-q^{-1})R_1-\mathds{1}=0\label{Reveq}
\ee Relation (\ref{Rev}) is a characteristic equation for the braid
$R$-matrix, whose eigenvalues were determined in sec.\ref{sec:Rev}.

In case of three-strand braid, the relevant extension of formulas
for the permutation group are less straightforward. Yet, they remain
rather simple for projectors on fully permutation and anti-permutation
representations. These projectors must satisfy \be SS_q^2=SS_q,\ \
AA_q^2=AA_q,\ \ SS_qAA_q=AA_qSS_q=0,\label{SSqdef} \ee and \be
R_1SS_q=R_2SS_q=qSS_q,\ \ R_1AA_q=R_2AA_q=-q^{-1}AA_q.\label{SSqev}
\ee Explicit expressions for $SS_q$ and $AA_q$ are straightforward
analogues of formulas (\ref{SSdef}) for $SS$ and $AA$; permutation
group generators are substituted by the corresponding $R$-matrices,
a factor of $q^k$ is put before a product of $k$ $R$-matrices, and
the normalization factors are changed so that the first and second
of equalities (\ref{SSqdef}) are satisfied: \be
SS_q&\equiv&\frac{1}{q^3[2]_q[3]_q}\left(\mathds{1}+qR_1+qR_2+q^2R_1R_2+q^2R_2R_1+q^3R_1R_2R_1\right),\\
AA_q&\equiv&-\frac{q^3}{[2]_q[3]_q}\left(\mathds{1}-q^{-1}R_1-q^{-1}R_2+q^{-2}R_1R_2+q^{-2}R_2R_1-q^{-3}R_1R_2R_1\right).
\ee Satisfying of (\ref{SSqdef}) and (\ref{SSqev}) is checked with
help of (\ref{Rev}), a similar identity for $R_2$, and the
Yang-Baxter equation, which in the case takes the from
$R_1R_2R_1=R_2R_1R_2$. An expression for the remaining projector
(denote it $AS_q$) is more involved; the easiest way to obtain it
is: \be
AS_q=\mathds{1}-SS_q-AA_q=\frac{1}{[3]_q}\big(R_1-R_2\big)^2; \ee
the equalities \be AS_q^2=AS_q,\ \
AS_qSS_q=SS_qAS_q=AS_qAA_q=AA_qAS_q=0, \ee follow then just from
(\ref{SSqdef}).

Relations (\ref{SSqev}) already imply that $SS_q$ and $AA_q$ are
projectors on the common eigenvectors of $R_1$ and $R_2$, i.e., that
for $X$ being any formal polynomial in $R$-matrices, $SS_qX$ and
$AA_qX$ are the eigenvectors, with the values $q$ and $-q^{-1}$,
correspondingly. It remains to verify that, in analogy with the
permutation group case, the expression $S_qAS_q$ yields a projector on
a two-dimensional common eigenspace. One immediately gets that \be
R_1S_qAS_q&=&S_qAS_q.\label{21qldop1} \ee An analog of
(\ref{21ldop}) is also derived straightforwardly, though a bit
lengthy. Using the eigenvalue equations for $R_1$ and $R_2$, and the
Yang-Baxter equation, one obtains that the operators $S_qAS_q$,
$R_2S_qAS_q$, and $R_1R_2S_qAS$ are expanded over the six basis
products of $R$-matrices as \be
\begin{array}{|c|c|c|c|c|c|c|}
\hline&&&&&&\\[-4mm]
&1&R_1&R_2&R_1R_2&R_2R_1&R_1R_2R_1\\[0.5mm]
\hline&&&&&&\\[-4mm]
S_qAS_q&q^2+1&q^3+q&q^{-1}&-1&-1&-q\\[0.5mm]
\hline&&&&&&\\[-4mm]
R_2S_qAS&-q^{-1}&-1&q^2+q^{-2}&-q&q^3+q^{-1}&-q^2\\[0.5mm]
\hline&&&&&&\\[-4mm]
R_1R_2S_qAS_q&-1&-q&-q&1+q^{-2}&-q^2&q+q^{-1}\\[0.5mm]
\hline
\end{array},
\ee wherefrom one derives the identity: \be
\left(1+qR_2+q^2R_1R_2\right)S_qAS_q=0.\label{21qldop2} \ee
Relations (\ref{21qldop1}, \ref{21qldop2}) imply that a basis in a
two-dimensional common eigenspace of the $R$-matrices is obtained
from an arbitrary formal polynomial in $R$-matrices $X$ as \be
S_qAS_qX\equiv\left(\begin{array}{c}1\\0\end{array}\right),\ \
R_2S_qAS_qX\equiv\left(\begin{array}{c}0\\1\end{array}\right). \ee
Alternatively, one can construct a basic of $R_1$ eigenvectors in
the same space. Writing down the corresponding condition \be
R_1\left(\alpha+\beta R_2\right)S_qAS_q
=\lambda\left(\alpha+\beta R_2\right)S_qAS_q \ee and solving the
system \be \left\{\begin{array}{rcl}
q\alpha-\beta q^{-2}&=&\lambda\alpha,\\
\beta q^{-1}&=&\lambda\beta,
\end{array}
\right. \Leftrightarrow \left[\begin{array}{ll}
\lambda=q,&\beta=0,\\
\lambda=-q^{-1},&\alpha=q^2[2]_q\beta,
\end{array}
\right. \ee one obtains the corresponding expressions for the basis
vectors, \be
SS_qAS_qX\equiv\left(\begin{array}{c}1\\0\end{array}\right),\ \
\frac{1}{\sqrt{[3]_q}}\big(\mathds{1}+q^2[2]_qR_2X\big)SS_qAS_qX\equiv\left(\begin{array}{c}0\\1\end{array}\right),
\ee and for acting on them $R$-matrices, \be
R_1=\left(\begin{array}{cc}1&0\\0&-1\end{array}\right),\ \ \
R_2=\left(\begin{array}{cc}-\frac{1}{q^2[2]_q}&\frac{\sqrt{[3]_q}}{[2]_q}
\\\frac{\sqrt{[3]_q}}{[2]_q}&\frac{1}{[2]_q}\end{array}\right).
\ee The result coincides with the formulas for the braid-crossing
matrices, derived in the framework of the permutation group approach
in sec.\ref{sec:3str}.

\

\paragraph{The summary}
Content of the present section may be reviewed as follows. The
contraction of the $\mathcal{R}$-matrix associated with a closure of
a three-strand braid can be presented as a composition of linear
operators of two types, $R_1\equiv R\otimes I$ and $R_2\equiv
I\otimes R$. Both operators acts on the space $V\otimes V\otimes V$,
where one can chose a basis of tensor monomials
$\xi_i\otimes\xi_j\otimes\xi_k\ \in\ V$ composed of basis vectors
$\xi$ of the space $V$, the subscripts running from $1$ to $N$.
Operators $R_1$ and $R_2$ can be considered then as rank six tensors
$\left(R_a\right)^{ijk}_{lmn}$ for $a=1,2$. If we substitute
(\ref{slnf}) for $R$, all the non-zero elements have the form
$(R_a)^{ijk}_{\sigma(ijk)}$, with the subscripts being a permutation
of the superscripts. In other words, a linear space
$\mathcal{V}_{ijk}$ spanned by all the monomials
$(\xi\otimes\xi\otimes\xi)_{\sigma(ijk)}$, where $\sigma$ runs over
all distinct permutations of $ijk$, is a common eigenspace of $R_1$
and $R_2$. The dimension of this space equals one for $i=j=k$, two
for $i=j\ne k$, and six for $i\ne j\ne k$. As we verified above, the
subspace $\mathcal{V}_{ijk}$ further decomposes into irreducible
common eigenspaces of $R_1$ and $R_2$, which are hence the common
eigenspaces of all the three-strand operators. The irreducible
common eigenspaces are either one- or two-dimensional, so that the
operators $R_1$ and $R_2$ acting on these spaces are represented
either by $1\times1$ or by $2\times 2$ matrices. In a certain basis,
the matrices reproduce the braid crossing matrices of
sec.\ref{sec:3str}.

The smallest common eigenspaces described above turn out to be nothing but the spaces of
the permutation group irreducible representations, as the $R$-matrices
eigenvalues $q$ and $-q^{-1}$ tend to $1$ and $-1$, respectively.
The same spaces for a generic $q$ are the spaces of Hecke algebra
(or $q$-permutation group) irreducible representations.

\section{Conclusion}
As a closing remark, let us note that this text does not apply in no sense for the full presentation of the questions we referred to. Although the text is a historical review, we have encountered with many subtleties of the discussed notions and constructions, some of them leading to open essential problem. As a conclusion, we formulate two such problem, which seem especially interesting to us.

The first problem is the already mentioned in the introduction
problem of relating of the $\mathcal{R}$-matrix formalism to the
perturbation theory for the Chern-Simons theory in the temporial
gauge \cite{MorRos,MorSm}. We add now that a major difficulty on
this way is that the $\mathcal{R}$-matrix is related (if does) to a
Wilson line, the classical Chern-Simons fields at different points
of the one not commuting with each other. Hence, examining the
simple explicit examples of such fields, like the example presented
in sec.\ref{sec:CSex}, might spread some light to the problem.

The second problem is to develop the covariant version of the $\mathcal{R}$-matrix approach, which would involve the turn-over operators instead of the extremum point operators. The problem has at least two applications. First, a QFT interpretation of the $\mathcal{R}$-matrix formalism (if any) might be more natural for this version of the covariant version of the formalism. The second application concerns the attempts of involving elements of the  $\mathcal{R}$-matrix approach in evaluating the superpolynomials of the knots \cite{AM2}. Namely, the morphism actin on the Seifert cycles on the resolved knot diagrams in the modified Khovanov contraction \cite{DM3} might be expressed in terms of the turn-over operators.

Although the above problems thus far attracted undeservedly few attention, we hope for they being studied properly in the nearest future.

\section{Acknowledgements}
The author would like to thank G.Aminov, D.Vasiliev, D.Galakhov,
A.Sleptsov, Sh.Shakirov, and all participants of the ITEP
math.phys.group seminar for numerous fruitful discussions. Author is
especially grateful to I.Daninlenko, P.Dunin-Barkowski, And.Morozov,
A.Popolitov,  for reading the preprint and making important remarks. Author also thanks G.Galakhov, S.Mironov and A.Sleptsov for useful comments to the second version.
Author is indebted to E.Vyrodov for discussing the physical
analogues, and to A.Mironov and A.Morozov for supervising the
authors work on $\mathcal{R}$-matrices approach to knot invariants. The work
was partially supported by RFBR grants
14-01-31492\_mol\_a, 14-01-92691\_Ind\_a, 15-31-20832\_mol\_a\_ved, 15-51-50034\_Yaf, NSh-1500.2014.2.

\bibliographystyle{utphys}
\bibliography{Hist_bibl,Knots_our,Ind}

\appendix

\section{Obtaining of a non-trivial solution of classical equations for the $SU(2)$ Chern-Simons\label{App:SU2sol}}
Here we explain how explicit form (\ref{SU2cl}) of a classical $SU(2)$ Chern-Simons field with a line-like singularity can be obtained.

The simplest non-trivial solution for the classical $SU(2)$
Chern-Simons equations can be obtained with help of the fact the
there-dimensional sphere is the group variety of the $SU(2)$ group.
Because the three-dimensional sphere is obtained from the
three-dimensional flat space by adding the infinitely distant point,
and because the $SU(2)$ group is a double covering of the rotation
group, this fact can be presented visually. Namely, a point $\vec
r=(x,y,z)$ in the flat three-dimensional space corresponds to the
rotation around the unity vector $\vec n=\frac{r}{r}$ on the angle
$\psi=4\arctan\frac{r}{a}$, a being an arbitrary parameter; the pair
of points $(-\vec r, \vec r)$ corresponds to the same rotation, and
the point $\vec r=(0,0,0)$ together with the infinitely distant
point correspond to the unity transformation. In turn, a rotation
around a unity vector $\vec n$ on the angle $\psi$ can be related to
the $SU(2)$ matrix \be \Omega\left(\vec n=\frac{\vec
r}{r},\psi=\arctan\frac{a}{r}\right)=\left(\begin{array}{cc}\cos
t+in_z\sin t&(n_x+in_i)\sin t\\(n_x-in_y)\sin t&\cos
t-in_z\sin t\end{array}\right)=\nn\\= \frac{1}{r\sqrt{r^2+a^2}}
\left(\begin{array}{cc}r^2+iaz&a(ix+y)\\a(ix-y)&r^2-iaz\end{array}\right),\ \ \
\vec r\equiv(x,y,z), \label{quatrep}\ee composition of two rotations
corresponding then to the product of the matrices (this
correspondence is known as the quaternionic representation of the
rotation group), and an $SU(2)$ matrix corresponds to a rotation,
since it is a generic $SU(2)$ matrix standing in (\ref{quatrep}).

The described one to one correspondence is even excessive for our
purposes. We will use (\ref{quatrep}) just as an explicit example of
a non-trivial, regular everywhere but the coordinate origin,  distribution of
the unitary matrices $\Omega$ in the three-dimensional space. The
corresponding components of the Chern-Simons field are
\be
\arraycolsep=0.5mm
A_x=\Omega^{-1}\p_x\Omega=\frac{a}{r^2\left(r^2+a^2\right)}
\left(\begin{array}{cc}-i(ay+2xz)&i\left(-x^2+y^2+z^2\right)+(az-2xy)\\i\left(-x^2+y^2+z^2\right)-(az-2xy)&i(ay+2xz)\end{array}\right),\nn\\
A_y=\Omega^{-1}\p_y\Omega=\frac{a}{r^2\left(r^2+a^2\right)}
\left(\begin{array}{cc}i(ax-2yz)&-i(az+xy)+\left(x^2-y^2+z^2\right)\\-i(az+xy)-\left(x^2-y^2+z^2\right)&-i(ax-2yz)\end{array}\right),\nn\\
A_z=\Omega^{-1}\p_z\Omega=\frac{a}{r^2\left(r^2+a^2\right)}
\left(\begin{array}{cc}i\left(x^2+y^2-z^2\right)&(ix+y)(ia-2z)\\(ix-y)(ia+2z)&-i\left(x^2+y^2-z^2\right)\end{array}\right).
\ee

Solution (\ref{SU2cl}) with a line-like singularity is obtained from (\ref{quatrep}) as \be
\tilde{\Omega}=\Omega(x,y,z=0)=\frac{1}{\sqrt{r^2+a^2}}
\left(\begin{array}{cc}a&ix+y\\ix-y&a\end{array}\right), \ee the
components of the Chern-Simons fields being obtained then as \be
\tilde{A}_x=A_x(x,y,z=0),\ \ \tilde{A}_y=A_y(x,y,z=0),\ \ \
\tilde{A}_z=0. \ee

\section{On the form of the Green functions for the abelian Chern-Simons theory in Lorenz and holomorphic gauges.\label{App:CSGreen}}
Here we clarify some subtleties of deriving the explicit expressions for the Chern-Simons Green functions in the Lorentz and holomorphic gauge.

\paragraph{Lorentz gauge.} First, we derive the expression
for the Green function for the Euclidian Chern-Simons in the covariant Lorentz gauge $\p_kA_k=0$, which we used when discussing the QFT interpretation of the linking number in sec.\ref{sec:lnum}. First, let us notice that the Green function by
definition appears in the integral form of the classical equations
of motions,
\be\epsilon^{kji}\p_{j}A_{i}(x)&=&4\pi J^k(x)\mbox{ + boundary conditions }\label{cldiff}\\
&\Updownarrow\nn\\
A_i(x)&=&\int d^3y G_{ij}(x-y)J^j(y)=\int d^3y
J^j(y)G_{ji}(y-x).\label{clint}\ee The gauge and boundary conditions
hold for $A_i(x)$, if $G_{ij}(x-y)$ satisfy similar conditions,
i.e., $\p^iG_{ij}(x-y)=\p^jG_{ij}(x-y)=0$, and $G_{ij}$ is regular
in the whole space but the field sources and vanishes at the
infinity. One can rewrite then (\ref{cldiff}) as
$\p_iA_j-\p_jA_i=4\pi\epsilon_{ijk}J^k$ and take the divergency of
both parts to obtain $\p^2A_i=4\pi\epsilon_{ijk}\p^jJ^k$ provided
that $\p_iA^i=0$. The same follows from (\ref{clint}) if the Green
function satisfy $\p^2G_{ij}(x-y)=4\pi\epsilon_{ijk}\p^k\delta(x-y)$
($J^k(y)$ is unaffected by $\p^2$ since $\p\equiv \frac{d}{dx}$).
The latter condition together with the gauge and boundary conditions
define $G_{ij}(x-y)$ unambiguously, hence the solution \be
G_{ij}(x-y)=4\pi\epsilon_{ijk}\p^k\frac{1}{|\vec x-\vec
y|},\label{Grab} \ee which satisfies all those conditions, should
satisfy (\ref{clint}) as well. Note that the Green function is
\emph{not} required to be a solution of the classical equation with
the delta-function in the r.h.s.;
definition (\ref{clint}) requires only that \be
\epsilon^{kji}\p_{j}G_{il}(x-y)= 4\pi\delta^k_l\delta(x-y)+\p^kf_l
\ee for a function $f_l$ vanishing fast enough, since $\p_kJ^k=0$
due to the classical equations of motion.

\paragraph{Holomorphic gauge.} Now demonstrate how one can verify that each factor in the kernel of Kontsevich integral (\ref{Konts2},\ref{Konts4}) is indeed the Green function of the classical equation of motions for Chern-Simons action in the holomorphic gauge (\ref{holac}). Namely, one may can either imply  the definition \be \p_zf(z,\bar
z)\equiv\lim_{|\mathcal{C}|\rightarrow
0}\frac{1}{|\mathcal{C}|}\oint_{z\in\mathcal{C}}f(z,\bar z)\ dz, \ee
or pass to the real vector fields: \be
\p_zu\equiv(\p_x+i\p_y)(u_x-iu_y)=\mathbf{div}\vec u+i\mathbf{rot}\vec u,\nn\\
\frac{1}{z}=\frac{\bar{z}}{|z|^2}=\frac{x-iy}{x^2+y^2}=\frac{\vec r}{r^2}\nn\\
\Gamma_{\partial G}\frac{\vec r}{r^2}=0,\ \ \Phi_{\partial
G}\frac{\vec r}{r^2}=\delta_{0\in G}. \label{Ghol}\ee

\section{Plat representation of knot and presenting Kontsevich integral as a contraction of the ``elementary contribution'' operators\label{App:eldec}}
\begin{wrapfigure}{r}{80pt}
\begin{picture}(80,120)(-10,-60)
\qbezier(0,48)(0,36)(6,30)
\qbezier(12,48)(12,36)(8,32)
\qbezier(6,30)(12,24)(12,12)
\qbezier(4,28)(0,24)(0,12)
\qbezier(12,12)(12,0)(24,-6)
\qbezier(24,-6)(36,-12)(36,-24) \qbezier(48,48)(48,0)(48,-24)
\qbezier(0,12)(0,0)(0,-48) \qbezier(0,0)(0,-12)(0,-24)
\qbezier(36,-24)(36,-48)(42,-48)
\qbezier(48,-18)(48,-48)(42,-48)
\put(-10,12){\line(1,0){80}}\put(-10,-24){\line(1,0){80}}
\put(60,30){I}\put(60,-8){II}\put(60,-48){III}
\qbezier(0,12)(-2,12)(-4,14)\qbezier(0,12)(-2,12)(-4,10)
\qbezier(12,12)(14,12)(16,14)\qbezier(12,12)(14,12)(16,10)
\qbezier(48,12)(46,12)(44,14)\qbezier(48,12)(46,12)(44,10)
\qbezier(0,-24)(2,-24)(4,-26)\qbezier(0,-24)(2,-24)(4,-22)
\qbezier(36,-24)(34,-24)(32,-26)\qbezier(36,-24)(34,-24)(32,-22)
\qbezier(48,-24)(50,-24)(52,-26)\qbezier(48,-24)(50,-24)(52,-22)
\put(3,3){$\delta$}\put(30,3){$\rho$}
\put(39,-20){$\delta^{\prime}$}\put(12,-20){$\rho^{\prime}$}
\put(0,-60){$\rho\gg\delta,\ \ \rho^{\prime}\gg\delta^{\prime}$}
\end{picture}
\caption{``Non-trivial events'' in a plat representation of knot.
\label{fig:Konz}}
\end{wrapfigure}
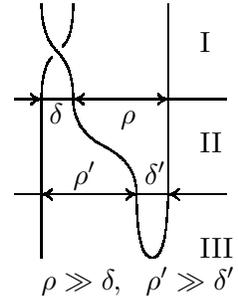

Here we outline the Wilson average with the knot as a contour, defined as the Gaussian average w.r.t. the Chern-Simons action in the holomorphic gauge, can be reduced to the Kontsevich integral presented as a contraction of certain elementary constituents.

\paragraph{Structure of the Kontsevich integral.} A knot, which is by definition a curve embedded in the three-dimensional space, is cut into layers by the horizontal, i.e., normal to the selected direction (the $t$ axes) with  the  planes tangent to the curve at its various critical points. Each layer contains several (even number) disconnected
pieces of the original curve. The Kontsevich integral is given then
by an infinite series of integrals of the increasing multiplicity. A
term with a certain multiplicity is, in turn, a sum of integrals
with a certain distribution of the integration variables over the
layers. Finally, each summand is a product of ``elementary
factors'' with all the variables running in the same segment
for each factor.

\paragraph{Plat representation of knot.} To match the Wilson average with the Kontsevich
integral, one has to bring a knot to a special form, performing the
proper continuous deformation. This form of a knot is a called a
\emph{plat} representation of knot \cite{KauffTB} and may be
described as follows.

A knot is by definition a closed curve embedded into the
three-dimensional space. If one selects a space direction, the
contours have critical points. Select then two horizontal planes
normal to the selected direction, so that no critical points are
contained between the planes. If the $t$ axis passes in the selected
direction, the plains are given by the equations $t=t_0$ and
$t=t_n>t_0$, and we will refer them to as the lower one and the
upper one, respectively. The curve crosses the upper plane in $2m$
points, which are pairwise connected above the plane by $m$ arcs of
the contour. One should then continuously transform the curve
arranging all the arcs in $m$ mutually parallel vertical planes, all
maximum points having the same hight $t=t_{max}$. The arcs of the
contour below the lower plane (section III in fig.\ref{fig:Konz})
should be arranged similarly, the hight of the minimum points being
$t=t_{min}$. The arcs of the contours between the selected planes
have no critical points. Therefore, one can arrange vertically each
one, but generally not all of them are the same time. For instance,
if two arcs intertwine (section I in fig.\ref{fig:Konz}), at least
one of them can not be everywhere vertical. Apart from that, one arc
may first intertwine with an other arc and with one more arc
(section II in fig.\ref{fig:Konz}); the first arc can not be then
vertical in the intermediate region. It turns out that all
``non-trivial events'' are exhausted by these two cases
\cite{ChumDuzMost}. One can complete then the procedure by selecting
$n-1$ more horizontal planes $t=t_i$ with $i=\overline{1,n-1}$ and
$t_0\le t_1\le\ldots\le t_2\le t_n$ such that no more than one
elementary event happens between any two neighboring planes, i.e.
all but no more than three arcs are vertical in each layer $t_i\le
t\le t_{i+1}$, $i=\overline{1,n}$.

\bigskip

\paragraph{Elementary segments of the integration contour and elementary constituent of the integral.}
After the knot is brought to a plat representation, it is split into
a collection of subsequent segments, \be \gamma=
\bigcup_{i=1}^m\bar\gamma_{2i-1}\gamma_{2i-1}\bar\gamma_{2i}\gamma_{2i},\ee
where $\gamma_{2i-1}$ and $\gamma_{2i}$  for $i= \overline{1,m}$
are, respectively, ascending and descending arcs with $t_1\le t\le
t_2$, while $\bar\gamma_{2i-1}$ and $\bar\gamma_{2i}$ with
$i=\overline{1,m}$ are, respectively, upper and lower closing arcs
with $t_n\le t$ and $t\le t_0$. Composition property of the path
exponential (\ref{Pexp:comp}) enables to write then  \be
\mathrm{Pexp}\oint dx^{\mu}A_{\mu}=
\prod_i\mathrm{Pexp}\int_{\gamma_i}\prod_jdx^{\mu}A_{\mu}
\mathrm{Pexp}\int_{\gamma_{i,j}} dx^{\mu}A_{\mu}.\ee

Taking the trace and introducing the operators \be
S(\gamma_{i,j})=\bigoplus_{i=1}^k\mathrm{Pexp}\int_{\gamma_{i,j}}dx^{\mu}A_{\mu},\nn\\
Q(\bar\gamma_1,\bar\gamma_3, \ldots)=\bigoplus_{i=0}^{
m}\mathrm{Pexp}\int_{\bar\gamma_{2i+1}}dx^{\mu}A_{\mu},\ \ \
\tilde{Q}(\bar\gamma_2,\bar\gamma_4, \ldots)=\bigoplus_{i=0}^{
m}\mathrm{Pexp}\int_{\bar\gamma_{2i}}dx^{\mu}A_{\mu}, \ee one can
identically rewrite the same decomposition as
\be\mathrm{Tr\,Pexp}\oint dx^{\mu}A_{\mu}= Q^{k^0_1\ldots
k^0_{2m}}(\bar\gamma_1,\bar\gamma_3, \ldots)
\tilde{Q}_{{k^{n+1}_1\ldots
k^{n+1}_{2m}}}(\bar\gamma_2,\bar\gamma_4, \ldots)\prod_j
S^{k^j_1\ldots k^j_{2m}}_{k^{j+1}_1\ldots
k^{j+1}_{2m}}(\gamma_{i,j})\ee Due to the presence of the
delta-function $\delta(t-t^{\prime})$ in the Green function, the
pairing, and hence all Gaussian correlators vanish for the fields
taken at the points lying in different layers, so that coordinates
of the points satisfy $t< t_i <t^{\prime}$ at least for one $i$.
Therefore, properties (\ref{Gav:fac}) hold for the selected
splitting of the contour, and decomposition (\ref{Gav:fac}) still
takes place after taking the average. One writes,
\be&\ckl\mathrm{Tr\,Pexp}\oint
dx^{\mu}A_{\mu}\brr=\\&=\mathrm{Tr}\left\{\ckl Q^{k^0_1\ldots
k^0_{2m}}(\bar\gamma_1,\bar\gamma_3, \ldots)\brr\ckl
\tilde{Q}_{{k^{n+1}_1\ldots
k^{n+1}_{2m}}}(\bar\gamma_2,\bar\gamma_4, \ldots)\brr\prod_j \ckl
S^{k^j_1\ldots k^j_{2m}}_{k^{j+1}_1\ldots
k^{j+1}_{2m}}(\gamma_{i,j})\brr\right\},\ee Further observation
concerns the arcs placed in the same layer. First, the Green
function being the decreasing function of the distance between arcs,
and the value of the integral being independent of the arcs
positions in the space, a contribution that contains the pairing of
fields from the points in different arcs from the same layer must
vanish if the arcs can be separated in the layer, i.e., if they do
not intertwine neither in this layer nor in the two neighboring
ones. In this case, \be \ckl S(\gamma_1,\gamma_2)\brr\equiv
\ckl\mathrm{Pexp} \int_{\gamma_1}dx^{\mu}A_{\mu}\otimes
\mathrm{Pexp} \int_{\gamma_2}dx^{\mu}A_{\mu}\brr=\ckl\mathrm{Pexp}
\int_{\gamma_1}dx^{\mu}A_{\mu}\brr\otimes \ckl\mathrm{Pexp}
\int_{\gamma_2}dx^{\mu}A_{\mu}\brr\equiv\\\equiv \ckl
S(\gamma_1)\brr\otimes \ckl S(\gamma_2)\brr,\mbox{ if
}\gamma_1\mbox{ is separated from }\gamma_2.\nn\ee


\section{Examples of calculating the elementary non-trivial contributions to the Kontsevich integral\label{App:elcalc}}

In the above section, we sketched the prove of the Kontsevich integral taking form of the tensor contraction of the operators associated with the specially selected parts of the integration contour (which is the knot). To complete our discussion on properties of the Kontsevich integral,
we enumerate the distinct elementary operators, discussing their properties and calculating explicitly the lowest order contributions to each one.

\paragraph{Trivial contributions}
Due to the form of the Green function, all integrals over the
vertical arcs vanish.  Integrals over the closing arcs (section III in fig.\ref{fig:Konz}) vanish as
well, since the points of any two arcs can be pairwise matched by
the horizontal segments. Hence, the corresponding elementary
operators contain only the zero-order term of the perturbative series, being equal \be
S_{\gamma_1,\ldots,\gamma_l}=\underbrace{\mathds{1}\otimes\ldots\otimes\mathds{1}}_{l\mbox{
 factors}},\ \ \ \mbox{if}\ \ \ \dot{z}(t)=0\mbox{ for }(t,z,\bar{z})\in\gamma_k\mbox{ with }k=\overline{1,l},\ee
and
\be
Q=\tilde{Q}=\underbrace{\mathds{1}\otimes\ldots\otimes\mathds{1}}_{m\mbox{
 factors}},\ee
respectively.
\paragraph{Crossing point contributions}
First of the two non-trivial operators corresponds to an
intertwining of two arcs, which contains a crossing point when
projected on a plane (section I in fig.\ref{fig:Konz}). This
operator can be evaluated explicitly and takes a rather simple form,
\be
R\equiv\ckl\mathrm{Pexp}\int_{\gamma}dx^{\mu}A_{\mu}(x)\otimes\mathrm{Pexp}\int_{\gamma^{\prime}}dx^{\prime
\nu}A_{\nu}(x^{\prime})\brr=q^{T_a\otimes T_a}\label{holcr}\ee The
last term of the equality contains a somewhat symbolic notation for
the result, and the exact sense of which is clarified below.

We recall that modulo discussion in sec.\ref{sec:CSfram}, only
pairings of the fields at the points from different arcs should be
left in the Wick theorem, for instance, up to the fourth order one
writes, \be\ckl\!\!\left(\mathds{1}\!\!+\!\! \int_{\gamma}\!\!
dx^{\mu}A^{\mu}(x)\!\!+\!\!\int_{\gamma}\!\!
dx^{\mu}\!\!\int_0^x\!\!\!\!
dy^{\nu}A_{\mu}(x)A_{\nu}(y)\!\!+\!\!\ldots\right)\!\!\otimes
\!\!\left(\mathds{1}\!\!+\!\! \int_{\gamma^{\prime}}\!\!
dx^{\prime\mu}A_{\mu}(x^{\prime})\!\!+\!\!\int_{\gamma^{\prime}}\!\!\!\!
dx^{\prime\mu}\!\!\int_0^{x^{\prime}}\!\!\!\!
dy^{\prime\nu}A_{\mu}(x^{\prime})A_{\nu}(y^{\prime})\!\!+\!\!\ldots\right)\!\!\brr\nn\\
\!\!=\!\!\mathds{1}\otimes\mathds{1}+\nn\\+\int_{\gamma}\!\!
dx^{\prime\mu}\!\!\int_{\gamma^{\prime}} \!\!dx^{\mu}\ckl
A^a_{\mu}(x)A^b_{\nu}(x^{\prime})\brr T^a\otimes
T^a+\int_{\gamma}\!\!
dx^{\mu}\!\!\int_0^x\!\!\!dy^{\nu}\!\!\!\int_{\gamma^{\prime}}\!\!
dx^{\prime\mu}\!\!\int_0^{x^{\prime}}\!\!\!dy^{\prime\nu}\!\!\left\{\!\!\ckl
A^a_{\mu}(x)A^{c\phantom{d}\!\!\!\!}_{\rho}(x^{\prime})\brr\!\!\ckl
A^b_{\nu}(y)A^d_{\sigma}(y^{\prime})\brr\right.\!\!+\!\!\nn\\\left.+\ckl
A^b_{\nu}(y)A^c_{\rho}(x^{\prime})\brr\!\!\ckl
A^a_{\mu}(x)A^d_{\sigma}(y^{\prime})\brr\!\!\right\}T^aT^b\otimes
T^cT^d+\ldots\ee

Now, it is important that the Green function decomposes into the
product of the group factor and coordinate dependent scalar factor,
\be \ckl
A^a_{\mu}(x)A^c_{\rho}(x^{\prime})\brr=t^{ac}g_{\mu\rho}(x-x^{\prime}),\ee
in particular,
$\frac{A^1_{\mu}(x)A^2_{\rho}(x^{\prime})}{A^2_{\mu}(x)A^3_{\rho}(x^{\prime})}=const$
and
$\frac{A^1_{\mu}(x)A^2_{\rho}(x^{\prime})}{A^1_{\nu}(y)A^2_{\rho}(y^{\prime})}=
\frac{A^2_{\mu}(x)A^3_{\rho}(x^{\prime})}{A^2_{\nu}(y)A^3_{\rho}(y^{\prime})}$.
As a result, all terms with the same distribution of the Green
function arguments over the contour arcs have assemble into the sum
over various parings of the group generators multiplied on the
common coordinate dependent factor, e.g., the second order term
takes form \be ,\label{holcr:2gr}\ee while the forth order term
becomes \be \int_{\gamma} dx^{\mu}\int_0^x
dy^{\nu}\int_{\gamma^{\prime}} dx^{\prime\mu}\int_0^{x^{\prime}}
dy^{\prime\nu} g_{\mu\rho}(x-x^{\prime})
g_{\nu\sigma}(y-y^{\prime})\left\{\tau^{ac}\tau^{bd}+\tau^{ad}\tau^{bc}\right\}T^aT^b\otimes
T^cT^d,\label{holcr:4gr}\ee where four out of six possible parings
remained, after the self-interaction contributions were excluded.
Generally, one can demonstrate that the group factor in a $k$ order
term has the form \be
R^{(k)}=\sum_{\sigma\in\mathrm{perm}(k)}T_{a_1\ldots a_k}\otimes
T_{\sigma(a_1\ldots a_k)},\ee where the sum is over all permutation
of $k$ elements. Moreover, the scalar coefficients, which arise as
the integrals of the coordinate dependent factors, turn out to be
proportional to the subsequent powers of the same quantity for the
subsequent terms of the expansion. Finally, each term enters the sum
with a numeric factor of $\frac{1}{k!}$. These three observations
are concentrated in the notation for the crossing operator used in
(\ref{holcr}).

\par\smallskip\noindent

Let us now take the integrals of the coordinate dependent factors
explicitly. We use not plain and prime variables for one and the
arcs, \be\gamma:\ \ x=x(t)\equiv\big(t,z(t),\bar{z}(t)\big),\ \ \
\gamma^{\prime}:\ \
x^{\prime}=x^{\prime}(t)\big(t,z^{\prime}(t),\bar{z}^{\prime}(t)\big),\ee
and we recall that the coordinate factor in the Green function is
equal to
$g_{0z}\big(x(t)-x^{\prime}(t^{\prime})\big)=\frac{\delta(t-t^{\prime})}{z(t)-z^{\prime}(t)}$.
The fourth order coordinate factor is then  \be \int_0^1 dt\int_0^t
ds \int_0^1 dt^{\prime} \int_0^{t^{\prime}} ds^{\prime}\ \dot
x^{\mu}(t)\dot
x^{\prime\rho}(t^{\prime})g_{\mu\rho}\big(x(t)-x^{\prime}(t^{\prime})\big)\
\dot
y^{\nu}(s)y^{\prime\sigma}(s^{\prime})g_{\nu\sigma}\big(y(s)-y^{\prime}(s^{\prime})\big)=\nn\\=
\int_0^1 dt\int_0^t ds\ h_{\mu\rho}\dot x^{\mu}(t)\dot
x^{\prime\rho}(t)f\big(x(t)-x^{\prime}(t)\big)\
h_{\nu\sigma}\dot y^{\nu}(s)y^{\prime\sigma}(s)f\big(y(s)-y^{\prime}(s)\big)=\nn\\
\int_0^1 dt\int_0^t ds\ \frac{\big(\dot z(t)-\dot
z^{\prime}(t)\big)\big(\dot z(s)-\dot
z^{\prime}(s)\big)}{\big(z(t)-z^{\prime}(t)\big)\big(z(s)-z^{\prime}(s)\big)}=
\frac{1}{2}\log^2\frac{z(0)-z^{\prime}(0)}{z(1)-z^{\prime}(1)}, \ee
where $\log$ stands for the multi-valued function
$\log\vartheta(t)\equiv \int_0^1
dt\frac{\dot\vartheta(t)}{\vartheta(t)}$. Higher order factors are
evaluated similarly, \be \int_{\gamma}\!
dx_1^{\mu_1}\!\int_0^{x_1}\!\!\!\!dx^{\mu_2}_2\ldots\!\int_0^{x_{k-1}}\!\!\!\!dx_k^{\mu_k}
\!\int_{\gamma^{\prime}}\!
dx_1^{\prime\nu_1}\!\int_0^{x\prime_1}\!\!\!\!dx_2^{\prime\nu_2}\ldots\!\int_0^{x\prime_{k-1}}\!\!\!\!dx_k^{\prime\nu_k}
g_{\mu_1\nu_{m_1}}(x_1-x_1^{\prime})g_{\mu_2\nu_{m_2}}(x_2-x_2^{\prime})\ldots
g_{\mu_k\nu_{m_k}}(x_k-x_k^{\prime})\nn\\=
\int_0^1dt_1\int_0^{t_1}dt_2\ldots\int_0^{t_{k-1}}dt_k
\frac{\big(\dot z(t_1)-\dot z^{\prime}(t_1)\big)\big(\dot
z(t_2)-\dot z^{\prime}(t_2)\big)\ldots\big(\dot z(t_k)-\dot
z^{\prime}(t_k)\big)}
{\big(z(t_1)-z^{\prime}(t_1)\big)\big(z(t_2)-z^{\prime}(t_2)\big)\ldots\big(z(t_k)-z^{\prime}(t_k)\big)}
=\frac{1}{k!}\log^k\frac{z(0)-z^{\prime}(0)}{z(1)-z^{\prime}(1)}.\nn\ee
Finally, the answer is \be R=e^{\hbar T_a\otimes
T^a}\equiv\mathds{1}+\sum_{k=1}^{\infty}\frac{\hbar^k}{k!}\sum_{\sigma\in
\mathrm{perm}_k}T_{a_1\ldots a_k}\otimes T_{\sigma(a_1\ldots a_k)}.
\ee
\paragraph{Associators}
The second non-trivial elementary contribution operator is known
under the name of \emph{Drinfeld associator}. It involves already
three arcs placed as in sec.II in fig.\ref{fig:Konz}, \be \Phi\equiv
\ckl \mathrm{Pexp}\int_{\gamma}dx^{\mu}A_{\mu}(x)\otimes
\mathrm{Pexp}\int_{\gamma^{\prime}}dx^{\prime\nu}A_{\nu}(x^{\prime})
\otimes\mathrm{Pexp}\int_{\gamma^{\prime\prime}}dx^{\prime\prime\rho}A_{\rho}(x^{\prime\prime})\brr,\label{holac}\ee
and it has a much more involved structure than the crossing point
operator. To demonstrate the difference between the two operators,
we calculate one of the fourth order contributions to the Drinfeld
associator  explicitly.

Namely, we consider the contribution arising from the pairing \be
\int_{\gamma} dx^{\mu}\int_0^x dy^{\nu}\int_{\gamma^{\prime}}
dx^{\prime\mu}\int_{\gamma^{\prime\prime}} dy^{\prime\nu}\ckl
A^a_{\mu}(x)A^{c\phantom{d}\!\!\!\!}_{\rho}(x^{\prime})\brr\ckl
A^b_{\nu}(y)A^d_{\sigma}(y^{\prime})\brr T^a\otimes T^bT^c\otimes
T^d,\ee which, in turn, arises from the term \be
\left(\int_{\gamma}\!\! dx^{\mu}A_{\mu}(x)\right)\otimes
\left(\int_{\gamma^{\prime}}\!\!
dx^{\prime\nu}\!\!\int_0^{x^{\prime}}\!\!\!\!
dy^{\prime\rho}A_{\nu}(x^{\prime})A_{\rho}(y^{\prime})\right)\otimes
\left(\int_{\gamma}\!\! dx^{\sigma}A_{\sigma}(y)\right) \ee in the
expansion of (\ref{holac}). Selecting the variables on the
connection components as \be\gamma:x=x(t)=\big(t,z(t),\bar
z(t)\big),\ \ \ \gamma^{\prime}:x=x(t)=\big(t,z^{\prime}(t),\bar
z^{\prime}(t)\big),\ \ \
\gamma^{\prime\prime}:x=x(t)=\big(t,z^{\prime\prime}(t),\bar
z^{\prime\prime}(t)\big), \ee we obtain that the integral of the
coordinate dependent factor of the considered contribution takes the
explicit form \be \int_0^1 \int_0^t dsdt\ \frac{\big(\dot z(t)-\dot
z^{\prime}(t)\big)\big(\dot z^{\prime}(s)-\dot
z^{\prime\prime}(s)\big)}{\big(z(t)-z^{\prime}(t)\big)\big(z^{\prime}(s)-z^{\prime\prime}(s)\big)}.
\ee Taking into account that the first and the last arcs being
vertical and supposing  that \be z^{\prime\prime}(t)=1+z(t),\ee one
can present the result in the form \be \int_0^1 \int_0^t dsdt\
\frac{\big(\dot z(t)-\dot z^{\prime}(t)\big)\big(\dot z(s)-\dot
z^{\prime}(s)\big)}{\big(z(t)-z^{\prime}(t)\big)\big(1+z(s)-z^{\prime}(s)\big)}
= \int_0^1 dt\
\frac{\dot z(t)-\dot z^{\prime}(t)}{z(t)-z^{\prime}(t)}\log\frac{1+z(t)-z^{\prime}(t)}{1+z(0)-z^{\prime}(0)}=\nn\\
=\mathrm{Li}_2\big(z^{\prime}(0)-z(0)\big)-\mathrm{Li}_2(z^{\prime}(1)-z(1))-
\log(1+z(0)-z^{\prime}(0))\log\frac{z(1)-z^{\prime}(1)}{z(0)-z^{\prime}(0)},
\label{holac:exp} \ee where the dilogarithm function is by
definition \be \mathrm{Li}_2(x)\equiv-\int_0^x
\frac{dy}{y}\log(1-y)=\int_0^x\sum_{k=0}^{\infty}\frac{y^{k-1}}{k}=\sum_{k=0}^{\infty}\frac{y^k}{k^2}
\ee One usually suppose that the second contour is infinitely close
to the first one on the upper boundary of the selected area,
becoming infinitely close the third one on the lower boundary.
Expression $(\ref{holac:exp})$ then notably simplifies, \be
\boxed{z^{\prime}(0)\rightarrow z(0),\ \ z^{\prime}(1)\rightarrow
z^{\prime\prime}(1),\ \ \Rightarrow\  \ \mbox{the answer}\rightarrow
\zeta(2)=\frac{\pi^2}{6}},\ee where we used that
$\mathrm{Li}_2(0)=0$, $\mathrm{Li}_2(1)=\zeta(2)$, and the Riemann
zeta-function is defined as the series
$\zeta(m)\equiv\sum\limits_{k=0}^{\infty}\frac{1}{k^m}$ and can be
evaluated explicitly with help of the expansion
 by definition \be
\frac{\sin(\pi x)}{\pi
x}=\prod_{n=1}^{\infty}\left(1-\frac{x^2}{n^2}\right)=1-\zeta(2)x^2+\ldots.
\ee

A remarkable property of the Drinfeld associator is that it can be
obtained from a solution of the Knizhnick-Zamolodchikov equation for
the WZW conformal blocks \cite{KZ:WZW}. This property gives a link
between the Kontsevich integral, with is associated with
perturbative expansion for the Chern-Simons Wilson average, and the
axiomatic definition of the exact Wilson average, which we briefly
outlined in sec.\ref{sec:WZW}.

\end{document}